\documentclass[aps,twocolumn,superscriptaddress,nofootinbib,longbibliography]{revtex4-1}
\usepackage{graphics}
\usepackage[hidelinks]{hyperref}
\hypersetup{
    colorlinks,
    linkcolor={blue},
    citecolor={blue},
    urlcolor={blue}
}
\usepackage{epsfig}
 \usepackage[normalem]{ulem}
\usepackage{float}
\usepackage[usenames]{color}
\usepackage{amsmath}
\usepackage{booktabs}
 \usepackage{multirow}
\usepackage{blindtext}
\usepackage{longtable}
\usepackage{color}
\usepackage{array}
\usepackage{url}

\newcommand{\del}[1]{{  }}

\newcommand*\Laplace{\mathop{}\!\mathbin\bigtriangleup}

\newcommand{\be}{\begin{equation}}
\newcommand{\ee}{\end{equation}}
\newcommand{\bea}{\begin{eqnarray}}
\newcommand{\eea}{\end{eqnarray}}

\def\inbar{\,\vrule height1.5ex width.4pt depth0pt}
\def\IR{\relax{\rm I\kern-.18em R}}
\def\IC{\relax\hbox{$\inbar\kern-.3em{\rm C}$}}

\begin{document}
\title{Broken detailed balance and non-equilibrium dynamics in living systems}

\author{F. Gnesotto}
 \thanks{These authors contributed equally}
\affiliation{Arnold-Sommerfeld-Center for Theoretical Physics and Center for
  NanoScience, Ludwig-Maximilians-Universit\"at M\"unchen,
   D-80333 M\"unchen, Germany.}

\author{F. Mura}
 \thanks{These authors contributed equally}
\affiliation{Arnold-Sommerfeld-Center for Theoretical Physics and Center for
  NanoScience, Ludwig-Maximilians-Universit\"at M\"unchen,
   D-80333 M\"unchen, Germany.}

\author{J. Gladrow}
 \thanks{These authors contributed equally}
\affiliation{Cavendish Laboratory, University of Cambridge, Cambridge CB3 0HE, United Kingdom}

\author{C.P. Broedersz}
\email{C.broedersz@lmu.de}
\affiliation{Arnold-Sommerfeld-Center for  Theoretical Physics and Center for
  NanoScience, Ludwig-Maximilians-Universit\"at M\"unchen,
   D-80333 M\"unchen, Germany.}

\begin{abstract}  
Living systems operate far from thermodynamic equilibrium. Enzymatic activity can induce broken detailed balance at the molecular scale. This molecular scale breaking of detailed balance is crucial to achieve  biological functions such as high-fidelity transcription and translation, sensing, adaptation, biochemical patterning, and force generation. While biological systems such as motor enzymes violate detailed balance at the molecular scale, it remains unclear how non-equilibrium dynamics manifests at the mesoscale in systems that are driven through the collective activity of many motors. Indeed, in several cellular systems the presence of non-equilibrium dynamics is not always evident at large scales.  For example, in the cytoskeleton or in chromosomes one can observe stationary stochastic processes that appear at first glance thermally driven. This raises the question how non-equilibrium fluctuations can be discerned from thermal noise.  We  discuss approaches that have recently been developed to address this question, including methods based on  measuring the extent to which the system violates the fluctuation-dissipation theorem. We also review applications of this approach to reconstituted cytoskeletal networks, the cytoplasm of living cells, and cell membranes.  Furthermore, we discuss a more recent approach to detect actively driven dynamics, which is based on inferring broken detailed balance. This constitutes a non-invasive method that uses time-lapse microscopy data, and can be applied to a broad range of systems in cells and tissue. We discuss the ideas underlying this method and its application to several examples including flagella, primary cilia, and cytoskeletal networks. Finally, we briefly discuss recent developments in stochastic thermodynamics and non-equilibrium statistical mechanics, which offer new perspectives to understand the physics of living systems.
\end{abstract}                                                                 
\date{\today}
\maketitle
\tableofcontents

\section{Introduction}
Living organisms are inherently out of equilibrium. A constant consumption and dissipation of energy results in non-equilibrium activity, which lies at the heart of  biological functionality:  internal activity enables cells to accurately sense and adapt in noisy environments~\cite{Lan2012,Mehta2012}, and it is crucial for high-fidelity DNA transcription and  for replication~\cite{Hopfield1974a,Murugan2012}. Non-equilibrium processes also enable subcellular systems to generate forces for internal transport, structural organization and directional motion~\cite{Needleman2014,Fletcher2010a,brangwynne2008cytoplasmic,juelicher2007active,Cates2012}.  Moreover, active dynamics can also guide spatial organization, for instance, through nonlinear reaction-diffusion patterning systems~\cite{Huang2003,Frey2017,Halatek2012}.
Thus, non-equilibrium dynamics is essential to maintain life in cells~\cite{Bialek}.

Physically, cells and tissue constitute a class of non-equilibrium many-body systems termed \emph{active living matter}. However, cellular systems are not driven out of equilibrium by external forces, as in conventional active condensed matter, but rather internally by enzymatic processes. While much progress has been made to understand active behavior in individual cases, the common physical principles underlying emergent active behavior in living systems remain  unclear. In this review, we primarily focus on research efforts that combine recent developments in non-equilibrium statistical mechanics and stochastic thermodynamics~\cite{Seifert_review,Ritort_review,Esposito_review} (see Section~\ref{StochThermo}) together with techniques  for detecting and quantifying non-equilibrium behavior~\cite{mackintosh2010active} (see Section~\ref{FDTsec} and \ref{BDB}). For  phenomenological and hydrodynamic approaches to active matter, we refer the reader to several excellent reviews~\cite{Marchetti2013,Ramaswamy2010, joanny2009active,prost2015active}.

\begin{figure}
\centering
  \includegraphics[width=9cm]{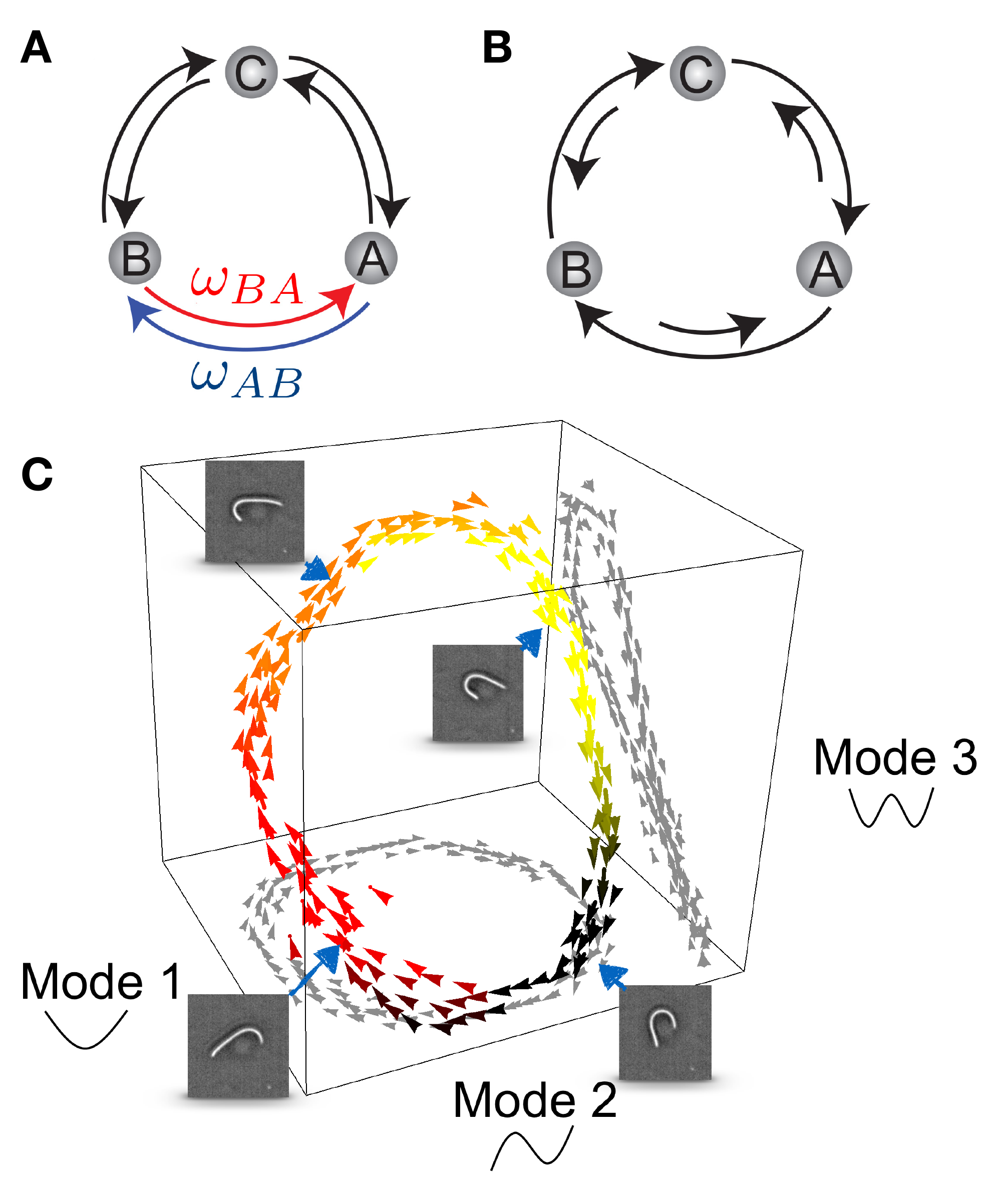}
  \caption{A) In thermodynamic equilibrium, transitions between microscopic states are pairwise-balanced, precluding net flux among states. B) Non-equilibrium steady states can break detailed balance and exhibit flux loops. C) Snapshots of an isolated \emph{Chlamydomonas} flagellum's beat cycle together with the 3D probability flux map of flagellar dynamics in a coarse grained phase space spanned by the first three modes. Adapted from \cite{Battle2016BDB}.
}
    \label{fig:Fig1}
\end{figure}

A characteristic feature of  living systems is that they are driven out of equilibrium at the molecular scale. For instance, metabolic processes, such as the citric acid cycle in animals and the Calvin cycle for carbon fixation in plants, generally involve driven molecular reaction cycles. 
Such closed-loop fluxes break detailed balance, and are thus forbidden in thermodynamic equilibrium (Fig.~\ref{fig:Fig1}A, B)~\cite{Zia2007}. Similar directed chemical cycles also power reaction-diffusion patterning systems in cells \cite{Frey2017} and molecular motors, including myosins or kinesins~\cite{Ajdari1997}.
 Indeed, such molecular motors can generate mechanical force by coupling the hydrolysis of adenosine triphosphate (ATP) to conformational changes in a mechano-chemical cycle~\cite{Ajdari1997,Howard2001}. The dissipation of this chemical energy drives unidirectional transitions between molecular states in this cycle. Such unbalanced transitions break detailed balance and result in directional motion of an individual motor. 

One of the central theoretical challenge in the field of active living matter is to understand how the non-equilibrium dynamics of individual molecular components act in concert to drive collective non-equilibrium behavior in large interacting systems, which in general is made of both active and passive constitutents. Motor activity may drive sub-components of cells and tissue\cite{mackintosh2010active,Bausch2006,Glaser2010}, but it remains unclear to what extent this activity manifests in the dynamics at large scales. Interestingly, even for systems out of equilibrium, broken detailed balance, for instance, does not need to be apparent at the supramolecular scale. In fact, at large scales, specific driven systems may even effectively regain thermodynamic equilibrium and obey detailed balance~\cite{Egolf2000,Rupprecht2016}.

There are, of course, ample examples where the dynamics of a living system is manifestly out of equilibrium, such as cell division or cell migration. In many cellular systems, however,  one can observe stationary stochastic processes that appear at first glance thermally driven.  Indeed, for many macromolecular assemblies in cells such as chromosomes~\cite{Weber},  the nucleus \cite{Almonacid2015}, the cytoplasm~\cite{brangwynne2009intracellular,Brangwynne2011,fakhri2014high}, membranes~\cite{Betz2009,Turlier2016,Ben_Isaac,Tuvia1997,Monzel2015}, primary cilia~\cite{Battle2015,Battle2016BDB}, and tissue \cite{Fodor2015} it has been debated to what extent non-equilibrium processes dominate their dynamics. Such observations  raise the  fundamental and practical question  how one can distinguish non-equilibrium dynamics from dynamics at thermal equilibrium. To address this question, a variety of methods and approaches have been developed to detect and quantify non-equilibrium in biological systems. When active and passive microrheology are combined, one can compare spontaneous fluctuations to linear response functions, which are related to each other  through the Fluctuation-Dissipation theorem (FDT) when the system is  at thermal equilibrium~\cite{Lau2003,Mizuno370,Mizuno2008,Guo2014}. Thus, the extent to which a system violates the FDT can provide insight into the non-equilibrium activity in a system. We will discuss this approach in detail in section~\ref{FDTsec}. Other methods employ temperature or chemical perturbations to test the extent to which thermal or enzymatic activities primarily drive the behavior of a system, but such experiments are invasive and are often difficult to interpret. More recently, a  non-invasive method to discriminate active and thermal fluctuations based on detecting broken detailed balance was proposed to study the dynamics of  mesoscopic systems. This new approach has been demonstrated for isolated flagella (see Fig.~\ref{fig:Fig1}C) and primary cilia on membranes of living cells \cite{Battle2016BDB}. The ideas underlying this method will be detailed in section~\ref{BDB} after briefly reviewing related work in stochastic thermodynamics in Sec.~\ref{StochThermo}.

Additional important insights on the collective effects of internal activity  came from studies on a host of simple reconstituted biological systems. Prominent examples include a variety of filamentous actin assemblies, which are driven internally by myosin molecular motors. Two-dimensional actin-myosin assays have been employed to study emergent phenomena, such as self-organization and pattern formation~\cite{Schaller2010polar, Schaller2011frozen}.
 Moreover, actin-myosin gels have been used as model systems to study the influence of microscopic forces on macroscopic network properties in cellular components~\cite{Mizuno370, Silva2011, murrell2012f, Alvarado2013molecular, Lenz2014}.
  Microrheology experiments in such reconstituted actin cytoskeletal networks have revealed that motor activity can drastically alter the rigidity of actin networks~\cite{Koenderink2009, sheinman2012actively, broedersz2011molecular} and significantly enhance fluctuations~\cite{brangwynne2008nonequilibrium, Mizuno370}.
Importantly, effects of motor forces observed \emph{in-vitro}, have now also been recovered in their native context, the cytoplasm~\cite{brangwynne2008nonequilibrium,Guo2014,fakhri2014high} and membranes~\cite{Betz2009, Turlier2016}. Further experimental and theoretical developments have employed fluorescent filaments as multiscale tracers, which offer a spectrum of simultaneously observable variables: their bending modes~\cite{aragon1985dynami, Gittes1993, Brangwynne2007}. The stochastic dynamics of these bending modes can be exploited to study non-equilibrium behavior by looking for breaking of detailed balance or breaking of Onsager symmetry of the corresponding correlations functions~\cite{Gladrow2016,Gladrow2017}. This approach will be discussed further in section~\ref{sec:filamentBDB}.

\label{sec:intro}
\section{Non-equilibrium activity in biological systems and the Fluctuation-Dissipation theorem}
\label{FDTsec}
Over the last decades, a broad variety of microrheological methods have been developed to study the stochastic dynamics and mechanical response of soft systems. Examples of such systems include synthetic soft matter~\cite{Cicuta2007,Mason97,Mackintosh99,Waigh05,Levine00}, reconstituted biological networks ~\cite{Jensen_Rev,Bausch2006,Lieleg2007,Lieleg_Rev_2010,Mahaffy00,Gardel03,Tseng02,Keller03,Uhde04},
as well as cells, tissue, cilia and flagella~\cite{Mizuno370,prost2015active,Battle2016BDB,Wilhelm08, Fabry2001,Tseng02, Bausch1999,ma2014active}.
In this section, we discuss how the combination of passive and active microrheology can be used to probe non-equilibrium activity in soft living matter. After briefly introducing the basic framework and the most commonly used microrheological techniques, we will discuss a selection of recent studies employing these approaches in conjunction with the fluctuation-dissipation theorem to quantify non-equilibrium dynamics.
\begin{figure}[htb]
\centering
  \includegraphics[width=8.75cm]{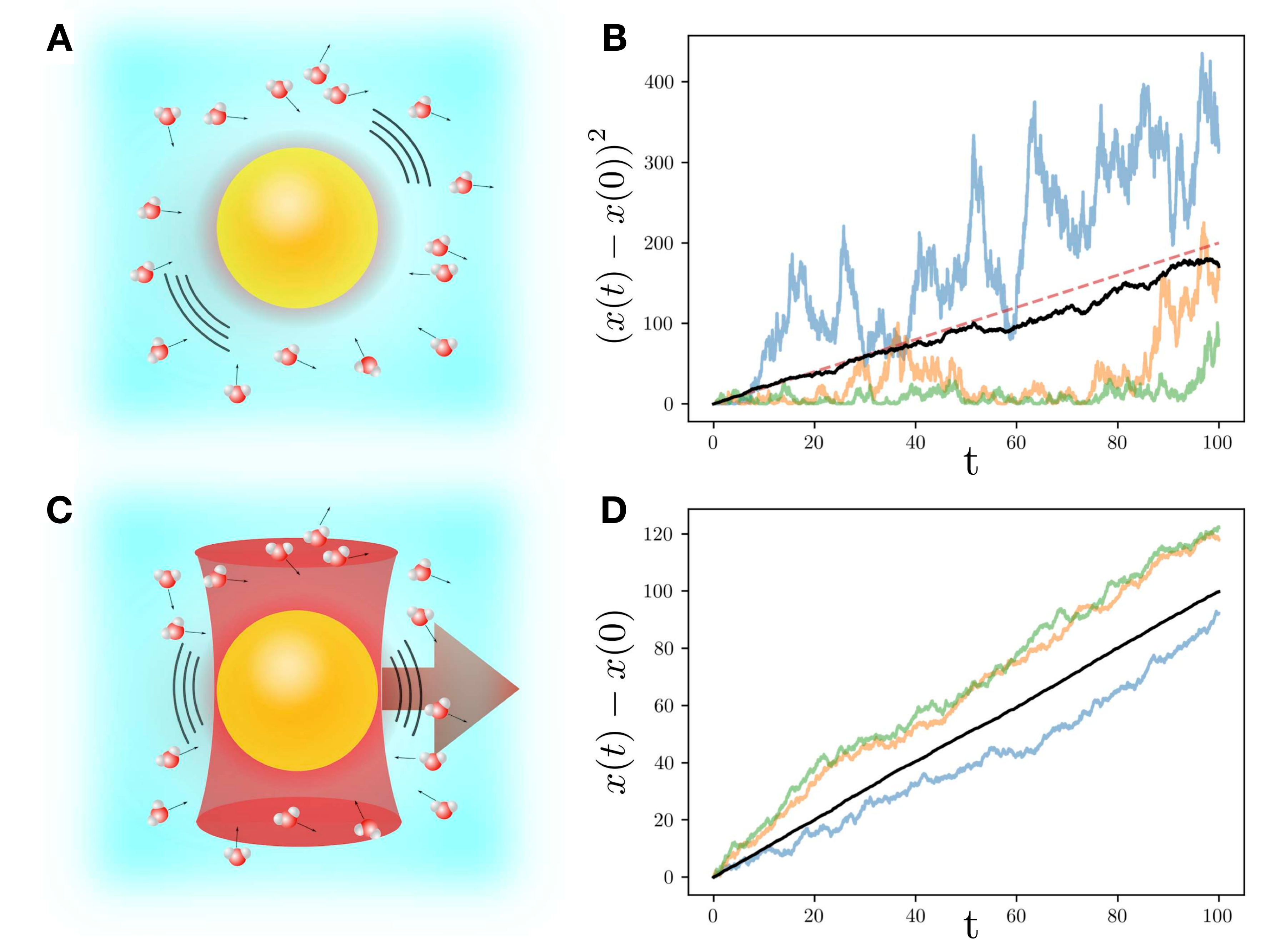}
  \caption{The fluctuation dissipation theorem implies a relation between thermal forces exerted by the molecules of the fluid on a Brownian bead and drag forces due to the viscosity of the fluid. A) Cartoon of a freely diffusing Brownian particle. B) Mean square displacement of the particle obtained by performing a Brownian simulation (black), and comparison with the analytical value $<(x(t)-x(0))^2>=2Dt$ (red). C)  Schematic of an external force $f$ in the positive $x$ direction applied on the particle via an optical tweezer. D) The average displacement for the driven particle (black), obtained from Brownian simulation, increases linearly with time, as $<x(t)-x(0)>=\mu ft$, where $\mu$ is the mobility. In this simple cases, the FDT reduces to the Einstein relation: $D=\mu k_{\rm B}T$.}
    \label{fig:Fig2}
\end{figure}

Microscopic probes embedded in soft viscoelastic environments can not only be used to retrieve data about the spontaneous fluctuations of the surrounding medium, but can also be employed to measure the mechanical response of this medium to a weak external force. In the absence of an applied force, the average power spectrum $S_{x}(\omega)=\langle |\Delta x^2(\omega)|\rangle$ of fluctuations in the bead position $x(t)$ can be directly measured. The brackets here indicate an ensemble average. The same bead can, in principle, be used to extract the linear response function $\chi_x(\omega)=\langle \Delta x(\omega) \rangle /f(\omega)$ by measuring the average displacement induced by a small applied force $f(\omega)$. In systems at thermal equilibrium, these two quantities are related through the Fluctuation-Dissipation theorem (FDT), derived in the context of \emph{linear response theory}~\cite{CallenFDT,Kubo_FDT} (see Fig.~\ref{fig:Fig2}). In frequency space, the FDT relates the autocorrelation function of position fluctuations of an embedded probe particle, in the absence of external forces, to the imaginary part of the associated response function:
\begin{equation}
S_{x}(\omega)=\frac{2k_B T}{\omega}\chi''_x(\omega) \,,
\label{eq:FDT}
\end{equation}
Importantly, a  system that is actively driven into a non-equilibrium steady-state will typically not satisfy this equality; this fact can be used to our advantage to study activity in such a system. Indeed, the violation of the FDT has proven to be a useful method to assess the stochastic non-equilibrium nature of biological systems, for instance, by providing direct access to the active force spectrum in cells~\cite{Guo2014}.

One of the first efforts to investigate deviations from the FDT in a biological system was performed on hair bundles present in the aural canal of a frog \cite{Juelicher2001}. Hair bundles are thought to be primarily responsible for the capability of the ear to actively filter external inputs and emit sound \cite{Juelicher2001, vanDijk_EarMechanics_2011}. 
To trace the dynamics of the hair bundle,  a flexible glass fiber was attached to the bundle's tip to measure both the position autocorrelation function  and the associated response to periodic external stimuli. Interestingly, the magnitude of position fluctuations was observed to largely exceed the linear-reponse-based levels for a purely thermal system. This violation of the FDT indicates the presence of an internal energy source driving the system out of equilibrium.

A suggested measure of the degree of violation of the FDT is a frequency-dependent ``effective temperature" $T_{\rm eff}(\omega)$\cite{Cugliandolo_EffT_97,cugliandolo2011effective,loi2008effective,bursac2005cytoskeletal,Juelicher2001,prost2009generalized,Wilhelm2009}, defined as the ratio between fluctuations and dissipation: $T_{\rm eff}(\omega)\equiv \omega \, S_{x}(\omega)/2k_B\chi_x''(\omega)$.
For a system at thermal equilibrium $T_{\rm eff}=T$. However, this quantity can be drastically modified for an actively driven bundle: Close to its spontaneous oscillation frequency of $\omega_0$, the imaginary part of the response function of the hair bundle becomes negative. This implies that $T_{\rm eff}$ is frequency dependent and can also assume negative values.

Even though this example illustrates how the dimensionless quantity $T_{\rm eff}/T$ provides a simple metric for non-equilibrium, the concept of an effective temperature in this context remains a topic of debate ~\cite{Ben_Isaac,cates_how_far,Turlier2016,Juelicher2001,fodor2015activity}.  Note, the existence of an effective temperature should not be mistaken for the existence of a physical mapping between an active system and an equilibrium at a temperature  $T_{\rm eff}$.  While there certainly are examples where such a mapping exists, this will not be the case in general. Furthermore, it is not obvious how to interpret negative or frequency dependent effective temperatures, but an interesting perspective is offered by Cugliandolo \emph{et al}.~\cite{Cugliandolo_EffT_97}. They demonstrated for a class of systems that  the effective temperature can indicate the direction of heat flow and acts as a criterion for thermalization~\cite{Cugliandolo_EffT_97}. In a more recent study, the conditions were derived for systems in non-equilibrium steady states to be governed by quasi-FDT: a ralation similar to the equilibrium FDT, but with the temperature replaced by a constant $T_{\rm eff}>T$\cite{Dieterich2015}. These conditions entail that the intrinsic relaxation time of the system is much longer than the characteristic time scale of the active forces. However, these conditions may become more complicated by systems with a viscoelastic response governed by a spectrum of timescales for which the thermal force spectrum is colored~\cite{chaikin1995}. Beyond  being a simple way of measuring deviations from the FDT, the concept of an effective temperatures may thus provide insight into active systems, but this certainly requires further investigation.  Alternative measures for non-equilibrium have been the subject of more recent developments based on phase spaces currents and entropy productions rates, which are discussed in sections \ref{StochThermo} and \ref{BDB}.

\subsection{Active and passive microrheology}
The successful application of the FDT in an active unidimensional context like the case of the hair bundle described above, paved the road for new approaches: microscopic probes were embedded into increasingly more complex biological environments to study the mechanics and detect activity inside reconstituted cytoskeletal systems~\cite{Gardel03,Bausch2006,Lau2003,Mizuno370} and living cells~\cite{Fabry2001,Yamada2000,Lau2003}.

Probing violations of the FDT in such soft biological systems relies on high-precision microrheological approaches. Conventional single particle microrheology is divided into two categories: passive microrheology (PMR)~\cite{Weitz95} and active microrheology (AMR)~\cite{Ziemann94,Amblard96,Schmidt1996}.
PMR depends on the basic assumption that both the FDT and the generalized Stokes-Einstein relationship apply. This assumption ensures that a measurement of the position fluctuation spectrum directly yields the rheological properties of the medium. Indeed, the generalized Stokes-Einstein relation connects the force-response function to the viscoelastic response of the medium~\cite{Weitz95},
\be
\chi_x(\omega)=\frac{1}{6\pi a G(\omega)} \, ,
\label{eq:generalized_stokes}
\ee
where $a$ is the radius of the bead.
This equation is valid in the limit of Stokes' assumptions, \emph{i.e.} overdamped spherical particle embedded in a homogeneous incompressible continuum medium with no slip boundary conditions at the particle's surface. Here, $G(\omega)=G'(\omega)+i G''(\omega)$ describes the complex shear modulus, where the the real part is the storage modulus $G'$  describing the elastic component of the rheological response, and the imaginary part, $G''$, is the loss modulus accounting for the dissipative contribution. Under equilibrium conditions, the imaginary part of the response function $\chi_x''$ is  also related to the position power spectral density via the FDT (Eq.~\eqref{eq:FDT}). Thus, in PMR, the response function and the shear modulus are measured by monitoring the mean square displacement (MSD) $\langle\Delta x^2\rangle(t) \equiv \langle (x(t)-\langle x \rangle)^2 \rangle$ of the embedded beads.
By contrast, in AMR the mechanical response is directly assessed by applying an external force on an embedded probe particle, usually by means of optical traps or magnetic tweezers. 
Within the linear response regime, the response function can be measured as $\chi _{x}=\langle \Delta x (\omega) \rangle / f(\omega)$, and the complex shear modulus can then be determined from the generalized Stokes-Einstein relation (Eq.~\eqref{eq:generalized_stokes}).

Although one-particle PMR has proven to be a useful tool to determine the equilibrium properties of homogeneous systems, biological environments are typically inhomogeneous. Such intrinsic inhomogeneity can strongly affect the local mechanical properties ~\cite{beroz2017physical, jones2015micromechanics}, posing  a challenge to determine the global mechanical properties using microrheology. To circumvent this issue, two-point particle microrheology is employed~\cite{Crocker2000,Lau2003}. This method is conceptually similar to one-point microrheology, but it is based on a generalized Stokes-Einstein relation for the cross-correlation of two particles at positions ${\bf r}_1$ and ${\bf r}_2$ with a corresponding power spectral density $S_{{r}_1,{r}_2}(R,\omega)$ with $R=|{\bf r}_2-{\bf r}_1|$. This correlation function depends only on the distance between the two particles and on the macroscopic shear modulus of the medium. Thus, $S_{{r}_1,{r}_2}$ is expected to be less sensitive to local inhomogeneities of the medium \cite{Crocker2000}. 

PMR has been extensively employed to assess the rheology of thermally driven soft materials in equilibrium, such as polymer networks~\cite{Schnurr1997,Mizuno2008,Mason97,Addas2004,Gittes1998,GittesMicro1997,Weitz95,chen2003rheological,mason2000estimating},  membranes and biopolymer-membrane complexes ~\cite{helfer2000microrheology,fedosov2010multiscale,Turlier2016}, as well as foams and interfaces~\cite{lee2009combined,prasad2006two,ortega2010interfacial}. However, a PMR approach cannot be employed by itself to establish the mechanical properties of non-equilibrium systems, for which the FDT generally does not apply. If the rheological properties of the active system are known, the power spectrum of microscopic stochastic forces $\Delta(\omega)$--- with both thermal and active contributions --- can be extracted directly from PMR data for a single sphere of radius $a$~\cite{Lau2003,Mizuno2008,caspi2000enhanced}
\begin {equation}
\Delta (\omega)=   6\pi a \, S_{x}(\omega) |G (\omega)|^2 \, .
\label{Lau_theory}
\end{equation}
The expression for the power spectrum of force fluctuations  was justified theoretically \cite{Lau2003,MacKintosh2008}, considering the medium as a continuous, incompressible, and viscoelastic continuum at large length scales. The results discussed above laid out the foundations for a variety of studies that employed microrheological approaches to investigate active dynamics in reconstituted cytoskeletal networks and live cells, which will be discussed next.

\subsection{Activity in reconstituted gels}
The cytoskeleton of a cell is a composite network of semiflexible polymers that include microtubules, intermediate filaments, F-actin, as well as associated proteins for cross-linking and force generation~\cite{Fletcher2010a,alberts1994molecular,kasza2007cell,Bausch2006}. The actin filament network is constantly stretched and displaced by collections of molecular motors such as Myosin II. These motors are able to convert ATP into directed mechanical motion and play a major role in the active dynamics of the cytoskeleton~\cite{Mizuno370,kohler2012contraction,stricker2010mechanics,juelicher2007active,fakhri2014high}. 

To develop a systematic and highly controlled platform for studying this complex environment, simplified cytoskeletal modules with a limited number of components were reconstituted \emph{in vitro}, opening up a new field of study~\cite{Bausch2006,Jensen_Rev,Lin07,Kasza2010}. Among these reconstituted systems, F-actin networks are perhaps the most thoroughly examined~\cite{Lieleg_Rev_2010,Mizuno370,gardel2008mechanical,joanny2009active,pelletier2009microrheology,Kasza2010,murrell2015forcing}. Indeed, in the presence of motor activity, these networks display a host of intriguing non-equilibrium behaviors, including pattern formation~\cite{Silva2011,Schaller2010polar,Schaller2011frozen,Schaller2011polar}, active contractiliy and nonlinear elasticity ~\cite{Koenderink2009,Bendix2008quantitative,ronceray2016fiber,murrell2012f,lenz2012contractile,wang2012active}, as well as motor-induced critical behavior~\cite{Alvarado2013molecular,sheinman2012actively}.

\begin{figure}
\centering
  \includegraphics[width=8.5cm]{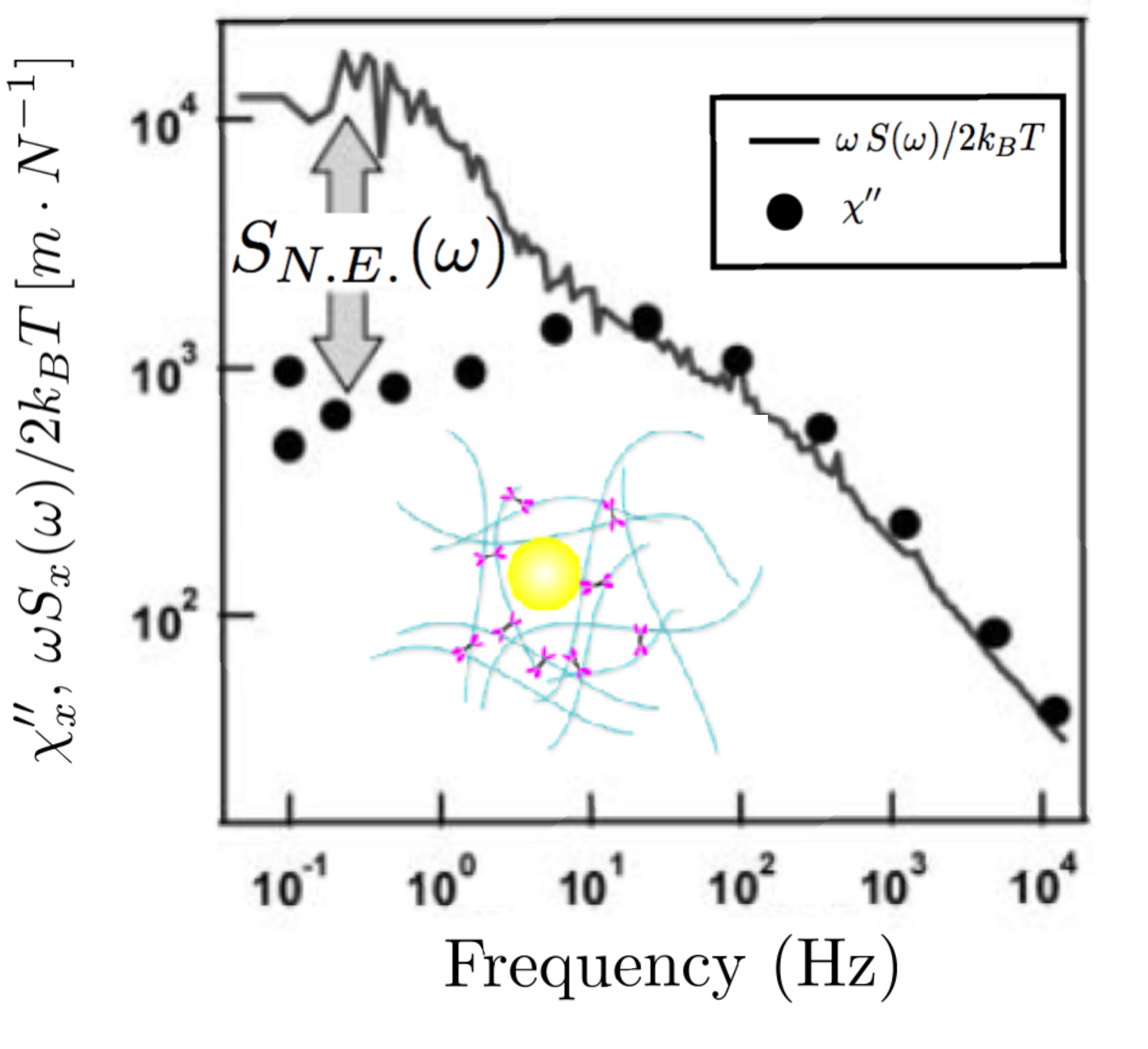}

  \caption{Violation of the FDT in reconstituted actin-myosin networks (inset). At frequencies below 10 Hz the response function estimated using the FDT from spontaneous fluctuations of a probe bead  deviates significantly from the response $\chi''$ measured directly using active microrheology (full circles).  Adapted from \cite{Mizuno370}.}
    \label{fig:Fig3}
\end{figure}

To study the steady state non-equilibrium dynamics of motor-activated gels, Mizuno \emph{et al.} constructed a three-component \emph{in vitro} model of a cytoskeleton, including filamentous actin, an actin crosslinker, and Myosin II molecular motors~\cite{Mizuno370}. The mechanical properties of the network were determined via AMR, while the activity-induced motion of an embedded particle was tracked via PMR.  The measured imaginary component of the mechanical compliance, $\chi_x''(\omega)$, was compared to the response predicted via the FDT, i.e. $\omega S_x(\omega)/2 k_B T$, as shown in Fig.~\ref{fig:Fig3}. In the presence of myosin, the fluctuations in the low-frequency regime were observed to be considerably larger than expected from the the measured response function and the FDT, indicating that myosin motors generate non-equilibrium stress fluctuations that rise well above thermally generated fluctuations at low frequencies. 

These observations raise the question why motor-driven active fluctuations only dominate at low frequencies.  This can be understood from a simple physical picture in which myosin motor filaments bind to the actin network and steadily build up a contractile force during a  characteristic processivity time $\tau_p$ ~\cite{howard2002mechanics}. After this processivity time, the motor filament detaches from the actin polymers to which they are bound, producing a sudden drop in the force that is exerted locally on the network. Such dynamics generically generate a force spectrum $\Delta(\omega) \sim \omega^{-2}$~\cite{MacKintosh2008,levine2009mechanics}, which can dominate over thermally driven fluctuations in an elastic network on time scales shorter than the processivity time (See Sec. \ref{sec:omega2} for a more detailed discussion). 

In addition to the appearance of non-equilibrium fluctuations, the presence of motors in the network led to a substantial ATP-dependent stiffening. It is well known that crosslinked semiflexible polymer networks stiffen under an external strain ~\cite{Storm2005,Gardel2004,Lieleg2007,Kasza2009,Lin2010}. Motors can effectively crosslink the network leading to stiffening, but they can also generate local contractile forces, and it is less clear how internal stress generation from such motor activity can induce large scale stresses and control network stiffness~\cite{ronceray2016fiber,broedersz2011molecular,shokef2012scaling,ronceray2015connecting,hawkins2014stress,xu2015nonlinearities,MacKintosh2008,chen2011strain,wang2012active}.
In a more recent experimental study, it was shown that motor generated stresses can induce a dramatic stiffening behavior of semiflexible networks~\cite{Koenderink2009}. This mechanism could be employed by cells and tissues to actively regulate their stiffness ~\cite{shokef2012scaling,tee2009mechanical,lam2011mechanics,jansen2013cells}.

An ensemble of beads dispersed in an active gel can not only be used to obtain fluctuation spectra,  but also to infer the full probability distribution of the beads' displacements at a time-lag $\tau$ \cite{toyota2011non,stuhrmann2012nonequilibrium,fodor2015activity}. This distribution is typically observed to be Gaussian for a thermal systems, while non-Gaussian tails are often reported for an active system. In actin-myosin gels, for example, exponential tails in the particle position distributions are observed at timescales $\tau$ less than the processivity time of the motors. By contrast, at larger time lags, a Gaussian distributions is observed, in agreement with what was previously found for fluctuation spectra in frequency space \cite{Mizuno370}. Importantly however, non-Gaussianity is not a distinctive trait of non-equilibrium activity, since it can also appear in thermal systems with anharmonic potentials. Moreover, Gaussian distributions could also govern active systems (see Sec.~\ref{sec:coordinateInvariance}).

The hallmarks of activity discussed above for actin-myosin gels are also observed in synthesized biomimetic motor-driven filament assemblies. Betrand \emph{et al.} created a DNA-based gel composed of stiff DNA tubes with flexible DNA linkers~\cite{Bertrand2012}. As an active component, they injected FtsK50C, a bacterial motor protein that can exert forces on DNA. An important difference with the actin-based networks described above, is that here the motors do not directly exert forces on the DNA tubes, which constitute the filaments in the gel. Instead, the motors attach to long double-stranded DNA segments that were designed to act as a cross-linker between two stiff DNA tubes. Upon introduction of the motors, the MSD of tracer beads  that were embedded in the gel was strongly reduced, even though the motors act as an additional source of fluctuations. This observation suggests  a substantial stiffening of the gel upon motor activation.  Furthermore, the power spectrum of bead fluctuations exhibited $\sim \omega^{-2}$ behavior, similar to results for \emph{in vitro} actin-myosin systems and even for live cells, which we discuss next.

\subsection{Activity in cells}
The extensive variety of biological functions performed by living cells places daunting demands on their mechanical properties. The cellular cytoskeleton needs to be capable of resisting external stresses like an elastic system to maintain its structural integrity, while still permitting remodelling like a fluid-like system to enable internal transport as well as migration of the cell as a whole~\cite{deng2006fast,kasza2007cell}. The optimal mechanical response clearly depends on the context.  An appealing idea is that the cell can use active forces and remodelling to dynamically adapt its (nonlinear) viscoelastic properties in response to internal and external cues~\cite{ahmed2015dynamic,ehrlicher2015alpha,yao2013stress}. In light of this, it is interesting to note that experiments on reconstituted networks suggest that activity and stresses can lead to responses varying from fluidization to actual stiffening~\cite{Humphrey2002,Koenderink2009,brangwynne2008cytoplasmic}.
Currently, however, it remains unclear how such a mechanical response plays a role in controlling the complex mechanical response of living cells~\cite{Fletcher2010a,Fernandez2008, Wolff2012,Krishnan2009,deng2006fast,Trepat2007,ehrlicher2015alpha}.

Important insights into the mechanical response of cell were provided by experiments conducted by Fabry \emph{et al.} via beads attached to focal adhesions near the cortex of human airway muscle cells. Their data indicate a rheological response where the loss and storage moduli are comparable, with a magnitude roughly in the range 100-1000 Pa around 1 Hz, and depend on frequency as a power law $|G(\omega)|\sim \omega^{x}$ with a small exponent $0.1\le x \le0.3$~\cite{Fabry2001}, reminiscent of  soft glassy rheology~\cite{Sollich1997,Semmrich2007,hoffman2009cell,bursac2005cytoskeletal,balland2006power}.

The studies conducted by Lau et al. \cite{Lau2003} and Fabry \cite{Fabry2001}  employed different probes at different cell sites for active and passive measurements, and determined a diffusive-like spectrum $\langle \Delta x^{2} \rangle\sim \omega^{-2}$.  A more recent assessment \cite{Wilhelm08} was able to measure the cellular response and the fluctuation spectrum with the same probe and at the same cellular location. The rheological measurement of $G$ was found to depend critically on the size of the engulfed magnetic beads and yielded a power law dependence on the applied torque-frequency $G(\omega)\sim \omega^{0.5-0.6}$. Furthermore, the conjuncted PMR and AMR assessments revealed a clear violation of the FDT, with the MSD of the beads increasing super diffusively with time. Measurements of the MSD of micron-size beads located around the nucleus of a living fibroblast also exhibited super-diffusive spectra, with a $\sim t^{3/2}$ dependence \cite{Sakaue2017}. Upon depolymerization of the microtubule network, diffusive behavior was restored suggesting that the rectifying action of microtubule-related molecular motors might be responsible for the super diffusive behavior. Furthermore, when the motors were inhibited without perturbing the polymer network, subdiffusive behavior was observed, in accordance to what is expected in equilibrium for a Brownian particle diffusing in a viscoelastic environment~\cite{caspi2000enhanced} 

\begin{figure}
\centering
  \includegraphics[width=8.5cm]{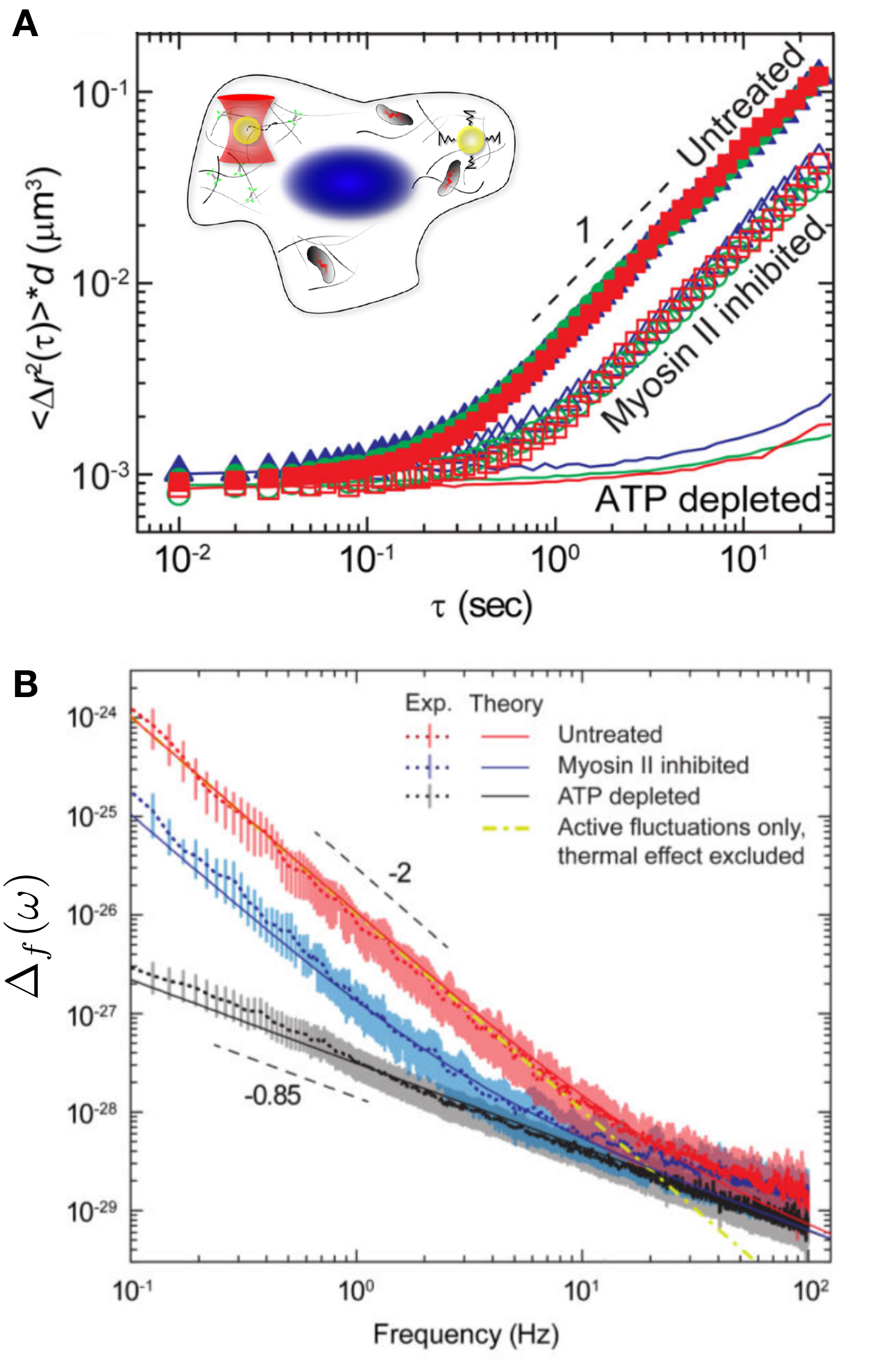}
  \caption{Fluctuations of probe particles inside living cells. A) The MSD, $\langle \Delta x^2(\tau)\rangle$, of tracer beads rescaled by  the particle diameter $d$, for untreated, Myosin inhibited, and ATP depleted cells. For untreated cells the MSD shows a plateau at  short time scales, after which the MSD increases linearly with time. When Myosin is inhibited by blebbistatin, the power law does not change but the magnitude of the MSD is reduced.  By depleting ATP in the cytoplasm, the dependence of the MSD on time becomes consistent with  thermal motion in a viscoelastic environment at short time. A cartoon of AMR and PMR performed inside the cytoplasm is shown in the inset. B) Measured force spectrum in the cytoplasm of untreated (red), blebbistatin treated (blue) and ATP-depleted (black) A7 cells.  Adapted from \cite{Guo2014}.}
  \label{fig:Fig4}
\end{figure}

A systematic measurement of both active and passive cytoplasmic properties was carried out by Guo \emph{et al.} via sub-micron colloidal beads injected into the cytoplasm of live A7 Melanoma cells. The probe beads were conveniently employed to perform both PMR and AMR with the use of optical tweezers. The active microrheology experiments indicated a  response with a shear modulus around 1 Pa, softer than measured near the cortex in \cite{Fabry2001}, but with a  similar power-law dependence of the complex shear modulus on frequency $|G(\omega)|\sim \omega^{0.15}$ \cite{Guo2014}. Passive microrheology was employed  to measure the mean square displacement (MSD) of position fluctuations under the same conditions in the cytoplasm (Fig. \ref{fig:Fig4}A). At short time-scales, the MSD is almost constant, as  expected for a particle embedded in a simple elastic medium. By contrast, at long time scales, the system can relax, resulting in an MSD that increases linearly with time, as would be expected for simple diffusion-like behavior of a probe particle in a viscous liquid as was also observed in earlier studies~\cite{Yamada2000,Alcaraz2003} 

Although these observations are deceptively close to the features of simple Brownian motion, this is clearly not the correct explanation for this phenomenon, given that the mechanical response of the system measured by AMR is predominantly elastic at these time scales. Furthermore, by treating cells with blebbistatin, an inhibitor of Myosin II, the magnitude of fluctuations notably decreased in the long time regime. While this suggests an important role for motor generated activity in driving the fluctuations of the probe particle, Myosin inhibition could also affect the mechanical properties of the cytoplasm, and thereby also the passive, thermally driven fluctuations of the probe particle. Nonetheless, by combining AMR and PMR it became clear that the system violates the FDT at these long time scales, implying that the system is not only out of equilibrium, but also that non-equilibrium activity can strongly alter the spectrum of force fluctuations.

The combination of AMR and PMR measurements was employed to infer the spectrum of force fluctuations  using a  method called Force Spectrum Microscopy (FSM). This method makes use of the relation $\Delta(\omega)= |k(\omega)|^2\langle \Delta x^2\rangle(\omega)$, where the complex spring constant $k\equiv1/\chi_x$ is related to $G$ by $k=6\pi G a$ (see Eq. \eqref{eq:generalized_stokes}). The measured force spectrum exhibited two different power-law regimes: at high frequencies $\Delta(\omega) \sim\omega^{-0.85}$, while at low frequencies ($\omega \lesssim 10$ Hz),  $\Delta(\omega) \propto \omega ^{-2}$, in agreement with what is expected for typical molecular motor power spectra, as depicted in Fig. \ref{fig:Fig4}B.

The observed high-frequency behavior  is in accordance with predictions for particle fluctuations driven by thermal forces in a nearly elastic medium. In fact, if 
$G\sim \omega^{\beta}$, then $\langle \Delta x^2\rangle(\omega) \sim \omega^{-(\beta +1)}$ at thermal equilibrium \cite{Lau2003}.  This implies that $ \Delta(\omega) \sim\omega^{-0.85}$, with the measured $\beta=0.15$.  By contrast, an active model predicts $\langle \Delta x^{2} \rangle(t)  \sim t^{1+2\beta}$ if $\Delta (\omega)\sim \omega ^{-2}$, which is consistent with what is observed in reconstituted motorized gels at times shorter than the processivity time $\tau_p$ ~\cite{Koenderink2009,brangwynne2008nonequilibrium}. These experiments and others~\cite{fakhri2014high,brangwynne2007force} have thus established the active nature and the characteristics of force spectra in the cytoplasm using embedded beads.

Various experiments employing PMR in live cells have been performed using alternative synthetic probes, such as nanotubes or  embedded intracellular entities, including microtubules, vesicles, and fluorescently labeled chromosomal loci.
In a recent study, Fakhri \emph{et al.} developed a new technology to investigate the stochastic dynamics of motor proteins along cytoskeletal tracks~\cite{fakhri2014high}. This cutting-edge method consists of imaging the near-infrared luminescence of single-walled carbon nanotubes (SWNT) targeted to kinesin-1 motors in live cells. Although traces of moving SWNT show long and relatively straight unidirectional runs, the dependence of the MSD of the tracer particle on time
exhibits several powerlaw regimes with an exponent that depends on the time range: At $t\approx 0.1 sec$ the exponent transitions from a value around 0.25
at short times to a value of 1 at larger times. By decomposing the MSD in motion along and perpendicular to the microtubule axis, it was shown that the dynamics of SWNT tracers originates from two distinct contributions: directed motion along the microtubules together with transverse non-directed fluctuations. The transverse fluctuations were attributed to bending fluctuations of the stiff microtubules, owing to motor-generated activity in the surrounding cytoskeleton, consistent with prior observations\cite{brangwynne2007force}. Indeed, the full time dependence of the MSD of traced kinesin motors could be described quantitatively with a model that assumes cytoskeletal stress fluctuations with long correlation times and sudden jumps. This is in agreement with a physical picture in which myosin mini-filaments locally contract the actin network during an attachment time set by the processivity time of the motors, followed by a sudden release.

Active bursts generated by Myosin-V are fundamental for nuclear positioning in mouse oocytes. In fact, active diffusion is here thought to create pressure gradient and directional forces strong enough to induce nuclear displacements~\cite{Almonacid2015,Razin2017,Razin2017a}. As in the earlier studies discussed above, the FDT is sharply violated at low frequencies, while it is recovered at large ones \cite{ahmed2015active}.

To study the steady-state stochastic dynamics of chromosomes in bacteria, novel fluorescence-labelling techniques were employed on chromosomal loci in \emph{E.Coli} cells. These experiments yielded sub-diffusive MSD behavior: $\langle \Delta x^2\rangle(t) \sim t^{0.4}$~\cite{weber2010bacterial,Weber,Sakaue2017,Vandebroek2015}. Although purely thermal forces in a viscoelastic system, such as the cytoplasm or a nucleoid, can also generate sub-diffusive motion~\cite{mackintosh2012active}, Weber \emph{et al.} demonstrated a clear dependence of the MSD on ATP levels: When ATP was depleted from the cell, the MSD magnitude was reduced. Surprisingly however, the exponent, $\alpha=0.4$, was not affected by varying ATP levels. Under the assumption that a change in the ATP level does not effect the dynamic shear modulus of the cytoplasm, this could be interpreted as resulting from active forces with a white noise spectrum and shear modulus that scales with frequency as  $G\sim \omega^{0.7}$. While these results provide   evidence for the existence of active diffusion by chromosomal loci, less invasive and more direct approaches are required to confirm and further study non-equilibrium behavior in the bacterial cytoplasm~\cite{Parry2014} and to understand dynamics of the chromosome. 

\subsection{ATP-dependent elastic properties and membrane fluctuations in red blood cells} 
\label{sec:betzSection}
The elastic properties of cells play an important role in many biological systems. The unusually high deformability of red blood cells (RBCs) is a prominent example in this respect, lying at the heart of the cardiovascular system. RBCs have the astonishing capability to squeeze through micron-sized holes, which ensures seamless blood flow through tight capillaries. To explore how these astonishing properties emerge, a detailed understanding of passive and active behavior of the membrane enclosing RBCs and its connection to the underlying cytoskeleton is required.

The bending dynamics of membranes are largely determined by their curvature and their response to bending forces thus depends on their local geometry~\cite{Evans1983, granek1997, granek2011,Lin2006}. In flat membranes, the power spectral density of bending fluctuations is expected to scale as $\omega^{-5/3}$ for large $\omega$~\cite{Milner1987, Betz2009,Lin2006}. A spectrum close to a $-5/3$-decay has indeed been reported in measurements of red blood cell membrane fluctuations~\cite{Betz2009}. Interestingly, the same experiments showed decreasing fluctuation amplitudes upon ATP-depletion, possibly indicating the role of non-equilibrium processes. The precise origin and nature of these processes, however, is difficult to determine due to the composite, ATP-dependent structure of erythrocyte membranes and cytoskeleton.

In addition, a flickering motion of RBC membranes observed in in microscopy experiments has sparked a discussion about the origin of these fluctuations. Indeed, the extent to which active processes determine the properties of RBCs is subject of intense research activity ~\cite{Brochard1975, Strey1995, Gov2004, Gov2005, Betz2009, Park2010, yoon2011red, Ben_Isaac, Rodriguez-Garcia2016}.

\begin{figure}
 \centering
 \includegraphics[width=9cm]{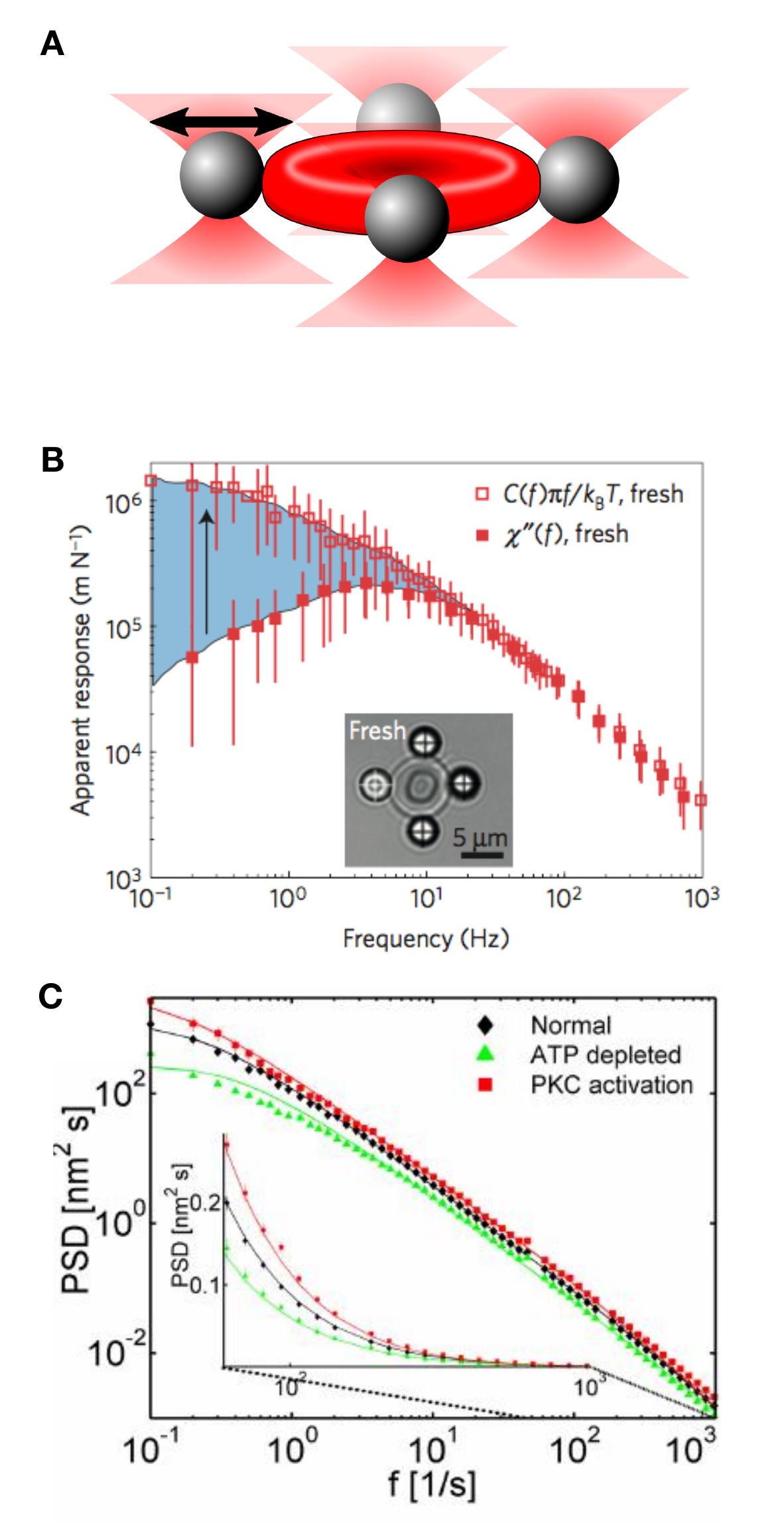}
 \caption{A) Cartoon of a red blood cell whose membrane conformations and response are tracked via four attached microscopic beads. B) The response and flickering spectrum of a red blood cell differ below 10 Hz, indicating a clear violation of the FDT. Adapted from \cite{Turlier2016}. C) Power spectrum of RBC membrane fluctuations under normal conditions (black), after ATP-depletion (green) and after addition of a PKC (red). PKC stands for {\em protein kinase C}, which catalyzes the phosphorylation of 4.1R, leading to increased dissociation of actin-spectrin structures. Adapted from~\cite{Betz2009}.}
 \label{fig:Fig5}
\end{figure}

Although myosin is present in the cytoskeleton of human erythrocytes, mechano-chemical motors are not the only source of active forces in the cell. In the membrane of RBCs, actin forms triangular structures with another filamentous protein called spectrin. These structures are linked together by a protein known as 4.1R. Phosphorylation of 4.1R, an ATP-consuming process, causes the spectrin-actin complex to dissociate, which could lead to a softening of the cell. In accordance with this model, ATP-depletion was found to increase cell stiffness~\cite{Tuvia1997}, and at the same time reduce membrane fluctuations on the $1-10$~s time scale. This is exemplified by the comparison between the green (ATP-depleted) and black (normal conditions) curves in Fig.~\ref{fig:Fig5}C. 

In order to relate the magnitude of fluctuations to membrane stiffness $\kappa$ and tension $\sigma$, Betz \emph{et al}.~\cite{Betz2009} employed a classical bending free-energy~\cite{Lin2006}
\begin{align}
 \mathcal{F}\left[h(r)\right] &= \int \mathrm{d}^2 r \, \left [\frac{\kappa}{2}\left(\Laplace h \right)^2+ \frac{\sigma}{2}\left(\nabla h \right)^2\right]. \label{eq:freeEnergy}
\end{align}
Here, $\kappa$ represents the membrane stiffness and $\sigma$ is the surface tension. A mode decomposition of the transverse displacement $h(\vec{q})$, evolving under thermal equilibrium dynamics of this energy functional leads to the correlator,
\begin{align}
  \langle h(\vec{q},t)h(\vec{q} \ ',t')) \rangle &= \frac{\left(2 \pi\right)^2 k_B T}{\kappa q^4+\sigma q^2}\delta(\vec{q}+\vec{q} \ ') e^{-\frac{|t-t'|}{\tau_q}}, \label{eq:membraneCorrelator} 
\end{align}
which is reminiscent of the correlator derived for semiflexible filaments (See section \ref{sec:filamentBDB}).
The decorrelation time $\tau_q$ is given by $\tau_q = 4\eta q /(\kappa q^4+\sigma q^2)$. A Fourier transformation of the correlator yields the theoretical prediction for the power spectral density shown in Fig.~\ref{fig:Fig5}. This model was also generalized to consider membrane fluctuations in the presence of active forces~\cite{Gov2005, Rodriguez-Garcia2016,Lin2006}.

The observed stiffening of the membrane upon ATP-depletion, presented a dilemma: membrane stiffening at low ATP could be the cause of the reduction of thermally driven membrane flickering, as apposed to a picture in which membrane flickering is primarily due to stochastic ATP-driven processes. This conundrum was resolved in a subsequent study, in which RBC flickering motion was shown to violate the equilibrium FDT, providing strong evidence for an active origin of the flickering~\cite{Turlier2016}. To demonstrate this, Turlier \emph{et al}.~\cite{Turlier2016} attached four beads to live erythrocytes, three of them serving as a handle, while the remaining bead can either be driven by a force exerted by optical tweezers or the unforced bead motion can be observed to monitor spontaneous fluctuations. The complex response $\chi_x(\omega)$ is then obtained from the ratio of Fourier transformations of the position $x(\omega)$ and force $F(\omega)$. The equilibrium FDT in Eq.~(\ref{eq:FDT}) relates these two quantities. 
The measured imaginary response $ \chi_x''(\omega)$ is plotted together with the response calculated from Eq.~(\ref{eq:FDT}) in Fig.~\ref{fig:Fig5}B. While the two curves exhibit stark differences at low frequencies, they become comparable for frequencies above 10~$\rm Hz$. Thus, whatever the precise nature of active processes in erythocyte membranes is, the intrinsic timescales of these processes appear to be on the order of $1 - 10$~Hz.

To explore the contributions to the mechanical properties of the membrane that arise specifically due to phosphorylation of 4.1R (and other molecules) in erythrocytes, the authors devised a semi-analytical non-equilibrium model for the elastic response of the membrane. Phosphorylation events are here modelled as on-off telegraph processes, which are added to an equilibrium description of membrane bending, such as in Eq.~(\ref{eq:freeEnergy}). The authors then decompose the membrane shape into spherical harmonic modes and calculate the single-mode power spectral density, which reads 
\begin{align}
 S_x^{lm}(\omega) &= \frac{2k_B T}{\omega} \chi_x^{lm} (\omega) ''+ \frac{2 \langle n_a \rangle \left(1-\langle n_a \rangle \right)\tau_a}{1+\omega \tau_a } \left |N^{lm}(\omega) \right | ^2, \label{eq:sphericalHarmonicsFDT} 
\end{align}
with $\tau_a=(k_a+k_i)^{-1}$ being the timescale, $n_a=k_a/(k_a+k_i)$ being the phosphorylation activity, and $N^{lm}(\omega)$ capturing the effects of tangential active noise on the membrane shape. The rate coefficients $k_a$ and $k_i$ characterize the simplified activate-inactivate ($a$-$i$) telegraph model, that the authors employ. Again, the expression in Eq.~(\ref{eq:sphericalHarmonicsFDT}) bears interesting similarities with the power spectrum of filament fluctuations (see Sec.~\ref{sec:filamentBDB}, Eqs.~(\ref{eq:thermalModeCorrelation}), and (\ref{eq:motorModeCorrelation})). The mode response here in Eq.~(\ref{eq:sphericalHarmonicsFDT})  is also composed of independent thermal and non-equilibrium contributions. Interestingly, the model shows that the curvature of the membrane is crucial for it to sustain active flickering motions. Only a curved surface allows fluctuations of tangential stress to result in transversal motion. Modes that correspond to wavelengths too short to couple to tangential stresses also do not seem to be affected by non-equilibrium processes. The flickering therefore appears to be caused by a coupling of tangential stresses to transversal motion only within a certain window of spherical modes $2\leq l\leq l*$. 

ATP-dependent fluctuations seem to contribute directly to the extraordinary mechanic properties of erythrocytes and may even help maintain their characteristic biconcave shape~\cite{Park2010}. Recently, bending fluctuations of membranes have been implicated in general cell-to-cell adhesion~\cite{Fenz2017}. The satisfactory agreement of theoretical and experimental fluctuation spectra in the examples discussed above highlights the merit of non-equilibrium statistical approaches to model and indeed explain properties of living biological matter.

The violation of the FDT is an elegant tool for the detection of activity in biological systems, as illustrated by the many examples discussed in the section above. That being said, for such a method to be applicable, the simultaneous measurement of fluctuations \emph{and} response is required. Even though this method gives information on the rheological properties of the system, its applicability can be challenging in contexts where the system is particularly delicate or poorly accessible such as chromosomes, the cytoskeleton, intracellular organelles, and membranes. Thus, in many cases a less invasive approach might be desired. These alternative approaches are further discussed in section~\ref{BDB}.

\subsection{Simple model for $\omega^{-2}$ active force spectra in biological systems}\label{sec:omega2}
As illustrated by the many examples discussed above, the mean square displacement of a probe particle in the cytoskeleton or in a reconstituted motor-activated gel has been widely observed to be surprisingly similar to a diffusive spectrum in a viscous medium: $\langle \Delta x^2\rangle \sim t$. In a purely viscous environment, with only Brownian thermal forces, the force spectrum is well-described by white noise, which has a flat power spectrum over the whole frequency range by definition. The magnitude of the complex shear modulus for a purely viscous fluid is $|G|^2 \sim \omega^{2}$. Such a simple rheological response, taken together with a white noise force spectrum, yields a displacement spectrum $\langle \Delta x^2\rangle \sim \Delta /|G|^2 \sim \omega^{-2} $ \emph{at all frequencies}. This mechanism however does not explain the effective diffusive behavior measured in cells below 10 Hz ~\cite{Guo2014,fodor2015activity,brangwynne2009intracellular,brangwynne2008cytoplasmic,gallet2009power,fakhri2014high}.  Below, we illustrate with a simple model~\cite{MacKintosh2008,levine2009mechanics,gallet2009power,Osmanovic2017,Samanta2016} that any active force with a sufficiently rapid decorrelation time will produce effective diffusive behavior of a bead in an elastic medium. 

 Consider a particle moving in a simple viscoelastic solid with both active forces, $f_A$, and thermal forces, $f_T$. The stochastic motion of such a particle can be described by an overdamped Langevin equation~\cite{Romanczuk2012,Lau2003,Mizuno2008,gardel2005microrheology,Levine00,Samanta2016,Ben_Isaac,ben2015modeling}:
\begin{equation}
\label{eq:EOM}
\gamma \dot{x}(t)=-k x(t) +f_{T}(t)+f_A(t) \, ,
\end{equation}
where $k$ is the elastic stiffness and $\gamma$ the friction coefficient of the gel, which is modelled as a Kelvin-Voigt medium~\cite{Doi2013a}. For such a system, the thermal noise is described by:
\begin{align*}
\langle f_T(t) \rangle &= 0 \, ,\\
\langle f_T(t')f_T(t) \rangle&=2\gamma k_BT\delta(t'-t)\,.
\end{align*}
By contrast, the independent active contribution, $f_A$, is modelled as a zero-average random telegraph process of amplitude $f_0$\cite{gardiner1985stochastic,Samanta2016}, whose autocorrelation function is 
\begin{align*}
\langle f_A(t)f_A(s) \rangle = \frac{f_0^2}{4}e^{-|t-s|/\tau}.
\end{align*} 
The inverse time constant $\tau^{-1}=\tau_{on}^{-1}+\tau_{off}^{-1}\ll k/\gamma$ is the sum of the switching rates of the motors between  \emph{on} and \emph{off} states.

Suppose we perform a PMR experiment in which we only have access to the power spectral density of the position, we would measure
\be
S_{x}(\omega) = \frac{\langle f_T^2 \rangle +\langle f_A^2 \rangle }{k^2+\gamma^2\omega^2} = \frac{2\gamma k_B T + \frac{f_0^2}{2}\frac{\tau}{(\omega \tau)^2+1}}{k^2+\gamma^2\omega^2} \, .
\ee
If we consider frequencies  $\tau^{-1}\ll\omega\ll k/\gamma$, the spectrum reduces to $S_{x} \approx f_0^2 \tau^{-1}/2(k\omega)^2 $. Note that the functional dependence on frequency in this limit is identical to the case of purely Brownian motion in a simple liquid. Thus, this simple model illustrates how active forces with a characteristic correlation time can  account for the characteristic features of active particle motion in viscoelastic solids.\\ \\

\section{Entropy production and stochastic thermodynamics}
\label{StochThermo}

Put colloquially, entropy is about disorder and irreversibility: transitions that increase the entropy of the universe are associated with an exchange of heat and should not be expected to spontaneously occur in reverse. Historically, this picture was shaped by experiments on the macroscopic scale, where temperature and pressure are well-defined variables. However, on length scales ranging from  nanometers to microns, where most cellular processes occur, fluctuations matter. Entropy, once thought to increase incessantly, here becomes a stochastic variable with fluctuations around its norm. These ideas sparked many new developments in stochastic thermodynamics~\cite{Seifert_review,Ritort_review,Esposito_review}. 

In this section, we briefly introduce and motivate several recent theoretical and experimental advances of this stochastic approach, which has extended thermodynamics to the realm of
small systems. In particular, we will discuss a class of results known as ``{\em fluctuation theorems}'' (FTs), together with a selection of general developments that highlight the applications of these results to living systems. In Section~\ref{sec:coordinateInvariance} we discuss aspects of entropy production that are specific to  linear multidimensional system, and in Section~\ref{sec:energySpeedAccuracy}, we review a recent study that demonstrates how these concepts can be used to understand  noisy control systems in cells. Finally, in Section~\ref{sec:englandBound}, we discuss a recently introduced fundamental lower bound for fluctuations around the currents of probability, which are associated with out-of-equilibrium systems. 

A key idea of stochastic thermodynamics is to extend the classical notion of ensembles and define ensemble averages of variables, such as heat, work , and entropy  over specific stochastic time trajectories of the system~\cite{Jarzynski2017}. These trajectories can be seen as realizations of a common generating process, associated with a particular thermodynamic state. The distribution $P( \delta )$ of fluctuations $\delta$ is often of interest. Fluctuation theorems are usually applicable far from equilibrium and constrain the shape of this distribution. Most FTs derived so far adhere to the following form
\begin{equation}
\label{FR}
\frac{P(\delta)}{P(-\delta)}= e^{\delta},
\end{equation}
which is always fulfilled for Gaussian probability distributions $P(\delta) \propto e^{-1/2 (\delta-\theta)^2/\sigma^2}$ with a mean $\theta$ that equals the variance $\theta=\sigma^2/2$. Other distributions may of course also fulfill this theorem. The fluctuation theorem governing the amount of entropy produced after a time $\Delta t$, $S(\Delta t)=\pm \omega$, $P(\omega)/P(-\omega) = e^{\omega}$ has received particular attention.
This result underlines the statistical nature of the second law of thermodynamics: a spontaneous decrease in the entropy of an isolated system is not prohibited, but becomes exponentially unlikely. However, since the entropy is an extensive quantity, negative fluctuations only become relevant when dealing with small systems, such as molecular machines.
  
The first fluctuation theorems were derived in a deterministic context~\cite{GallavPRL}, then extended to finite time transitions between two equilibrium states~\cite{Jarzynski1997},  and finally to microscopically reversible stochastic systems~\cite{CrooksPRE}. Later, mesoscopic stochastic approaches  based on a Langevin descriptions were proposed. These descriptions turn out to be ecpecially suitable in an experimental biological context were typically only mesoscopic degrees of freedom are tracked~\cite{KurchanFT, Sekimoto1, Sekimoto2, Seifert_stoch_FT}.  

Further physical intuition for entropy production can be obtained in the description provided by Seifert~\cite{Seifert_stoch_FT}. Here, the one-dimensional overdamped motion of a colloidal particle is treated as a model system. The particle moves in a medium at fixed temperature $T$ and is subject to an external force $F(x,\lambda)$ at position $x$, which evolves according to a protocol $\lambda$.  The entropy production associated with individual trajectories, $\Delta s_{\rm tot}=\Delta s_{m}+ \Delta s$, is given by the sum of two distinct contributions: the change of entropy of the medium $\Delta s_m $ and the change of entropy of the system $\Delta s$. The former is related to the amount of heat dissipated into the medium, $\dot{q}=F(x,\lambda)\dot{x}$, as $\Delta s_m=\int dt' \, \dot{q}/T$. The entropy change of the system is obtained from a trajectory-dependent entropy:
\begin{equation}
s(t)=- k_B \ln (P(x(t),t)).
 \end{equation}
Taking the average of $s(t)$ naturally leads to the Gibbs entropy, $S= -k_B \langle \ln (P(x(t),t)\rangle$. Within this framework, the integral fluctuation theorem (IFT) for $\Delta s_{\rm tot}$ can be derived~\cite{Seifert_stoch_FT}, which reads
\begin{equation}
\label{IFT}
 \left\langle e^{-\frac{\Delta s_{\rm tot}}{k_{\rm B}}}\right\rangle=1.
  \end{equation}
The IFT expresses a universal property of entropy production, which is valid if the process can be captured by a Langevin or master equation description. Note, that in this context this theorem also implies the second law, since it implies $\langle\Delta s_{\rm tot} \rangle\geq 0$. In steady-state, a similar approach leads to the steady-state fluctuation theorem (SSFT)
\begin{equation}
P(-\Delta s_{\rm tot})/P(\Delta s_{\rm tot})= e^{-\frac{\Delta s_{\rm tot}}{k_{\rm B}}},
\end{equation}
which is a strongler relation from which Eq.(\ref{IFT}) follows directly.
In early studies~\cite{KurchanFT,Lebowitz1999} this theorem was obtained only in the long time limit, but it has been now extended to shorter timescales~\cite{Seifert_stoch_FT}.
To experimentally validate the fluctuation theorems discussed, Speck \emph{et al}. studied a silica bead maintained in a NESS by an optical tweezer. In this study, a single silica bead is driven along a circular path by an optical tweezer~\cite{Speck_stoch_coll}. The forces felt by the bead fluctuate fast enough to result in an effective force $f$, which is constant along the entire circular path. 
The entropy production calculated directly from trajectories indeed adhered to the SSFT described above.

The development of fluctuation theorems has given a fresh boost to the field of stochastic thermodynamics and has led to a number of interesting studies. For example, several conditions for thermodynamic optimal paths have been established~\cite{Schmiedl2007,Machta2015,Sivak2012}. These optimal paths represent a protocol for an external control parameter that minimizes the mean work required to drive the system between two equilibrium states in a given amount of time. These results could provide insight into thermodynamic control of small biological systems. Recently, a fundamental trade-off between the amount of entropy produced and the degree of uncertainty in probability currents has been derived, which was considered in the context of sensory adaptation in bacteria. This trade-off is discussed in Section~\ref{sec:englandBound}.

Another important connection between energy dissipation and the spontaneous fluctuations of a system in a non-equilibrium steady-state was found by Harada and Sasa~\cite{Harada_Sasa}. 
When a system is driven out of equilibrium, the fluctuation dissipation theorem (FDT) is violated (See section \ref{FDTsec}). A natural question to ask is what the violation of the FDT teaches us about the non-equilibrium state of a system. Starting from a Langevin description for a system of colloidal particles in a non-equilibrium steady state, a relation was derived between the energy dissipation rate and the extent of violation of the equilibrium FDT~\cite{Harada_Sasa},
\begin{equation}
\langle \dot{W} \rangle = \sum _{i=0}^{N-1} \gamma  _i \left\{\overline{v}^2_i + \int_{-\infty}^{+\infty}\left[\tilde{S}_{v,ii}( \omega)- 2T\tilde{\chi}_{v,ii}'( \omega)\right]\frac{d\omega}{2\pi}\right\}
\label{H-S}
\end{equation}
where $\langle \dot{W} \rangle$  is the average rate of energy dissipation and $\gamma _i$ denotes the friction coefficient for the $i_{th}$-coordinate; $\tilde{S}_{v,ii}( \omega)$ and $\tilde{\chi}_{v,ii}( \omega)$ are  the Fourier transform of the velocity correlation function and response function respectively. A remarkable feature of this relation is that it involves experimentally measurable quantities such as the correlation function and the response function, thereby allowing a direct estimate of the rate of energy dissipation. The violation of FDT has been measured, for instance, for molecular motors such as $\text{F}_1$ ATP-ase or Kinesin. Using the Harada-Sasa relation, it has been possible to infer information on the dissipated energy and efficiencies of  such biological engines~\cite{Toyabe_ATPase,Ariga_Kinesin}. 

Intuitively, any experimental estimate of the entropy production rate will be affected by the temporal and spatial resolution of the observation. In \cite{Esposito2016stochcoarsgran} a coarse-grained description of a system in terms of mesostates was considered. With this approach, it was shown how the entropy production obtained from the mesoscopic dynamics, gives a lower bound on the total rate of entropy production. Interestingly,  in systems characterized by a large separation of timescales~\cite{wang2016timescalesep} where only the slow variables are monitored, the hidden entropy production arising from the coupling between slow and fast degrees of freedom, can be recovered using Eq.(\ref{H-S}). 

The entropy production rate appears to be a good way of quantifying the breakdown of time reversal symmetry and energy dissipation. However, it is still unclear how this quantity is affected by the timescales that characterize the system. To address this, a system of Active Ornstein-Uhlenbeck particles was considered~\cite{cates_how_far}. This system can be driven out of equilibrium by requiring the self-propulsion velocity of each particle to be a persistent Gaussian stochastic variable with decorrelation time $\tau$, thereby providing a simple, yet rich theoretical framework to study non-equilibrium processes. Interestingly, to linear order in $\tau$, an effective equilibrium regime can be recovered: This regime is characterized by an effective Boltzmann distribution and a generalized FDT, even though the system is still being driven out of equilibrium. Indeed, the leading order contribution of the entropy production rate only sets in at $\sim \tau^2$. 

In general, how much information is needed to safely say if a system is evolving forward or backward in time? In complex  systems, we may sometimes face limited information about local or global thermodynamic forces. In such situations, the direction in which processes evolve, that is, the direction of time itself may in principle become unclear. Due to micro-reversibility, individual backward and forward trajectories are indistinguishable in a steady-state. Thus, it is natural how much information is needed to tell if a given trajectory runs forward or backward in time? 

This question was studied by Roldan \emph{et al}.~\cite{Roldan2015} using decision-theory, a natural bridge between thermodynamic and information-theoretic perspectives. Entropy production is here defined as $ S_{\text{tot}}(t) = k_B \ln (P(X_t)/P(\tilde{X}_t))$ with $X_t$ and $\tilde{X}_t$ denoting a forward trajectory and its time-reversed counterpart . The unitless entropy production, $S(t)/k_B$ assumes the role of a log-likelihood ratio $\mathcal{L}(t)$ of the probability associated with the forward-hypothesis $P(X_t | H_\rightarrow)$ and the backward-hypothesis $P(X_t | H_\leftarrow)$, that is, $\mathcal{L}(t)= \ln P(X_t | H_\rightarrow)/P(X_t | H_\leftarrow)$.
In a sequential-probability ratio test, $\mathcal{L}(t)$ is required to exceed a pre-defined threshold $L_1$ or subceed a lower threshold $L_0$, to decide which of the respective hypotheses $H_1, \, H_0$ is to be rejected. The log-likelihood ratio $\mathcal{L}(t)$ evolves over time as more and more information is gathered from the trajectory under scrutiny.

Interestingly, for decision-thresholds placed symmetrically around the origin $L_0 =-L, \, L_1=L$, the observation time $\tau_{\text{dec}}$ required for $\mathcal{L}(t)$ to pass either threshold turns out to be distributed independently of the sign of $L$, i.e.
\begin{align}
 P(\tau_\text{dec} | \leftarrow ) & = P(\tau_\text{dec} | \rightarrow ).  \label{eq:decisionTimeEquivalence}
\end{align}

\begin{figure}
 
 \centering
 \includegraphics[width=9cm]{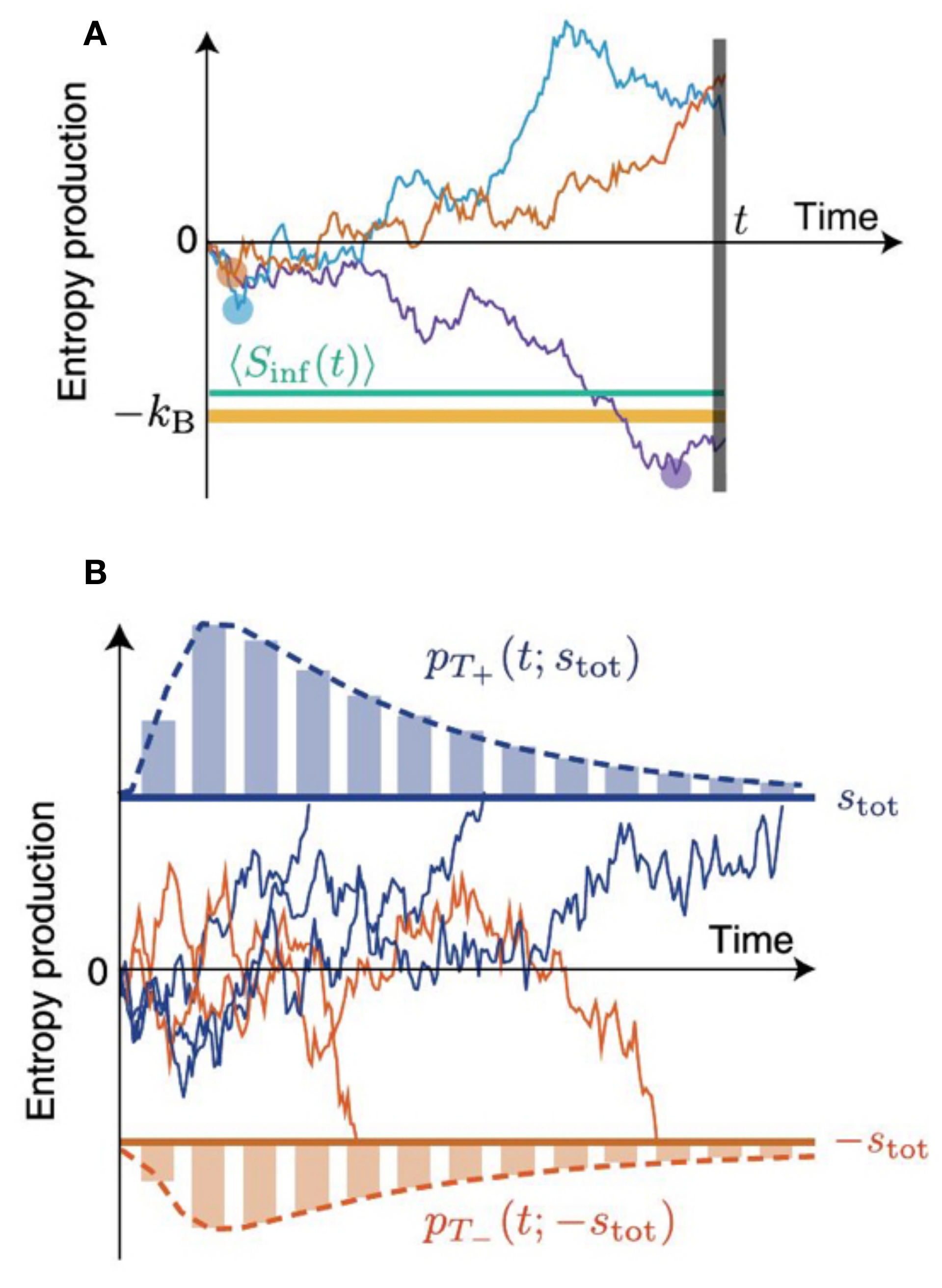}
 \caption{Entropy as a stochastic variable: Illustration of the mean infimum inequality and the equivalence of entropy production stopping times. A) The minimum of the average of an ensemble of entropy-trajectories (purple, red and blue) $\langle S_{\text{inf}}(t)\rangle$ (cyan) is bound from below by $k_B$ (thick yellow). B) For entropy-bounds $\pm s_\text{tot}$ that are symmetrically placed around $0$ (thick red and blue), the stopping times $T_+$ and $T_-$ share the same probability distribution (the figure shows unnormalized histograms). The stopping time $T_+$ ($T_-$) here is defined as the first-passage time of the entropy past the upper (lower) bound. Adapted from~\cite{Neri2016}\label{fig:Fig6}}

\end{figure}

From a thermodynamics perspective, this insight, implies that the average time it takes for a given process to produce a certain amount of entropy, must equal the average time it takes the same process to {\em consume} this amount of entropy. The latter process, of course, would be less likely to occur. In a related recent study, Neri \emph{et al}.~\cite{Neri2016} discuss the properties of ``{\em stopping times}'' of entropy production processes using a rigorous mathematical approach.
The stopping time here is defined as the time a process on average takes to produce or consume a certain amount of entropy relative to time $t_0$. This stopping time equivalence is sketched in Fig.~\ref{fig:Fig6}B. Importantly, stopping times are first passage times conditioned on the process actually reaching the threshold. The distribution of stopping times, therefore does not say anything about how probable it is for an observer to witness the process of reaching the threshold at all. Only {\em if} a trajectory reaches the threshold, the conditional first passage time can be measured. Fig.~\ref{fig:Fig6}A depicts another property of the entropy $S(t)$: the average entropy is bounded from below by $k_B$. 

These ideas were further illustrated by a few examples. The time a discrete molecular stepper, similar to the one illustrated in Fig.~\ref{fig:Fig9}, would spend making $N$ steps forward in a row, on average, is the same as it would spent making $N$ steps backwards. This results from the way the entropy production for this system scales with the position, $S_\text{tot}(t) = -N(t) Fl/(k_B T)$, where $F$  is the driving force and $l$ denotes the step length. Related first-passage-time equivalences have been discussed in the context of transport~\cite{Berezhkovskii2006}, enzymes~\cite{Qian2006a}, molecular motors~\cite{Kolomeisky2005}, and drift-diffusion processes~\cite{Stern1977}. Entropy stopping times, however, provide a unifying and fruitful perspective on first passage times of thermodynamic processes.

Living systems form one of the most intriguing candidates to apply key concepts of stochastic thermodynamic. Several fluctuation relations have been experimentally verified for various biological processes~\cite{Loutchko2017,Berg2008,Liphardt2002} and a stochastic thermodynamic description for chemical reaction networks have been developed~\cite{Seifert_networks} and applied, for instance, in catalytic enzymatic cycles~\cite{Loutchko2017}. A multitude of  thermodynamic equalities and lower bound inequalities involving the entropy production have been used to investigate the efficiency of biological systems. This provides insight into the energy dissipation required for a system to perform its biological function at some degree of accuracy. Important contributions in this direction can be found, for instance,  in~\cite{Seifert_motors} where the efficiency of molecular motors in transforming ATP-derived chemical energy into mechanical work is discussed. Following this line, we could ask how precise cells can sense  their  environment and use this information for their internal regulation. This was addressed in several works~\cite{Seifert_sensing,Lan2012,Mehta2012}, highlighting a close connection between the amount of entropy produced by the cellular reaction network responsible for performing the ``measurement", and the accuracy of the final measured information (See~\ref{sec:energySpeedAccuracy}). In~\cite{England2013,England2015}  these concepts were further expanded and applied to more complex macroscopic systems, such as the self-replication of bacteria, whose description is not captured by a simple system of chemical reaction networks. Despite the system's complexity, insightful results were obtained by deriving the more general inequality:
\begin{equation}
\label{England2law}
\beta \langle \Delta Q \rangle _{I \rightarrow II} + \Delta S_{int} \geq -\ln \frac{\pi(II\rightarrow I)}{\pi(I\rightarrow II)}
\end{equation}
where the system's irreversibility, i.e the ratio between the probability of transition between two macrostates $\pi(II\rightarrow I)$ and its reverse, represents a lower bound for the total entropy production $\beta \langle \Delta Q \rangle _{I \rightarrow II} + \Delta S_{int}$, $\beta=1/k_{\rm B}T$, and $\Delta Q$ is the heat dissipated in the environment. One can now identify the two macrostates $I$ and $II$ with an environment containing  one and two bacterial cells respectively.
Using probabilistic arguments it is then possible to express the probability ratio in Eq.~\eqref{England2law} as a function of measurable parameters, which  characterize the system's dynamics. With this approach, one can make a quantitative comparison between the actual heat produced by \textit{E.coli} bacteria during a self-replication event and the physical lower bound imposed by thermodynamics constraints. These results may also have implications for the adaptation of internally driven systems, which are discussed in\cite{England2015,Perunov2016}.

\subsection{Coordinate invariance in multivariate stochastic systems}
\label{sec:coordinateInvariance}
Energy dissipation, variability, unpredictability are traits not exclusively found in biological systems. In fact, it was a meterologist, Edward Lorenz, who coined the term 'butterfly effect' to describe an unusually high sensitivity on initial conditions in what are now known as 'chaotic systems' ~\cite{Rouvas-Nicolis2009}. In a fresh attempt to explain their large variability, stochastic models have been applied to periodically recurring meteorological systems. El-Ni$\tilde{\rm n}$o, for example, is characterized by a slow oscillation of the sea surface temperature, which can cause violent weather patterns when the temperature is close to its maximum. Such a change in temperature can lead to new steady-states, in which the system is permanently driven out-of equilibrium under constant dissipation of energy and exhibits a rich diversity of weather 'states'. Out of equilibrium, transitions between states are still random, but certain transitions clearly unfold in a preferred temporal sequence.

 Interestingly, in an effort to model meteorological systems stochastically, Weiss uncovered a direct link between energy dissipation and variability, which is intimately related to broken detailed balance~\cite{Weiss2003}. 
More specifically, he found that out-of equilibrium systems can react more violently to perturbations than their more well-behaved equilibrium counterparts. This finding may be relevant in a much broader context, including biology, and we will therefore briefly summarize the main points here.
Specifically, we will briefly explore this phenomenon of {\em noise amplification} from a perspective of coordinate-invariant properties~\cite{Weiss2003}.

In an open thermodynamic system in equilibrium, all state variables $\vec{x}$, are subject to the dialectic interplay of random forcing (noise) $\vec{\xi}$, relaxation, and dissipation. Consider an overdamped two-bead toy system at equilibrium, for example (see Fig.~\ref{fig:Fig12}A and Sec.~\ref{sec:2beads}), where the two beads are coupled by springs and are placed in contact with independent heat baths. Energy stored in the springs is permanently released and refuelled by the thermal bath, and flows back and forth between the two colloids in a balanced way. 
A sustained difference in temperature between the beads, $T_1\neq  T_2$, however, will permanently rectify the flow of energy and break this balance.
Crucially, this temperature difference is a matter of perspective. If we set, for example, $T_1=0$, then bead $1$ will not receive any noise any more and energy will flow to it from bead number $2$. Interestingly, if we look at the normal coordinates of the beads $u_1(t)=(x_1(t)-x_2(t))/2$ and $u_2(t)=(x_1(t)+x_2(t))/2$, we find that their respective thermal noise has exactly the same temperature $T_2/2$. However, if we could measure the fluctuations of noise in these coordinates, we would find that both noise terms correlate. Thus, in this case, mode $1$ and $2$ are driven not only by the same temperature, but by the very same white noise process.

Correlations amongst noise processes $\xi_1(t)$ and $\xi_2(t)$ exciting different variables $x_1(t)$ and $x_2(t)$ can, in principle, break detailed balance, even if the overall variance of the noise is equal in all directions, i.e. $\langle \xi_1^2 \rangle=\langle \xi^2_2\rangle$. In other words, correlations in random forces in one coordinate system, result in differences in temperature in other coordinates and vice-versa. The simple temperature criterion $T_1=T_2$ is thus insufficient to rule out broken detailed balance (see Section~\ref{subsection:DB}); a comprehensive coordinate-invariant criterion is required.

Consider variables $\vec{x}(t)$ of a generic system, evolving stochastically according to a Langevin equation (\ref{eq:langevinPicture}), 
\begin{align}
 \frac{\mathrm{d} \vec{x}}{\mathrm{d}t}(t) & = \mathbf{A}\vec{x}(t)+\mathbf{F}\vec{\xi}(t) \label{eq:langevinPicture}
 \end{align}
while the dynamics of the associated probability density $\rho(\vec{x},t)$ is given by the corresponding Fokker-Planck equation (\ref{eq:fpePicture}). 
\begin{align}
\frac{\partial \rho}{\partial t} \left(\vec{x},t \right) &= - \nabla\cdot \left(\mathbf{A}\vec{x}\rho\left(\vec{x},t \right) - \mathbf{D}\nabla\rho\left(\vec{x},t \right) \right) \label{eq:fpePicture}
\end{align}
In the equations above, $\mathbf{F}$ denotes the forcing matrix, in which any noise variance is absorbed, such that $\vec{\xi}$ here has unit variance $\langle \vec{\xi}(t)\vec{\xi}^T(t') \rangle = \mathbf{1}\delta(t-t')$. The forcing matrix is directly related to the diffusion matrix $\mathbf{D}= \frac{1}{2}\mathbf{F}^T \mathbf{F}$, and the term $\mathbf{A}\vec{x}$ describes deterministic forces, and the matrix $\mathbf{A}$ therefore contains all relaxational timescales. Any linear system with additive, state-independent white noise $\vec{\xi}$ can be mapped onto these generic equations.

In an equilibrium system with independent noise processes, $\mathbf{D}$ is diagonal and fulfills the standard fluctuation-dissipation theorem $\mathbf{D} = \frac{k_B T}{\gamma} \mathbf{1}$, where $\gamma$ denotes the friction coefficient. In steady-state, the correlation matrix $\mathbf{C}=\langle \vec{x} \cdot \vec{x}^T \rangle$ in and out of equilibrium obeys the Lyanpunov equation $\mathbf{A}\mathbf{C}+\mathbf{C}\mathbf{A}^T=- 2\mathbf{D}$, which can be thought of as a multidimensional FDT. The density $\rho$ can therefore always be written as a multivariate Gaussian distribution $\rho(\vec{x},t)=1/\sqrt{|\mathbf{C}|} e^{-\frac{1}{2} \vec{x}^T\mathbf{C}^{-1}\vec{x}}$. 

Apart from systems with temperature gradients, detailed balance is also broken in systems with non-conservative forces $\mathbf{A}\vec{x}$, which have a non-zero rotation $ \partial_{i}\left( \mathbf{A}\vec{x} \right)_j \neq  \partial_{j}\left( \mathbf{A}\vec{x} \right)_i $. Within our matrix framework, this condition simplifies to $ A_{i,j} \neq A_{j,i}$ and thus requires $\mathbf{A}$ to be symmetric in equilibrium. For instance, if we consider the two bead model in Sec.~\ref{sec:2beads}, $\mathbf{A}$ would represent a product between a mobility matrix and a stiffness matrix, both of which are symmetric resulting in a symmetric $\mathbf{A}$. Note, this framework only  applies to systems with dissipative coupling; reactive currents require a separate analysis. The two ways of breaking detailed balance in our case (temperature gradients and non-conservative forces) are reflected by a coordinate-independent commutation criterion for equilibrium~\cite{Weiss2003}: 
\begin{align}
\label{eq:weissCriterion}
\mathbf{A}\mathbf{D}-\mathbf{D}\mathbf{A}^T =0.
\end{align}
It was also argued that a system with broken detailed balance will sustain a larger variance than a similar system with the same level of noise, which is in equilibrium. This effect, referred to by Weiss as {\em noise amplification}, had previously been attributed to non-normality of the matrix $\mathbf{A}$, which is only true for diagonal $\mathbf{D}$. This type of noise amplification is now understood to be caused by broken detailed balance.  

Although this amplification can be captured by different metrics, we here focus on the gain matrix $\mathbf{G}=\mathbf{1}+\mathbf{A}\mathbf{C}\mathbf{D}^{-1}$. The gain matrix is a measure of the variance of the system normalized by the amplitude of the noise input. To obtain a scalar measure, one can take, for example, the determinant of $\mathbf{G}$ which yields  the gain $g$. It can be shown, that $g\geq g_0$ when detailed balance is broken, where $g_0$ is the gain of the same system in equilibrium. Finally, it is interesting to note, that the noise amplification matrix $\mathbf{G}$ is related to the average production of entropy in our generic model system. Let $\Pi$ denote the production of entropy, then
\begin{align}
 \Pi &= k_B\rm{tr}\left(\mathbf{A}\mathbf{G} \right), \label{eq:entropyAndNoise}
\end{align}
providing a direct link between dissipation and increased variability in multivariate systems out of equilibrium.

\subsection{Energy-speed-accuracy trade-off in sensory adaption}
\label{sec:energySpeedAccuracy}
Energy dissipation is essential to various control circuits found in living organisms. Faced with the noise inherent to small systems, cells are believed to have evolved strategies to increase the accuracy, efficiency, and robustness of their chemical reaction networks~\cite{Barkai1997, Leibler1999, Qian2006}.  Implementing these strategies, however, comes at an energetic price, as is exemplified by Lan \emph{et al}. in the case of the energy-speed-accuracy (ESA) trade-off in sensory adaption~\cite{Lan2012}. This particular circuit is, of course, not the only active control in cell biology. The canonical example of molecular `{\em quality control}' is the kinetic proofreading process, in which chemical energy is used to ensure low error rates in gene transcription and translation~\cite{Hopfield1974a}. Furthermore, fast and accurate learning and inference processes, which form the basis of sensing and adaptation, require some energy due to the inherent cost of information processing~\cite{Mehta2012, Ito2013, Sartori2014, Lang2014}. 

\begin{figure}[h!]
 \centering
 \includegraphics[width=9cm]{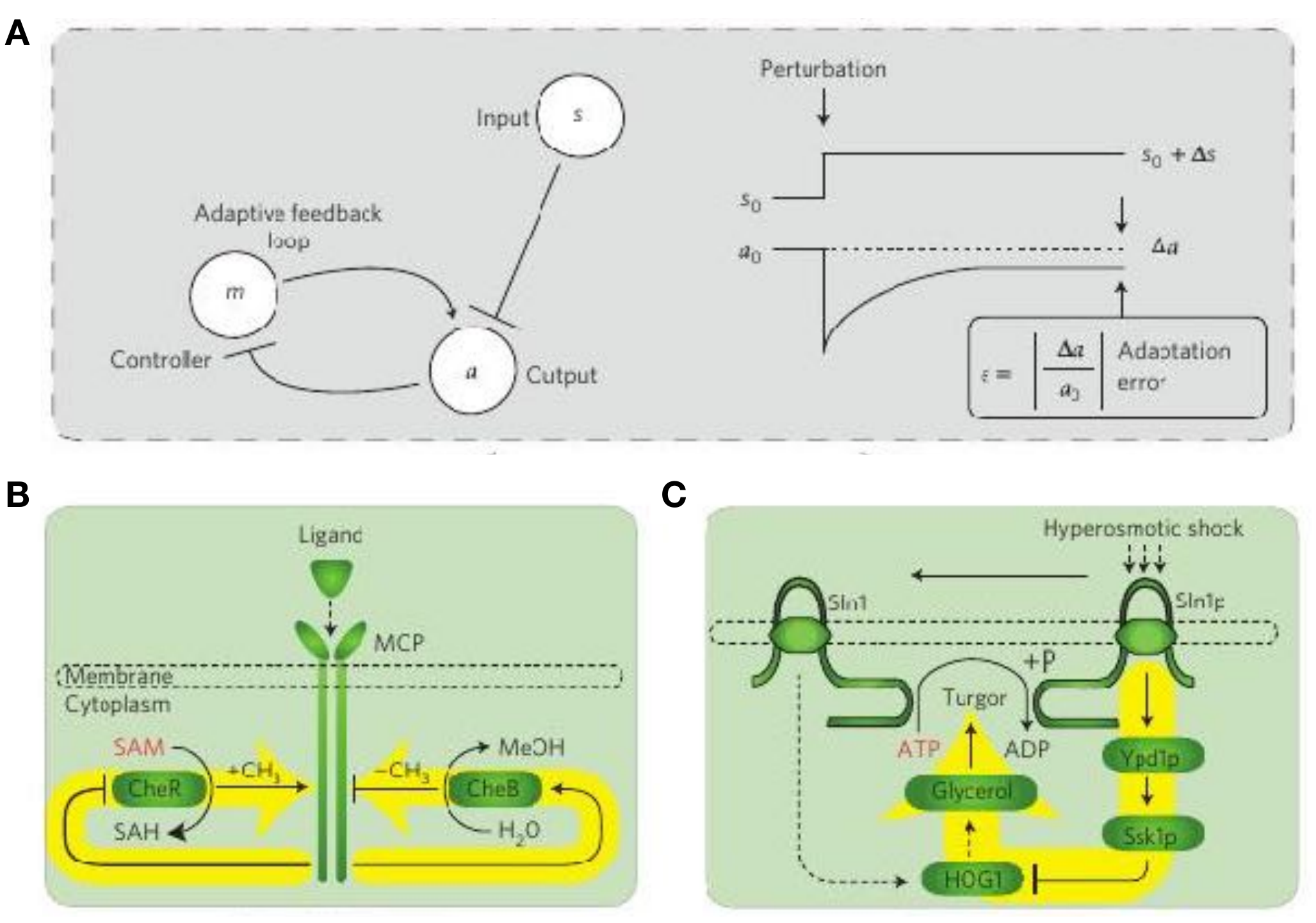}
 \caption{Models of adaptive feedback systems. A) Simplified topology of a feedback circuit. The input $s$ here is chosen to have an inhibitory effect.  On the right the response of the output $a$ is shown following a step in the input $s$. B) Chemotactic circuit in \emph{E. coli}. Ligand binding to a methyl-accepting-protein (MCP) causes further addition (mediated by CheR) or deletion (mediated by CheB) of methylgroups to MCP. This methylation counteracts the effects of ligand binding.  C) Osmotic sensing circuit in yeast. A reduction of osmolarity results in dephosphorylation of Sln1p$\to$ Sln1, which activates the HOG1 (High osmolarity glycerol) mechanism. This mechanism acts to restore the tugor pressure inside the cell and eventually phosphorylates Sln1$\to$Sln1p. 
Adapted from Lan \emph{et al}~\cite{Lan2012}.}
 \label{fig:Fig7}
\end{figure}

Sensory learning and adaptation at the cellular level involves chemical feedback circuits that are directly or indirectly driven by ATP hydrolysis, which provides energy input to break detailed balance. Examples of adaptation circuits are shown schematically in Fig.~\ref{fig:Fig7}B,C. These examples include the chemotactic adaption mechanism in \emph{E. coli } (panel (B)), a well-established model system for environmental sensing. Common to all circuits is a three-node feedback structure, as depicted in Fig.~\ref{fig:Fig7}A.   
Conceptually, this negative feedback circuit aims to sustain a given level of activity $a_0$, independent of the steady amplitude of an external stimulus $s$, which here is assumed to be inhibitory. This adaptive behavior allows the circuit to respond sensitively to changes to the external stimulus over a large dynamic range in $s$.

The authors condense the dynamics of such a chemical network into a simple model (Fig.~\ref{fig:Fig7}A) with abstract control $m(t)$ and activity $a(t)$ variables described by two coupled Langevin equations,
\begin{align}
 \dot{a}&= F_a(a,m,s) + \xi_a(t) \label{eq:adapt1} \\
 \dot{m}&= F_m(a,m,s) + \xi_m(t) \label{eq:adapt2}
\end{align}
with $F_a$, $F_m$ denoting the coarse-grained biochemical response and $\xi_a$, $\xi_m$ being white-noise processes with different variances $2\Delta_a$ and $2\Delta_m$, respectively.  Importantly, these biochemical responses do not fulfil the condition for conservative forces discussed in the previous section (above Eq.~(\ref{eq:weissCriterion})). To function as an adaptive system with negative feedback, $\partial_m F_a$ and $\partial_a F_m$ must have different signs, which implies a breaking of detailed balance. Indeed, adaptation manifests in a sustained probability current $J=(J_a, J_m)$ in the phase space spanned by $a\times m$; the energetic cost to maintain this non-equilibrium steady-state is given by the amount of heat exchanged with the environment per unit time, which must equal the entropy production rate $\Pi$ multiplied by the temperature $T$ of the heatbath to which the system is coupled.
 
In general, a non-equilibrium system at steady-state that adheres to a Fokker-Planck equation produces entropy at a rate~\cite{Tome2012, Seifert_review},
\begin{align}
 \Pi &= k_B \int \,\mathrm{d}\vec{x}\sum\limits_i \frac{J_i(\vec{x},t)^2}{D_i \rho(\vec{x},t)}. \label{eq:entropyProduction}
\end{align}
where $\rho(\vec{x},t)$ is the probability density in phase space, $D_i$ is the diffusion coefficient, and the sum runs over all phase space variables.
Applying Eq.~(\ref{eq:entropyProduction}) to the model above for sensory adaption, yields the heat exchange rate $\dot{W}~=~\int~\int~\mathrm{d}m\mathrm{d}a\left[J_a^2/(\Delta_a \rho )~+~J_m^2/(\Delta_m \rho )\right]$. An assumed separation of timescales that govern the fast activity $a$ and the slower control $m$, allows the authors to derive an {\em Energy-Speed-Accuracy} (ESA) relation, which reads
\begin{align}
 \dot{W} &\approx (c_0\sigma_a^2) \omega_m \log\left(\frac{\epsilon_0}{\epsilon} \right), \label{eq:esaRelation}
\end{align}
where,  $\sigma_a^2$ represents the variance of the activity, and $\epsilon$ denotes the adaptation error defined as $\epsilon \equiv \left |1- \langle a\rangle /a_0 \right |$, while $c_0$ and $\epsilon_0$ are constants that depend on details of the model. Here, $\omega_m$  parametrizes the rate of the control variable $m$. Therefore, an increase in $\omega_m$ or a reduction in $\epsilon$ requires an increased dissipation $\dot{W}$; put simply, swift and accurate adaptation can only be achieved at high energetic cost. 

\begin{figure}
 \centering
 \includegraphics[width=9cm]{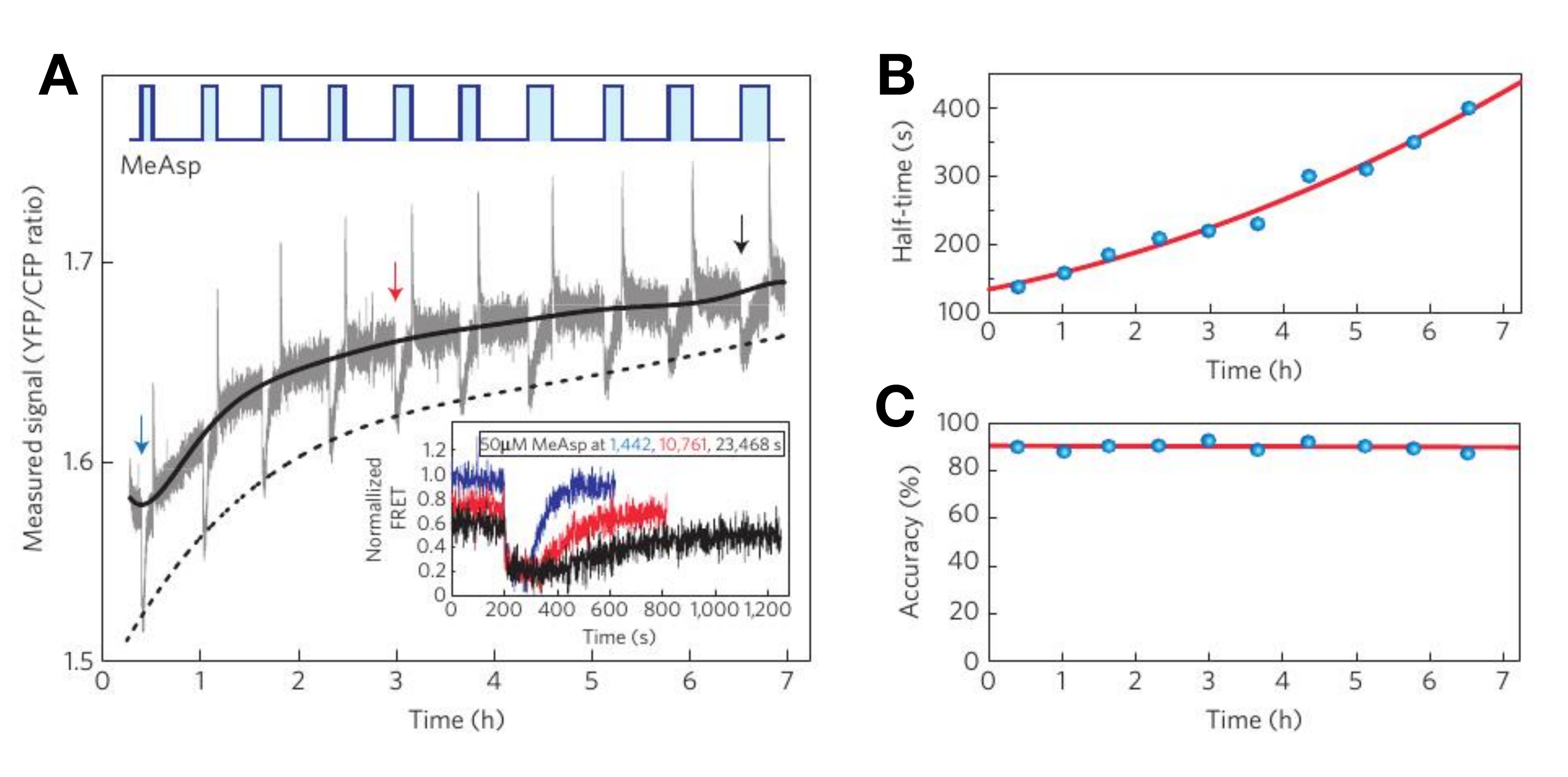}
 \caption{Experimental evidence for an energy-speed-accuracy (ESA) trade-off in \emph{E. coli} chemotaxis. A) Ratio of intensity of fluorescent reporters of adaptation. Changes in this signal are indicative of adaptation in the chemotatic circuit to external stimuli presented by the addition/removal of MeAsp. The inset illustrates the reduction of the FRET signal at the three different points in time indicated by arrows. B) Half-times inferred from the responses to addition/removal cycles shown in A). C) Relative accuracies of adaptation. Adapted from \cite{Lan2012}.}
 \label{fig:Fig8}
\end{figure}
The authors argue that a dilution of chemical energy in living bacteria will mainly affect the adaptation rate, but leave the adaptation error unchanged. Starvation should therefore lead to lower adaptation rates to uphold the ESA relation. This prediction was tested in starving \emph{E. coli} colonies under repeated addition and removal of MeAsp (see Fig.~\ref{fig:Fig8}), an attractant which stimulates the chemotactic system shown in Fig.~\ref{fig:Fig7}B. The cells in this study were engineered to express fluorescent markers attached to two proteins involved in adaptation. Physical proximity between any of these two molecules is an indicator of ongoing chemosensing, and was measured using Foerster-resonance-energy transfer (FRET). Since the donor-acceptor distance correlates with the acceptor intensity, but anticorrelates with the donor intensity, the ratio of YFP (acceptor) and CFP (donor) intensities lends itself as a readout signal to monitor adaptation. 
Indeed, after each addition/removal cycle of MeAsp, the signal recovers, albeit at a gradually decreasing pace, as is shown in the inset in Fig.~\ref{fig:Fig8}A. The decrease in the speed of adaptation is attributed to the progressing depletion of nutrition in the colony. In panels b and c, the adaptation half-time and relative accuracy are plotted. The graph in panel c clearly demonstrates the constancy of the accuracy of chemotatic system as nutrients are depleted over time, which is argued to be close to optimality.  

\subsection{Current fluctuations in non-equilibrium systems}
\label{sec:englandBound}
 Directed and chemically-specific transport of proteins, RNA, ions, and other molecules across the various membranes that foliate the cell is often achieved by active processes. A library of active membrane channel proteins has been described, which `{\em pump}' ions into and out of cells to control osmolarity, the electrical potential or the pH~\cite{Stein2014}. Furthermore, in eukaryotic cells, a concentration gradient of signalling molecules across the nuclear envelope causes messenger RNA (mRNA) molecules, expressed within the nucleus, to diffuse outwards through channels known as nuclear pore complexes (NPC)~\cite{alberts1994molecular}. Outside of the nucleus, the mRNA is translated into proteins by the ribosomes, which are too large to traverse the NPCs. All these directed transport processes are essential to the cell. Thus, this raises the question of reliability of such processes~\cite{Bezrukov2000, Berezhkovskii2008}. For example, how steady should we expect the supply of mRNA to the ribosomes to be~\cite{Grunwald2011}? Or, more generally, how predictable is the output rate of any given non-equilibrium process? Even active processes still endure fluctuations: molecular motors, at times, make a step backwards, or stall. Polymerizing filaments will undergo brief periods of sluggish growth or even shrinkage. Similarly, active membrane channels will sometimes transport more, and in other times fewer molecules. To illustrate this, an abstract example of such current fluctuations is depicted in Fig.~\ref{fig:Fig10}, which will be further discussed below.

\begin{figure}[t]
 \includegraphics[width=9cm]{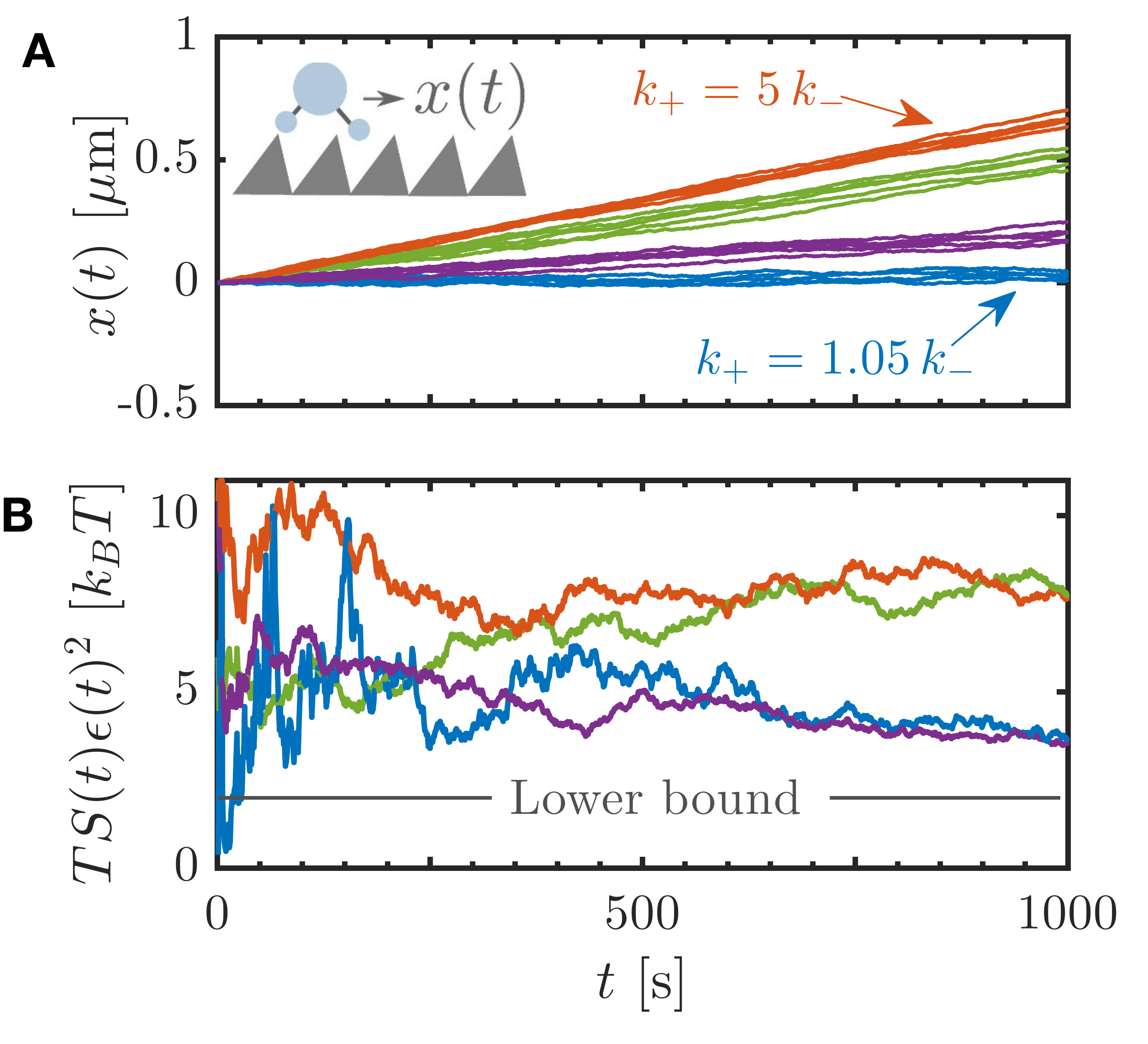}
  \caption{Variability of non-equilibrium steady states: A) Example trajectories to show the spread in the average position $\langle x(t) \rangle$ after $t$ steps. The inset depicts a simple model for a molecular motor in a sawtooth potential. B) The products  $T S(t) \epsilon(t)^2$ calculated over an ensemble of trajectories are bounded from below by the uncertainty relation. Despite the small size of the ensemble (100), Eq.~(\ref{eq:uncertaintyRelationBarato}) is fulfilled.}
  \label{fig:Fig9}
\end{figure}

\begin{figure}
 \includegraphics[width=9cm]{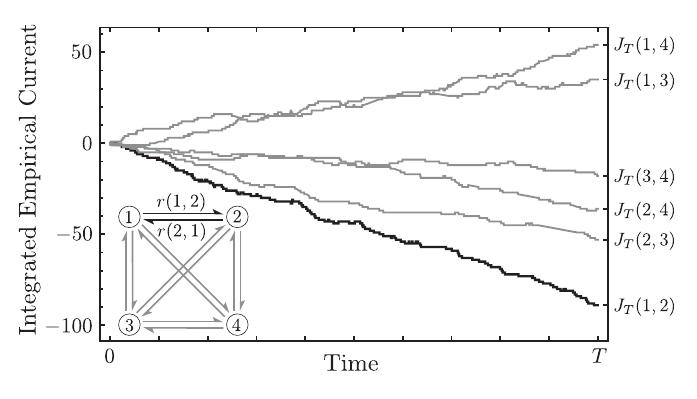}
  \caption{Variability of non-equilibrium steady states: Fluctuations of the cumulative probability current $J_T(m,n)=\int ^T j_{m,n}(t) \mathrm{d}t$ along all nodes in the four state system shown in the inset. Fluctuations result in perturbations of the currents around their intrinsic rates $r(m,n)$. Adapted from~\cite{Gingrich2016}. }
  \label{fig:Fig10}
\end{figure}

It seems intuitive, that predictability on the microscale always comes with an energy-price tag. In recent years, significant progress has been made to calculate the level of deviations from the average rate of a non-equilibrium process that is to be expected over finite times~\cite{Barato2015, Machta2015, Gingrich2016, Pietzonka2016, Pietzonka2017, Horowitz2017}. More formally, a universal bound for finite-time fluctuations of a probability current in steady-state has been established. Such an uncertainty relation is perhaps best illustrated by the simple motor model discussed by Barato \emph{et al}.~\cite{Barato2015}: A molecular motor moves to the right at a rate $k_+$, and to the left at a rate $k_-$. The movement is biased, i.e. $k_+> k_-$, driven by a free energy gradient $\Delta \mathcal{F}=k_B T \log \left( k_+/k_- \right)$. A few trajectories for various values of $k_+$ are depicted in Fig.~\ref{fig:Fig9}A. As can be seen, the walker (shown in the inset), on average, moves with a constant drift $\langle x (t) \rangle = t(k_+-k_-)$. Associated with this drift is a constant rate of entropy production $\Pi = (k_+-k_-)\Delta \mathcal{F}/T$.
Barato \emph{et al}. showed that the product of the total entropy produced $S(t)=\Pi t$ and the squared uncertainty $\epsilon^2=\langle (x(t) -\langle x(t) \rangle)^2 \rangle /\langle x(t)\rangle ^2$  always fulfils the bound
\begin{align}
 T S(t)\epsilon(t)^2  \ge  2k _B T. \label{eq:uncertaintyRelationBarato}
\end{align}
For this particular model, the square uncertainty reads $\epsilon(t)^2 = (k_++k_-)/[(k_+-k_-)t]$, such that the product $T S(t)\epsilon(t)^2 $ is constant in time. To further illustrate this point, we plotted the quantity $T S(t)\epsilon(t)^2$ for each choice of $k_+$ in Fig.~\ref{fig:Fig9}A, averaged over an ensemble of a hundred simulated trajectories in Fig.~\ref{fig:Fig9}B. Due to the finite ensemble size, the graphs fluctuate, but stay well above the universal lower bound of $2 k_B T$ for longer times $t$. So far, the theory underlying uncertainty relations was shown to be valid in the long time limit. Only recently, its validity has been extended to finite time scales~\cite{Pietzonka2017, Horowitz2017}.

The bound in Eq.~(\ref{eq:uncertaintyRelationBarato}) can be generalized to any non-equilibrium steady-state~\cite{Gingrich2016, Pietzonka2016}.
For a Markov system, such as the four-node system in the inset in Fig.~\ref{fig:Fig10}, the integrated current  $J_t=\int^t j(t') \mathrm{d}t'$ between any two nodes is distributed as $P(J_t=t j ) \sim e^{-t I(j)}$, with $I(j)$ denoting the large deviation function. This function therefore controls the variability of  $J_t$. Interestingly, it can be shown that the large deviation function obtained in the linear response regime $I_\text{LR}$, is never exceeded by $I$, even far away from equilibrium~\cite{Gingrich2016}. Thus, an increase in currents is accompanied by an increase in the variability of these currents when a system is driven further away from equilibrium. 

This general lower bound for the squared uncertainty $\epsilon^2$ is shown to be given by $2/\Sigma$, i.e. $\epsilon^2 \ge 2/\Sigma$, where  $\Sigma~=~\sum_{m<n} j (m,n)\log[ (\rho (m)r(m,n))/(\rho(n)r(n,m))]$ is the average dissipation rate in the system in steady-state. The steady-state probability distribution is denoted here by $\rho(m)$. The relation above is a bound for the uncertainty of the entire system. A similar relation also applies to any individual edge between two nodes $n$ and $m$, $\langle ( j(m,n) -\langle j(m,n) \rangle)^2 \rangle/\langle j(m,n)\rangle^2 \ge  2/\Pi(m,n)$, where $\Pi(m,n)$ denotes the entropy production associated with the edge $(m,n)$. 

While the uncertainty relations discussed above appear abstract at first, they may soon prove useful in studying transport or control systems in cellular biology due to their general applicability. 
Reminiscent of Carnot's efficiency for macroscopic engines, one implication of Eq.~(\ref{eq:uncertaintyRelationBarato}) is that a reduction in uncertainty can only be achieved by dissipating more energy when the system is close to optimality.

\section{Detecting broken detailed balance in living systems}
\label{BDB}
Up to this point we discussed intrinsically invasive methods to probe biological systems for non-equilibrium dynamics. For instance, to determine violations of the fluctuation-dissipation theorem a response function is required, which can only be measured by performing a perturbation in non-equilibrium systems (see Section~\ref{FDTsec}). Other methods that are used to probe for non-equilibrium involve thermal or chemical perturbations, and are therefore also inherently invasive. Such approaches are not ideal for investigating  the stochastic dynamics of  delicate sub-cellular system. Performing a controlled perturbation of such a system might not only be technically  challenging,  it may also be undesirable because of potential effects on the behavior or function of such a fragile system. 

Ideally, we would like to avoid the technical and conceptual difficulties of invasive protocols to probe for non-equilibrium behavior. This raises the question: Could we perhaps measure a system's non-equilibrium behavior simply by looking at it? With this purpose in mind, we recently developed a method that indeed uses conventional video microscopy data of cellular and subcellular systems~\cite{Battle2016BDB}. Detecting non-equilibrium behavior in the stochastic dynamics of mesoscopic coordinates of such systems can be accomplished by demonstrating that these dynamics break detailed balance. In this section, we will illustrate these ideas and discuss some recent related theoretical developments.

\subsection{Equilibrium, steady state, and detailed balance}
\label{subsection:DB}
Suppose we can describe a system on a mesoscopic level by dividing phase space into small cells, such that the state of the system can be described by a state variable $n$. If the system is ergodic and irreducible, it will evolve towards a unique stationary solution $p_n^{(s)}$, which is constant in time. A necessary  and sufficient requirement for such steady-state conditions is that the rate of transitions into any particular microstate, $m$, is balanced by the total rate of transitions from $m$ to other microstates $n$:
\begin{equation}
\label{eq:EqTrans}
\sum_n W_{n,m}  = \sum_n W_{m,n},
\end{equation}
where $W_{n,m}$ describes the rate of transitions from state $m$ to $n$. This result must hold for any system, at equilibrium or far from equilibrium, that has reached steady state conditions. When the system is Markovian, Eq.(\ref{eq:EqTrans}) reduces to
\begin{equation}
\sum_n w_{nm} p_m^{(s)} = \sum_n w_{mn}p_n^{(s)},
\end{equation}
where $w_{nm}$ describes the rate of transitions from state $m$ to $n$, given that the system is in state $m$. 

In thermodynamics equilibrium, it can be shown that a system must obey an even stronger condition: detailed balance. Classical closed ergodic systems are characterized by a time-independent Hamiltonian, which we will here restrict to be an even function of the momenta and independent of magnetic fields. The microscopic degrees of freedom of such a system obey deterministic dynamics described by Hamilton's equations, which are time reversal invariant. This  has important implications also for the probability distribution of mesoscopic observables, which characterize the systems states at thermodynamic equilibrium. Consider, for instance,  a mesoscopic variable $y$, which represents a generalized coordinate that either does not depend on the microscopic momenta, or that is an even functions of the microscopic momenta. Then, the transition between states must obey~\cite{gardiner1985stochastic}
\begin{equation}
\label{eq:detBalEvenVar}
p_2^{(e)}(y_2,\tau;y_1,0) = p_2^{(e)}(y_1,\tau;y_2,0) \, .
\end{equation}
Here we indicate with $p_2^{(e)}$ the two-point joint probability distribution.
This result is referred to as the principle of detailed balance. Put simply, it means that the transitions between any two mesostates are pairwise balanced, and this result derives from the transition rates between any two microstates are also being pairwise balanced. For Markovian systems we can  write detailed balance more conveniently as 
\begin{equation}
w(y_2|y_1)p^{(e)}(y_1) = w(y_1|y_2)p^{(e)}(y_2)\,,
\end{equation}
where the $w$'s indicate the conditional rates between states. 
Finally,  we note that if we add observables $z$, which are odd functions of the momenta, Eq.~\ref{eq:detBalEvenVar} needs to be  generalized to 
\begin{equation}
p_2^{(e)}(y_2,z_2,\tau;y_1,z_1,0) = p_2^{(e)}(y_1,-z_1,\tau;y_2,-z_2,0)\, .
\end{equation}

It is important to note that  for a system in steady state dynamics, broken detailed balance is direct evidence of non-equilibrium, but showing that a system obeys detailed balance in a subspace of  coordinates is insufficient to prove equilibrium. Indeed, even for systems out of equilibrium, broken detailed balance is not necessarily apparent at the supramolecular scale~\cite{Egolf2000,Rupprecht2016,Gladrow2016, Gladrow2017}. One can also often observe stationary stochastic processes in cells that, at first glance, appear to be thermally driven. Examples include the fluctuations of cytoskeletal filaments such as microtubuli, F-actin filaments or the fluctuations of intracellular organelles. These cases should be contrasted with obvious examples of mesoscopic non-equilibrium, non-stationary, irreversible  processes such as cell growth, locomotion and mitosis. Thus, in general, it is unclear how and when broken detailed balance that realized on the molecular level also manifests at larger scales.

\subsection{Probability Flux Analysis}

In this section, we  describe the basis and methodology that can be used to infer broken detailed balance from microscopy data. We consider a system, which is assumed to evolve according to stationary dynamics. This could, for instance, be a primary cilium or a flagellum~\cite{Battle2016BDB}. In general, these system exhibits stochastic dynamics, comprised of both a deterministic and a stochastic component. The dynamics of such a system can be captured by conventional video microscopy. To quantify this measured stochastic dynamics, we first need to parameterize the configuration of the system. The shape of a flagellum, for instance, could be conveniently decomposed into the dynamic normal modes of an elastic beam. In this example, the corresponding mode amplitudes represent time-dependent generalized coordinates of the system. Note, these mode amplitudes can be extracted from a single time frame and strictly represent configurational coordinates, which are independent of the microscopic momenta.

In general, a video microsocopy experiment can be used to extract time traces of $D$ mesoscopic tracked coordinates $x_1,...,x_D$, which represent the instantaneous configuration of the system. Clearly, this only represents a chosen subset of all coordinates that completely specify the whole system. Furthermore, only spatial or conformational degrees of freedom are considered in this discussion here.  Indeed, fluctuations in momenta in a typical overdamped biological or soft-matter systems relax on very short time-scales, which are not resolved in typical video microscopy experiments. However, the basic methodology described below can readily be generalized to also include momentum-like variables. 

We define a probability density, $\rho (x_1,...,x_D,t)$, in terms of only the tracked degrees of freedom. This probability density can be obtained from the full joint probability density in terms of a complete set of variables, by integrating out all the untracked degrees of freedom. In the reduced configurational phase space of the tracked degrees of freedom, the dynamics of the system still obeys a continuity equation:

\begin{equation}
\frac{\partial \rho (x_1,...,x_D,t)}{\partial t}=- \nabla\cdot\vec{j}(x_1,...,x_D,t)			
\end{equation}
where  $\vec{j}(x_1,...,x_D,t)$  is the current density describing the net flow of transitions of the system in the D-dimensional configurational phase space. Here, we only consider systems with dissipative currents~\cite{chaikin1995}. While at steady state the divergence of the current needs to vanish, in equilibrium any dissipative current itself must be identically zero.

\subsubsection{Estimating phase space currents}
\label{sec:expBDB}
\begin{figure}
 \centering
 \includegraphics[width=9cm]{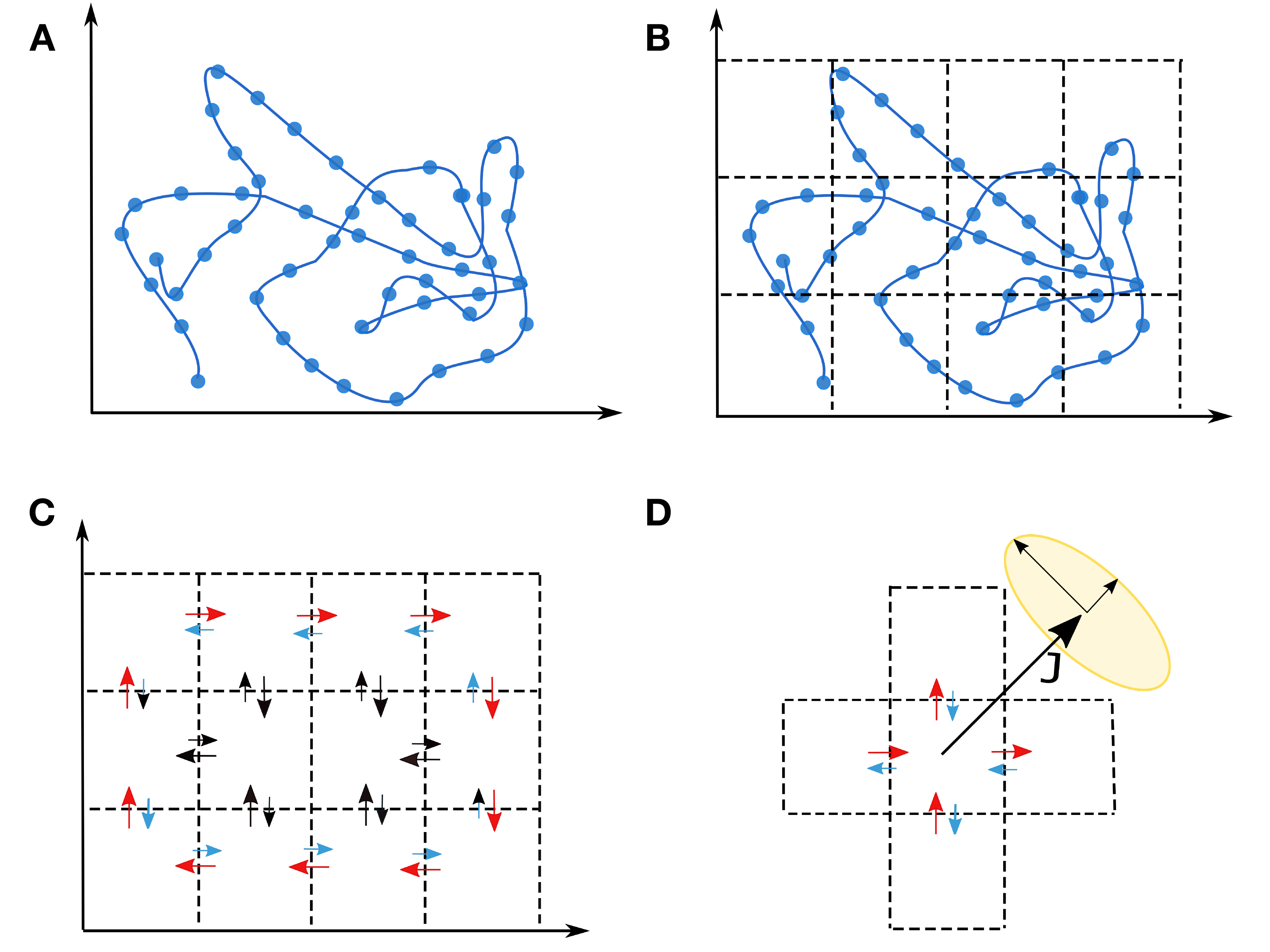}
 \caption{Schematic illustrating  the coarse-graining procedure for the estimation of phase space current and corresponding errorbars. A) Trajectory in continuous phase space. B) Grid illustrating the discretization of the continuous phase space. C) By counting transitions between  first neighbour discrete states it is possible to estimate the currents (indicated by the arrows) across the boundaries.  D) Current errorbars representation obtained through the bootstrapping procedure. Adapted from~ \cite{Battle2016BDB}.}
 \label{fig:Fig11}
\end{figure}

Here we discuss one way of estimating currents from a set of time-traces.  To provide a simple illustration of this approach, we consider a system with a two-dimensional configurational phase space,  as illustrated in Fig.~\ref{fig:Fig11}A. The dynamics of the system is captured by a time trace in this configurational phase space.
It is convenient to analyze these trajectories using a discretized coarse-grained representation of the two-dimensional phase space. This Coarse-Grained Phase Space (CGPS) consists of a collection of  equally sized, rectangular boxes, each of which represents a discrete state  Fig.~\ref{fig:Fig11}B. Such a discrete state in CGPS encompasses a continuous set of microstates, each of which belongs to a unique, discrete state. The primary reason for using this discretized representation of phase space is to be able to obtain informative results on experimental data with limited statistics.  

In this two-dimensional CGPS, a discrete state $\alpha$  has two neighboring states, respectively $\alpha _+$ (larger $x_i$) and $\alpha _-$  (smaller $x_i$), along each direction $x_i$, resulting in four possible transitions. The dynamics of the system indeed satisfies the continuity equation
\begin{equation}
\frac{d p_\alpha}{dt}=-\tilde{W}_{\alpha^-,\alpha}^{(x_1)}+\tilde{W}_{\alpha,\alpha^+}^{(x_1 )} -\tilde{W}_{\alpha^-,\alpha}^{(x_2)}+\tilde{W}_{\alpha,\alpha^+}^{(x_2)},
\end{equation}
where $\tilde{W}_{\alpha,\beta}= W_{\alpha,\beta} - W_{\beta,\alpha}$ is the net rate of transitions from state $\beta$ to $\alpha$ and $p_{\alpha}$ is the probability to be in discrete state $\alpha$, which will become time independent when the system reaches steady-state conditions. 

This probability $p_{\alpha}$ is related to the probability density $\rho (x_1,...,x_2,t)$ defined above, and can be obtained by integrating $\rho$ over the volume of state $\alpha$ in CGPS. We can estimate this probability from a measured trajectory by using
\begin{equation}
p_\alpha=t_\alpha/t_{\rm total},
\end{equation}
where $t_\alpha$ is the accumulated time that the system spends in state $\alpha$ and $t_{\rm total}$  is the total duration of the experiment.

The net rates $\tilde{W}$ in CGPS can be estimated  from the measured trajectories simply by counting the net number of transitions per unit time:
\begin{equation}
\label{netrates}
\tilde{W}_{\alpha,\beta}^{(x_i)}=\frac{N_{\alpha,\beta}^{(x_i)}-N_{\beta,\alpha}^{(x_i)}}{t_{\rm total}}
\end{equation}
Here  $N_{\alpha,\beta}^{(x_i)}$ is the number of transitions from state $\beta$ to state $\alpha$ along the direction $x_i$. In a mechanical system, the trajectories   through phase space are continuous such that there can be only transitions between neighboring states.
However, due to the discreteness in a measured time trajectory, it is possible that a transition between neighboring states is ``skipped", resulting in an apparent transition between non-neighboring states. In these cases, it is convenient to perform an interpolation of the time trace to estimate the intermediate transitions. It is important that this interpolation is performed in a time-symmetric way, so that the interpolation filter preserves time-reversal symmetry. In fact, this should be taken into account with any kind of filtering that is performed on measured time traces.

The currents in CGPS that describes back-and-forth transitions through all four boundaries of the box associated with a discrete state (Fig.~\ref{fig:Fig11}C), can be defined by:
\begin{equation}
\vec{j}(\vec{x}_\alpha)=\frac{1}{2} \left(\tilde{W}_{\alpha^-,\alpha}^{(x_1)}+\tilde{W}_{\alpha,\alpha^+}^{(x_1)},\tilde{W}_{\alpha^-,\alpha}^{(x_2)}+\tilde{W}_{\alpha,\alpha^+}^{(x_2)}\right).
\end{equation}
Here, $\vec{x}_\alpha$ is the center position of the box associated with state $\alpha$. 

With this approach, prominent examples such as an isolated beating flagella of \emph{Chlamydomonas reinhardtii} were examined~\cite{Battle2016BDB} (see Fig.~\ref{fig:Fig1}). Dynein motors drive relative axial sliding of microtubules inside the axoneme of the flagellum ~\cite{ma2014active,RiedelKruse2007,Wan2014}. To quantify the non-equilibrium dynamics of this system, we decomposed the axoneme shapes measured using time-lapse microscopy into the dynamic normal modes of an elastic filament freely suspended in a liquid. Using this approach, we obtained the amplitudes of the projections coefficients for the first 3 modes. These amplitude time series were used to construct a trajectory in a phase space spanned by the three lowest-order modes, which were analyzed using PFA, as shown in Fig.~\ref{fig:Fig1}C. Here, the vector fields indicate the fluxes for the first three modes. Thus, this method can be used to quantify the non-equilibrium dynamics of the flagellum in a phase space of configurational degrees of freedom.

In addition, we considered  primary cilia of Madin-Darby Canine Kidney (MDCK II) epithelial. Primary cilia are hair-like mechano and chemosensive organelles that grow from the periphery of certain eukaryotic cells~\cite{Singla2006,Battle2015,Barnes1961}. At first glance the dynamics of the deflection angle and curvature of primary cilia appear to exhibit random fluctuations. Using Probability Flux Analysis (PFA), however, it was demonstrated that there are significant circulating probability fluxes in a configurational phase space of angle and curvature, providing evidence for the non-equilibrium nature of primary cilia~\cite{Battle2016BDB}. This approach is now gaining traction in variety of systems, ranging from the post translation Kai circadian clock ~\cite{Paijmans2017} to motility phenotypes~\cite{JacobC.Kimmeletal.2017}. When the mobility of a system is known, a related approach can be used to estimate the heat dissipation~\cite{Lander2012}. However, in a non-equilibrium system, the mobility must be obtained by a perturbative measurement. 

\subsubsection{Bootstrapping} 
In practice, the finite length of experimental or simulated trajectories limits the accuracy with which we can estimate fluxes in phase space. This has an important implication: even when considering a system at thermodynamic equilibrium, a  measurement from finite  data will typically result in apparent non-zero currents. In such a case we can not statistically distinguish the measured apparent current from a zero current. Therefore, it is important to asses if the estimated currents are statistically significant. Moreover, these current fluctuations may also be interesting to study in and of themselves (see Section~\ref{sec:englandBound}). In this section we briefly describe ``bootstrapping", a method that can be used to associate errorbars to the measured currents.

The error bars on the probability flux can be determined by counting statistics of the number of transitions in Eq.(\ref{netrates}). In general, however, there may be correlations between in-and-outward transitions for a given state, which renders it difficult to perform a simple estimate of the errorbar. A possible way around this, which naturally takes correlations into account, is to bootstrap trajectories from the experimentally measured or simulated trajectories. 

To perform this bootstrapping procedure, we first determine all the transitions between discrete states  in the CGPS from the measured trajectories. From this data, we construct a set $A$ of $n$ events, describing specific transitions of the system  between two states, including the transition time. Given $A$, we can generate a new set of  transitions, $A'$, by randomly sampling $n$ \emph{single} events (with replacements) from $A$. This procedure, however, ignores possible correlations. To capture the effects of  correlations on the accuracy of our current estimator, we bootstrap trajectories by randomly sampling a group of $m$ consecutive events  from $A$ to construct a new set of transitions $A'_i(m)$ ~\cite{Shannon}.

For each bootstrapped trajectory we calculate the current field and by averaging over all the realizations, we estimate the covariance matrix. To visualize the errorbars (standard error of the mean) on the estimated currents, we depict an ellipse aligned with the principle components of this covariance matrix. The short and long axes of these error-ellipses are defined by the square roots of the small and large eigenvalues, respectively, of the covariance matrix, Fig.~\ref{fig:Fig11}D. Empirically, we found that the estimated error bars reduce substantially by including pairwise correlations, i.e. in going from $m=1$ to $m=2$, after which the errorbars became largely insensitive to $m$. Such correlations can arise because of the coarse graining of phase space, which can introduce a degree of non-Markovianity.

\begin{figure*}
\centering
  \includegraphics[width=18cm]{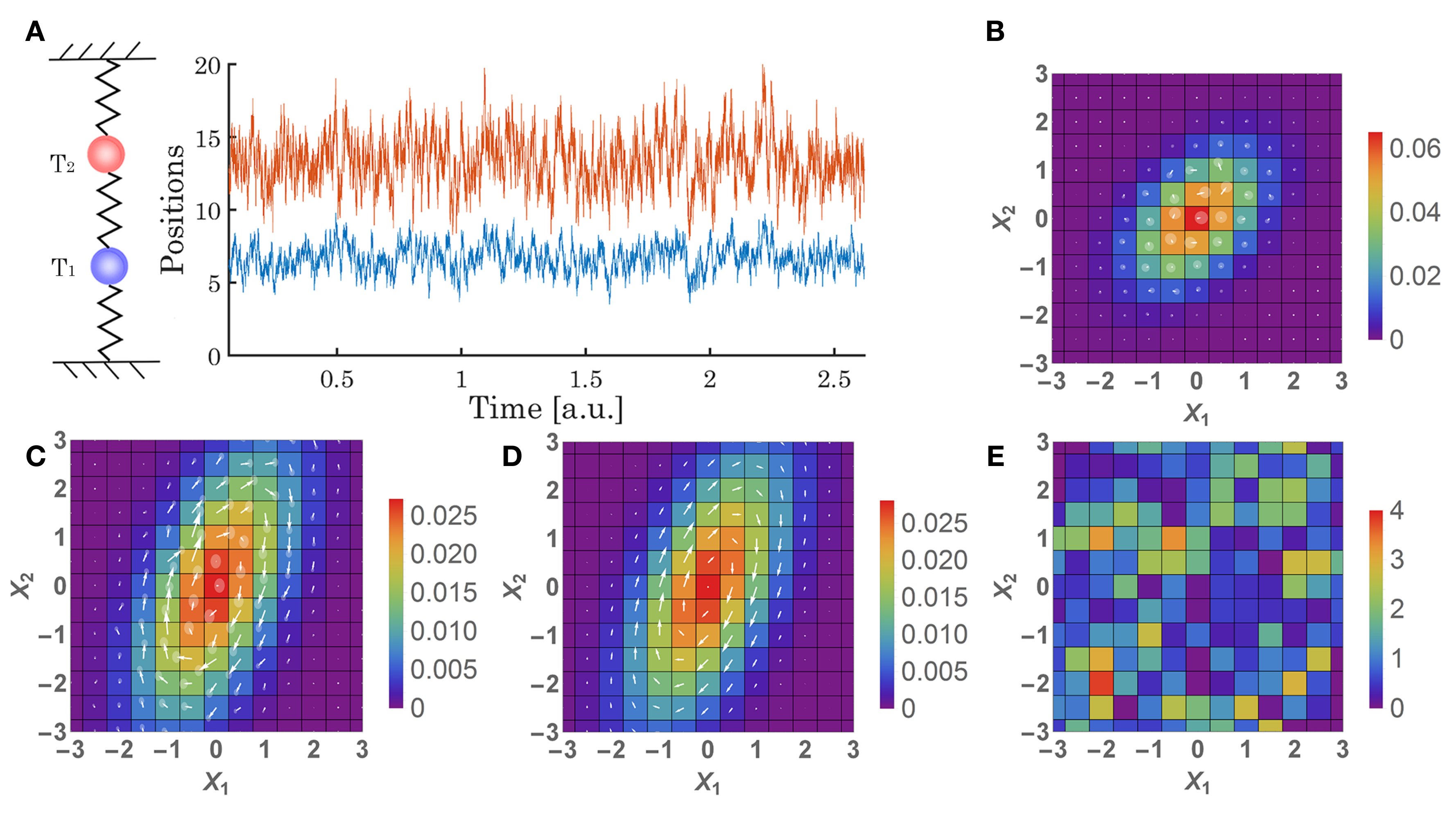}
  \caption{A) Schematic of the two coupled beads system and simulated time series of the beads positions for $T_2=5T_1$. B-C)Probability distribution (color) and flux map (white arrows) obtained by  Brownian dynamics simulations at equilibrium ($T_1=T_2)$ (B) and non-equilibrium $(T_2=5T_1)$ (C). Translucent discs represent a 2$\sigma$ confidence interval for fluxes. D) Analytical result for the probability distribution (color) and flux map (white arrows) obtained for a non-equilibrium case, ($T_2=5T_1$). Adapted from~\cite{Battle2016BDB} E) Compatibility estimated from Eq.(\ref{comp}) between the estimated and theoretical second components of the currents.}
  \label{fig:Fig12}
\end{figure*}

\subsubsection{Toy model: two stochastically driven  coupled beads} \label{sec:2beads}
To provide some basic intuition for stochastic non-equilibrium systems we next discuss a simple model, which can  easily be solved both analytically and numerically. With this model, we  illustrate how probability flux analysis (PFA) can be used on simulated data to obtain current densities in coarse grained phase space. The results are shown to be consistent with analytical calculations within errorbars.

Consider a system consisting of two microscopic over-damped beads in a liquid connected to each  other and to a rigid boundary by springs with elastic constant $k$, as depicted in  Fig.~\ref{fig:Fig12}A. The two beads are assumed to be in contact with two independent heat baths, respectively at temperatures $T_1$ and $T_2$. 
The stochastic dynamics of this system is described by the overdamped equation of motion
\begin{equation}
\frac{d \vec{x}}{d t} = \mathbf{A} \vec{x}+\mathbf{F}\vec{\xi},
\label{Langevin}
\end{equation}
where $\vec{x}=(x_1,x_2)^T$ represents the beads positions. The deterministic dynamics is captured by the matrix
\begin{equation}
\mathbf{A} =\frac{k}{\gamma} \begin{pmatrix} -2 & 1 \\
1 & -2\\
\end{pmatrix}.
\end{equation}
The drag coefficient $\gamma$, characterizing the viscous interactions between the beads and the liquid, is assumed to be identical for the two beads.
The stochastic contribution, $\xi _i $, in the equation of motion is defined by
\begin{equation}
\langle \vec{\xi} \rangle = 0, \ \ \ \ \  \langle \vec{\xi}(t) \otimes \vec{\xi}(t') \rangle = I\delta(t-t'),
\end{equation}
and the amplitude of the noise is captured by the matrix
\begin{equation}
\mathbf{F} =\sqrt{\frac{2 k_{\rm B}}{\gamma}} \begin{pmatrix} \sqrt{ T_1} &  0 \\ 0  & \sqrt{ T_2} \end{pmatrix}.
\end{equation}

We can generate simulated trajectories for this system by numerically integrating Eq.~\eqref{Langevin}. We will consider two exemplificative cases: i) thermal equilibrium with $T_1=T_2$, and ii) non-equilibrium with $T_2=5T_1$. An example of the two simulated trajectories for this last case is shown in Fig.~\ref{fig:Fig12}A, where we note that  the dynamics of individual trajectories appears to be, at first glance, indistinguishable from equilibrium dynamics. Interestingly however, the nonequilibrium nature of this system is revealed by applying PFA to these data, which gives coherently circulating probability fluxes in the phase space (Fig.~\ref{fig:Fig12}C).
By contrast, in the case of thermal equilibrium ($T_1=T_2$) we find, as expected, that the flux vanishes, as shown in Fig.~\ref{fig:Fig12}B. \\

To compare these results of the estimated fluxes from simulations with analytical calculations, we next consider the time evolution for the probability density function $p(\vec{x},t)$ of the system, which is described by the Fokker Planck equation:
\begin{equation}
\frac{\partial p(\vec{x},t)}{\partial t} = - \nabla \cdot [\mathbf{A}\vec{x}p(\vec{x},t)]+\nabla \cdot \mathbf{D}\nabla p(\vec{x},t),
\label{Fokker}
\end{equation}
where $ \mathbf{D}=\frac{1}{2} \mathbf{F} \mathbf{F}^T$ is the diffusion matrix. The steady-state solution of this equation is a Gaussian distribution, with a covariance matrix, $\mathbf{C},$ which is found by  solving the Lyapunov equation
\begin{align}
\mathbf{AC+C}\mathbf{A}^T= -2\mathbf{D}. \label{eq:matrixFDT}
\end{align}
The steady state probability flux density is given by $\vec{J}~=~\mathbf{\Omega} \vec{x} p(\vec{x})$, where 
\begin{equation}
\mathbf{\Omega} =\frac{k(T_1-T_2)}{\gamma c} \begin{pmatrix} 2(T_1+T_2)&-(7T_1+T_2)\\
 (T_1+7T_2)&-2(T_1+T_2)
\end{pmatrix}
\end{equation}
with $c= (T_1^2+14T_1T_2+T_2^2)$. As expected, the flux vanishes at thermal equilibrium when $T_1=T_2$. 
In the near equilibrium regime, we can consider $T_1=T$ and $T_2=T+\epsilon$ with $\epsilon$ small. Within this limit, the current field can be written as
\begin{equation}
\label{neareqcurrent}
\vec{J} \propto \frac{\epsilon}{T^2}e^{-\frac{k(x_1^2 - x_1x_2 +x_2^2)}{k_{\rm B}T}}\begin{pmatrix} x_1-2x_2\\
-2x_1+x_2
\end{pmatrix}
+ \mathcal{O}(\epsilon)^2
\end{equation}
where we note how the amplitude and the direction of the flux are set by the ratio $\frac{\epsilon}{T^2}$, which vanishes at equilibrium. To gain some intuition on how the current decays with the distance in phase space, we can for example constrain Eq.\eqref{neareqcurrent}  along the vertical direction ($x_1=0$), 
\begin{equation}
\label{currentvert}
\vec{J} \propto \frac{\epsilon}{T^2}e^{-\frac{kx_2^2}{k_{\rm B}T}}\begin{pmatrix} -2x_2\\
x_2
\end{pmatrix}
+ \mathcal{O}(\epsilon)^2
\end{equation}
 From Eq.\eqref{currentvert} we can notice two opposite contributions to the amplitude, the linear dependence, dominant for small $x_2$ and the exponential dependence, dominant for larger $x_2$. This indicates an optimal distance from the origin at which the flux is maximum.

To compare the analytical expectation for the flux $\vec{J}$ with the results obtained using PFA on simulated trajectories, we calculate the compatibility $c_{ij{,l}}$ between the estimated $\hat{J}$ and the theoretical $J^{th}$ values of the flux field in  cell $i,j$, and in direction $x_l$:
\begin{equation}
c_{ij,l}=\frac{|\hat{J}_{ij,x_l}-J^{th}_{ij,x_l}|}{\sigma},
\label{comp}
\end{equation}
where $\sigma$ is the error obtained from the bootstrapping analysis in PFA. The results for the second component of $\vec{J}$ yield an average compatibility of $\langle c_{ij}\rangle\simeq 1.02$ (Fig.~\ref{fig:Fig12}E), indicating a good quantitative agreement  between our estimation and the exact currents. A similar result is obtained  in the equilibrium case $(T_1=T_2)$, for which the average compatibility is  $\langle c_{ij}\rangle\simeq 0.95$. This concludes our analysis of probability fluxes in phase space for stochastic trajectories. These results illustrate how PFA can be used to infer accurate currents in coarse grained phase space from stochastic trajectories.

\subsection{Probe filaments to study broken detailed balance across scales in motor-activated gels}
\label{sec:filamentBDB}
While mesoscopic objects, such as cilia or flagella, can often be directly imaged, detecting non-equilibrium dynamics inside live cells on the microscale and below is more challenging. The cellular cytoskeleton, discussed in  Sec.\ref{FDTsec}, is a prominent example of active matter, which can best be described as a viscoelastic meshwork of biopolymers, activated by myosin motors~\cite{mackintosh2010active, prost2015active}. Random contractions of these myosin proteins fuelled by ATP hydrolysis can drive vigorous steady-state fluctuations in this polymer network. Such fluctuations can be quantified experimentally by embedding fluorescent probe particles. This technique has revealed multiple scaling regimes of the time dependence of the mean-squared displacement~\cite{brangwynne2009intracellular}, which were attributed to a combination of the viscoelastic behavior of the network and the temporal dynamics of motor activity. In particular, endogenous embedded filaments such as microtubules, or added filaments such as single-walled carbon nanotubes have proved to be convenient probes~\cite{brangwynne2008nonequilibrium, fakhri2014high}. 

These experiments and others~\cite{toyota2011non, Guo2014, Weber2015} have sparked a host of theoretical efforts~\cite{Everaers1999, Liverpool2003, Levine2004, MacKintosh2008, Kikuchi24112009, Loi2011, Ghosh20141065, ben2015modeling, Eisenstecken2017} to elucidate the stochastic dynamics of probe particles and filaments in an active motorized gel. More recently, it has been suggested that probe filaments can be also used as a multi-variable probe to discriminate active from thermal fluctuations using detailed balance~\cite{Gladrow2016, Gladrow2017}, and could be used to detect correlations in the profile of active forces along its backbone~\cite{Brangwynne2007}.
\begin{figure}
\centering
\includegraphics[width=9cm]{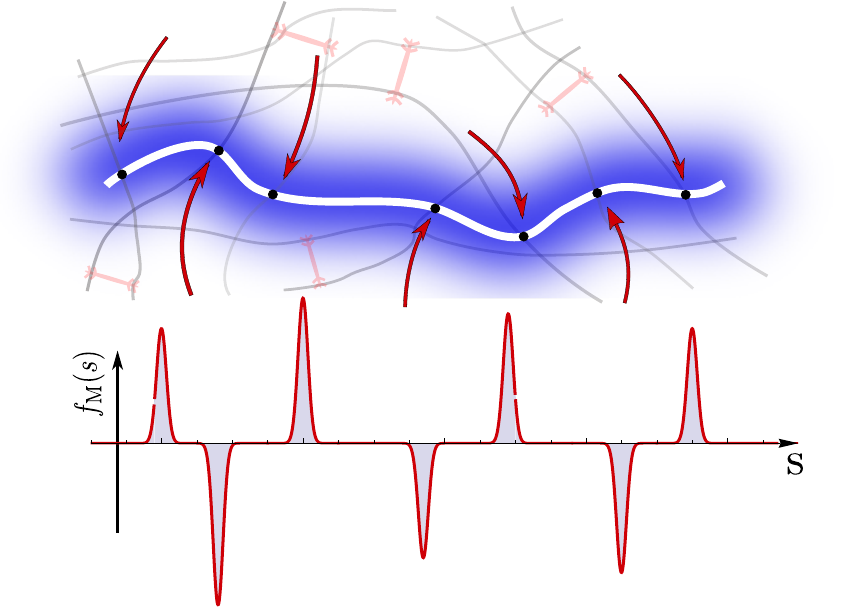}
\caption{Fluctuations of a probe filament (blue) embedded in a viscoelastic actin (grey) network, driven out of equilibrium by random contractions of myosin (red, red arrows). Adapted from ~\cite{Gladrow2016}. }
\label{fig:Fig13}
\end{figure}

In the following, we lay out a framework to describe fluctuations of a semiflexible probe filament~\cite{kratkyPorod, aragon1985dynami, Goldstein1995, Hallatschek2007}, which is embedded in a motor-activated network~\cite{MacKintosh2008, levine2009mechanics,Osmanovic2017}. We assume the probe filament to be weakly-bending, such that we can focus on the transverse coordinate $r_\perp(s,t)$, where the arclength $0< s < L$ parametrizes the backbone, as shown in Fig.~\ref{fig:Fig13}.
The overdamped dynamics of such a probe filament is governed by a balance of (i) viscous and elastic forces of the surrounding viscoelastic medium, (ii) bending forces, (iii) thermal agitation, and (iv) motor-induced fluctuations, which read in this order as
\begin{align}
  \label{eq:equation of motion}
  \int \limits_{-\infty}^t \,\mathrm{d}t'\,  \alpha(t-t')  r_\perp(s,t')  +\kappa \frac{\partial^4 r_\perp}{\partial t^4}(s,t)& = \xi(s,t) + f_\text{M}(s,t).
\end{align}
Terms on the left describe relaxation, while terms on the right contain stochastic contributions. For a predominantly elastic network, we can use the generalized Stokes equation, $\hat{\alpha}(\omega)= k_0  \hat{G}(\omega)$ to approximate the viscoelastic kernel on the left hand side as $\hat{G} = G_0+i\eta \omega$, i.e. as a Kelvin-Voigt-type viscoelastic solid. The factor $k_0$ has a geometrical origin, and is given by $k_0\approx 4 \pi/\ln(L/d) $ for an infinitesimal rod segment of diameter $d$~\cite{howard2002mechanics}. In a crosslinked actin network this approximation is reasonable for low frequencies typically below roughly $100\, \text{Hz}$, beyond which the network modulus exhibits a characteristic stiffening with frequency~\cite{Koenderink2006, Mizuno2009, Guo2014}. When the network is described as such a simple viscoelastic solid, the thermal noise is given by a Gaussian white-noise process $\xi(s,t)$, to which we add independent actively induced forces $f_\text{M}(s,t)$, specified in detail further below.

Bending forces can be conveniently studied from the perspective of bending modes of the probe filament. Following the approach in~\cite{Gladrow2016, Gladrow2017}, a description in terms of bending modes can be obtained from a decomposition of the backbone coordinates into orthogonal dynamic modes $r_\perp(s,t)= L \sum_q a_q(t) y_q(s)$~\cite{aragon1985dynami, Gittes1993, Brangwynne2007}. In this coordinate system, the multiscale character of probe filaments becomes apparent: each bending mode amplitude $a_q(t)$ is sensitive to a lengthscale corresponding to its wavelength. The precise form of bending modes, however, depends on the boundary conditions of the filament. 
The simplest case is a filament with zero transverse deflections at its end, where classical sine-modes $y_{q_m}(s)=\sqrt{2/L} \sin(q_m s)$ form an orthonormal set. Importantly, these modes are independent in equilibrium, due to their orthogonality. For fixed-end modes, mode number $m\in\{1,2, 3,...\}$ and wave-vector $q$ are related via  $q(m)= m \pi/L$. The relaxational timescale of each mode is set by a balance between both elastic and viscous forces of the network and the bending rigidity of the filament. For inextensible filaments in purely viscous environments, this results in a strongly length-dependent decay
\begin{align}
\tau_q &= \frac{\eta}{\kappa q^{4}/k_0+G_0}, \label{eq:relaxationTimeDecay}
\end{align}
In the linear-response regime, we obtain the mode-response function to transverse deflections, $\chi_q(t)$, in fourier space  $\hat{\chi}_q(\omega) = (\hat{\alpha}(\omega) + \kappa q^4)^{-1}$. This response function is related to mode variances in equilibrium via the mode fluctuation-dissipation theorem
\begin{align}
\label{eq:modeFDT}
 \langle \left | \hat{a}_q (\omega) \right| ^2 \rangle &= \frac{2k_BT}{L^2 \omega} \hat{\chi}''_q(\omega).
\end{align}
Bending modes are thus ideally suited to not only detect motor activity, but also to measure their spatial and temporal characteristics. Perhaps for these reasons, bending mode fluctuations have been the subject of a number of studies in biological non-equilibrium systems.	

In a study by Brangwynne \emph{et al}.~\cite{brangwynne2008nonequilibrium}, fluorescently labelled microtubuli were used to probe the active fluctuations in actin-myosin gels. The persistence lengths of microtubuli is on the order of millimeters \cite{Gittes1993}, such that these filaments can be treated effectively as rigid on microscopic lengthscales under thermal conditions. By contrast, in actin-myosin gels, microtubuli exhibit significant fluctuations, caused by contractions of myosin, which deform the network in which the microtubules are embedded. A quantitative analysis of thermal bending mode fluctuations reveals a $q^{-4}$-decay in actin networks (without myosin). By contrast, adding myosin not only increases the amplitudes of fluctuations, but also results in a breakdown of the standard mode decay (see Eq.~(\ref{eq:relaxationTimeDecay}). The spatial extent of individual indentations in motor-agitated microtubuli can be used to extract forces induced by myosing. These force range between $0-30$ pN, in accord with more recent studies in live cells~\cite{Guo2014, fakhri2014high}. Importantly, the results also suggest a very narrow profile of the force exerted on the microtubules. Furthermore, in the cell cortex, microtubules often appear considerably more curved, despite their rigidity. Indeed, this curved microtubule structure is not due to temporal bending fluctuations of the microtubule, but rather results from geometrical constraints that randomly deflect the microtubule tip during polymerization \cite{brangwynne2007force}.

Motivated by these experimental observations, we can model the motor-induced force, exerted on the probe at the points where it is coupled elastically to the network, as a superposition of all active forces in the environment:
\begin{align}
  \label{eq:motorForce}
  f_\text{M}(s,t) = \sum\limits_{n}\, f_n(s,t),
\end{align}
where each $f_n(s,t)$ denotes the force contribution from active motors, which affect the filament at the $n_{\rm th}$ entanglement point. Active forces have a characteristic spatial decay, since myosin motors exert forces in dipoles rather than in single directions~\cite{Yuval2013}. The model in Eq.~(\ref{eq:motorForce}) does not account for such details; its main purpose is to provide a non-uniform force background $f(s)$ along the backbone $s$.

Measurements of myosin dynamics have revealed a Lorentzian power spectrum~\cite{MacKintosh2008, Mizuno2009}. A simple on-off telegraph process $\mathcal{T}(t)$ is in accord with these observations and appears to be adequate to model the stochastic force dynamics of individual motors.
Taking furthermore into account the narrow profile of motor forces inferred from experiments~\cite{brangwynne2008nonequilibrium}, we arrive at a model for motor-induced forces, which reads $f_n(s,t)= f_n \delta(s-s_n) \mathcal{T}_n(t)$. Here $\mathcal{T}_n(t)$ is a telegraph process with exponential decorrelation $\langle \mathcal{T}(t)\mathcal{T}(t') \rangle = C_2 \rm{exp}(-|t-t'|/\tau _{M})$. 

\begin{figure}
 \centering
 \includegraphics[width=8cm]{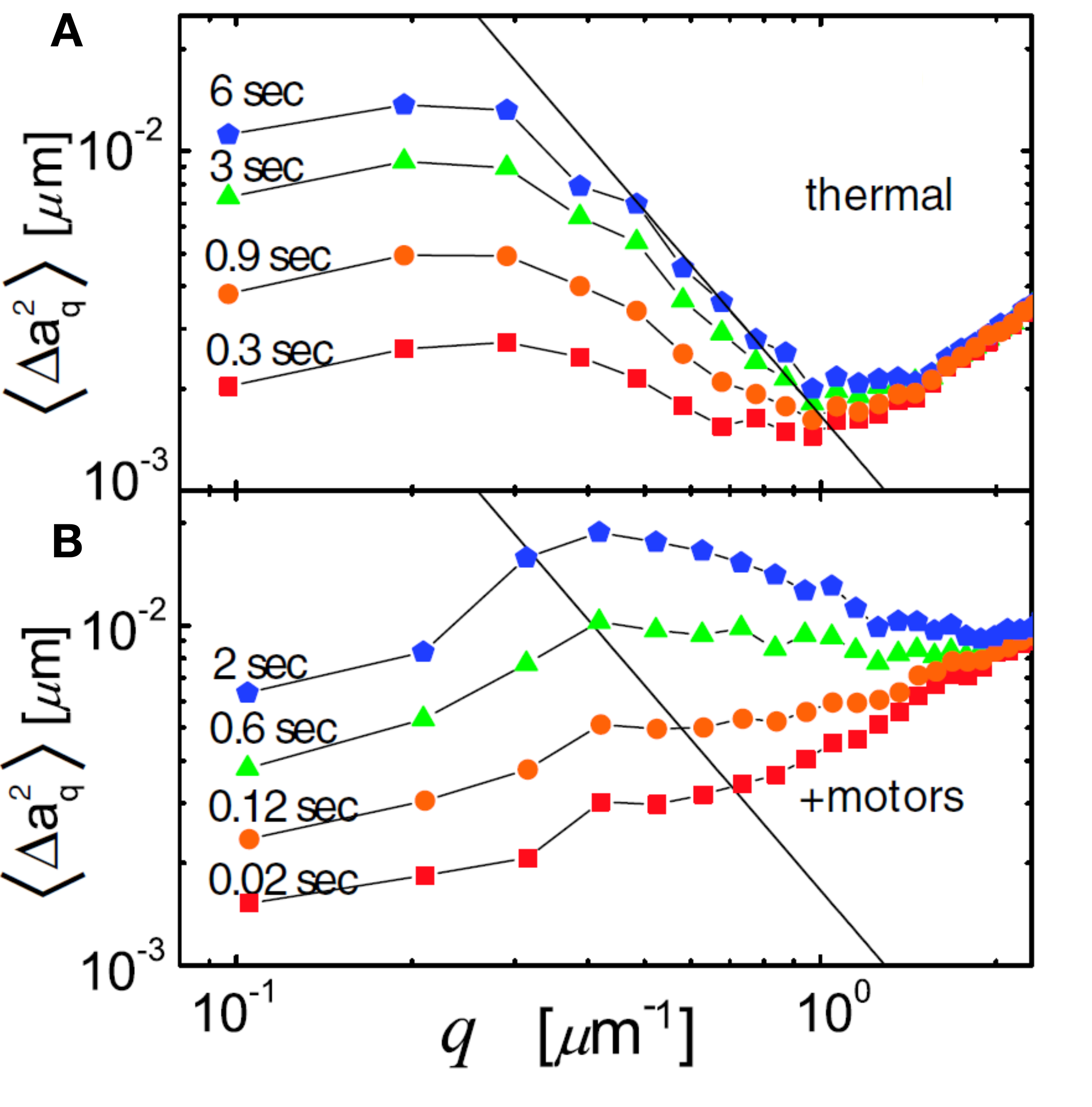}
 \caption{Mode amplitude variations in fluctuating microtubules embedded in actin-myosin gels over mode number $q$. A) In ATP-depleted gels (purely thermal noise), mode variances follow a power law decay. B) Active fluctuations result in enhanced mode variances in accord with the theory In Fig.~\ref{fig:Fig15}. At high $q$-values measurement noise leads to an increase in mode variances. Adapted from~\cite{brangwynne2008nonequilibrium}.}
 \label{fig:Fig14}
\end{figure}

Using this simple description for the stochastic behavior of motor-generated forces together with Eq.~(\ref{eq:modeFDT}), we compute the mode correlator, which decomposes into active and thermal contributions: $\langle a_q(t)a_w(t')\rangle= \langle a_q(t)a_w(t')\rangle^{\text{Th}}+\langle a_q(t)a_w(t')\rangle^{\text{M}}$, given by 
\begin{align}
  \langle a_q(t)a_w(t')\rangle^{\text{Th}} &=  \frac{k_B T \tau_q}{L^2\gamma}\delta_{q,w}e^{-\frac{\left | t-t'\right | }{\tau_q}}  \label{eq:thermalModeCorrelation}\\
  \langle a_q(t)a_w(t')\rangle^{\text{M}} &= \frac{1}{L^2\gamma^2}\mathbf{F}_{q,w}C_2\mathcal{C}_{q,w}\left( t-t'\right). \label{eq:motorModeCorrelation}
\end{align}
$\mathbf{F}_{q,w}$ specifies the geometry of motor-induced forces in mode space and is defined by $\mathbf{F}_{q,w}=\sum_n f_n^2 y_q\left (s_n \right)y_w\left(s_n\right)$, where the sum runs over the filament-network contacts. The function $\mathcal{C}_{q,w}(\Delta t)$ denotes the temporal decorrelation of active mode fluctuations.
In contrast to thermal equilibrium, active fluctuations decay as a double exponential 
\begin{align}
    \mathcal{C}_{q,w}( t-t')&= \tau_q\tau_w\left(\frac{e^{-\frac{\lvert  t-t' \rvert }{\tau_\text{M}}}}{\left(1-\frac{\tau_q}{\tau_\text{M}}\right)\left(1+\frac{\tau_w}{\tau_\text{M}}\right)} \right.\nonumber \\
&\left. -2 \frac{\tau_q}{\tau_\text{M}}\frac{e^{-\frac{\lvert t'-t\rvert }{\tau_q}}}{\left(1-\left(\frac{\tau_q}{\tau_{\rm M}}\right)^2\right)\left(1+\frac{\tau_w}{\tau_q}\right)}\right),\label{eq:modeMotorCrossCorrelator}
\end{align}
which indicates a competition between two decorrelating processes: mode relaxation and the internal decorrelation of the motor state. The correlator is not symmetric in the indices $q$ and $w$ as can be seen from Eq.~(\ref{eq:modeMotorCrossCorrelator}), which results in a breaking of Onsager's time-reversal symmetry~\cite{Gladrow2017}. 

\begin{figure}
 \centering
 \includegraphics[width=8cm]{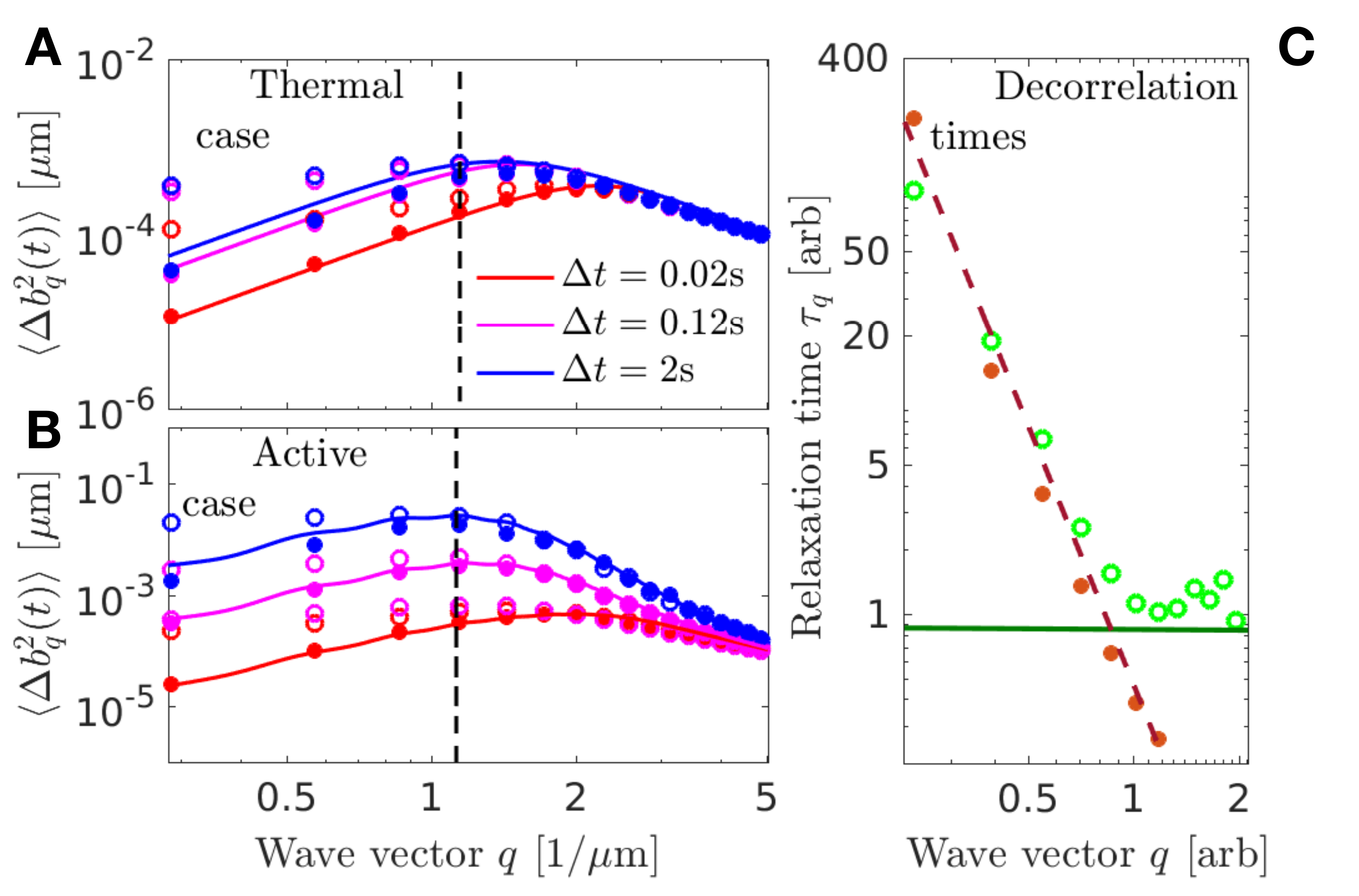}
 \caption{Mode fluctuations under (A) purely thermal agitated and (B) under additional influence of motor-induced forces. C) Convergence of mode decorrelation times onto the motor timescale. Different colors correspond to the different times $\Delta t$ shown in the legend in panel (A). Adapted from~\cite{Gladrow2016}.}
  \label{fig:Fig15}
 \end{figure}

This double exponential in Eq.~\eqref{eq:modeMotorCrossCorrelator} is the footprint of colour of the noise process, which we use to describe motor-induced forces. The $q^{-4}$-decay of mode amplitudes relaxation times, $\tau_q $, levels off around $\tau_\text{M}$ as shown in Fig.~\ref{fig:Fig15}C. This saturation occurs because modes cannot decorrelate faster than the force that is driving them. Under coloured noise, the relaxation times cannot be directly inferred from decorrelation. This is indeed confirmed in Brownian dynamics simulations of filaments subject to active fluctuations~\cite{Gladrow2016}. 
To further illustrate these results, simulations of mode variances over mode vector in passive (Fig.\ref{fig:Fig15}A) and  active (Fig.\ref{fig:Fig15}B) networks are shown together with theoretical predictions from Eqs.~(\ref{eq:thermalModeCorrelation},\ref{eq:motorModeCorrelation}). For comparison, experimentally obtained mode variances~\cite{brangwynne2008nonequilibrium} are plotted over $q$ in Fig.~\ref{fig:Fig14}. As one would expect, in both cases mode variances are elevated in active environments. 

\begin{figure}
 \centering
 \includegraphics[width=1\columnwidth]{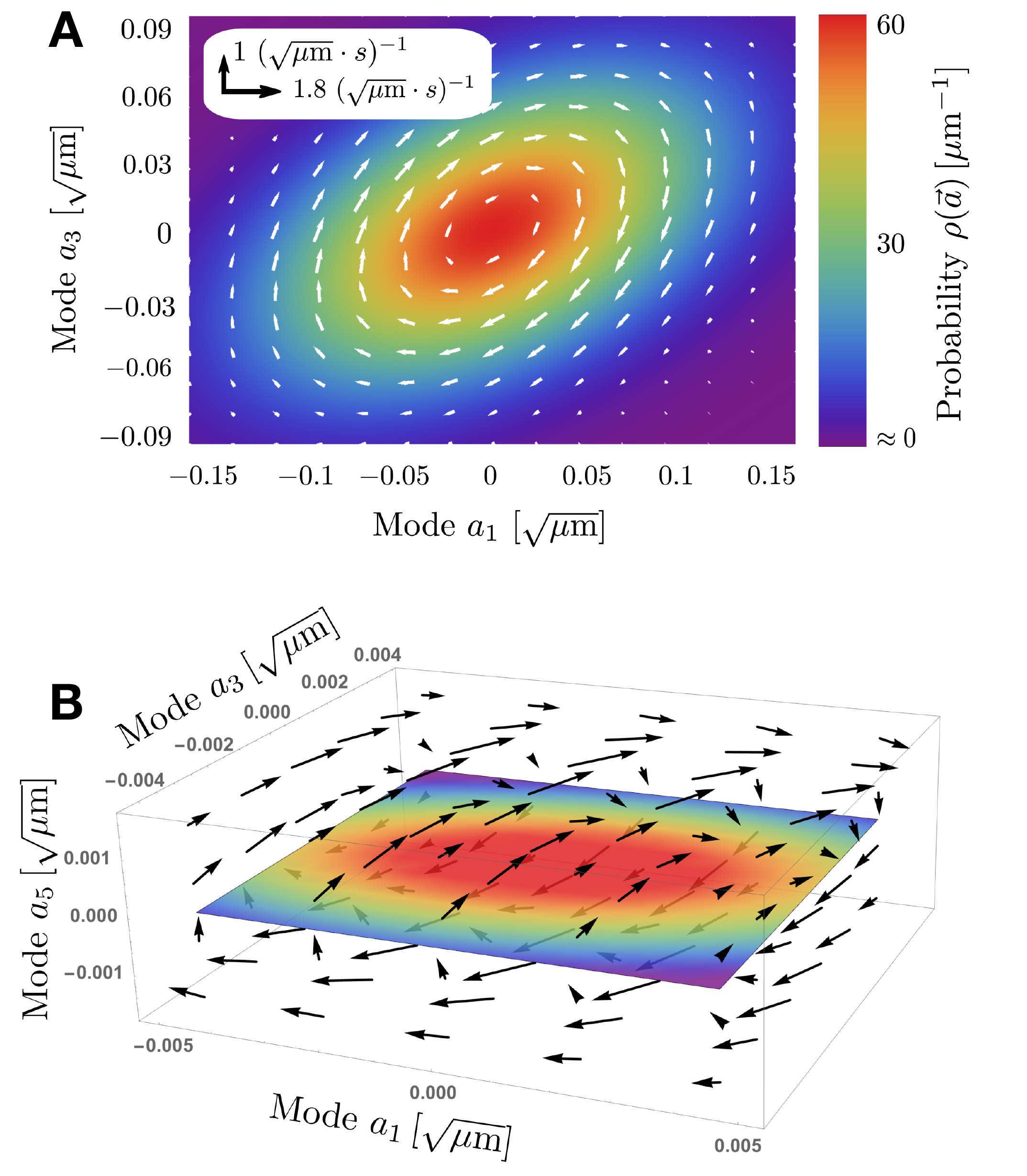}
 \caption{Steady-state probability currents in mode space. A) Projection of the multidimensional current on the mode amplitude pair $a_1$ and $a_3$. B) The same current in three dimensions $a_1$, $a_3$, and $a_5$. Due to the geometry of probe-network interactions in this example, only modes of similar number parity (e.g. odd-odd) couple. Adapted from~\cite{Gladrow2016}.}
 \label{fig:Fig16}
\end{figure}

In the long time regime $t\gg \tau_\text{M}$, motor forces effectively appear as sources of white-noise. In this `{\em white-noise limit}', the motor correlator converges to a $\delta$-function, $\langle\Delta \mathcal{T}(t)\Delta \mathcal{T}(t') \rangle \to C_2/2 \tau_\text{M}\delta(t-t')$ with a factor $\tau_\text{M}$, which remains only as a scale of the variance of the motor process. The mode correlation function in the white-noise limit can be derived by a series expansion of $\mathcal{C}_{q,w}(t)$, which yields
\begin{align}
 \langle a_q(t)a_w(t') \rangle^\text{M} &= \frac{C_2\tau_M}{L^2 \gamma^2}\mathbf{F}_{q,w} \frac{\tau_q\tau_w}{\tau_q+\tau_W} \delta(t-t'). \label{eq:whiteNoiseCorrelator}
\end{align}
We can now contract the thermal and motor-white noise processes into a single process $\psi(t)$, with a correlator $ \langle \psi_q(t)\psi_w(t')\rangle =\left(4k_BT\gamma\delta_{q,w} +  C_2\tau_\text{M} \mathbf{F}_{q,w} \right) \frac{\delta \left( t-t'\right)}{2L^2}$. 

It is useful at this stage to compare this scenario with that described in Sec.~\ref{sec:coordinateInvariance}: Here, mode variables are independent, but are subject to noise with a cross-mode correlations, such that different modes are simultanously excited by a motor event. By constrast, thermal noise is uniform in amplitude and uncorrelated throughout the system, giving rise to independent stochastic forces in mode-space. As we discussed in Sec.~\ref{sec:coordinateInvariance}, a correlation in the external noise in one coordinate system, may appear as a ``temperature'' gradient in different coordinates. It is this mechanism, which gives rise to a probability flux in mode space, which breaks detailed balance in the fluctuations of the probe filament. In other words, a motor-induced force background $\langle f_\text{M}(s,t)\rangle_{\text{temporal}}=\lim_{t_1 \to \infty} \int_{t_0}^{t_1} \mathrm{d}t f_\text{M}(s,t) $ (see lower panel in Fig.~\ref{fig:Fig13}), which varies along the filament, will lead to a breaking of detailed balance in a hyperplane spanned by the affected modes. 

The magnitude and structure of this probability current, is given as a solution of the multivariate Fokker-Planck equation $\partial_t \rho(\vec{a},t)= - \vec{\nabla} \cdot\vec{j}(\vec{a},t)$ in mode space. The probability current, $\vec{j}(\vec{a})$, can be written in steady-state as 
\begin{align}
 \vec{j}(\vec{a}) = (\mathbf{K}+\mathbf{D}\mathbf{C}^{-1})\vec{a}\rho(\vec{a}) \equiv \mathbf{\Omega} \vec{a}\rho(\vec{a}), \label{eq:current}
\end{align}
where $\mathbf{K}_{q,w}=-1/\tau_q \delta_{q,w}$ is the deterministic matrix which defines the linear force field, and $\mathbf{D}$ and $\mathbf{C}$ represent respectively the  diffusion and covariance matrices. Within this linear description,  $\mathbf{\Omega}$ is the matrix that captures the structure of the current~\cite{Weiss2003, WeissFluctuationProperties}. A rotational probability current in the Fokker-Planck picture is associated with a net rotation of variables in the Langevin description: On average, mode amplitudes cycle around the origin, when detailed balance is broken in steady-state, as illustrated in Fig.~\ref{fig:Fig16}. 

The circular character of the current is reflected mathematically by the skew-symmetry of $\mathbf{\Omega}^T=-\mathbf{\Omega}$ in the coordinate system where $\tilde{\mathbf{C}} = \mathbf{1}$ (``{\em correlation-identity coordinates}''). This can be seen from Eq.~(\ref{eq:matrixFDT}), which dictates that $\tilde{\mathbf{K}}+\tilde{\mathbf{K}}^T=-2\tilde{\mathbf{D}}$ in this system, such that $\tilde{\mathbf{\Omega}} = \frac{1}{2} \left( \tilde{\mathbf{K}}-\tilde{\mathbf{K}}^T\right)$. The eigenvalues of any skew-symmetric matrix $\mathbf{M}$ are either zero or purely imaginary, with the latter leading to rotational currents in a hypothetical dynamical system described by $\dot{\vec{x}} = \mathbf{M}\vec{x}$. Moreover, since $\text{tr}\left(\mathbf{\Omega} \right)=0$, in two dimensions, this implies that the eigenvalues can be rewritten as $\lambda_{1,2} = \pm i \omega$. In a steady-state, the probability current $\vec{j}$ (in any dimension) has to be orthogonal to the gradient of the density $\rho(\vec{a}, t)$, since $\vec{\nabla}\cdot \vec{j}(\vec{a}) = \rho(\vec{a})\vec{\nabla}\cdot (\mathbf{\Omega}\vec{a})+ \left(\mathbf{\Omega} \vec{a} \right)\cdot \vec{\nabla}\vec{\rho}(\vec{a})=\partial_t \rho =0$. The first term must be zero, since it is proportional to $\text{tr}\left(\mathbf{\Omega}\right)\vec{a}$, so that the second term has to vanish as well. This, however, implies that $\vec{\nabla}\rho \perp \mathbf{\Omega}\vec{a}$: the gradient of the density must be perpendicular to the flow field. In a linear system, the probability density is always Gaussian $\rho \propto e^{-\frac{1}{2} \vec{a}^T \mathbf{C}^{-1} \vec{a}}$, such that the flow field must have an ellipsoidal structure~\cite{Weiss2003}. 
In correlation-identity coordinates, where the density has a radial symmetry, the profile of $\tilde{\vec{j}}$ would thus be purely azimuthal, and its magnitude would represent an angular velocity. An average over the angular movements $\langle \dot{\varphi}\rangle$ of the mode vector $\vec{a}(t)$ in the plane will yield the cycling frequency. The imaginary part of the positive eigenvalue of $\mathbf{\Omega}$ must therefore represent the average cycling frequency of the mode vector in the plane. 

In a reduced two-dimensional system consisting only of $a_q(t)$ and $a_w(t)$, the cycling frequency can be calculated analytically and reads
\begin{equation}
\omega_{q,w}^\text{2D} =  \frac{  \left (\tau_q -\tau_w\right)\mathbf{F}_{q,w} }{\sqrt{\tau_q\tau_w\left( \left(\tau_q+\tau_w\right)^2 \beta  -4\tau_q\tau_w \mathbf{F}_{q,w}^2\right) }}.\label{eq:netCyclingFrequency}
\end{equation}
where $\beta= ((\frac{2k_B T\gamma}{C_2\tau_M})^2+\frac{2k_B T \gamma}{C_2 \tau_\text{M}}\mathbf{F}_{q,q}+\mathbf{F}_{w,w} )+\mathbf{F}_{w,w}\mathbf{F}_{q,q}  )$.

Interestingly, Eq.~(\ref{eq:netCyclingFrequency}) shows that in the case of equal relaxation times $\tau_q=\tau_w$, the cycling frequency would be zero and thus, detailed balance would be restored, regardless of differences between the modes in motor-induced fluctuations. This hints at an important role of relaxation times in determining the shape of the current in multidimensional systems. Furthermore, the denominator of Eq.~(\ref{eq:netCyclingFrequency}) shows how an increase in overall temperature $T$ could mask broken detailed balance by reducing the cycling frequency. 

In summary, filaments as multi-scale and multi-variable probes offer a novel perspective on non-equilibrium phenomena in active matter and could be used in the future as ``non-equilibrium antennae". As we illustrated in this section, a heterogeneous force background $f_\text{M}(s,t)$ created by motor-induced fluctuations leads to a breaking of detailed balance in mode space of embedded filaments. The intricate structure of the probability current in steady-state may contain a wealth of information about the geometric and, perhaps, temporal structure of impinging active forces. 

The theory laid out in this section can be generalized to other objects, such as membranes~\cite{Strey1995, granek1997, Betz2009, Turlier2016}. In principle, the membrane mode decomposition described in Sec.~\ref{FDTsec} could be used to detect a breaking of detailed balance, since active processes in the cortex of red blood cells might result in noise input that correlates over different membrane modes. 

\section{Outlook}
The examples discussed in this review illustrate how  experimental measurements of nonequilibrium activity and irreversibility can provide a deeper conceptual understanding of active biological assemblies and non-equilibrium processes in cells. In many cases, non-equilibrium fluctuations have successfully been identified and quantified using the combination of active and passive microrheology techniques to study the violation of the fluctuation-dissipation theorem,~\cite{Mizuno370,Turlier2016}. Such studies can for instance reveal the force spectrum inside cells, \cite{Guo2014}. However, these approaches require invasive micromechanical manipulation. Furthermore, a complete generalization of the fluctuation-dissipation theorem for non-equilibrium system is still lacking, such that the response of a non-equilibrium system can not be inferred from it's spontaneous fluctuations. However, this does not mean that the fluctuations of a non-equilibrium steady state do not contain valuable information about the nature of the system. Indeed, non-invasive approaches  to measure broken detailed balance from stochastic dynamics have now been establishes to reveal phase space currents in mesoscopic degrees of freedom of biological systems, \cite{Battle2016BDB}. It remains an open question what information can be inferred about the underlying system from such phase space currents, \cite{Gladrow2016,Gladrow2017}. However, recently derived theoretical relations for energy dissipation and entropy production to characterize non-equilibrium activity are  finding traction in various biological systems such as molecular motors and chemical control systems \cite{Seifert_sensing,Lan2012,Mehta2012,Barato2015}. 

Taken together, the research discussed in this review illustrate that the gap between fundamental approaches in stochastic thermodynamics and its application to real biological system is slowly closing. Indeed, studies of biological active matter are not only yielding insights in non-equilibrium physics, they have also suggested conceptually novel mechanisms in cell biology.
For instance, the collective effect of forces exerted by molecular motors has been implicated in intracellular transport and positioning of the nucleus~\cite{caspi2000enhanced, brangwynne2009intracellular,fakhri2014high,Guo2014, Almonacid2015}. 
This novel mode of transport, known as {\em active diffusion}, is thought to complement thermal diffusion and directed, motor-driven modes of transport in cells.  Another intriguing example is the role of DNA-binding ATPases, which have been suggested to be capable of generating forces on the chromosomes through a DNA-relay mechanism~\cite{Lim2014} or loop extrusion~\cite{Wang2013,Wilhelm2015}. ATP- or GTPases  can also interact with membranes or DNA to play a role in pattern forming systems~\cite{Huang2003,Frey2017,Halatek2012}, for instance in the Min system in \emph{E. coli} and CDC42 in yeast. In these systems, certain proteins can switch irreversibly between different conformational states, affecting their affinity to be in the cytosol or the membrane. This, together with nonlinear interactions between these different proteins, can result in non-equilibrium dynamic pattern formation. 

Another important example in this respect is how cells break symmetry to form a polarity axes. Intracellular myosin activity has been implicated in establishing a sense of direction ('polarity') in cells. In order to divide, cells must 'decide' on the axis of the mytotic spindle, which is a crucial part of the cell division apparatus~\cite{Grill2001, Grill2003, Pecreaux2006, Ou2010}. Cortical flows resulting from asymmetries in myosin activity have been shown to effectively polarize \emph{C.elegans} cells and break the initial cellular symmetry~\cite{Mayer2010}.

Non-equilibrium phenomena also emerge at the multicellular scale: Groups of motile cells  exhibit collective active dynamics, such as flocking, swarming, non-equilibrium phase transitions or the coordinated movements of cells during embryonic developments ~\cite{Thutupalli2015,Peruani2012}.
More broadly, non-equilibrium physics is  emerging as a guiding framework to understand phenomena related to self-replication and adaptation~\cite{England2013,England2015,Perunov2016}, the origin of life (see for example~\cite{Frauenfelder1999,Agerschou2017}), as well as synthetic life-like systems (See \cite{Schwille2009} and references therein).

\begin{acknowledgments}
This work was supported by the German Excellence Initiative via
 the program NanoSystems Initiative Munich (NIM) and by the German Research Council (DFG) within the framework of the Transregio 174 ``Spatiotemporal dynamics of bacterial cells". F.M. is supported by a DFG Fellowship through the Graduate School of Quantitative Biosciences Munich (QBM). This project has received funding from the European Union's Horizon 2020 research and innovation programme under European Training Network (ETN) grant 674979-NANOTRANS (J.G.)  and was performed in part at the Aspen Center for Physics (C.P.B), which is supported by National Science Foundation grant PHY-1607611. We thank 
T.\ Betz,
G.\ Berman,
W.\ Bialek, 
D.\ Braun,
D.\ Brueckner, 
C.\ Brangwynne, 
G.\ Crooks,
N.\ Fakhri,
B.\ Fabry, 
E.\ Frey, 
N.\ Gov,
M.\ Guo,
G.\ Gradziuk,
R.\ Granek,
L.\ Jawerth, 
F.\ Jülicher,
G.\ Koenderink, 
K.\ Kroy,
M.\ Lenz, 
T.\ Liverpool, 
T.\ Lubensky, 
B.\ Machta, 
F.\ MacKintosh, 
J.\ Messelink, 
K.\ Miermans,
J.\ Raedler, 
P.\ Ronceray,
J.\ Shaevitz, 
D.\ Schwab,
M.\ Sheinman, 
Y.\ Shokef,
C.\ Schmidt, 
C.\ Storm,
J.\ Tailleur,
M.\ Tikhonov, 
D.\ Weitz,  
M.\ Wigbers, and
N.\ Wingreen, 
for many stimulating discussions.
\end{acknowledgments}


\begin{thebibliography}{274}%
\makeatletter
\providecommand \@ifxundefined [1]{%
 \@ifx{#1\undefined}
}%
\providecommand \@ifnum [1]{%
 \ifnum #1\expandafter \@firstoftwo
 \else \expandafter \@secondoftwo
 \fi
}%
\providecommand \@ifx [1]{%
 \ifx #1\expandafter \@firstoftwo
 \else \expandafter \@secondoftwo
 \fi
}%
\providecommand \natexlab [1]{#1}%
\providecommand \enquote  [1]{``#1''}%
\providecommand \bibnamefont  [1]{#1}%
\providecommand \bibfnamefont [1]{#1}%
\providecommand \citenamefont [1]{#1}%
\providecommand \href@noop [0]{\@secondoftwo}%
\providecommand \href [0]{\begingroup \@sanitize@url \@href}%
\providecommand \@href[1]{\@@startlink{#1}\@@href}%
\providecommand \@@href[1]{\endgroup#1\@@endlink}%
\providecommand \@sanitize@url [0]{\catcode `\\12\catcode `\$12\catcode
  `\&12\catcode `\#12\catcode `\^12\catcode `\_12\catcode `\%12\relax}%
\providecommand \@@startlink[1]{}%
\providecommand \@@endlink[0]{}%
\providecommand \url  [0]{\begingroup\@sanitize@url \@url }%
\providecommand \@url [1]{\endgroup\@href {#1}{\urlprefix }}%
\providecommand \urlprefix  [0]{URL }%
\providecommand \Eprint [0]{\href }%
\providecommand \doibase [0]{http://dx.doi.org/}%
\providecommand \selectlanguage [0]{\@gobble}%
\providecommand \bibinfo  [0]{\@secondoftwo}%
\providecommand \bibfield  [0]{\@secondoftwo}%
\providecommand \translation [1]{[#1]}%
\providecommand \BibitemOpen [0]{}%
\providecommand \bibitemStop [0]{}%
\providecommand \bibitemNoStop [0]{.\EOS\space}%
\providecommand \EOS [0]{\spacefactor3000\relax}%
\providecommand \BibitemShut  [1]{\csname bibitem#1\endcsname}%
\let\auto@bib@innerbib\@empty
\bibitem [{\citenamefont {Lan}\ \emph {et~al.}(2012)\citenamefont {Lan},
  \citenamefont {Sartori}, \citenamefont {Neumann}, \citenamefont {Sourjik},\
  and\ \citenamefont {Tu}}]{Lan2012}%
  \BibitemOpen
  \bibfield  {author} {\bibinfo {author} {\bibfnamefont {Ganhui}\ \bibnamefont
  {Lan}}, \bibinfo {author} {\bibfnamefont {Pablo}\ \bibnamefont {Sartori}},
  \bibinfo {author} {\bibfnamefont {Silke}\ \bibnamefont {Neumann}}, \bibinfo
  {author} {\bibfnamefont {Victor}\ \bibnamefont {Sourjik}}, \ and\ \bibinfo
  {author} {\bibfnamefont {Yuhai}\ \bibnamefont {Tu}},\ }\bibfield  {title}
  {\enquote {\bibinfo {title} {{The energy ? speed ? accuracy trade-off in
  sensory adaptation}},}\ }\href {\doibase 10.1038/nphys2276} {\bibfield
  {journal} {\bibinfo  {journal} {Nat. Phys.}\ }\textbf {\bibinfo {volume}
  {8}},\ \bibinfo {pages} {422--428} (\bibinfo {year} {2012})},\ \Eprint
  {http://arxiv.org/abs/0402594v3} {arXiv:0402594v3 [arXiv:cond-mat]}
  \BibitemShut {NoStop}%
\bibitem [{\citenamefont {Mehta}\ and\ \citenamefont
  {Schwab}(2012)}]{Mehta2012}%
  \BibitemOpen
  \bibfield  {author} {\bibinfo {author} {\bibfnamefont {Pankaj}\ \bibnamefont
  {Mehta}}\ and\ \bibinfo {author} {\bibfnamefont {David~J.}\ \bibnamefont
  {Schwab}},\ }\bibfield  {title} {\enquote {\bibinfo {title} {{Energetic costs
  of cellular computation}},}\ }\href {\doibase 10.1073/pnas.1207814109}
  {\bibfield  {journal} {\bibinfo  {journal} {Proc. Natl. Acad. Sci.}\ }\textbf
  {\bibinfo {volume} {109}},\ \bibinfo {pages} {17978--17982} (\bibinfo {year}
  {2012})},\ \Eprint {http://arxiv.org/abs/1203.5426} {arXiv:1203.5426}
  \BibitemShut {NoStop}%
\bibitem [{\citenamefont {Hopfield}(1974)}]{Hopfield1974a}%
  \BibitemOpen
  \bibfield  {author} {\bibinfo {author} {\bibfnamefont {J.~J.}\ \bibnamefont
  {Hopfield}},\ }\bibfield  {title} {\enquote {\bibinfo {title} {{Kinetic
  Proofreading: A New Mechanism for Reducing Errors in Biosynthetic Processes
  Requiring High Specificity}},}\ }\href {\doibase 10.1073/pnas.71.10.4135}
  {\bibfield  {journal} {\bibinfo  {journal} {Proc. Natl. Acad. Sci.}\ }\textbf
  {\bibinfo {volume} {71}},\ \bibinfo {pages} {4135--4139} (\bibinfo {year}
  {1974})}\BibitemShut {NoStop}%
\bibitem [{\citenamefont {Murugan}\ \emph {et~al.}(2012)\citenamefont
  {Murugan}, \citenamefont {Huse},\ and\ \citenamefont
  {Leibler}}]{Murugan2012}%
  \BibitemOpen
  \bibfield  {author} {\bibinfo {author} {\bibfnamefont {Arvind}\ \bibnamefont
  {Murugan}}, \bibinfo {author} {\bibfnamefont {David~A.}\ \bibnamefont
  {Huse}}, \ and\ \bibinfo {author} {\bibfnamefont {Stanislas}\ \bibnamefont
  {Leibler}},\ }\bibfield  {title} {\enquote {\bibinfo {title} {{Speed,
  dissipation, and error in kinetic proofreading}},}\ }\href {\doibase
  10.1073/pnas.1119911109} {\bibfield  {journal} {\bibinfo  {journal} {Proc.
  Natl. Acad. Sci.}\ }\textbf {\bibinfo {volume} {109}},\ \bibinfo {pages}
  {12034--12039} (\bibinfo {year} {2012})}\BibitemShut {NoStop}%
\bibitem [{\citenamefont {Needleman}\ and\ \citenamefont
  {Brugues}(2014)}]{Needleman2014}%
  \BibitemOpen
  \bibfield  {author} {\bibinfo {author} {\bibfnamefont {Dan}\ \bibnamefont
  {Needleman}}\ and\ \bibinfo {author} {\bibfnamefont {Jan}\ \bibnamefont
  {Brugues}},\ }\bibfield  {title} {\enquote {\bibinfo {title} {{Determining
  physical principles of subcellular organization}},}\ }\href {\doibase
  10.1016/j.devcel.2014.04.018} {\bibfield  {journal} {\bibinfo  {journal}
  {Dev. Cell}\ }\textbf {\bibinfo {volume} {29}},\ \bibinfo {pages} {135--138}
  (\bibinfo {year} {2014})}\BibitemShut {NoStop}%
\bibitem [{\citenamefont {Fletcher}\ and\ \citenamefont
  {Mullins}(2010)}]{Fletcher2010a}%
  \BibitemOpen
  \bibfield  {author} {\bibinfo {author} {\bibfnamefont {Daniel~A.}\
  \bibnamefont {Fletcher}}\ and\ \bibinfo {author} {\bibfnamefont {R.~Dyche}\
  \bibnamefont {Mullins}},\ }\bibfield  {title} {\enquote {\bibinfo {title}
  {{Cell mechanics and the cytoskeleton}},}\ }\href {\doibase
  10.1038/nature08908} {\bibfield  {journal} {\bibinfo  {journal} {Nature}\
  }\textbf {\bibinfo {volume} {463}},\ \bibinfo {pages} {485--492} (\bibinfo
  {year} {2010})}\BibitemShut {NoStop}%
\bibitem [{\citenamefont {Brangwynne}\ \emph
  {et~al.}(2008{\natexlab{a}})\citenamefont {Brangwynne}, \citenamefont
  {Koenderink}, \citenamefont {MacKintosh},\ and\ \citenamefont
  {Weitz}}]{brangwynne2008cytoplasmic}%
  \BibitemOpen
  \bibfield  {author} {\bibinfo {author} {\bibfnamefont {Clifford~P}\
  \bibnamefont {Brangwynne}}, \bibinfo {author} {\bibfnamefont {Gijsje~H}\
  \bibnamefont {Koenderink}}, \bibinfo {author} {\bibfnamefont {Frederick~C}\
  \bibnamefont {MacKintosh}}, \ and\ \bibinfo {author} {\bibfnamefont
  {David~A}\ \bibnamefont {Weitz}},\ }\bibfield  {title} {\enquote {\bibinfo
  {title} {{Cytoplasmic diffusion: molecular motors mix it up}},}\ }\href
  {\doibase 10.1083/jcb.200806149} {\bibfield  {journal} {\bibinfo  {journal}
  {J. Cell Biol.}\ }\textbf {\bibinfo {volume} {183}},\ \bibinfo {pages}
  {583--587} (\bibinfo {year} {2008}{\natexlab{a}})}\BibitemShut {NoStop}%
\bibitem [{\citenamefont {Juelicher}\ \emph {et~al.}(2007)\citenamefont
  {Juelicher}, \citenamefont {Kruse}, \citenamefont {Prost},\ and\
  \citenamefont {Joanny}}]{juelicher2007active}%
  \BibitemOpen
  \bibfield  {author} {\bibinfo {author} {\bibfnamefont {Frank}\ \bibnamefont
  {Juelicher}}, \bibinfo {author} {\bibfnamefont {K}~\bibnamefont {Kruse}},
  \bibinfo {author} {\bibfnamefont {J}~\bibnamefont {Prost}}, \ and\ \bibinfo
  {author} {\bibfnamefont {J-F}\ \bibnamefont {Joanny}},\ }\bibfield  {title}
  {\enquote {\bibinfo {title} {{Active behavior of the cytoskeleton}},}\
  }\href@noop {} {\bibfield  {journal} {\bibinfo  {journal} {Phys. Rep.}\
  }\textbf {\bibinfo {volume} {449}},\ \bibinfo {pages} {3--28} (\bibinfo
  {year} {2007})}\BibitemShut {NoStop}%
\bibitem [{\citenamefont {Cates}(2012)}]{Cates2012}%
  \BibitemOpen
  \bibfield  {author} {\bibinfo {author} {\bibfnamefont {M~E}\ \bibnamefont
  {Cates}},\ }\bibfield  {title} {\enquote {\bibinfo {title} {{Diffusive
  transport without detailed balance in motile bacteria: does microbiology need
  statistical physics?}}}\ }\href {\doibase 10.1088/0034-4885/75/4/042601}
  {\bibfield  {journal} {\bibinfo  {journal} {Reports Prog. Phys.}\ }\textbf
  {\bibinfo {volume} {75}},\ \bibinfo {pages} {042601} (\bibinfo {year}
  {2012})},\ \Eprint {http://arxiv.org/abs/1208.3957} {arXiv:1208.3957}
  \BibitemShut {NoStop}%
\bibitem [{\citenamefont {Huang}\ \emph {et~al.}(2003)\citenamefont {Huang},
  \citenamefont {Meir},\ and\ \citenamefont {Wingreen}}]{Huang2003}%
  \BibitemOpen
  \bibfield  {author} {\bibinfo {author} {\bibfnamefont {Kerwyn~Casey}\
  \bibnamefont {Huang}}, \bibinfo {author} {\bibfnamefont {Yigal}\ \bibnamefont
  {Meir}}, \ and\ \bibinfo {author} {\bibfnamefont {Ned~S}\ \bibnamefont
  {Wingreen}},\ }\bibfield  {title} {\enquote {\bibinfo {title} {{Dynamic
  structures in Escherichia coli: Spontaneous formation of MinE rings and MinD
  polar zones}},}\ }\href {\doibase 10.1073/pnas.2135445100} {\bibfield
  {journal} {\bibinfo  {journal} {Proc. Natl. Acad. Sci.}\ }\textbf {\bibinfo
  {volume} {100}},\ \bibinfo {pages} {12724--12728} (\bibinfo {year}
  {2003})}\BibitemShut {NoStop}%
\bibitem [{\citenamefont {Frey}\ \emph {et~al.}(2017)\citenamefont {Frey},
  \citenamefont {Halatek}, \citenamefont {Kretschmer}, ,\ and\ \citenamefont
  {Schwille}}]{Frey2017}%
  \BibitemOpen
  \bibfield  {author} {\bibinfo {author} {\bibfnamefont {E.}~\bibnamefont
  {Frey}}, \bibinfo {author} {\bibfnamefont {J.}~\bibnamefont {Halatek}},
  \bibinfo {author} {\bibfnamefont {S.}~\bibnamefont {Kretschmer}}, , \ and\
  \bibinfo {author} {\bibfnamefont {P.}~\bibnamefont {Schwille}},\ }\bibfield
  {title} {\enquote {\bibinfo {title} {{Protein Pattern Formation}},}\ }in\
  \href
  {https://www.theorie.physik.uni-muenchen.de/lsfrey/publications/articles/springer{\_}protein{\_}patterns.pdf}
  {\emph {\bibinfo {booktitle} {Phys. Biol. Membr.}}},\ \bibinfo {editor}
  {edited by\ \bibinfo {editor} {\bibfnamefont {P.}~\bibnamefont {Bassereau}}\
  and\ \bibinfo {editor} {\bibfnamefont {P.~C.~A.}\ \bibnamefont {Sens}}}\
  (\bibinfo  {publisher} {Springer-Verlag GmbH},\ \bibinfo {address}
  {Heidelberg},\ \bibinfo {year} {2017})\BibitemShut {NoStop}%
\bibitem [{\citenamefont {Halatek}\ and\ \citenamefont
  {Frey}(2012)}]{Halatek2012}%
  \BibitemOpen
  \bibfield  {author} {\bibinfo {author} {\bibfnamefont {Jacob}\ \bibnamefont
  {Halatek}}\ and\ \bibinfo {author} {\bibfnamefont {Erwin}\ \bibnamefont
  {Frey}},\ }\bibfield  {title} {\enquote {\bibinfo {title} {{Highly Canalized
  MinD Transfer and MinE Sequestration Explain the Origin of Robust
  MinCDE-Protein Dynamics}},}\ }\href {\doibase 10.1016/j.celrep.2012.04.005}
  {\bibfield  {journal} {\bibinfo  {journal} {Cell Rep.}\ }\textbf {\bibinfo
  {volume} {1}},\ \bibinfo {pages} {741--752} (\bibinfo {year}
  {2012})}\BibitemShut {NoStop}%
\bibitem [{\citenamefont {Bialek}(2012)}]{Bialek}%
  \BibitemOpen
  \bibfield  {author} {\bibinfo {author} {\bibfnamefont {William}\ \bibnamefont
  {Bialek}},\ }\href@noop {} {\emph {\bibinfo {title} {{Biophysics:Searching
  for Principles}}}}\ (\bibinfo  {publisher} {Princeton University Press},\
  \bibinfo {year} {2012})\BibitemShut {NoStop}%
\bibitem [{\citenamefont {Seifert}(2012)}]{Seifert_review}%
  \BibitemOpen
  \bibfield  {author} {\bibinfo {author} {\bibfnamefont {Udo}\ \bibnamefont
  {Seifert}},\ }\bibfield  {title} {\enquote {\bibinfo {title} {{Stochastic
  thermodynamics, fluctuation theorems and molecular machines}},}\ }\href
  {\doibase 10.1088/0034-4885/75/12/126001} {\bibfield  {journal} {\bibinfo
  {journal} {Reports Prog. Phys.}\ }\textbf {\bibinfo {volume} {75}},\ \bibinfo
  {pages} {126001} (\bibinfo {year} {2012})}\BibitemShut {NoStop}%
\bibitem [{\citenamefont {Ritort}(2008)}]{Ritort_review}%
  \BibitemOpen
  \bibfield  {author} {\bibinfo {author} {\bibfnamefont {F}~\bibnamefont
  {Ritort}},\ }\bibfield  {title} {\enquote {\bibinfo {title} {{Nonequilibrium
  fluctuations in small systems: From physics to biology}},}\ }\href@noop {}
  {\bibfield  {journal} {\bibinfo  {journal} {Adv. Chem. Phys.}\ }\textbf
  {\bibinfo {volume} {137}} (\bibinfo {year} {2008})}\BibitemShut {NoStop}%
\bibitem [{\citenamefont {{Van den Broeck}}\ and\ \citenamefont
  {Esposito}(2015)}]{Esposito_review}%
  \BibitemOpen
  \bibfield  {author} {\bibinfo {author} {\bibfnamefont {C.}~\bibnamefont {{Van
  den Broeck}}}\ and\ \bibinfo {author} {\bibfnamefont {M}~\bibnamefont
  {Esposito}},\ }\bibfield  {title} {\enquote {\bibinfo {title} {{Ensemble and
  trajectory thermodynamics: A brief introduction}},}\ }\href {\doibase
  10.1016/j.physa.2014.04.035} {\bibfield  {journal} {\bibinfo  {journal}
  {Phys. A Stat. Mech. its Appl.}\ }\textbf {\bibinfo {volume} {418}},\
  \bibinfo {pages} {6--16} (\bibinfo {year} {2015})}\BibitemShut {NoStop}%
\bibitem [{\citenamefont {MacKintosh}\ and\ \citenamefont
  {Schmidt}(2010)}]{mackintosh2010active}%
  \BibitemOpen
  \bibfield  {author} {\bibinfo {author} {\bibfnamefont {Frederick~C}\
  \bibnamefont {MacKintosh}}\ and\ \bibinfo {author} {\bibfnamefont
  {Christoph~F}\ \bibnamefont {Schmidt}},\ }\bibfield  {title} {\enquote
  {\bibinfo {title} {{Active cellular materials}},}\ }\href {\doibase
  10.1016/j.ceb.2010.01.002} {\bibfield  {journal} {\bibinfo  {journal} {Curr.
  Opin. Cell Biol.}\ }\textbf {\bibinfo {volume} {22}},\ \bibinfo {pages}
  {29--35} (\bibinfo {year} {2010})}\BibitemShut {NoStop}%
\bibitem [{\citenamefont {Marchetti}\ \emph {et~al.}(2013)\citenamefont
  {Marchetti}, \citenamefont {Joanny}, \citenamefont {Ramaswamy}, \citenamefont
  {Liverpool}, \citenamefont {Prost}, \citenamefont {Rao},\ and\ \citenamefont
  {Simha}}]{Marchetti2013}%
  \BibitemOpen
  \bibfield  {author} {\bibinfo {author} {\bibfnamefont {M.~C.}\ \bibnamefont
  {Marchetti}}, \bibinfo {author} {\bibfnamefont {J.~F.}\ \bibnamefont
  {Joanny}}, \bibinfo {author} {\bibfnamefont {S.}~\bibnamefont {Ramaswamy}},
  \bibinfo {author} {\bibfnamefont {T.~B.}\ \bibnamefont {Liverpool}}, \bibinfo
  {author} {\bibfnamefont {J.}~\bibnamefont {Prost}}, \bibinfo {author}
  {\bibfnamefont {Madan}\ \bibnamefont {Rao}}, \ and\ \bibinfo {author}
  {\bibfnamefont {R.~Aditi}\ \bibnamefont {Simha}},\ }\bibfield  {title}
  {\enquote {\bibinfo {title} {{Hydrodynamics of soft active matter}},}\ }\href
  {\doibase 10.1103/RevModPhys.85.1143} {\bibfield  {journal} {\bibinfo
  {journal} {Rev. Mod. Phys.}\ }\textbf {\bibinfo {volume} {85}},\ \bibinfo
  {pages} {1143--1189} (\bibinfo {year} {2013})},\ \Eprint
  {http://arxiv.org/abs/1207.2929} {arXiv:1207.2929} \BibitemShut {NoStop}%
\bibitem [{\citenamefont {Ramaswamy}(2010)}]{Ramaswamy2010}%
  \BibitemOpen
  \bibfield  {author} {\bibinfo {author} {\bibfnamefont {Sriram}\ \bibnamefont
  {Ramaswamy}},\ }\bibfield  {title} {\enquote {\bibinfo {title} {{The
  Mechanics and Statistics of Active Matter}},}\ }\href {\doibase
  10.1146/annurev-conmatphys-070909-104101} {\  (\bibinfo {year} {2010}),\
  10.1146/annurev-conmatphys-070909-104101},\ \Eprint
  {http://arxiv.org/abs/1004.1933} {arXiv:1004.1933} \BibitemShut {NoStop}%
\bibitem [{\citenamefont {Joanny}\ and\ \citenamefont
  {Prost}(2009)}]{joanny2009active}%
  \BibitemOpen
  \bibfield  {author} {\bibinfo {author} {\bibfnamefont {Jean-Fran{\c{c}}ois}\
  \bibnamefont {Joanny}}\ and\ \bibinfo {author} {\bibfnamefont {Jacques}\
  \bibnamefont {Prost}},\ }\bibfield  {title} {\enquote {\bibinfo {title}
  {{Active gels as a description of the actin-myosin cytoskeleton}},}\
  }\href@noop {} {\bibfield  {journal} {\bibinfo  {journal} {HFSP J.}\ }\textbf
  {\bibinfo {volume} {3}},\ \bibinfo {pages} {94--104} (\bibinfo {year}
  {2009})}\BibitemShut {NoStop}%
\bibitem [{\citenamefont {Prost}\ \emph {et~al.}(2015)\citenamefont {Prost},
  \citenamefont {Julicher},\ and\ \citenamefont {Joanny}}]{prost2015active}%
  \BibitemOpen
  \bibfield  {author} {\bibinfo {author} {\bibfnamefont {J}~\bibnamefont
  {Prost}}, \bibinfo {author} {\bibfnamefont {F}~\bibnamefont {Julicher}}, \
  and\ \bibinfo {author} {\bibfnamefont {J-F.~F}\ \bibnamefont {Joanny}},\
  }\bibfield  {title} {\enquote {\bibinfo {title} {{Active gel physics}},}\
  }\href {http://dx.doi.org/10.1038/nphys3224} {\bibfield  {journal} {\bibinfo
  {journal} {Nat. Phys.}\ }\textbf {\bibinfo {volume} {11}},\ \bibinfo {pages}
  {111--117} (\bibinfo {year} {2015})}\BibitemShut {NoStop}%
\bibitem [{\citenamefont {Battle}\ \emph {et~al.}(2016)\citenamefont {Battle},
  \citenamefont {Broedersz}, \citenamefont {Fakhri}, \citenamefont {Geyer},
  \citenamefont {Howard}, \citenamefont {Schmidt},\ and\ \citenamefont
  {MacKintosh}}]{Battle2016BDB}%
  \BibitemOpen
  \bibfield  {author} {\bibinfo {author} {\bibfnamefont {Christopher}\
  \bibnamefont {Battle}}, \bibinfo {author} {\bibfnamefont {Chase~P}\
  \bibnamefont {Broedersz}}, \bibinfo {author} {\bibfnamefont {Nikta}\
  \bibnamefont {Fakhri}}, \bibinfo {author} {\bibfnamefont {Veikko~F}\
  \bibnamefont {Geyer}}, \bibinfo {author} {\bibfnamefont {Jonathon}\
  \bibnamefont {Howard}}, \bibinfo {author} {\bibfnamefont {Christoph~F}\
  \bibnamefont {Schmidt}}, \ and\ \bibinfo {author} {\bibfnamefont {Fred~C}\
  \bibnamefont {MacKintosh}},\ }\bibfield  {title} {\enquote {\bibinfo {title}
  {{Broken detailed balance at mesoscopic scales in active biological
  systems}},}\ }\href {\doibase 10.1126/science.aac8167} {\bibfield  {journal}
  {\bibinfo  {journal} {Science}\ }\textbf {\bibinfo {volume} {352}},\ \bibinfo
  {pages} {604--607} (\bibinfo {year} {2016})}\BibitemShut {NoStop}%
\bibitem [{\citenamefont {Zia}\ and\ \citenamefont
  {Schmittmann}(2007)}]{Zia2007}%
  \BibitemOpen
  \bibfield  {author} {\bibinfo {author} {\bibfnamefont {R.~K.~P.}\
  \bibnamefont {Zia}}\ and\ \bibinfo {author} {\bibfnamefont {B.}~\bibnamefont
  {Schmittmann}},\ }\bibfield  {title} {\enquote {\bibinfo {title}
  {{Probability currents as principal characteristics in the statistical
  mechanics of non-equilibrium steady states}},}\ }\href {\doibase
  10.1088/1742-5468/2007/07/P07012} {\bibfield  {journal} {\bibinfo  {journal}
  {J. Stat. Mech. Theory Exp.}\ }\textbf {\bibinfo {volume} {2007}},\ \bibinfo
  {pages} {P07012--P07012} (\bibinfo {year} {2007})},\ \Eprint
  {http://arxiv.org/abs/0701763} {arXiv:0701763 [cond-mat]} \BibitemShut
  {NoStop}%
\bibitem [{\citenamefont {Ajdari}\ \emph {et~al.}(1997)\citenamefont {Ajdari},
  \citenamefont {Prost}, \citenamefont {Ju},\ and\ \citenamefont
  {J{\"{u}}licher}}]{Ajdari1997}%
  \BibitemOpen
  \bibfield  {author} {\bibinfo {author} {\bibfnamefont {Armand}\ \bibnamefont
  {Ajdari}}, \bibinfo {author} {\bibfnamefont {Jacques}\ \bibnamefont {Prost}},
  \bibinfo {author} {\bibfnamefont {Frank}\ \bibnamefont {Ju}}, \ and\ \bibinfo
  {author} {\bibfnamefont {Frank}\ \bibnamefont {J{\"{u}}licher}},\ }\bibfield
  {title} {\enquote {\bibinfo {title} {{Modeling molecular motors}},}\ }\href
  {\doibase 10.1103/RevModPhys.69.1269} {\bibfield  {journal} {\bibinfo
  {journal} {Rev. Mod. Phys.}\ }\textbf {\bibinfo {volume} {69}},\ \bibinfo
  {pages} {1269--1281} (\bibinfo {year} {1997})}\BibitemShut {NoStop}%
\bibitem [{\citenamefont {Howard}(2001)}]{Howard2001}%
  \BibitemOpen
  \bibfield  {author} {\bibinfo {author} {\bibfnamefont {Jonathon}\
  \bibnamefont {Howard}},\ }\bibfield  {title} {\enquote {\bibinfo {title}
  {{Mechanics of Motor Proteins and the Cytoskeleton}},}\ }\href {\doibase
  10.1017/CBO9781107415324.004} {\bibfield  {journal} {\bibinfo  {journal}
  {Sinauer}\ }\textbf {\bibinfo {volume} {7}},\ \bibinfo {pages} {384}
  (\bibinfo {year} {2001})},\ \Eprint {http://arxiv.org/abs/arXiv:1011.1669v3}
  {arXiv:arXiv:1011.1669v3} \BibitemShut {NoStop}%
\bibitem [{\citenamefont {Bausch}\ and\ \citenamefont
  {Kroy}(2006)}]{Bausch2006}%
  \BibitemOpen
  \bibfield  {author} {\bibinfo {author} {\bibfnamefont {A~R}\ \bibnamefont
  {Bausch}}\ and\ \bibinfo {author} {\bibfnamefont {K}~\bibnamefont {Kroy}},\
  }\bibfield  {title} {\enquote {\bibinfo {title} {{A bottom-up approach to
  cell mechanics}},}\ }\href@noop {} {\bibfield  {journal} {\bibinfo  {journal}
  {Nat Phys}\ }\textbf {\bibinfo {volume} {2}},\ \bibinfo {pages} {231--238}
  (\bibinfo {year} {2006})}\BibitemShut {NoStop}%
\bibitem [{\citenamefont {Glaser}\ and\ \citenamefont
  {Kroy}(2010)}]{Glaser2010}%
  \BibitemOpen
  \bibfield  {author} {\bibinfo {author} {\bibfnamefont {Jens}\ \bibnamefont
  {Glaser}}\ and\ \bibinfo {author} {\bibfnamefont {Klaus}\ \bibnamefont
  {Kroy}},\ }\bibfield  {title} {\enquote {\bibinfo {title} {{Fluctuations of
  stiff polymers and cell mechanics}},}\ }\href
  {http://www.intechopen.com/books/biopolymers/fluctuations-of-stiff-polymers-and-cell-mechanics-}
  {\bibfield  {journal} {\bibinfo  {journal} {Biopolymers}\ ,\ \bibinfo {pages}
  {509--534}} (\bibinfo {year} {2010})}\BibitemShut {NoStop}%
\bibitem [{\citenamefont {Egolf}(2000)}]{Egolf2000}%
  \BibitemOpen
  \bibfield  {author} {\bibinfo {author} {\bibfnamefont {D.~A.}\ \bibnamefont
  {Egolf}},\ }\bibfield  {title} {\enquote {\bibinfo {title} {{Equilibrium
  Regained: From Nonequilibrium Chaos to Statistical Mechanics}},}\ }\href
  {\doibase 10.1126/science.287.5450.101} {\bibfield  {journal} {\bibinfo
  {journal} {Science}\ }\textbf {\bibinfo {volume} {287}},\ \bibinfo {pages}
  {101--104} (\bibinfo {year} {2000})}\BibitemShut {NoStop}%
\bibitem [{\citenamefont {Rupprecht}\ and\ \citenamefont
  {Prost}(2016)}]{Rupprecht2016}%
  \BibitemOpen
  \bibfield  {author} {\bibinfo {author} {\bibfnamefont {Jean-Francois}\
  \bibnamefont {Rupprecht}}\ and\ \bibinfo {author} {\bibfnamefont {Jacques}\
  \bibnamefont {Prost}},\ }\bibfield  {title} {\enquote {\bibinfo {title} {{A
  fresh eye on nonequilibrium systems}},}\ }\href {\doibase
  10.1126/science.aaf4611} {\bibfield  {journal} {\bibinfo  {journal}
  {Science}\ }\textbf {\bibinfo {volume} {352}},\ \bibinfo {pages} {514--515}
  (\bibinfo {year} {2016})}\BibitemShut {NoStop}%
\bibitem [{\citenamefont {Weber}\ \emph {et~al.}(2012)\citenamefont {Weber},
  \citenamefont {Spakowitz},\ and\ \citenamefont {Theriot}}]{Weber}%
  \BibitemOpen
  \bibfield  {author} {\bibinfo {author} {\bibfnamefont {Stephanie~C}\
  \bibnamefont {Weber}}, \bibinfo {author} {\bibfnamefont {Andrew~J}\
  \bibnamefont {Spakowitz}}, \ and\ \bibinfo {author} {\bibfnamefont {Julie~A}\
  \bibnamefont {Theriot}},\ }\bibfield  {title} {\enquote {\bibinfo {title}
  {{Nonthermal ATP-dependent fluctuations contribute to the in vivo motion of
  chromosomal loci}},}\ }\href {\doibase 10.1073/pnas.1119505109} {\bibfield
  {journal} {\bibinfo  {journal} {Proc. Natl. Acad. Sci.}\ }\textbf {\bibinfo
  {volume} {109}},\ \bibinfo {pages} {7338--7343} (\bibinfo {year}
  {2012})}\BibitemShut {NoStop}%
\bibitem [{\citenamefont {Almonacid}\ \emph {et~al.}(2015)\citenamefont
  {Almonacid}, \citenamefont {Ahmed}, \citenamefont {Bussonnier}, \citenamefont
  {Mailly}, \citenamefont {Betz}, \citenamefont {Voituriez}, \citenamefont
  {Gov},\ and\ \citenamefont {Verlhac}}]{Almonacid2015}%
  \BibitemOpen
  \bibfield  {author} {\bibinfo {author} {\bibfnamefont {Maria}\ \bibnamefont
  {Almonacid}}, \bibinfo {author} {\bibfnamefont {Wylie~W.}\ \bibnamefont
  {Ahmed}}, \bibinfo {author} {\bibfnamefont {Matthias}\ \bibnamefont
  {Bussonnier}}, \bibinfo {author} {\bibfnamefont {Philippe}\ \bibnamefont
  {Mailly}}, \bibinfo {author} {\bibfnamefont {Timo}\ \bibnamefont {Betz}},
  \bibinfo {author} {\bibfnamefont {Rapha{\"{e}}l}\ \bibnamefont {Voituriez}},
  \bibinfo {author} {\bibfnamefont {Nir~S.}\ \bibnamefont {Gov}}, \ and\
  \bibinfo {author} {\bibfnamefont {Marie-H{\'{e}}l{\`{e}}ne}\ \bibnamefont
  {Verlhac}},\ }\bibfield  {title} {\enquote {\bibinfo {title} {{Active
  diffusion positions the nucleus in mouse oocytes}},}\ }\href {\doibase
  10.1038/ncb3131} {\bibfield  {journal} {\bibinfo  {journal} {Nat. Cell
  Biol.}\ }\textbf {\bibinfo {volume} {17}},\ \bibinfo {pages} {470--479}
  (\bibinfo {year} {2015})}\BibitemShut {NoStop}%
\bibitem [{\citenamefont {Brangwynne}\ \emph {et~al.}(2009)\citenamefont
  {Brangwynne}, \citenamefont {Koenderink}, \citenamefont {MacKintosh},\ and\
  \citenamefont {Weitz}}]{brangwynne2009intracellular}%
  \BibitemOpen
  \bibfield  {author} {\bibinfo {author} {\bibfnamefont {Clifford~P}\
  \bibnamefont {Brangwynne}}, \bibinfo {author} {\bibfnamefont {Gijsje~H}\
  \bibnamefont {Koenderink}}, \bibinfo {author} {\bibfnamefont {Frederick~C}\
  \bibnamefont {MacKintosh}}, \ and\ \bibinfo {author} {\bibfnamefont
  {David~A}\ \bibnamefont {Weitz}},\ }\bibfield  {title} {\enquote {\bibinfo
  {title} {{Intracellular transport by active diffusion}},}\ }\href {\doibase
  10.1016/j.tcb.2009.04.004} {\bibfield  {journal} {\bibinfo  {journal} {Trends
  Cell Biol.}\ }\textbf {\bibinfo {volume} {19}},\ \bibinfo {pages} {423--427}
  (\bibinfo {year} {2009})}\BibitemShut {NoStop}%
\bibitem [{\citenamefont {Brangwynne}\ \emph {et~al.}(2011)\citenamefont
  {Brangwynne}, \citenamefont {Mitchison},\ and\ \citenamefont
  {Hyman}}]{Brangwynne2011}%
  \BibitemOpen
  \bibfield  {author} {\bibinfo {author} {\bibfnamefont {C.~P.}\ \bibnamefont
  {Brangwynne}}, \bibinfo {author} {\bibfnamefont {T.~J.}\ \bibnamefont
  {Mitchison}}, \ and\ \bibinfo {author} {\bibfnamefont {A.~A.}\ \bibnamefont
  {Hyman}},\ }\bibfield  {title} {\enquote {\bibinfo {title} {{Active
  liquid-like behavior of nucleoli determines their size and shape in Xenopus
  laevis oocytes}},}\ }\href {\doibase 10.1073/pnas.1017150108} {\bibfield
  {journal} {\bibinfo  {journal} {Proc. Natl. Acad. Sci.}\ }\textbf {\bibinfo
  {volume} {108}},\ \bibinfo {pages} {4334--4339} (\bibinfo {year}
  {2011})}\BibitemShut {NoStop}%
\bibitem [{\citenamefont {Fakhri}\ \emph {et~al.}(2014)\citenamefont {Fakhri},
  \citenamefont {Wessel}, \citenamefont {Willms}, \citenamefont {Pasquali},
  \citenamefont {Klopfenstein}, \citenamefont {MacKintosh},\ and\ \citenamefont
  {Schmidt}}]{fakhri2014high}%
  \BibitemOpen
  \bibfield  {author} {\bibinfo {author} {\bibfnamefont {Nikta}\ \bibnamefont
  {Fakhri}}, \bibinfo {author} {\bibfnamefont {Alok~D}\ \bibnamefont {Wessel}},
  \bibinfo {author} {\bibfnamefont {Charlotte}\ \bibnamefont {Willms}},
  \bibinfo {author} {\bibfnamefont {Matteo}\ \bibnamefont {Pasquali}}, \bibinfo
  {author} {\bibfnamefont {Dieter~R}\ \bibnamefont {Klopfenstein}}, \bibinfo
  {author} {\bibfnamefont {Frederick~C}\ \bibnamefont {MacKintosh}}, \ and\
  \bibinfo {author} {\bibfnamefont {Christoph~F}\ \bibnamefont {Schmidt}},\
  }\bibfield  {title} {\enquote {\bibinfo {title} {{High-resolution mapping of
  intracellular fluctuations using carbon nanotubes}},}\ }\href@noop {}
  {\bibfield  {journal} {\bibinfo  {journal} {Science}\ }\textbf {\bibinfo
  {volume} {344}},\ \bibinfo {pages} {1031--1035} (\bibinfo {year}
  {2014})}\BibitemShut {NoStop}%
\bibitem [{\citenamefont {Betz}\ \emph {et~al.}(2009)\citenamefont {Betz},
  \citenamefont {Lenz}, \citenamefont {Joanny},\ and\ \citenamefont
  {Sykes}}]{Betz2009}%
  \BibitemOpen
  \bibfield  {author} {\bibinfo {author} {\bibfnamefont {T.}~\bibnamefont
  {Betz}}, \bibinfo {author} {\bibfnamefont {M.}~\bibnamefont {Lenz}}, \bibinfo
  {author} {\bibfnamefont {J.-F.}\ \bibnamefont {Joanny}}, \ and\ \bibinfo
  {author} {\bibfnamefont {C.}~\bibnamefont {Sykes}},\ }\bibfield  {title}
  {\enquote {\bibinfo {title} {{ATP-dependent mechanics of red blood cells}},}\
  }\href {\doibase 10.1073/pnas.0904614106} {\bibfield  {journal} {\bibinfo
  {journal} {Proc. Natl. Acad. Sci.}\ }\textbf {\bibinfo {volume} {106}},\
  \bibinfo {pages} {15320--15325} (\bibinfo {year} {2009})}\BibitemShut
  {NoStop}%
\bibitem [{\citenamefont {Turlier}\ \emph {et~al.}(2016)\citenamefont
  {Turlier}, \citenamefont {Fedosov}, \citenamefont {Audoly}, \citenamefont
  {Auth}, \citenamefont {Gov}, \citenamefont {Sylkes}, \citenamefont {Joanny},
  \citenamefont {Gompper},\ and\ \citenamefont {Betz}}]{Turlier2016}%
  \BibitemOpen
  \bibfield  {author} {\bibinfo {author} {\bibfnamefont {H}~\bibnamefont
  {Turlier}}, \bibinfo {author} {\bibfnamefont {D~A}\ \bibnamefont {Fedosov}},
  \bibinfo {author} {\bibfnamefont {B}~\bibnamefont {Audoly}}, \bibinfo
  {author} {\bibfnamefont {T}~\bibnamefont {Auth}}, \bibinfo {author}
  {\bibfnamefont {N~S}\ \bibnamefont {Gov}}, \bibinfo {author} {\bibfnamefont
  {C}~\bibnamefont {Sylkes}}, \bibinfo {author} {\bibfnamefont {J.-F.}\
  \bibnamefont {Joanny}}, \bibinfo {author} {\bibfnamefont {G}~\bibnamefont
  {Gompper}}, \ and\ \bibinfo {author} {\bibfnamefont {T}~\bibnamefont
  {Betz}},\ }\bibfield  {title} {\enquote {\bibinfo {title} {{Equilibrium
  physics breakdown reveals the active nature of red blood cell flickering}},}\
  }\href {\doibase 10.1038/nphys3621} {\bibfield  {journal} {\bibinfo
  {journal} {Nat. Phys.}\ }\textbf {\bibinfo {volume} {12}},\ \bibinfo {pages}
  {513--519} (\bibinfo {year} {2016})}\BibitemShut {NoStop}%
\bibitem [{\citenamefont {Ben-Isaac}\ \emph {et~al.}(2011)\citenamefont
  {Ben-Isaac}, \citenamefont {Park}, \citenamefont {Popescu}, \citenamefont
  {Brown}, \citenamefont {Gov},\ and\ \citenamefont {Shokef}}]{Ben_Isaac}%
  \BibitemOpen
  \bibfield  {author} {\bibinfo {author} {\bibfnamefont {Eyal}\ \bibnamefont
  {Ben-Isaac}}, \bibinfo {author} {\bibfnamefont {YongKeun}\ \bibnamefont
  {Park}}, \bibinfo {author} {\bibfnamefont {Gabriel}\ \bibnamefont {Popescu}},
  \bibinfo {author} {\bibfnamefont {Frank L~H}\ \bibnamefont {Brown}}, \bibinfo
  {author} {\bibfnamefont {Nir~S}\ \bibnamefont {Gov}}, \ and\ \bibinfo
  {author} {\bibfnamefont {Yair}\ \bibnamefont {Shokef}},\ }\bibfield  {title}
  {\enquote {\bibinfo {title} {{Effective Temperature of Red-Blood-Cell
  Membrane Fluctuations}},}\ }\href {\doibase 10.1103/PhysRevLett.106.238103}
  {\bibfield  {journal} {\bibinfo  {journal} {Phys. Rev. Lett.}\ }\textbf
  {\bibinfo {volume} {106}},\ \bibinfo {pages} {238103} (\bibinfo {year}
  {2011})}\BibitemShut {NoStop}%
\bibitem [{\citenamefont {Tuvia}\ \emph {et~al.}(1997)\citenamefont {Tuvia},
  \citenamefont {Almagor}, \citenamefont {Bitler}, \citenamefont {Levin},
  \citenamefont {Korenstein},\ and\ \citenamefont {Yedgar}}]{Tuvia1997}%
  \BibitemOpen
  \bibfield  {author} {\bibinfo {author} {\bibfnamefont {Shmuel}\ \bibnamefont
  {Tuvia}}, \bibinfo {author} {\bibfnamefont {Ada}\ \bibnamefont {Almagor}},
  \bibinfo {author} {\bibfnamefont {Arkady}\ \bibnamefont {Bitler}}, \bibinfo
  {author} {\bibfnamefont {Shlomo}\ \bibnamefont {Levin}}, \bibinfo {author}
  {\bibfnamefont {Rafi}\ \bibnamefont {Korenstein}}, \ and\ \bibinfo {author}
  {\bibfnamefont {Saul}\ \bibnamefont {Yedgar}},\ }\bibfield  {title} {\enquote
  {\bibinfo {title} {{Cell membrane fluctuations are regulated by medium
  macroviscosity: Evidence for a metabolic driving force}},}\ }\href {\doibase
  10.1073/pnas.94.10.5045} {\bibfield  {journal} {\bibinfo  {journal} {Proc.
  Natl. Acad. Sci.}\ }\textbf {\bibinfo {volume} {94}},\ \bibinfo {pages}
  {5045--5049} (\bibinfo {year} {1997})}\BibitemShut {NoStop}%
\bibitem [{\citenamefont {Monzel}\ \emph {et~al.}(2015)\citenamefont {Monzel},
  \citenamefont {Schmidt}, \citenamefont {Kleusch}, \citenamefont
  {Kirchenb{\"{u}}chler}, \citenamefont {Seifert}, \citenamefont {Smith},
  \citenamefont {Sengupta},\ and\ \citenamefont {Merkel}}]{Monzel2015}%
  \BibitemOpen
  \bibfield  {author} {\bibinfo {author} {\bibfnamefont {C}~\bibnamefont
  {Monzel}}, \bibinfo {author} {\bibfnamefont {D}~\bibnamefont {Schmidt}},
  \bibinfo {author} {\bibfnamefont {C}~\bibnamefont {Kleusch}}, \bibinfo
  {author} {\bibfnamefont {D}~\bibnamefont {Kirchenb{\"{u}}chler}}, \bibinfo
  {author} {\bibfnamefont {U}~\bibnamefont {Seifert}}, \bibinfo {author}
  {\bibfnamefont {A-S}\ \bibnamefont {Smith}}, \bibinfo {author} {\bibfnamefont
  {K}~\bibnamefont {Sengupta}}, \ and\ \bibinfo {author} {\bibfnamefont
  {R}~\bibnamefont {Merkel}},\ }\bibfield  {title} {\enquote {\bibinfo {title}
  {{Measuring fast stochastic displacements of bio-membranes with dynamic
  optical displacement spectroscopy.}}}\ }\href {\doibase 10.1038/ncomms9162}
  {\bibfield  {journal} {\bibinfo  {journal} {Nat. Commun.}\ }\textbf {\bibinfo
  {volume} {6}},\ \bibinfo {pages} {8162} (\bibinfo {year} {2015})}\BibitemShut
  {NoStop}%
\bibitem [{\citenamefont {Battle}\ \emph {et~al.}(2015)\citenamefont {Battle},
  \citenamefont {Ott}, \citenamefont {Burnette}, \citenamefont
  {Lippincott-Schwartz},\ and\ \citenamefont {Schmidt}}]{Battle2015}%
  \BibitemOpen
  \bibfield  {author} {\bibinfo {author} {\bibfnamefont {Christopher}\
  \bibnamefont {Battle}}, \bibinfo {author} {\bibfnamefont {Carolyn~M.}\
  \bibnamefont {Ott}}, \bibinfo {author} {\bibfnamefont {Dylan~T.}\
  \bibnamefont {Burnette}}, \bibinfo {author} {\bibfnamefont {Jennifer}\
  \bibnamefont {Lippincott-Schwartz}}, \ and\ \bibinfo {author} {\bibfnamefont
  {Christoph~F.}\ \bibnamefont {Schmidt}},\ }\bibfield  {title} {\enquote
  {\bibinfo {title} {{Intracellular and extracellular forces drive primary
  cilia movement}},}\ }\href {\doibase 10.1073/pnas.1421845112} {\bibfield
  {journal} {\bibinfo  {journal} {Proc. Natl. Acad. Sci.}\ }\textbf {\bibinfo
  {volume} {112}},\ \bibinfo {pages} {1410--1415} (\bibinfo {year}
  {2015})}\BibitemShut {NoStop}%
\bibitem [{\citenamefont {Fodor}\ \emph
  {et~al.}(2015{\natexlab{a}})\citenamefont {Fodor}, \citenamefont {Mehandia},
  \citenamefont {Comelles}, \citenamefont {Thiagarajan}, \citenamefont {Gov},
  \citenamefont {Visco}, \citenamefont {van Wijland},\ and\ \citenamefont
  {Riveline}}]{Fodor2015}%
  \BibitemOpen
  \bibfield  {author} {\bibinfo {author} {\bibfnamefont {{\'{E}}.}~\bibnamefont
  {Fodor}}, \bibinfo {author} {\bibfnamefont {V.}~\bibnamefont {Mehandia}},
  \bibinfo {author} {\bibfnamefont {J.}~\bibnamefont {Comelles}}, \bibinfo
  {author} {\bibfnamefont {R.}~\bibnamefont {Thiagarajan}}, \bibinfo {author}
  {\bibfnamefont {N.~S.}\ \bibnamefont {Gov}}, \bibinfo {author} {\bibfnamefont
  {P.}~\bibnamefont {Visco}}, \bibinfo {author} {\bibfnamefont
  {F.}~\bibnamefont {van Wijland}}, \ and\ \bibinfo {author} {\bibfnamefont
  {D.}~\bibnamefont {Riveline}},\ }\bibfield  {title} {\enquote {\bibinfo
  {title} {{From motor-induced fluctuations to mesoscopic dynamics in
  epithelial tissues}},}\ }\href {http://arxiv.org/abs/1512.01476} {\bibfield
  {journal} {\bibinfo  {journal} {arXiv Prepr.}\ }\textbf {\bibinfo {volume}
  {1}},\ \bibinfo {pages} {1--5} (\bibinfo {year} {2015}{\natexlab{a}})},\
  \Eprint {http://arxiv.org/abs/1512.01476} {arXiv:1512.01476} \BibitemShut
  {NoStop}%
\bibitem [{\citenamefont {Lau}\ \emph {et~al.}(2003)\citenamefont {Lau},
  \citenamefont {Hoffman}, \citenamefont {Davies}, \citenamefont {Crocker},\
  and\ \citenamefont {Lubensky}}]{Lau2003}%
  \BibitemOpen
  \bibfield  {author} {\bibinfo {author} {\bibfnamefont {A~W~C}\ \bibnamefont
  {Lau}}, \bibinfo {author} {\bibfnamefont {B~D}\ \bibnamefont {Hoffman}},
  \bibinfo {author} {\bibfnamefont {A}~\bibnamefont {Davies}}, \bibinfo
  {author} {\bibfnamefont {J~C}\ \bibnamefont {Crocker}}, \ and\ \bibinfo
  {author} {\bibfnamefont {T~C}\ \bibnamefont {Lubensky}},\ }\bibfield  {title}
  {\enquote {\bibinfo {title} {{Microrheology, Stress Fluctuations, and Active
  Behavior of Living Cells}},}\ }\href {\doibase 10.1103/PhysRevLett.91.198101}
  {\bibfield  {journal} {\bibinfo  {journal} {Phys. Rev. Lett.}\ }\textbf
  {\bibinfo {volume} {91}},\ \bibinfo {pages} {198101} (\bibinfo {year}
  {2003})}\BibitemShut {NoStop}%
\bibitem [{\citenamefont {Mizuno}\ \emph {et~al.}(2007)\citenamefont {Mizuno},
  \citenamefont {Tardin}, \citenamefont {Schmidt},\ and\ \citenamefont
  {MacKintosh}}]{Mizuno370}%
  \BibitemOpen
  \bibfield  {author} {\bibinfo {author} {\bibfnamefont {Daisuke}\ \bibnamefont
  {Mizuno}}, \bibinfo {author} {\bibfnamefont {Catherine}\ \bibnamefont
  {Tardin}}, \bibinfo {author} {\bibfnamefont {C~F}\ \bibnamefont {Schmidt}}, \
  and\ \bibinfo {author} {\bibfnamefont {F~C}\ \bibnamefont {MacKintosh}},\
  }\bibfield  {title} {\enquote {\bibinfo {title} {{Nonequilibrium Mechanics of
  Active Cytoskeletal Networks}},}\ }\href {\doibase 10.1126/science.1134404}
  {\bibfield  {journal} {\bibinfo  {journal} {Science}\ }\textbf {\bibinfo
  {volume} {315}},\ \bibinfo {pages} {370--373} (\bibinfo {year}
  {2007})}\BibitemShut {NoStop}%
\bibitem [{\citenamefont {Mizuno}\ \emph {et~al.}(2008)\citenamefont {Mizuno},
  \citenamefont {Head}, \citenamefont {MacKintosh},\ and\ \citenamefont
  {Schmidt}}]{Mizuno2008}%
  \BibitemOpen
  \bibfield  {author} {\bibinfo {author} {\bibfnamefont {D}~\bibnamefont
  {Mizuno}}, \bibinfo {author} {\bibfnamefont {D~A}\ \bibnamefont {Head}},
  \bibinfo {author} {\bibfnamefont {F~C}\ \bibnamefont {MacKintosh}}, \ and\
  \bibinfo {author} {\bibfnamefont {C~F}\ \bibnamefont {Schmidt}},\ }\bibfield
  {title} {\enquote {\bibinfo {title} {{Active and Passive Microrheology in
  Equilibrium and Nonequilibrium Systems}},}\ }\href {\doibase
  10.1021/ma801218z} {\bibfield  {journal} {\bibinfo  {journal}
  {Macromolecules}\ }\textbf {\bibinfo {volume} {41}},\ \bibinfo {pages}
  {7194--7202} (\bibinfo {year} {2008})}\BibitemShut {NoStop}%
\bibitem [{\citenamefont {Guo}\ \emph {et~al.}(2014)\citenamefont {Guo},
  \citenamefont {Ehrlicher}, \citenamefont {Jensen}, \citenamefont {Renz},
  \citenamefont {Moore}, \citenamefont {Goldman}, \citenamefont
  {Lippincott-Schwartz}, \citenamefont {Mackintosh},\ and\ \citenamefont
  {Weitz}}]{Guo2014}%
  \BibitemOpen
  \bibfield  {author} {\bibinfo {author} {\bibfnamefont {Ming}\ \bibnamefont
  {Guo}}, \bibinfo {author} {\bibfnamefont {Allen~J}\ \bibnamefont
  {Ehrlicher}}, \bibinfo {author} {\bibfnamefont {Mikkel~H}\ \bibnamefont
  {Jensen}}, \bibinfo {author} {\bibfnamefont {Malte}\ \bibnamefont {Renz}},
  \bibinfo {author} {\bibfnamefont {Jeffrey~R}\ \bibnamefont {Moore}}, \bibinfo
  {author} {\bibfnamefont {Robert~D}\ \bibnamefont {Goldman}}, \bibinfo
  {author} {\bibfnamefont {Jennifer}\ \bibnamefont {Lippincott-Schwartz}},
  \bibinfo {author} {\bibfnamefont {Frederick~C}\ \bibnamefont {Mackintosh}}, \
  and\ \bibinfo {author} {\bibfnamefont {David~A}\ \bibnamefont {Weitz}},\
  }\bibfield  {title} {\enquote {\bibinfo {title} {{Probing the stochastic,
  motor-driven properties of the cytoplasm using force spectrum microscopy}},}\
  }\href@noop {} {\bibfield  {journal} {\bibinfo  {journal} {Cell}\ }\textbf
  {\bibinfo {volume} {158}},\ \bibinfo {pages} {822--832} (\bibinfo {year}
  {2014})}\BibitemShut {NoStop}%
\bibitem [{\citenamefont {Schaller}\ \emph {et~al.}(2010)\citenamefont
  {Schaller}, \citenamefont {Weber}, \citenamefont {Semmrich}, \citenamefont
  {Frey},\ and\ \citenamefont {Bausch}}]{Schaller2010polar}%
  \BibitemOpen
  \bibfield  {author} {\bibinfo {author} {\bibfnamefont {Volker}\ \bibnamefont
  {Schaller}}, \bibinfo {author} {\bibfnamefont {Christoph}\ \bibnamefont
  {Weber}}, \bibinfo {author} {\bibfnamefont {Christine}\ \bibnamefont
  {Semmrich}}, \bibinfo {author} {\bibfnamefont {Erwin}\ \bibnamefont {Frey}},
  \ and\ \bibinfo {author} {\bibfnamefont {Andreas~R}\ \bibnamefont {Bausch}},\
  }\bibfield  {title} {\enquote {\bibinfo {title} {{Polar patterns of driven
  filaments}},}\ }\href {\doibase 10.1038/nature09312} {\bibfield  {journal}
  {\bibinfo  {journal} {Nature}\ }\textbf {\bibinfo {volume} {467}},\ \bibinfo
  {pages} {73--77} (\bibinfo {year} {2010})}\BibitemShut {NoStop}%
\bibitem [{\citenamefont {Schaller}\ \emph
  {et~al.}(2011{\natexlab{a}})\citenamefont {Schaller}, \citenamefont {Weber},
  \citenamefont {Hammerich}, \citenamefont {Frey},\ and\ \citenamefont
  {Bausch}}]{Schaller2011frozen}%
  \BibitemOpen
  \bibfield  {author} {\bibinfo {author} {\bibfnamefont {Volker}\ \bibnamefont
  {Schaller}}, \bibinfo {author} {\bibfnamefont {Christoph~A}\ \bibnamefont
  {Weber}}, \bibinfo {author} {\bibfnamefont {Benjamin}\ \bibnamefont
  {Hammerich}}, \bibinfo {author} {\bibfnamefont {Erwin}\ \bibnamefont {Frey}},
  \ and\ \bibinfo {author} {\bibfnamefont {Andreas~R}\ \bibnamefont {Bausch}},\
  }\bibfield  {title} {\enquote {\bibinfo {title} {{Frozen steady states in
  active systems}},}\ }\href@noop {} {\bibfield  {journal} {\bibinfo  {journal}
  {Proc. Natl. Acad. Sci.}\ }\textbf {\bibinfo {volume} {108}},\ \bibinfo
  {pages} {19183--19188} (\bibinfo {year} {2011}{\natexlab{a}})}\BibitemShut
  {NoStop}%
\bibitem [{\citenamefont {{Soares e Silva}}\ \emph {et~al.}(2011)\citenamefont
  {{Soares e Silva}}, \citenamefont {Depken}, \citenamefont {Stuhrmann},
  \citenamefont {Korsten}, \citenamefont {MacKintosh}, \citenamefont
  {Koenderink}, \citenamefont {e~Silva}, \citenamefont {Depken}, \citenamefont
  {Stuhrmann}, \citenamefont {Korsten}, \citenamefont {MacKintosh},\ and\
  \citenamefont {Koenderink}}]{Silva2011}%
  \BibitemOpen
  \bibfield  {author} {\bibinfo {author} {\bibfnamefont {M.}~\bibnamefont
  {{Soares e Silva}}}, \bibinfo {author} {\bibfnamefont {Martin}\ \bibnamefont
  {Depken}}, \bibinfo {author} {\bibfnamefont {Bj{\"{o}}rn}\ \bibnamefont
  {Stuhrmann}}, \bibinfo {author} {\bibfnamefont {Marijn}\ \bibnamefont
  {Korsten}}, \bibinfo {author} {\bibfnamefont {Fred~C.}\ \bibnamefont
  {MacKintosh}}, \bibinfo {author} {\bibfnamefont {Gijsje~H.}\ \bibnamefont
  {Koenderink}}, \bibinfo {author} {\bibfnamefont {Marina~Soares}\ \bibnamefont
  {e~Silva}}, \bibinfo {author} {\bibfnamefont {Martin}\ \bibnamefont
  {Depken}}, \bibinfo {author} {\bibfnamefont {Bj{\"{o}}rn}\ \bibnamefont
  {Stuhrmann}}, \bibinfo {author} {\bibfnamefont {Marijn}\ \bibnamefont
  {Korsten}}, \bibinfo {author} {\bibfnamefont {Fred~C.}\ \bibnamefont
  {MacKintosh}}, \ and\ \bibinfo {author} {\bibfnamefont {Gijsje~H.}\
  \bibnamefont {Koenderink}},\ }\bibfield  {title} {\enquote {\bibinfo {title}
  {{Active multistage coarsening of actin networks driven by myosin motors}},}\
  }\href {\doibase 10.1073/pnas.1016616108} {\bibfield  {journal} {\bibinfo
  {journal} {Proc. Natl. Acad. Sci.}\ }\textbf {\bibinfo {volume} {108}},\
  \bibinfo {pages} {9408--9413} (\bibinfo {year} {2011})}\BibitemShut {NoStop}%
\bibitem [{\citenamefont {Murrell}\ and\ \citenamefont
  {Gardel}(2012)}]{murrell2012f}%
  \BibitemOpen
  \bibfield  {author} {\bibinfo {author} {\bibfnamefont {Michael~P}\
  \bibnamefont {Murrell}}\ and\ \bibinfo {author} {\bibfnamefont {Margaret~L}\
  \bibnamefont {Gardel}},\ }\bibfield  {title} {\enquote {\bibinfo {title}
  {{F-actin buckling coordinates contractility and severing in a biomimetic
  actomyosin cortex}},}\ }\href@noop {} {\bibfield  {journal} {\bibinfo
  {journal} {Proc. Natl. Acad. Sci.}\ }\textbf {\bibinfo {volume} {109}},\
  \bibinfo {pages} {20820--20825} (\bibinfo {year} {2012})}\BibitemShut
  {NoStop}%
\bibitem [{\citenamefont {Alvarado}\ \emph {et~al.}(2013)\citenamefont
  {Alvarado}, \citenamefont {Sheinman}, \citenamefont {Sharma}, \citenamefont
  {MacKintosh},\ and\ \citenamefont {Koenderink}}]{Alvarado2013molecular}%
  \BibitemOpen
  \bibfield  {author} {\bibinfo {author} {\bibfnamefont {Jos{\'{e}}}\
  \bibnamefont {Alvarado}}, \bibinfo {author} {\bibfnamefont {Michael}\
  \bibnamefont {Sheinman}}, \bibinfo {author} {\bibfnamefont {Abhinav}\
  \bibnamefont {Sharma}}, \bibinfo {author} {\bibfnamefont {Fred~C}\
  \bibnamefont {MacKintosh}}, \ and\ \bibinfo {author} {\bibfnamefont
  {Gijsje~H}\ \bibnamefont {Koenderink}},\ }\bibfield  {title} {\enquote
  {\bibinfo {title} {{Molecular motors robustly drive active gels to a
  critically connected state}},}\ }\href {\doibase 10.1038/nphys2715}
  {\bibfield  {journal} {\bibinfo  {journal} {Nat. Phys.}\ }\textbf {\bibinfo
  {volume} {9}},\ \bibinfo {pages} {591--597} (\bibinfo {year}
  {2013})}\BibitemShut {NoStop}%
\bibitem [{\citenamefont {Lenz}(2014)}]{Lenz2014}%
  \BibitemOpen
  \bibfield  {author} {\bibinfo {author} {\bibfnamefont {Martin}\ \bibnamefont
  {Lenz}},\ }\bibfield  {title} {\enquote {\bibinfo {title} {{Geometrical
  Origins of Contractility in Disordered Actomyosin Networks}},}\ }\href
  {\doibase 10.1103/PhysRevX.4.041002} {\bibfield  {journal} {\bibinfo
  {journal} {Phys. Rev. X}\ }\textbf {\bibinfo {volume} {4}},\ \bibinfo {pages}
  {041002} (\bibinfo {year} {2014})},\ \Eprint {http://arxiv.org/abs/1407.6693}
  {arXiv:1407.6693} \BibitemShut {NoStop}%
\bibitem [{\citenamefont {Koenderink}\ \emph {et~al.}(2009)\citenamefont
  {Koenderink}, \citenamefont {Dogic}, \citenamefont {Nakamura}, \citenamefont
  {Bendix}, \citenamefont {MacKintosh}, \citenamefont {Hartwig}, \citenamefont
  {Stossel},\ and\ \citenamefont {Weitz}}]{Koenderink2009}%
  \BibitemOpen
  \bibfield  {author} {\bibinfo {author} {\bibfnamefont {Gijsje~H}\
  \bibnamefont {Koenderink}}, \bibinfo {author} {\bibfnamefont {Zvonimir}\
  \bibnamefont {Dogic}}, \bibinfo {author} {\bibfnamefont {Fumihiko}\
  \bibnamefont {Nakamura}}, \bibinfo {author} {\bibfnamefont {Poul~M}\
  \bibnamefont {Bendix}}, \bibinfo {author} {\bibfnamefont {Frederick~C}\
  \bibnamefont {MacKintosh}}, \bibinfo {author} {\bibfnamefont {John~H}\
  \bibnamefont {Hartwig}}, \bibinfo {author} {\bibfnamefont {Thomas~P}\
  \bibnamefont {Stossel}}, \ and\ \bibinfo {author} {\bibfnamefont {David~A}\
  \bibnamefont {Weitz}},\ }\bibfield  {title} {\enquote {\bibinfo {title} {{An
  active biopolymer network controlled by molecular motors}},}\ }\href
  {\doibase 10.1073/pnas.0903974106} {\bibfield  {journal} {\bibinfo  {journal}
  {Proc. Natl. Acad. Sci.}\ }\textbf {\bibinfo {volume} {106}},\ \bibinfo
  {pages} {15192--15197} (\bibinfo {year} {2009})}\BibitemShut {NoStop}%
\bibitem [{\citenamefont {Sheinman}\ \emph {et~al.}(2012)\citenamefont
  {Sheinman}, \citenamefont {Broedersz},\ and\ \citenamefont
  {MacKintosh}}]{sheinman2012actively}%
  \BibitemOpen
  \bibfield  {author} {\bibinfo {author} {\bibfnamefont {M}~\bibnamefont
  {Sheinman}}, \bibinfo {author} {\bibfnamefont {C~P}\ \bibnamefont
  {Broedersz}}, \ and\ \bibinfo {author} {\bibfnamefont {F~C}\ \bibnamefont
  {MacKintosh}},\ }\bibfield  {title} {\enquote {\bibinfo {title} {{Actively
  stressed marginal networks}},}\ }\href@noop {} {\bibfield  {journal}
  {\bibinfo  {journal} {Phys. Rev. Lett.}\ }\textbf {\bibinfo {volume} {109}},\
  \bibinfo {pages} {238101} (\bibinfo {year} {2012})}\BibitemShut {NoStop}%
\bibitem [{\citenamefont {Broedersz}\ and\ \citenamefont
  {MacKintosh}(2011)}]{broedersz2011molecular}%
  \BibitemOpen
  \bibfield  {author} {\bibinfo {author} {\bibfnamefont {C.~P.}\ \bibnamefont
  {Broedersz}}\ and\ \bibinfo {author} {\bibfnamefont {F.~C.}\ \bibnamefont
  {MacKintosh}},\ }\bibfield  {title} {\enquote {\bibinfo {title} {{Molecular
  motors stiffen non-affine semiflexible polymer networks}},}\ }\href {\doibase
  10.1039/c0sm01004a} {\bibfield  {journal} {\bibinfo  {journal} {Soft Matter}\
  }\textbf {\bibinfo {volume} {7}},\ \bibinfo {pages} {3186--3191} (\bibinfo
  {year} {2011})},\ \Eprint {http://arxiv.org/abs/1009.3848} {arXiv:1009.3848}
  \BibitemShut {NoStop}%
\bibitem [{\citenamefont {Brangwynne}\ \emph
  {et~al.}(2008{\natexlab{b}})\citenamefont {Brangwynne}, \citenamefont
  {Koenderink}, \citenamefont {MacKintosh},\ and\ \citenamefont
  {Weitz}}]{brangwynne2008nonequilibrium}%
  \BibitemOpen
  \bibfield  {author} {\bibinfo {author} {\bibfnamefont {Clifford~P}\
  \bibnamefont {Brangwynne}}, \bibinfo {author} {\bibfnamefont {Gijsje~H}\
  \bibnamefont {Koenderink}}, \bibinfo {author} {\bibfnamefont {Frederick~C}\
  \bibnamefont {MacKintosh}}, \ and\ \bibinfo {author} {\bibfnamefont
  {David~A}\ \bibnamefont {Weitz}},\ }\bibfield  {title} {\enquote {\bibinfo
  {title} {{Nonequilibrium microtubule fluctuations in a model
  cytoskeleton}},}\ }\href {\doibase 10.1103/PhysRevLett.100.118104} {\bibfield
   {journal} {\bibinfo  {journal} {Phys. Rev. Lett.}\ }\textbf {\bibinfo
  {volume} {100}},\ \bibinfo {pages} {118104} (\bibinfo {year}
  {2008}{\natexlab{b}})}\BibitemShut {NoStop}%
\bibitem [{\citenamefont {Aragon}\ and\ \citenamefont
  {Pecora}(1985)}]{aragon1985dynami}%
  \BibitemOpen
  \bibfield  {author} {\bibinfo {author} {\bibfnamefont {Sergio~R}\
  \bibnamefont {Aragon}}\ and\ \bibinfo {author} {\bibfnamefont
  {R.}~\bibnamefont {Pecora}},\ }\bibfield  {title} {\enquote {\bibinfo {title}
  {{Dynamics of wormlike chains}},}\ }\href {\doibase 10.1021/ma00152a014}
  {\bibfield  {journal} {\bibinfo  {journal} {Macromolecules}\ }\textbf
  {\bibinfo {volume} {18}},\ \bibinfo {pages} {1868--1875} (\bibinfo {year}
  {1985})}\BibitemShut {NoStop}%
\bibitem [{\citenamefont {Gittes}\ \emph {et~al.}(1993)\citenamefont {Gittes},
  \citenamefont {Mickey}, \citenamefont {Nettleton},\ and\ \citenamefont
  {Howard}}]{Gittes1993}%
  \BibitemOpen
  \bibfield  {author} {\bibinfo {author} {\bibfnamefont {F.}~\bibnamefont
  {Gittes}}, \bibinfo {author} {\bibfnamefont {B.}~\bibnamefont {Mickey}},
  \bibinfo {author} {\bibfnamefont {J.}~\bibnamefont {Nettleton}}, \ and\
  \bibinfo {author} {\bibfnamefont {J.}~\bibnamefont {Howard}},\ }\bibfield
  {title} {\enquote {\bibinfo {title} {{Flexural rigidity of microtubules and
  actin filaments measured from thermal fluctuations in shape}},}\ }\href
  {\doibase 10.1083/jcb.120.4.923} {\bibfield  {journal} {\bibinfo  {journal}
  {J. Cell Biol.}\ }\textbf {\bibinfo {volume} {120}},\ \bibinfo {pages}
  {923--934} (\bibinfo {year} {1993})},\ \Eprint
  {http://arxiv.org/abs/arXiv:1011.1669v3} {arXiv:arXiv:1011.1669v3}
  \BibitemShut {NoStop}%
\bibitem [{\citenamefont {Brangwynne}\ \emph
  {et~al.}(2007{\natexlab{a}})\citenamefont {Brangwynne}, \citenamefont
  {Koenderink}, \citenamefont {Barry}, \citenamefont {Dogic}, \citenamefont
  {MacKintosh},\ and\ \citenamefont {Weitz}}]{Brangwynne2007}%
  \BibitemOpen
  \bibfield  {author} {\bibinfo {author} {\bibfnamefont {Clifford~P.}\
  \bibnamefont {Brangwynne}}, \bibinfo {author} {\bibfnamefont {Gijsje~H.}\
  \bibnamefont {Koenderink}}, \bibinfo {author} {\bibfnamefont
  {Ed}~\bibnamefont {Barry}}, \bibinfo {author} {\bibfnamefont {Zvonimir}\
  \bibnamefont {Dogic}}, \bibinfo {author} {\bibfnamefont {Frederick~C.}\
  \bibnamefont {MacKintosh}}, \ and\ \bibinfo {author} {\bibfnamefont
  {David~A.}\ \bibnamefont {Weitz}},\ }\bibfield  {title} {\enquote {\bibinfo
  {title} {{Bending Dynamics of Fluctuating Biopolymers Probed by Automated
  High-Resolution Filament Tracking}},}\ }\href {\doibase
  10.1529/biophysj.106.096966} {\bibfield  {journal} {\bibinfo  {journal}
  {Biophys. J.}\ }\textbf {\bibinfo {volume} {93}},\ \bibinfo {pages}
  {346--359} (\bibinfo {year} {2007}{\natexlab{a}})}\BibitemShut {NoStop}%
\bibitem [{\citenamefont {Gladrow}\ \emph {et~al.}(2016)\citenamefont
  {Gladrow}, \citenamefont {Fakhri}, \citenamefont {MacKintosh}, \citenamefont
  {Schmidt},\ and\ \citenamefont {Broedersz}}]{Gladrow2016}%
  \BibitemOpen
  \bibfield  {author} {\bibinfo {author} {\bibfnamefont {J}~\bibnamefont
  {Gladrow}}, \bibinfo {author} {\bibfnamefont {N}~\bibnamefont {Fakhri}},
  \bibinfo {author} {\bibfnamefont {F~C}\ \bibnamefont {MacKintosh}}, \bibinfo
  {author} {\bibfnamefont {C~F}\ \bibnamefont {Schmidt}}, \ and\ \bibinfo
  {author} {\bibfnamefont {C~P}\ \bibnamefont {Broedersz}},\ }\bibfield
  {title} {\enquote {\bibinfo {title} {{Broken Detailed Balance of Filament
  Dynamics in Active Networks}},}\ }\href {\doibase
  10.1103/PhysRevLett.116.248301} {\bibfield  {journal} {\bibinfo  {journal}
  {Phys. Rev. Lett.}\ }\textbf {\bibinfo {volume} {116}},\ \bibinfo {pages}
  {248301} (\bibinfo {year} {2016})}\BibitemShut {NoStop}%
\bibitem [{\citenamefont {Gladrow}\ \emph {et~al.}(2017)\citenamefont
  {Gladrow}, \citenamefont {Broedersz},\ and\ \citenamefont
  {Schmidt}}]{Gladrow2017}%
  \BibitemOpen
  \bibfield  {author} {\bibinfo {author} {\bibfnamefont {J.}~\bibnamefont
  {Gladrow}}, \bibinfo {author} {\bibfnamefont {C.~P.}\ \bibnamefont
  {Broedersz}}, \ and\ \bibinfo {author} {\bibfnamefont {C.~F.}\ \bibnamefont
  {Schmidt}},\ }\bibfield  {title} {\enquote {\bibinfo {title} {{Nonequilibrium
  dynamics of probe filaments in actin-myosin networks}},}\ }\href {\doibase
  10.1103/PhysRevE.96.022408} {\bibfield  {journal} {\bibinfo  {journal} {Phys.
  Rev. E}\ }\textbf {\bibinfo {volume} {96}},\ \bibinfo {pages} {022408}
  (\bibinfo {year} {2017})},\ \Eprint {http://arxiv.org/abs/1704.06243}
  {arXiv:1704.06243} \BibitemShut {NoStop}%
\bibitem [{\citenamefont {Cicuta}\ and\ \citenamefont
  {Donald}(2007)}]{Cicuta2007}%
  \BibitemOpen
  \bibfield  {author} {\bibinfo {author} {\bibfnamefont {Pietro}\ \bibnamefont
  {Cicuta}}\ and\ \bibinfo {author} {\bibfnamefont {Athene~M}\ \bibnamefont
  {Donald}},\ }\bibfield  {title} {\enquote {\bibinfo {title} {{Microrheology:
  a review of the method and applications}},}\ }\href {\doibase
  10.1039/B706004C} {\bibfield  {journal} {\bibinfo  {journal} {Soft Matter}\
  }\textbf {\bibinfo {volume} {3}},\ \bibinfo {pages} {1449--1455} (\bibinfo
  {year} {2007})}\BibitemShut {NoStop}%
\bibitem [{\citenamefont {Mason}\ \emph {et~al.}(1997)\citenamefont {Mason},
  \citenamefont {Ganesan}, \citenamefont {van Zanten}, \citenamefont {Wirtz},\
  and\ \citenamefont {Kuo}}]{Mason97}%
  \BibitemOpen
  \bibfield  {author} {\bibinfo {author} {\bibfnamefont {T~G}\ \bibnamefont
  {Mason}}, \bibinfo {author} {\bibfnamefont {K}~\bibnamefont {Ganesan}},
  \bibinfo {author} {\bibfnamefont {J~H}\ \bibnamefont {van Zanten}}, \bibinfo
  {author} {\bibfnamefont {D}~\bibnamefont {Wirtz}}, \ and\ \bibinfo {author}
  {\bibfnamefont {S~C}\ \bibnamefont {Kuo}},\ }\bibfield  {title} {\enquote
  {\bibinfo {title} {{Particle Tracking Microrheology of Complex Fluids}},}\
  }\href {\doibase 10.1103/PhysRevLett.79.3282} {\bibfield  {journal} {\bibinfo
   {journal} {Phys. Rev. Lett.}\ }\textbf {\bibinfo {volume} {79}},\ \bibinfo
  {pages} {3282--3285} (\bibinfo {year} {1997})}\BibitemShut {NoStop}%
\bibitem [{\citenamefont {MacKintosh}\ and\ \citenamefont
  {Schmidt}(1999)}]{Mackintosh99}%
  \BibitemOpen
  \bibfield  {author} {\bibinfo {author} {\bibfnamefont {F.C.}\ \bibnamefont
  {MacKintosh}}\ and\ \bibinfo {author} {\bibfnamefont {C.F.}\ \bibnamefont
  {Schmidt}},\ }\bibfield  {title} {\enquote {\bibinfo {title}
  {{Microrheology}},}\ }\href {\doibase 10.1016/S1359-0294(99)90010-9}
  {\bibfield  {journal} {\bibinfo  {journal} {Curr. Opin. Colloid Interface
  Sci.}\ }\textbf {\bibinfo {volume} {4}},\ \bibinfo {pages} {300--307}
  (\bibinfo {year} {1999})}\BibitemShut {NoStop}%
\bibitem [{\citenamefont {Waigh}(2005)}]{Waigh05}%
  \BibitemOpen
  \bibfield  {author} {\bibinfo {author} {\bibfnamefont {T~A}\ \bibnamefont
  {Waigh}},\ }\bibfield  {title} {\enquote {\bibinfo {title} {{Microrheology of
  complex fluids}},}\ }\href {http://stacks.iop.org/0034-4885/68/i=3/a=R04}
  {\bibfield  {journal} {\bibinfo  {journal} {Reports Prog. Phys.}\ }\textbf
  {\bibinfo {volume} {68}},\ \bibinfo {pages} {685} (\bibinfo {year}
  {2005})}\BibitemShut {NoStop}%
\bibitem [{\citenamefont {Levine}\ and\ \citenamefont
  {Lubensky}(2000)}]{Levine00}%
  \BibitemOpen
  \bibfield  {author} {\bibinfo {author} {\bibfnamefont {Alex~J}\ \bibnamefont
  {Levine}}\ and\ \bibinfo {author} {\bibfnamefont {T~C}\ \bibnamefont
  {Lubensky}},\ }\bibfield  {title} {\enquote {\bibinfo {title} {{One- and
  Two-Particle Microrheology}},}\ }\href {\doibase 10.1103/PhysRevLett.85.1774}
  {\bibfield  {journal} {\bibinfo  {journal} {Phys. Rev. Lett.}\ }\textbf
  {\bibinfo {volume} {85}},\ \bibinfo {pages} {1774--1777} (\bibinfo {year}
  {2000})}\BibitemShut {NoStop}%
\bibitem [{\citenamefont {Jensen}\ \emph {et~al.}(2015)\citenamefont {Jensen},
  \citenamefont {Morris},\ and\ \citenamefont {Weitz}}]{Jensen_Rev}%
  \BibitemOpen
  \bibfield  {author} {\bibinfo {author} {\bibfnamefont {Mikkel~H}\
  \bibnamefont {Jensen}}, \bibinfo {author} {\bibfnamefont {Eliza~J}\
  \bibnamefont {Morris}}, \ and\ \bibinfo {author} {\bibfnamefont {David~A}\
  \bibnamefont {Weitz}},\ }\bibfield  {title} {\enquote {\bibinfo {title}
  {{Mechanics and dynamics of reconstituted cytoskeletal systems}},}\
  }\href@noop {} {\bibfield  {journal} {\bibinfo  {journal} {Biochim. Biophys.
  Acta}\ }\textbf {\bibinfo {volume} {1853}},\ \bibinfo {pages} {3038--3042}
  (\bibinfo {year} {2015})}\BibitemShut {NoStop}%
\bibitem [{\citenamefont {Lieleg}\ \emph {et~al.}(2007)\citenamefont {Lieleg},
  \citenamefont {Claessens}, \citenamefont {Heussinger}, \citenamefont {Frey},\
  and\ \citenamefont {Bausch}}]{Lieleg2007}%
  \BibitemOpen
  \bibfield  {author} {\bibinfo {author} {\bibfnamefont {O}~\bibnamefont
  {Lieleg}}, \bibinfo {author} {\bibfnamefont {M~M A~E}\ \bibnamefont
  {Claessens}}, \bibinfo {author} {\bibfnamefont {C}~\bibnamefont
  {Heussinger}}, \bibinfo {author} {\bibfnamefont {E}~\bibnamefont {Frey}}, \
  and\ \bibinfo {author} {\bibfnamefont {A~R}\ \bibnamefont {Bausch}},\
  }\bibfield  {title} {\enquote {\bibinfo {title} {{Mechanics of Bundled
  Semiflexible Polymer Networks}},}\ }\href {\doibase
  10.1103/PhysRevLett.99.088102} {\bibfield  {journal} {\bibinfo  {journal}
  {Phys. Rev. Lett.}\ }\textbf {\bibinfo {volume} {99}},\ \bibinfo {pages}
  {88102} (\bibinfo {year} {2007})}\BibitemShut {NoStop}%
\bibitem [{\citenamefont {Lieleg}\ \emph {et~al.}(2010)\citenamefont {Lieleg},
  \citenamefont {Claessens},\ and\ \citenamefont {Bausch}}]{Lieleg_Rev_2010}%
  \BibitemOpen
  \bibfield  {author} {\bibinfo {author} {\bibfnamefont {Oliver}\ \bibnamefont
  {Lieleg}}, \bibinfo {author} {\bibfnamefont {Mireille M A~E}\ \bibnamefont
  {Claessens}}, \ and\ \bibinfo {author} {\bibfnamefont {Andreas~R}\
  \bibnamefont {Bausch}},\ }\bibfield  {title} {\enquote {\bibinfo {title}
  {{Structure and dynamics of cross-linked actin networks}},}\ }\href {\doibase
  10.1039/B912163N} {\bibfield  {journal} {\bibinfo  {journal} {Soft Matter}\
  }\textbf {\bibinfo {volume} {6}},\ \bibinfo {pages} {218--225} (\bibinfo
  {year} {2010})}\BibitemShut {NoStop}%
\bibitem [{\citenamefont {Mahaffy}\ \emph {et~al.}(2000)\citenamefont
  {Mahaffy}, \citenamefont {Shih}, \citenamefont {MacKintosh},\ and\
  \citenamefont {K{\"{a}}s}}]{Mahaffy00}%
  \BibitemOpen
  \bibfield  {author} {\bibinfo {author} {\bibfnamefont {R~E}\ \bibnamefont
  {Mahaffy}}, \bibinfo {author} {\bibfnamefont {C~K}\ \bibnamefont {Shih}},
  \bibinfo {author} {\bibfnamefont {F~C}\ \bibnamefont {MacKintosh}}, \ and\
  \bibinfo {author} {\bibfnamefont {J}~\bibnamefont {K{\"{a}}s}},\ }\bibfield
  {title} {\enquote {\bibinfo {title} {{Scanning Probe-Based
  Frequency-Dependent Microrheology of Polymer Gels and Biological Cells}},}\
  }\href {\doibase 10.1103/PhysRevLett.85.880} {\bibfield  {journal} {\bibinfo
  {journal} {Phys. Rev. Lett.}\ }\textbf {\bibinfo {volume} {85}},\ \bibinfo
  {pages} {880--883} (\bibinfo {year} {2000})}\BibitemShut {NoStop}%
\bibitem [{\citenamefont {Gardel}\ \emph {et~al.}(2003)\citenamefont {Gardel},
  \citenamefont {Valentine}, \citenamefont {Crocker}, \citenamefont {Bausch},\
  and\ \citenamefont {Weitz}}]{Gardel03}%
  \BibitemOpen
  \bibfield  {author} {\bibinfo {author} {\bibfnamefont {M~L}\ \bibnamefont
  {Gardel}}, \bibinfo {author} {\bibfnamefont {M~T}\ \bibnamefont {Valentine}},
  \bibinfo {author} {\bibfnamefont {J~C}\ \bibnamefont {Crocker}}, \bibinfo
  {author} {\bibfnamefont {A~R}\ \bibnamefont {Bausch}}, \ and\ \bibinfo
  {author} {\bibfnamefont {D~A}\ \bibnamefont {Weitz}},\ }\bibfield  {title}
  {\enquote {\bibinfo {title} {{Microrheology of Entangled F-Actin
  Solutions}},}\ }\href {\doibase 10.1103/PhysRevLett.91.158302} {\bibfield
  {journal} {\bibinfo  {journal} {Phys. Rev. Lett.}\ }\textbf {\bibinfo
  {volume} {91}},\ \bibinfo {pages} {158302} (\bibinfo {year}
  {2003})}\BibitemShut {NoStop}%
\bibitem [{\citenamefont {Tseng}\ \emph {et~al.}(2002)\citenamefont {Tseng},
  \citenamefont {Kole},\ and\ \citenamefont {Wirtz}}]{Tseng02}%
  \BibitemOpen
  \bibfield  {author} {\bibinfo {author} {\bibfnamefont {Yiider}\ \bibnamefont
  {Tseng}}, \bibinfo {author} {\bibfnamefont {Thomas~P}\ \bibnamefont {Kole}},
  \ and\ \bibinfo {author} {\bibfnamefont {Denis}\ \bibnamefont {Wirtz}},\
  }\bibfield  {title} {\enquote {\bibinfo {title} {{Micromechanical mapping of
  live cells by multiple-particle-tracking microrheology.}}}\ }\href@noop {}
  {\bibfield  {journal} {\bibinfo  {journal} {Biophys. J.}\ }\textbf {\bibinfo
  {volume} {83}},\ \bibinfo {pages} {3162--3176} (\bibinfo {year}
  {2002})}\BibitemShut {NoStop}%
\bibitem [{\citenamefont {Keller}\ \emph {et~al.}(2003)\citenamefont {Keller},
  \citenamefont {Tharmann}, \citenamefont {Dichtl}, \citenamefont {Bausch},\
  and\ \citenamefont {Sackmann}}]{Keller03}%
  \BibitemOpen
  \bibfield  {author} {\bibinfo {author} {\bibfnamefont {Manfred}\ \bibnamefont
  {Keller}}, \bibinfo {author} {\bibfnamefont {Rainer}\ \bibnamefont
  {Tharmann}}, \bibinfo {author} {\bibfnamefont {Marius~A}\ \bibnamefont
  {Dichtl}}, \bibinfo {author} {\bibfnamefont {Andreas~R}\ \bibnamefont
  {Bausch}}, \ and\ \bibinfo {author} {\bibfnamefont {Erich}\ \bibnamefont
  {Sackmann}},\ }\bibfield  {title} {\enquote {\bibinfo {title} {{Slow filament
  dynamics and viscoelasticity in entangled and active actin networks}},}\
  }\href {\doibase 10.1098/rsta.2002.1158} {\ \textbf {\bibinfo {volume}
  {361}},\ \bibinfo {pages} {699--712} (\bibinfo {year} {2003})}\BibitemShut
  {NoStop}%
\bibitem [{\citenamefont {Uhde}\ \emph {et~al.}(2004)\citenamefont {Uhde},
  \citenamefont {Keller}, \citenamefont {Sackmann}, \citenamefont
  {Parmeggiani},\ and\ \citenamefont {Frey}}]{Uhde04}%
  \BibitemOpen
  \bibfield  {author} {\bibinfo {author} {\bibfnamefont {J{\"{o}}rg}\
  \bibnamefont {Uhde}}, \bibinfo {author} {\bibfnamefont {Manfred}\
  \bibnamefont {Keller}}, \bibinfo {author} {\bibfnamefont {Erich}\
  \bibnamefont {Sackmann}}, \bibinfo {author} {\bibfnamefont {Andrea}\
  \bibnamefont {Parmeggiani}}, \ and\ \bibinfo {author} {\bibfnamefont {Erwin}\
  \bibnamefont {Frey}},\ }\bibfield  {title} {\enquote {\bibinfo {title}
  {{Internal Motility in Stiffening Actin-Myosin Networks}},}\ }\href {\doibase
  10.1103/PhysRevLett.93.268101} {\bibfield  {journal} {\bibinfo  {journal}
  {Phys. Rev. Lett.}\ }\textbf {\bibinfo {volume} {93}},\ \bibinfo {pages}
  {268101} (\bibinfo {year} {2004})}\BibitemShut {NoStop}%
\bibitem [{\citenamefont {Wilhelm}(2008)}]{Wilhelm08}%
  \BibitemOpen
  \bibfield  {author} {\bibinfo {author} {\bibfnamefont {Claire}\ \bibnamefont
  {Wilhelm}},\ }\bibfield  {title} {\enquote {\bibinfo {title}
  {{Out-of-Equilibrium Microrheology inside Living Cells}},}\ }\href {\doibase
  10.1103/PhysRevLett.101.028101} {\bibfield  {journal} {\bibinfo  {journal}
  {Phys. Rev. Lett.}\ }\textbf {\bibinfo {volume} {101}},\ \bibinfo {pages}
  {28101} (\bibinfo {year} {2008})}\BibitemShut {NoStop}%
\bibitem [{\citenamefont {Fabry}\ \emph {et~al.}(2001)\citenamefont {Fabry},
  \citenamefont {Maksym}, \citenamefont {Butler}, \citenamefont {Glogauer},
  \citenamefont {Navajas},\ and\ \citenamefont {Fredberg}}]{Fabry2001}%
  \BibitemOpen
  \bibfield  {author} {\bibinfo {author} {\bibfnamefont {Ben}\ \bibnamefont
  {Fabry}}, \bibinfo {author} {\bibfnamefont {Geoffrey~N}\ \bibnamefont
  {Maksym}}, \bibinfo {author} {\bibfnamefont {James~P}\ \bibnamefont
  {Butler}}, \bibinfo {author} {\bibfnamefont {Michael}\ \bibnamefont
  {Glogauer}}, \bibinfo {author} {\bibfnamefont {Daniel}\ \bibnamefont
  {Navajas}}, \ and\ \bibinfo {author} {\bibfnamefont {Jeffrey~J}\ \bibnamefont
  {Fredberg}},\ }\bibfield  {title} {\enquote {\bibinfo {title} {{Scaling the
  Microrheology of Living Cells}},}\ }\href {\doibase
  10.1103/PhysRevLett.87.148102} {\bibfield  {journal} {\bibinfo  {journal}
  {Phys. Rev. Lett.}\ }\textbf {\bibinfo {volume} {87}},\ \bibinfo {pages}
  {148102} (\bibinfo {year} {2001})}\BibitemShut {NoStop}%
\bibitem [{\citenamefont {Bausch}\ \emph {et~al.}(1999)\citenamefont {Bausch},
  \citenamefont {M{\"{o}}ller},\ and\ \citenamefont {Sackmann}}]{Bausch1999}%
  \BibitemOpen
  \bibfield  {author} {\bibinfo {author} {\bibfnamefont {A~R}\ \bibnamefont
  {Bausch}}, \bibinfo {author} {\bibfnamefont {W}~\bibnamefont {M{\"{o}}ller}},
  \ and\ \bibinfo {author} {\bibfnamefont {E}~\bibnamefont {Sackmann}},\
  }\bibfield  {title} {\enquote {\bibinfo {title} {{Measurement of local
  viscoelasticity and forces in living cells by magnetic tweezers.}}}\
  }\href@noop {} {\bibfield  {journal} {\bibinfo  {journal} {Biophys. J.}\
  }\textbf {\bibinfo {volume} {76}},\ \bibinfo {pages} {573--579} (\bibinfo
  {year} {1999})}\BibitemShut {NoStop}%
\bibitem [{\citenamefont {Ma}\ \emph {et~al.}(2014)\citenamefont {Ma},
  \citenamefont {Klindt}, \citenamefont {Riedel-Kruse}, \citenamefont
  {J{\"{u}}licher},\ and\ \citenamefont {Friedrich}}]{ma2014active}%
  \BibitemOpen
  \bibfield  {author} {\bibinfo {author} {\bibfnamefont {Rui}\ \bibnamefont
  {Ma}}, \bibinfo {author} {\bibfnamefont {Gary~S}\ \bibnamefont {Klindt}},
  \bibinfo {author} {\bibfnamefont {Ingmar~H}\ \bibnamefont {Riedel-Kruse}},
  \bibinfo {author} {\bibfnamefont {Frank}\ \bibnamefont {J{\"{u}}licher}}, \
  and\ \bibinfo {author} {\bibfnamefont {Benjamin~M}\ \bibnamefont
  {Friedrich}},\ }\bibfield  {title} {\enquote {\bibinfo {title} {{Active Phase
  and Amplitude Fluctuations of Flagellar Beating}},}\ }\href {\doibase
  10.1103/PhysRevLett.113.048101} {\bibfield  {journal} {\bibinfo  {journal}
  {Phys. Rev. Lett.}\ }\textbf {\bibinfo {volume} {113}},\ \bibinfo {pages}
  {048101} (\bibinfo {year} {2014})}\BibitemShut {NoStop}%
\bibitem [{\citenamefont {Callen}\ and\ \citenamefont
  {Welton}(1951)}]{CallenFDT}%
  \BibitemOpen
  \bibfield  {author} {\bibinfo {author} {\bibfnamefont {Herbert~B}\
  \bibnamefont {Callen}}\ and\ \bibinfo {author} {\bibfnamefont {Theodore~A}\
  \bibnamefont {Welton}},\ }\bibfield  {title} {\enquote {\bibinfo {title}
  {{Irreversibility and Generalized Noise}},}\ }\href {\doibase
  10.1103/PhysRev.83.34} {\bibfield  {journal} {\bibinfo  {journal} {Phys.
  Rev.}\ }\textbf {\bibinfo {volume} {83}},\ \bibinfo {pages} {34--40}
  (\bibinfo {year} {1951})}\BibitemShut {NoStop}%
\bibitem [{\citenamefont {Kubo}(1966)}]{Kubo_FDT}%
  \BibitemOpen
  \bibfield  {author} {\bibinfo {author} {\bibfnamefont {Rep}\ \bibnamefont
  {Kubo}},\ }\bibfield  {title} {\enquote {\bibinfo {title} {{The
  fluctuation-dissipation theorem}},}\ }\href@noop {} {\bibfield  {journal}
  {\bibinfo  {journal} {Reports Prog. Phys.}\ }\textbf {\bibinfo {volume}
  {29}},\ \bibinfo {pages} {255} (\bibinfo {year} {1966})}\BibitemShut
  {NoStop}%
\bibitem [{\citenamefont {Martin}\ \emph {et~al.}(2001)\citenamefont {Martin},
  \citenamefont {Hudspeth},\ and\ \citenamefont
  {J{\"{u}}licher}}]{Juelicher2001}%
  \BibitemOpen
  \bibfield  {author} {\bibinfo {author} {\bibfnamefont {P}~\bibnamefont
  {Martin}}, \bibinfo {author} {\bibfnamefont {A~J}\ \bibnamefont {Hudspeth}},
  \ and\ \bibinfo {author} {\bibfnamefont {F}~\bibnamefont {J{\"{u}}licher}},\
  }\bibfield  {title} {\enquote {\bibinfo {title} {{Comparison of a hair
  bundle's spontaneous oscillations with its response to mechanical stimulation
  reveals the underlying active process}},}\ }\href {\doibase
  10.1073/pnas.251530598} {\bibfield  {journal} {\bibinfo  {journal} {Proc.
  Natl. Acad. Sci.}\ }\textbf {\bibinfo {volume} {98}},\ \bibinfo {pages}
  {14380--14385} (\bibinfo {year} {2001})}\BibitemShut {NoStop}%
\bibitem [{\citenamefont {{Van Dijk}}\ \emph {et~al.}(2011)\citenamefont {{Van
  Dijk}}, \citenamefont {Mason}, \citenamefont {Schoffelen}, \citenamefont
  {Narins},\ and\ \citenamefont {Meenderink}}]{vanDijk_EarMechanics_2011}%
  \BibitemOpen
  \bibfield  {author} {\bibinfo {author} {\bibfnamefont {Pim}\ \bibnamefont
  {{Van Dijk}}}, \bibinfo {author} {\bibfnamefont {Matthew~J}\ \bibnamefont
  {Mason}}, \bibinfo {author} {\bibfnamefont {Richard~L.M.}\ \bibnamefont
  {Schoffelen}}, \bibinfo {author} {\bibfnamefont {Peter~M}\ \bibnamefont
  {Narins}}, \ and\ \bibinfo {author} {\bibfnamefont {Sebastiaan~W.F.}\
  \bibnamefont {Meenderink}},\ }\bibfield  {title} {\enquote {\bibinfo {title}
  {{Mechanics of the frog ear}},}\ }\href {\doibase
  10.1016/j.heares.2010.02.004} {\bibfield  {journal} {\bibinfo  {journal}
  {Hear. Res.}\ }\textbf {\bibinfo {volume} {273}},\ \bibinfo {pages} {46--58}
  (\bibinfo {year} {2011})}\BibitemShut {NoStop}%
\bibitem [{\citenamefont {Cugliandolo}\ \emph {et~al.}(1997)\citenamefont
  {Cugliandolo}, \citenamefont {Kurchan},\ and\ \citenamefont
  {Peliti}}]{Cugliandolo_EffT_97}%
  \BibitemOpen
  \bibfield  {author} {\bibinfo {author} {\bibfnamefont {Leticia~F}\
  \bibnamefont {Cugliandolo}}, \bibinfo {author} {\bibfnamefont {Jorge}\
  \bibnamefont {Kurchan}}, \ and\ \bibinfo {author} {\bibfnamefont {Luca}\
  \bibnamefont {Peliti}},\ }\bibfield  {title} {\enquote {\bibinfo {title}
  {{Energy flow, partial equilibration, and effective temperatures in systems
  with slow dynamics}},}\ }\href {\doibase 10.1103/PhysRevE.55.3898} {\bibfield
   {journal} {\bibinfo  {journal} {Phys. Rev. E}\ }\textbf {\bibinfo {volume}
  {55}},\ \bibinfo {pages} {3898--3914} (\bibinfo {year} {1997})}\BibitemShut
  {NoStop}%
\bibitem [{\citenamefont {Cugliandolo}(2011)}]{cugliandolo2011effective}%
  \BibitemOpen
  \bibfield  {author} {\bibinfo {author} {\bibfnamefont {Leticia~F}\
  \bibnamefont {Cugliandolo}},\ }\bibfield  {title} {\enquote {\bibinfo {title}
  {{The effective temperature}},}\ }\href@noop {} {\bibfield  {journal}
  {\bibinfo  {journal} {J. Phys. A Math. Theor.}\ }\textbf {\bibinfo {volume}
  {44}},\ \bibinfo {pages} {483001} (\bibinfo {year} {2011})}\BibitemShut
  {NoStop}%
\bibitem [{\citenamefont {Loi}\ \emph {et~al.}(2008)\citenamefont {Loi},
  \citenamefont {Mossa},\ and\ \citenamefont {Cugliandolo}}]{loi2008effective}%
  \BibitemOpen
  \bibfield  {author} {\bibinfo {author} {\bibfnamefont {Davide}\ \bibnamefont
  {Loi}}, \bibinfo {author} {\bibfnamefont {Stefano}\ \bibnamefont {Mossa}}, \
  and\ \bibinfo {author} {\bibfnamefont {Leticia~F}\ \bibnamefont
  {Cugliandolo}},\ }\bibfield  {title} {\enquote {\bibinfo {title} {{Effective
  temperature of active matter}},}\ }\href@noop {} {\bibfield  {journal}
  {\bibinfo  {journal} {Phys. Rev. E}\ }\textbf {\bibinfo {volume} {77}},\
  \bibinfo {pages} {51111} (\bibinfo {year} {2008})}\BibitemShut {NoStop}%
\bibitem [{\citenamefont {Bursac}\ \emph {et~al.}(2005)\citenamefont {Bursac},
  \citenamefont {Lenormand}, \citenamefont {Fabry}, \citenamefont {Oliver},
  \citenamefont {Weitz}, \citenamefont {Viasnoff}, \citenamefont {Butler},\
  and\ \citenamefont {Fredberg}}]{bursac2005cytoskeletal}%
  \BibitemOpen
  \bibfield  {author} {\bibinfo {author} {\bibfnamefont {Predrag}\ \bibnamefont
  {Bursac}}, \bibinfo {author} {\bibfnamefont {Guillaume}\ \bibnamefont
  {Lenormand}}, \bibinfo {author} {\bibfnamefont {Ben}\ \bibnamefont {Fabry}},
  \bibinfo {author} {\bibfnamefont {Madavi}\ \bibnamefont {Oliver}}, \bibinfo
  {author} {\bibfnamefont {David~A}\ \bibnamefont {Weitz}}, \bibinfo {author}
  {\bibfnamefont {Virgile}\ \bibnamefont {Viasnoff}}, \bibinfo {author}
  {\bibfnamefont {James~P}\ \bibnamefont {Butler}}, \ and\ \bibinfo {author}
  {\bibfnamefont {Jeffrey~J}\ \bibnamefont {Fredberg}},\ }\bibfield  {title}
  {\enquote {\bibinfo {title} {{Cytoskeletal remodelling and slow dynamics in
  the living cell}},}\ }\href@noop {} {\bibfield  {journal} {\bibinfo
  {journal} {Nat. Mater.}\ }\textbf {\bibinfo {volume} {4}},\ \bibinfo {pages}
  {557--561} (\bibinfo {year} {2005})}\BibitemShut {NoStop}%
\bibitem [{\citenamefont {Prost}\ \emph {et~al.}(2009)\citenamefont {Prost},
  \citenamefont {Joanny},\ and\ \citenamefont
  {Parrondo}}]{prost2009generalized}%
  \BibitemOpen
  \bibfield  {author} {\bibinfo {author} {\bibfnamefont {J}~\bibnamefont
  {Prost}}, \bibinfo {author} {\bibfnamefont {J-F}\ \bibnamefont {Joanny}}, \
  and\ \bibinfo {author} {\bibfnamefont {J~M~R}\ \bibnamefont {Parrondo}},\
  }\bibfield  {title} {\enquote {\bibinfo {title} {{Generalized
  Fluctuation-Dissipation Theorem for Steady-State Systems}},}\ }\href
  {\doibase 10.1103/PhysRevLett.103.090601} {\bibfield  {journal} {\bibinfo
  {journal} {Phys. Rev. Lett.}\ }\textbf {\bibinfo {volume} {103}},\ \bibinfo
  {pages} {090601} (\bibinfo {year} {2009})}\BibitemShut {NoStop}%
\bibitem [{\citenamefont {Wilhelm}(2009)}]{Wilhelm2009}%
  \BibitemOpen
  \bibfield  {author} {\bibinfo {author} {\bibfnamefont {Claire}\ \bibnamefont
  {Wilhelm}},\ }\bibfield  {title} {\enquote {\bibinfo {title} {{Effective
  temperature inside living cells}},}\ }\href@noop {} {\bibfield  {journal}
  {\bibinfo  {journal} {MRS Online Proc. Libr. Arch.}\ }\textbf {\bibinfo
  {volume} {1227}} (\bibinfo {year} {2009})}\BibitemShut {NoStop}%
\bibitem [{\citenamefont {Fodor}\ \emph {et~al.}(2016)\citenamefont {Fodor},
  \citenamefont {Nardini}, \citenamefont {Cates}, \citenamefont {Tailleur},
  \citenamefont {Visco},\ and\ \citenamefont {van Wijland}}]{cates_how_far}%
  \BibitemOpen
  \bibfield  {author} {\bibinfo {author} {\bibfnamefont {{\'{E}}tienne}\
  \bibnamefont {Fodor}}, \bibinfo {author} {\bibfnamefont {Cesare}\
  \bibnamefont {Nardini}}, \bibinfo {author} {\bibfnamefont {Michael~E}\
  \bibnamefont {Cates}}, \bibinfo {author} {\bibfnamefont {Julien}\
  \bibnamefont {Tailleur}}, \bibinfo {author} {\bibfnamefont {Paolo}\
  \bibnamefont {Visco}}, \ and\ \bibinfo {author} {\bibfnamefont
  {Fr{\'{e}}d{\'{e}}ric}\ \bibnamefont {van Wijland}},\ }\bibfield  {title}
  {\enquote {\bibinfo {title} {{How Far from Equilibrium Is Active Matter?}}}\
  }\href {\doibase 10.1103/PhysRevLett.117.038103} {\bibfield  {journal}
  {\bibinfo  {journal} {Phys. Rev. Lett.}\ }\textbf {\bibinfo {volume} {117}},\
  \bibinfo {pages} {38103} (\bibinfo {year} {2016})}\BibitemShut {NoStop}%
\bibitem [{\citenamefont {Fodor}\ \emph
  {et~al.}(2015{\natexlab{b}})\citenamefont {Fodor}, \citenamefont {Guo},
  \citenamefont {Gov}, \citenamefont {Visco}, \citenamefont {Weitz},\ and\
  \citenamefont {van Wijland}}]{fodor2015activity}%
  \BibitemOpen
  \bibfield  {author} {\bibinfo {author} {\bibfnamefont {{\'{E}}}~\bibnamefont
  {Fodor}}, \bibinfo {author} {\bibfnamefont {M}~\bibnamefont {Guo}}, \bibinfo
  {author} {\bibfnamefont {N~S}\ \bibnamefont {Gov}}, \bibinfo {author}
  {\bibfnamefont {P}~\bibnamefont {Visco}}, \bibinfo {author} {\bibfnamefont
  {D~A}\ \bibnamefont {Weitz}}, \ and\ \bibinfo {author} {\bibfnamefont
  {F}~\bibnamefont {van Wijland}},\ }\bibfield  {title} {\enquote {\bibinfo
  {title} {{Activity-driven fluctuations in living cells}},}\ }\href@noop {}
  {\bibfield  {journal} {\bibinfo  {journal} {EPL (Europhysics Lett.}\ }\textbf
  {\bibinfo {volume} {110}},\ \bibinfo {pages} {48005} (\bibinfo {year}
  {2015}{\natexlab{b}})}\BibitemShut {NoStop}%
\bibitem [{\citenamefont {Dieterich}\ \emph {et~al.}(2015)\citenamefont
  {Dieterich}, \citenamefont {Seifert}, \citenamefont {Ritort}, \citenamefont
  {Camunas-Soler}, \citenamefont {Ribezzi-Crivellari}, \citenamefont
  {Seifert},\ and\ \citenamefont {Ritort}}]{Dieterich2015}%
  \BibitemOpen
  \bibfield  {author} {\bibinfo {author} {\bibfnamefont {E}~\bibnamefont
  {Dieterich}}, \bibinfo {author} {\bibfnamefont {U}~\bibnamefont {Seifert}},
  \bibinfo {author} {\bibfnamefont {F}~\bibnamefont {Ritort}}, \bibinfo
  {author} {\bibfnamefont {J.}~\bibnamefont {Camunas-Soler}}, \bibinfo {author}
  {\bibfnamefont {M.}~\bibnamefont {Ribezzi-Crivellari}}, \bibinfo {author}
  {\bibfnamefont {U}~\bibnamefont {Seifert}}, \ and\ \bibinfo {author}
  {\bibfnamefont {F}~\bibnamefont {Ritort}},\ }\bibfield  {title} {\enquote
  {\bibinfo {title} {{Single-molecule measurement of the effective temperature
  in non-equilibrium steady states}},}\ }\href {\doibase 10.1038/NPHYS3435}
  {\bibfield  {journal} {\bibinfo  {journal} {Nat. Phys.}\ }\textbf {\bibinfo
  {volume} {11}},\ \bibinfo {pages} {1--8} (\bibinfo {year}
  {2015})}\BibitemShut {NoStop}%
\bibitem [{\citenamefont {Chaikin}\ and\ \citenamefont
  {Lubensky}(1995)}]{chaikin1995}%
  \BibitemOpen
  \bibfield  {author} {\bibinfo {author} {\bibfnamefont {P.~M.}\ \bibnamefont
  {Chaikin}}\ and\ \bibinfo {author} {\bibfnamefont {T.~C.}\ \bibnamefont
  {Lubensky}},\ }\href@noop {} {\emph {\bibinfo {title} {{Principles of
  condensed matter physics}}}}\ (\bibinfo  {publisher} {Cambridge University
  Press},\ \bibinfo {year} {1995})\BibitemShut {NoStop}%
\bibitem [{\citenamefont {Yamada}\ \emph {et~al.}(2000)\citenamefont {Yamada},
  \citenamefont {Wirtz},\ and\ \citenamefont {Kuo}}]{Yamada2000}%
  \BibitemOpen
  \bibfield  {author} {\bibinfo {author} {\bibfnamefont {Soichiro}\
  \bibnamefont {Yamada}}, \bibinfo {author} {\bibfnamefont {Denis}\
  \bibnamefont {Wirtz}}, \ and\ \bibinfo {author} {\bibfnamefont {Scot~C}\
  \bibnamefont {Kuo}},\ }\bibfield  {title} {\enquote {\bibinfo {title}
  {{Mechanics of Living Cells Measured by Laser Tracking Microrheology}},}\
  }\href {http://www.sciencedirect.com/science/article/pii/S0006349500767257}
  {\bibfield  {journal} {\bibinfo  {journal} {Biophys. J.}\ }\textbf {\bibinfo
  {volume} {78}},\ \bibinfo {pages} {1736--1747} (\bibinfo {year}
  {2000})}\BibitemShut {NoStop}%
\bibitem [{\citenamefont {Mason}\ and\ \citenamefont {Weitz}(1995)}]{Weitz95}%
  \BibitemOpen
  \bibfield  {author} {\bibinfo {author} {\bibfnamefont {T~G}\ \bibnamefont
  {Mason}}\ and\ \bibinfo {author} {\bibfnamefont {D~A}\ \bibnamefont
  {Weitz}},\ }\bibfield  {title} {\enquote {\bibinfo {title} {{Optical
  Measurements of Frequency-Dependent Linear Viscoelastic Moduli of Complex
  Fluids}},}\ }\href {\doibase 10.1103/PhysRevLett.74.1250} {\bibfield
  {journal} {\bibinfo  {journal} {Phys. Rev. Lett.}\ }\textbf {\bibinfo
  {volume} {74}},\ \bibinfo {pages} {1250--1253} (\bibinfo {year}
  {1995})}\BibitemShut {NoStop}%
\bibitem [{\citenamefont {Ziemann}\ \emph {et~al.}(1994)\citenamefont
  {Ziemann}, \citenamefont {R{\"{a}}dler},\ and\ \citenamefont
  {Sackmann}}]{Ziemann94}%
  \BibitemOpen
  \bibfield  {author} {\bibinfo {author} {\bibfnamefont {F}~\bibnamefont
  {Ziemann}}, \bibinfo {author} {\bibfnamefont {J}~\bibnamefont
  {R{\"{a}}dler}}, \ and\ \bibinfo {author} {\bibfnamefont {E}~\bibnamefont
  {Sackmann}},\ }\bibfield  {title} {\enquote {\bibinfo {title} {{Local
  measurements of viscoelastic moduli of entangled actin networks using an
  oscillating magnetic bead micro-rheometer.}}}\ }\href@noop {} {\bibfield
  {journal} {\bibinfo  {journal} {Biophys. J.}\ }\textbf {\bibinfo {volume}
  {66}},\ \bibinfo {pages} {2210--2216} (\bibinfo {year} {1994})}\BibitemShut
  {NoStop}%
\bibitem [{\citenamefont {Amblard}\ \emph {et~al.}(1996)\citenamefont
  {Amblard}, \citenamefont {Maggs}, \citenamefont {Yurke}, \citenamefont
  {Pargellis},\ and\ \citenamefont {Leibler}}]{Amblard96}%
  \BibitemOpen
  \bibfield  {author} {\bibinfo {author} {\bibfnamefont {F}~\bibnamefont
  {Amblard}}, \bibinfo {author} {\bibfnamefont {A~C}\ \bibnamefont {Maggs}},
  \bibinfo {author} {\bibfnamefont {B}~\bibnamefont {Yurke}}, \bibinfo {author}
  {\bibfnamefont {A~N}\ \bibnamefont {Pargellis}}, \ and\ \bibinfo {author}
  {\bibfnamefont {S}~\bibnamefont {Leibler}},\ }\bibfield  {title} {\enquote
  {\bibinfo {title} {{Subdiffusion and Anomalous Local Viscoelasticity in Actin
  Networks}},}\ }\href {\doibase 10.1103/PhysRevLett.77.4470} {\bibfield
  {journal} {\bibinfo  {journal} {Phys. Rev. Lett.}\ }\textbf {\bibinfo
  {volume} {77}},\ \bibinfo {pages} {4470--4473} (\bibinfo {year}
  {1996})}\BibitemShut {NoStop}%
\bibitem [{\citenamefont {Schmidt}\ \emph {et~al.}(1996)\citenamefont
  {Schmidt}, \citenamefont {Ziemann},\ and\ \citenamefont
  {Sackmann}}]{Schmidt1996}%
  \BibitemOpen
  \bibfield  {author} {\bibinfo {author} {\bibfnamefont {Frank~G.}\
  \bibnamefont {Schmidt}}, \bibinfo {author} {\bibfnamefont {Florian}\
  \bibnamefont {Ziemann}}, \ and\ \bibinfo {author} {\bibfnamefont {Erich}\
  \bibnamefont {Sackmann}},\ }\bibfield  {title} {\enquote {\bibinfo {title}
  {{Shear field mapping in actin networks by using magnetic tweezers}},}\
  }\href {\doibase 10.1007/BF00180376} {\bibfield  {journal} {\bibinfo
  {journal} {Eur. Biophys. J.}\ }\textbf {\bibinfo {volume} {24}},\ \bibinfo
  {pages} {348--353} (\bibinfo {year} {1996})}\BibitemShut {NoStop}%
\bibitem [{\citenamefont {Beroz}\ \emph {et~al.}(2017)\citenamefont {Beroz},
  \citenamefont {Jawerth}, \citenamefont {M{\"{u}}nster}, \citenamefont
  {Weitz}, \citenamefont {Broedersz},\ and\ \citenamefont
  {Wingreen}}]{beroz2017physical}%
  \BibitemOpen
  \bibfield  {author} {\bibinfo {author} {\bibfnamefont {Farzan}\ \bibnamefont
  {Beroz}}, \bibinfo {author} {\bibfnamefont {Louise~M}\ \bibnamefont
  {Jawerth}}, \bibinfo {author} {\bibfnamefont {Stefan}\ \bibnamefont
  {M{\"{u}}nster}}, \bibinfo {author} {\bibfnamefont {David~A}\ \bibnamefont
  {Weitz}}, \bibinfo {author} {\bibfnamefont {Chase~P}\ \bibnamefont
  {Broedersz}}, \ and\ \bibinfo {author} {\bibfnamefont {Ned~S}\ \bibnamefont
  {Wingreen}},\ }\bibfield  {title} {\enquote {\bibinfo {title} {{Physical
  limits to biomechanical sensing in disordered fibre networks.}}}\ }\href@noop
  {} {\bibfield  {journal} {\bibinfo  {journal} {Nat. Commun.}\ }\textbf
  {\bibinfo {volume} {8}},\ \bibinfo {pages} {16096} (\bibinfo {year}
  {2017})}\BibitemShut {NoStop}%
\bibitem [{\citenamefont {Jones}\ \emph {et~al.}(2015)\citenamefont {Jones},
  \citenamefont {Cibula}, \citenamefont {Feng}, \citenamefont {Krnacik},
  \citenamefont {McIntyre}, \citenamefont {Levine},\ and\ \citenamefont
  {Sun}}]{jones2015micromechanics}%
  \BibitemOpen
  \bibfield  {author} {\bibinfo {author} {\bibfnamefont {Christopher A~R}\
  \bibnamefont {Jones}}, \bibinfo {author} {\bibfnamefont {Matthew}\
  \bibnamefont {Cibula}}, \bibinfo {author} {\bibfnamefont {Jingchen}\
  \bibnamefont {Feng}}, \bibinfo {author} {\bibfnamefont {Emma~A}\ \bibnamefont
  {Krnacik}}, \bibinfo {author} {\bibfnamefont {David~H}\ \bibnamefont
  {McIntyre}}, \bibinfo {author} {\bibfnamefont {Herbert}\ \bibnamefont
  {Levine}}, \ and\ \bibinfo {author} {\bibfnamefont {Bo}~\bibnamefont {Sun}},\
  }\bibfield  {title} {\enquote {\bibinfo {title} {{Micromechanics of
  cellularized biopolymer networks}},}\ }\href@noop {} {\bibfield  {journal}
  {\bibinfo  {journal} {Proc. Natl. Acad. Sci.}\ }\textbf {\bibinfo {volume}
  {112}},\ \bibinfo {pages} {E5117----E5122} (\bibinfo {year}
  {2015})}\BibitemShut {NoStop}%
\bibitem [{\citenamefont {Crocker}\ \emph {et~al.}(2000)\citenamefont
  {Crocker}, \citenamefont {Valentine}, \citenamefont {Weeks}, \citenamefont
  {Gisler}, \citenamefont {Kaplan}, \citenamefont {Yodh},\ and\ \citenamefont
  {Weitz}}]{Crocker2000}%
  \BibitemOpen
  \bibfield  {author} {\bibinfo {author} {\bibfnamefont {John~C}\ \bibnamefont
  {Crocker}}, \bibinfo {author} {\bibfnamefont {M~T}\ \bibnamefont
  {Valentine}}, \bibinfo {author} {\bibfnamefont {Eric~R}\ \bibnamefont
  {Weeks}}, \bibinfo {author} {\bibfnamefont {T}~\bibnamefont {Gisler}},
  \bibinfo {author} {\bibfnamefont {P~D}\ \bibnamefont {Kaplan}}, \bibinfo
  {author} {\bibfnamefont {A~G}\ \bibnamefont {Yodh}}, \ and\ \bibinfo {author}
  {\bibfnamefont {D~A}\ \bibnamefont {Weitz}},\ }\bibfield  {title} {\enquote
  {\bibinfo {title} {{Two-Point Microrheology of Inhomogeneous Soft
  Materials}},}\ }\href {\doibase 10.1103/PhysRevLett.85.888} {\bibfield
  {journal} {\bibinfo  {journal} {Phys. Rev. Lett.}\ }\textbf {\bibinfo
  {volume} {85}},\ \bibinfo {pages} {888--891} (\bibinfo {year}
  {2000})}\BibitemShut {NoStop}%
\bibitem [{\citenamefont {Schnurr}\ \emph {et~al.}(1997)\citenamefont
  {Schnurr}, \citenamefont {Gittes}, \citenamefont {MacKintosh},\ and\
  \citenamefont {Schmidt}}]{Schnurr1997}%
  \BibitemOpen
  \bibfield  {author} {\bibinfo {author} {\bibfnamefont {B}~\bibnamefont
  {Schnurr}}, \bibinfo {author} {\bibfnamefont {F}~\bibnamefont {Gittes}},
  \bibinfo {author} {\bibfnamefont {F~C}\ \bibnamefont {MacKintosh}}, \ and\
  \bibinfo {author} {\bibfnamefont {C~F}\ \bibnamefont {Schmidt}},\ }\bibfield
  {title} {\enquote {\bibinfo {title} {{Determining Microscopic Viscoelasticity
  in Flexible and Semiflexible Polymer Networks from Thermal Fluctuations}},}\
  }\href {\doibase 10.1021/ma970555n} {\bibfield  {journal} {\bibinfo
  {journal} {Macromolecules}\ }\textbf {\bibinfo {volume} {30}},\ \bibinfo
  {pages} {7781--7792} (\bibinfo {year} {1997})}\BibitemShut {NoStop}%
\bibitem [{\citenamefont {Addas}\ \emph {et~al.}(2004)\citenamefont {Addas},
  \citenamefont {Schmidt},\ and\ \citenamefont {Tang}}]{Addas2004}%
  \BibitemOpen
  \bibfield  {author} {\bibinfo {author} {\bibfnamefont {Karim~M}\ \bibnamefont
  {Addas}}, \bibinfo {author} {\bibfnamefont {Christoph~F}\ \bibnamefont
  {Schmidt}}, \ and\ \bibinfo {author} {\bibfnamefont {Jay~X}\ \bibnamefont
  {Tang}},\ }\bibfield  {title} {\enquote {\bibinfo {title} {{Microrheology of
  solutions of semiflexible biopolymer filaments using laser tweezers
  interferometry}},}\ }\href {\doibase 10.1103/PhysRevE.70.021503} {\bibfield
  {journal} {\bibinfo  {journal} {Phys. Rev. E}\ }\textbf {\bibinfo {volume}
  {70}},\ \bibinfo {pages} {021503} (\bibinfo {year} {2004})}\BibitemShut
  {NoStop}%
\bibitem [{\citenamefont {Gittes}\ and\ \citenamefont
  {MacKintosh}(1998)}]{Gittes1998}%
  \BibitemOpen
  \bibfield  {author} {\bibinfo {author} {\bibfnamefont {F}~\bibnamefont
  {Gittes}}\ and\ \bibinfo {author} {\bibfnamefont {F}~\bibnamefont
  {MacKintosh}},\ }\bibfield  {title} {\enquote {\bibinfo {title} {{Dynamic
  shear modulus of a semiflexible polymer network}},}\ }\href@noop {}
  {\bibfield  {journal} {\bibinfo  {journal} {Phys. Rev. E, Stat. physics,
  plasmas, fluids, Relat. Interdiscip. Top.}\ } (\bibinfo {year}
  {1998})}\BibitemShut {NoStop}%
\bibitem [{\citenamefont {Gittes}\ \emph {et~al.}(1997)\citenamefont {Gittes},
  \citenamefont {Schnurr}, \citenamefont {Olmsted}, \citenamefont
  {MacKintosh},\ and\ \citenamefont {Schmidt}}]{GittesMicro1997}%
  \BibitemOpen
  \bibfield  {author} {\bibinfo {author} {\bibfnamefont {F}~\bibnamefont
  {Gittes}}, \bibinfo {author} {\bibfnamefont {B}~\bibnamefont {Schnurr}},
  \bibinfo {author} {\bibfnamefont {P~D}\ \bibnamefont {Olmsted}}, \bibinfo
  {author} {\bibfnamefont {F~C}\ \bibnamefont {MacKintosh}}, \ and\ \bibinfo
  {author} {\bibfnamefont {C~F}\ \bibnamefont {Schmidt}},\ }\bibfield  {title}
  {\enquote {\bibinfo {title} {{Microscopic Viscoelasticity: Shear Moduli of
  Soft Materials Determined from Thermal Fluctuations}},}\ }\href {\doibase
  10.1103/PhysRevLett.79.3286} {\bibfield  {journal} {\bibinfo  {journal}
  {Phys. Rev. Lett.}\ }\textbf {\bibinfo {volume} {79}},\ \bibinfo {pages}
  {3286--3289} (\bibinfo {year} {1997})}\BibitemShut {NoStop}%
\bibitem [{\citenamefont {Chen}\ \emph {et~al.}(2003)\citenamefont {Chen},
  \citenamefont {Weeks}, \citenamefont {Crocker}, \citenamefont {Islam},
  \citenamefont {Verma}, \citenamefont {Gruber}, \citenamefont {Levine},
  \citenamefont {Lubensky},\ and\ \citenamefont {Yodh}}]{chen2003rheological}%
  \BibitemOpen
  \bibfield  {author} {\bibinfo {author} {\bibfnamefont {D~T}\ \bibnamefont
  {Chen}}, \bibinfo {author} {\bibfnamefont {E~R}\ \bibnamefont {Weeks}},
  \bibinfo {author} {\bibfnamefont {John~C}\ \bibnamefont {Crocker}}, \bibinfo
  {author} {\bibfnamefont {M~F}\ \bibnamefont {Islam}}, \bibinfo {author}
  {\bibfnamefont {R}~\bibnamefont {Verma}}, \bibinfo {author} {\bibfnamefont
  {J}~\bibnamefont {Gruber}}, \bibinfo {author} {\bibfnamefont {A~J}\
  \bibnamefont {Levine}}, \bibinfo {author} {\bibfnamefont {Thomas~C}\
  \bibnamefont {Lubensky}}, \ and\ \bibinfo {author} {\bibfnamefont {A~G}\
  \bibnamefont {Yodh}},\ }\bibfield  {title} {\enquote {\bibinfo {title}
  {{Rheological microscopy: local mechanical properties from microrheology}},}\
  }\href@noop {} {\bibfield  {journal} {\bibinfo  {journal} {Phys. Rev. Lett.}\
  }\textbf {\bibinfo {volume} {90}},\ \bibinfo {pages} {108301} (\bibinfo
  {year} {2003})}\BibitemShut {NoStop}%
\bibitem [{\citenamefont {Mason}(2000)}]{mason2000estimating}%
  \BibitemOpen
  \bibfield  {author} {\bibinfo {author} {\bibfnamefont {Thomas~G}\
  \bibnamefont {Mason}},\ }\bibfield  {title} {\enquote {\bibinfo {title}
  {{Estimating the viscoelastic moduli of complex fluids using the generalized
  Stokes--Einstein equation}},}\ }\href@noop {} {\bibfield  {journal} {\bibinfo
   {journal} {Rheol. Acta}\ }\textbf {\bibinfo {volume} {39}},\ \bibinfo
  {pages} {371--378} (\bibinfo {year} {2000})}\BibitemShut {NoStop}%
\bibitem [{\citenamefont {Helfer}\ \emph {et~al.}(2000)\citenamefont {Helfer},
  \citenamefont {Harlepp}, \citenamefont {Bourdieu}, \citenamefont {Robert},
  \citenamefont {MacKintosh},\ and\ \citenamefont
  {Chatenay}}]{helfer2000microrheology}%
  \BibitemOpen
  \bibfield  {author} {\bibinfo {author} {\bibfnamefont {E}~\bibnamefont
  {Helfer}}, \bibinfo {author} {\bibfnamefont {S}~\bibnamefont {Harlepp}},
  \bibinfo {author} {\bibfnamefont {L}~\bibnamefont {Bourdieu}}, \bibinfo
  {author} {\bibfnamefont {J}~\bibnamefont {Robert}}, \bibinfo {author}
  {\bibfnamefont {F~C}\ \bibnamefont {MacKintosh}}, \ and\ \bibinfo {author}
  {\bibfnamefont {D}~\bibnamefont {Chatenay}},\ }\bibfield  {title} {\enquote
  {\bibinfo {title} {{Microrheology of biopolymer-membrane complexes}},}\
  }\href@noop {} {\bibfield  {journal} {\bibinfo  {journal} {Phys. Rev. Lett.}\
  }\textbf {\bibinfo {volume} {85}},\ \bibinfo {pages} {457} (\bibinfo {year}
  {2000})}\BibitemShut {NoStop}%
\bibitem [{\citenamefont {Fedosov}\ \emph {et~al.}(2010)\citenamefont
  {Fedosov}, \citenamefont {Caswell},\ and\ \citenamefont
  {Karniadakis}}]{fedosov2010multiscale}%
  \BibitemOpen
  \bibfield  {author} {\bibinfo {author} {\bibfnamefont {Dmitry~A}\
  \bibnamefont {Fedosov}}, \bibinfo {author} {\bibfnamefont {Bruce}\
  \bibnamefont {Caswell}}, \ and\ \bibinfo {author} {\bibfnamefont {George~Em}\
  \bibnamefont {Karniadakis}},\ }\bibfield  {title} {\enquote {\bibinfo {title}
  {{A multiscale red blood cell model with accurate mechanics, rheology, and
  dynamics}},}\ }\href@noop {} {\bibfield  {journal} {\bibinfo  {journal}
  {Biophys. J.}\ }\textbf {\bibinfo {volume} {98}},\ \bibinfo {pages}
  {2215--2225} (\bibinfo {year} {2010})}\BibitemShut {NoStop}%
\bibitem [{\citenamefont {Lee}\ \emph {et~al.}(2010)\citenamefont {Lee},
  \citenamefont {Reich}, \citenamefont {Stebe},\ and\ \citenamefont
  {Leheny}}]{lee2009combined}%
  \BibitemOpen
  \bibfield  {author} {\bibinfo {author} {\bibfnamefont {Myung~Han}\
  \bibnamefont {Lee}}, \bibinfo {author} {\bibfnamefont {Daniel~H}\
  \bibnamefont {Reich}}, \bibinfo {author} {\bibfnamefont {Kathleen~J}\
  \bibnamefont {Stebe}}, \ and\ \bibinfo {author} {\bibfnamefont {Robert~L}\
  \bibnamefont {Leheny}},\ }\bibfield  {title} {\enquote {\bibinfo {title}
  {{Combined Passive and Active Microrheology Study of Protein-Layer Formation
  at an Air?Water Interface}},}\ }\href {\doibase 10.1021/la902881f}
  {\bibfield  {journal} {\bibinfo  {journal} {Langmuir}\ }\textbf {\bibinfo
  {volume} {26}},\ \bibinfo {pages} {2650--2658} (\bibinfo {year}
  {2010})}\BibitemShut {NoStop}%
\bibitem [{\citenamefont {Prasad}\ \emph {et~al.}(2006)\citenamefont {Prasad},
  \citenamefont {Koehler},\ and\ \citenamefont {Weeks}}]{prasad2006two}%
  \BibitemOpen
  \bibfield  {author} {\bibinfo {author} {\bibfnamefont {V}~\bibnamefont
  {Prasad}}, \bibinfo {author} {\bibfnamefont {S~A}\ \bibnamefont {Koehler}}, \
  and\ \bibinfo {author} {\bibfnamefont {Eric~R}\ \bibnamefont {Weeks}},\
  }\bibfield  {title} {\enquote {\bibinfo {title} {{Two-particle microrheology
  of quasi-2D viscous systems}},}\ }\href@noop {} {\bibfield  {journal}
  {\bibinfo  {journal} {Phys. Rev. Lett.}\ }\textbf {\bibinfo {volume} {97}},\
  \bibinfo {pages} {176001} (\bibinfo {year} {2006})}\BibitemShut {NoStop}%
\bibitem [{\citenamefont {Ortega}\ \emph {et~al.}(2010)\citenamefont {Ortega},
  \citenamefont {Ritacco},\ and\ \citenamefont
  {Rubio}}]{ortega2010interfacial}%
  \BibitemOpen
  \bibfield  {author} {\bibinfo {author} {\bibfnamefont {Francisco}\
  \bibnamefont {Ortega}}, \bibinfo {author} {\bibfnamefont {Hern{\'{a}}n}\
  \bibnamefont {Ritacco}}, \ and\ \bibinfo {author} {\bibfnamefont
  {Ram{\'{o}}n~G}\ \bibnamefont {Rubio}},\ }\bibfield  {title} {\enquote
  {\bibinfo {title} {{Interfacial microrheology: particle tracking and related
  techniques}},}\ }\href@noop {} {\bibfield  {journal} {\bibinfo  {journal}
  {Curr. Opin. Colloid Interface Sci.}\ }\textbf {\bibinfo {volume} {15}},\
  \bibinfo {pages} {237--245} (\bibinfo {year} {2010})}\BibitemShut {NoStop}%
\bibitem [{\citenamefont {Caspi}\ \emph {et~al.}(2000)\citenamefont {Caspi},
  \citenamefont {Granek},\ and\ \citenamefont {Elbaum}}]{caspi2000enhanced}%
  \BibitemOpen
  \bibfield  {author} {\bibinfo {author} {\bibfnamefont {Avi}\ \bibnamefont
  {Caspi}}, \bibinfo {author} {\bibfnamefont {Rony}\ \bibnamefont {Granek}}, \
  and\ \bibinfo {author} {\bibfnamefont {Michael}\ \bibnamefont {Elbaum}},\
  }\bibfield  {title} {\enquote {\bibinfo {title} {{Enhanced diffusion in
  active intracellular transport}},}\ }\href@noop {} {\bibfield  {journal}
  {\bibinfo  {journal} {Phys. Rev. Lett.}\ }\textbf {\bibinfo {volume} {85}},\
  \bibinfo {pages} {5655} (\bibinfo {year} {2000})}\BibitemShut {NoStop}%
\bibitem [{\citenamefont {MacKintosh}\ and\ \citenamefont
  {Levine}(2008)}]{MacKintosh2008}%
  \BibitemOpen
  \bibfield  {author} {\bibinfo {author} {\bibfnamefont {F.~C.}\ \bibnamefont
  {MacKintosh}}\ and\ \bibinfo {author} {\bibfnamefont {A.~J.}\ \bibnamefont
  {Levine}},\ }\bibfield  {title} {\enquote {\bibinfo {title} {{Nonequilibrium
  Mechanics and Dynamics of Motor-Activated Gels}},}\ }\href {\doibase
  10.1103/PhysRevLett.100.018104} {\bibfield  {journal} {\bibinfo  {journal}
  {Phys. Rev. Lett.}\ }\textbf {\bibinfo {volume} {100}},\ \bibinfo {pages}
  {18104} (\bibinfo {year} {2008})},\ \Eprint {http://arxiv.org/abs/0704.3794}
  {arXiv:0704.3794} \BibitemShut {NoStop}%
\bibitem [{\citenamefont {Alberts}\ \emph {et~al.}(1994)\citenamefont
  {Alberts}, \citenamefont {Bray}, \citenamefont {Lewis}, \citenamefont {Raff},
  \citenamefont {Roberts},\ and\ \citenamefont
  {Watson}}]{alberts1994molecular}%
  \BibitemOpen
  \bibfield  {author} {\bibinfo {author} {\bibfnamefont {Bruce}\ \bibnamefont
  {Alberts}}, \bibinfo {author} {\bibfnamefont {Dennis}\ \bibnamefont {Bray}},
  \bibinfo {author} {\bibfnamefont {Julian}\ \bibnamefont {Lewis}}, \bibinfo
  {author} {\bibfnamefont {Martin}\ \bibnamefont {Raff}}, \bibinfo {author}
  {\bibfnamefont {Keith}\ \bibnamefont {Roberts}}, \ and\ \bibinfo {author}
  {\bibfnamefont {James~D}\ \bibnamefont {Watson}},\ }\bibfield  {title}
  {\enquote {\bibinfo {title} {{Molecular biology of the cell. 3rd}},}\
  }\href@noop {} {\bibfield  {journal} {\bibinfo  {journal} {New York Garl.
  Pub}\ }\textbf {\bibinfo {volume} {43}},\ \bibinfo {pages} {67} (\bibinfo
  {year} {1994})}\BibitemShut {NoStop}%
\bibitem [{\citenamefont {Kasza}\ \emph {et~al.}(2007)\citenamefont {Kasza},
  \citenamefont {Rowat}, \citenamefont {Liu}, \citenamefont {Angelini},
  \citenamefont {Brangwynne}, \citenamefont {Koenderink},\ and\ \citenamefont
  {Weitz}}]{kasza2007cell}%
  \BibitemOpen
  \bibfield  {author} {\bibinfo {author} {\bibfnamefont {Karen~E}\ \bibnamefont
  {Kasza}}, \bibinfo {author} {\bibfnamefont {Amy~C}\ \bibnamefont {Rowat}},
  \bibinfo {author} {\bibfnamefont {Jiayu}\ \bibnamefont {Liu}}, \bibinfo
  {author} {\bibfnamefont {Thomas~E}\ \bibnamefont {Angelini}}, \bibinfo
  {author} {\bibfnamefont {Clifford~P}\ \bibnamefont {Brangwynne}}, \bibinfo
  {author} {\bibfnamefont {Gijsje~H}\ \bibnamefont {Koenderink}}, \ and\
  \bibinfo {author} {\bibfnamefont {David~A}\ \bibnamefont {Weitz}},\
  }\bibfield  {title} {\enquote {\bibinfo {title} {{The cell as a material}},}\
  }\href@noop {} {\bibfield  {journal} {\bibinfo  {journal} {Curr. Opin. Cell
  Biol.}\ }\textbf {\bibinfo {volume} {19}},\ \bibinfo {pages} {101--107}
  (\bibinfo {year} {2007})}\BibitemShut {NoStop}%
\bibitem [{\citenamefont {K{\"{o}}hler}\ and\ \citenamefont
  {Bausch}(2012)}]{kohler2012contraction}%
  \BibitemOpen
  \bibfield  {author} {\bibinfo {author} {\bibfnamefont {Simone}\ \bibnamefont
  {K{\"{o}}hler}}\ and\ \bibinfo {author} {\bibfnamefont {Andreas~R}\
  \bibnamefont {Bausch}},\ }\bibfield  {title} {\enquote {\bibinfo {title}
  {{Contraction mechanisms in composite active actin networks}},}\ }\href@noop
  {} {\bibfield  {journal} {\bibinfo  {journal} {PLoS One}\ }\textbf {\bibinfo
  {volume} {7}},\ \bibinfo {pages} {e39869} (\bibinfo {year}
  {2012})}\BibitemShut {NoStop}%
\bibitem [{\citenamefont {Stricker}\ \emph {et~al.}(2010)\citenamefont
  {Stricker}, \citenamefont {Falzone},\ and\ \citenamefont
  {Gardel}}]{stricker2010mechanics}%
  \BibitemOpen
  \bibfield  {author} {\bibinfo {author} {\bibfnamefont {Jonathan}\
  \bibnamefont {Stricker}}, \bibinfo {author} {\bibfnamefont {Tobias}\
  \bibnamefont {Falzone}}, \ and\ \bibinfo {author} {\bibfnamefont
  {Margaret~L}\ \bibnamefont {Gardel}},\ }\bibfield  {title} {\enquote
  {\bibinfo {title} {{Mechanics of the F-actin cytoskeleton}},}\ }\href
  {\doibase 10.1016/j.jbiomech.2009.09.003} {\bibfield  {journal} {\bibinfo
  {journal} {J. Biomech.}\ }\textbf {\bibinfo {volume} {43}},\ \bibinfo {pages}
  {9--14} (\bibinfo {year} {2010})}\BibitemShut {NoStop}%
\bibitem [{\citenamefont {Lin}\ \emph {et~al.}(2007)\citenamefont {Lin},
  \citenamefont {Koenderink}, \citenamefont {MacKintosh},\ and\ \citenamefont
  {Weitz}}]{Lin07}%
  \BibitemOpen
  \bibfield  {author} {\bibinfo {author} {\bibfnamefont {Yi-Chia}\ \bibnamefont
  {Lin}}, \bibinfo {author} {\bibfnamefont {Gijsje~H}\ \bibnamefont
  {Koenderink}}, \bibinfo {author} {\bibfnamefont {Frederick~C}\ \bibnamefont
  {MacKintosh}}, \ and\ \bibinfo {author} {\bibfnamefont {David~A}\
  \bibnamefont {Weitz}},\ }\bibfield  {title} {\enquote {\bibinfo {title}
  {{Viscoelastic properties of microtubule networks}},}\ }\href@noop {}
  {\bibfield  {journal} {\bibinfo  {journal} {Macromolecules}\ }\textbf
  {\bibinfo {volume} {40}},\ \bibinfo {pages} {7714--7720} (\bibinfo {year}
  {2007})}\BibitemShut {NoStop}%
\bibitem [{\citenamefont {Kasza}\ \emph {et~al.}()\citenamefont {Kasza},
  \citenamefont {Broedersz}, \citenamefont {Koenderink}, \citenamefont {Lin},
  \citenamefont {Messner}, \citenamefont {Millman}, \citenamefont {Nakamura},
  \citenamefont {Stossel}, \citenamefont {MacKintosh},\ and\ \citenamefont
  {Weitz}}]{Kasza2010}%
  \BibitemOpen
  \bibfield  {author} {\bibinfo {author} {\bibfnamefont {K~E}\ \bibnamefont
  {Kasza}}, \bibinfo {author} {\bibfnamefont {C~P}\ \bibnamefont {Broedersz}},
  \bibinfo {author} {\bibfnamefont {G~H}\ \bibnamefont {Koenderink}}, \bibinfo
  {author} {\bibfnamefont {Y~C}\ \bibnamefont {Lin}}, \bibinfo {author}
  {\bibfnamefont {W}~\bibnamefont {Messner}}, \bibinfo {author} {\bibfnamefont
  {E~A}\ \bibnamefont {Millman}}, \bibinfo {author} {\bibfnamefont
  {F}~\bibnamefont {Nakamura}}, \bibinfo {author} {\bibfnamefont {T~P}\
  \bibnamefont {Stossel}}, \bibinfo {author} {\bibfnamefont {F~C}\ \bibnamefont
  {MacKintosh}}, \ and\ \bibinfo {author} {\bibfnamefont {D~A}\ \bibnamefont
  {Weitz}},\ }\bibfield  {title} {\enquote {\bibinfo {title} {{Actin Filament
  Length Tunes Elasticity of Flexibly Cross-Linked Actin Networks}},}\ }\href
  {\doibase 10.1016/j.bpj.2010.06.025} {\bibfield  {journal} {\bibinfo
  {journal} {Biophys. J.}\ }\textbf {\bibinfo {volume} {99}},\ \bibinfo {pages}
  {1091--1100}}\BibitemShut {NoStop}%
\bibitem [{\citenamefont {Gardel}\ \emph {et~al.}(2008)\citenamefont {Gardel},
  \citenamefont {Kasza}, \citenamefont {Brangwynne}, \citenamefont {Liu},\ and\
  \citenamefont {Weitz}}]{gardel2008mechanical}%
  \BibitemOpen
  \bibfield  {author} {\bibinfo {author} {\bibfnamefont {Margaret~L}\
  \bibnamefont {Gardel}}, \bibinfo {author} {\bibfnamefont {Karen~E}\
  \bibnamefont {Kasza}}, \bibinfo {author} {\bibfnamefont {Clifford~P}\
  \bibnamefont {Brangwynne}}, \bibinfo {author} {\bibfnamefont {Jiayu}\
  \bibnamefont {Liu}}, \ and\ \bibinfo {author} {\bibfnamefont {David~A}\
  \bibnamefont {Weitz}},\ }\bibfield  {title} {\enquote {\bibinfo {title}
  {{Mechanical response of cytoskeletal networks}},}\ }\href@noop {} {\bibfield
   {journal} {\bibinfo  {journal} {Methods Cell Biol.}\ }\textbf {\bibinfo
  {volume} {89}},\ \bibinfo {pages} {487--519} (\bibinfo {year}
  {2008})}\BibitemShut {NoStop}%
\bibitem [{\citenamefont {Pelletier}\ \emph {et~al.}(2009)\citenamefont
  {Pelletier}, \citenamefont {Gal}, \citenamefont {Fournier},\ and\
  \citenamefont {Kilfoil}}]{pelletier2009microrheology}%
  \BibitemOpen
  \bibfield  {author} {\bibinfo {author} {\bibfnamefont {Vincent}\ \bibnamefont
  {Pelletier}}, \bibinfo {author} {\bibfnamefont {Naama}\ \bibnamefont {Gal}},
  \bibinfo {author} {\bibfnamefont {Paul}\ \bibnamefont {Fournier}}, \ and\
  \bibinfo {author} {\bibfnamefont {Maria~L}\ \bibnamefont {Kilfoil}},\
  }\bibfield  {title} {\enquote {\bibinfo {title} {{Microrheology of
  microtubule solutions and actin-microtubule composite networks}},}\
  }\href@noop {} {\bibfield  {journal} {\bibinfo  {journal} {Phys. Rev. Lett.}\
  }\textbf {\bibinfo {volume} {102}},\ \bibinfo {pages} {188303} (\bibinfo
  {year} {2009})}\BibitemShut {NoStop}%
\bibitem [{\citenamefont {Murrell}\ \emph {et~al.}(2015)\citenamefont
  {Murrell}, \citenamefont {Oakes}, \citenamefont {Lenz},\ and\ \citenamefont
  {Gardel}}]{murrell2015forcing}%
  \BibitemOpen
  \bibfield  {author} {\bibinfo {author} {\bibfnamefont {Michael}\ \bibnamefont
  {Murrell}}, \bibinfo {author} {\bibfnamefont {Patrick~W}\ \bibnamefont
  {Oakes}}, \bibinfo {author} {\bibfnamefont {Martin}\ \bibnamefont {Lenz}}, \
  and\ \bibinfo {author} {\bibfnamefont {Margaret~L}\ \bibnamefont {Gardel}},\
  }\bibfield  {title} {\enquote {\bibinfo {title} {{Forcing cells into shape:
  the mechanics of actomyosin contractility}},}\ }\href {\doibase
  10.1038/nrm4012} {\bibfield  {journal} {\bibinfo  {journal} {Nat. Rev. Mol.
  Cell Biol.}\ }\textbf {\bibinfo {volume} {16}},\ \bibinfo {pages} {486--498}
  (\bibinfo {year} {2015})}\BibitemShut {NoStop}%
\bibitem [{\citenamefont {Schaller}\ \emph
  {et~al.}(2011{\natexlab{b}})\citenamefont {Schaller}, \citenamefont {Weber},
  \citenamefont {Frey},\ and\ \citenamefont {Bausch}}]{Schaller2011polar}%
  \BibitemOpen
  \bibfield  {author} {\bibinfo {author} {\bibfnamefont {Volker}\ \bibnamefont
  {Schaller}}, \bibinfo {author} {\bibfnamefont {Christoph}\ \bibnamefont
  {Weber}}, \bibinfo {author} {\bibfnamefont {Erwin}\ \bibnamefont {Frey}}, \
  and\ \bibinfo {author} {\bibfnamefont {Andreas~R}\ \bibnamefont {Bausch}},\
  }\bibfield  {title} {\enquote {\bibinfo {title} {{Polar pattern formation:
  hydrodynamic coupling of driven filaments}},}\ }\href@noop {} {\bibfield
  {journal} {\bibinfo  {journal} {Soft Matter}\ }\textbf {\bibinfo {volume}
  {7}},\ \bibinfo {pages} {3213--3218} (\bibinfo {year}
  {2011}{\natexlab{b}})}\BibitemShut {NoStop}%
\bibitem [{\citenamefont {Bendix}\ \emph {et~al.}(2008)\citenamefont {Bendix},
  \citenamefont {Koenderink}, \citenamefont {Cuvelier}, \citenamefont {Dogic},
  \citenamefont {Koeleman}, \citenamefont {Brieher}, \citenamefont {Field},
  \citenamefont {Mahadevan},\ and\ \citenamefont
  {Weitz}}]{Bendix2008quantitative}%
  \BibitemOpen
  \bibfield  {author} {\bibinfo {author} {\bibfnamefont {Poul~M}\ \bibnamefont
  {Bendix}}, \bibinfo {author} {\bibfnamefont {Gijsje~H}\ \bibnamefont
  {Koenderink}}, \bibinfo {author} {\bibfnamefont {Damien}\ \bibnamefont
  {Cuvelier}}, \bibinfo {author} {\bibfnamefont {Zvonimir}\ \bibnamefont
  {Dogic}}, \bibinfo {author} {\bibfnamefont {Bernard~N}\ \bibnamefont
  {Koeleman}}, \bibinfo {author} {\bibfnamefont {William~M}\ \bibnamefont
  {Brieher}}, \bibinfo {author} {\bibfnamefont {Christine~M}\ \bibnamefont
  {Field}}, \bibinfo {author} {\bibfnamefont {L}~\bibnamefont {Mahadevan}}, \
  and\ \bibinfo {author} {\bibfnamefont {David~A}\ \bibnamefont {Weitz}},\
  }\bibfield  {title} {\enquote {\bibinfo {title} {{A Quantitative Analysis of
  Contractility in Active Cytoskeletal Protein Networks}},}\ }\href {\doibase
  10.1529/biophysj.107.117960} {\bibfield  {journal} {\bibinfo  {journal}
  {Biophys. J.}\ }\textbf {\bibinfo {volume} {94}},\ \bibinfo {pages}
  {3126--3136} (\bibinfo {year} {2008})}\BibitemShut {NoStop}%
\bibitem [{\citenamefont {Ronceray}\ \emph {et~al.}(2016)\citenamefont
  {Ronceray}, \citenamefont {Broedersz},\ and\ \citenamefont
  {Lenz}}]{ronceray2016fiber}%
  \BibitemOpen
  \bibfield  {author} {\bibinfo {author} {\bibfnamefont {Pierre}\ \bibnamefont
  {Ronceray}}, \bibinfo {author} {\bibfnamefont {Chase~P}\ \bibnamefont
  {Broedersz}}, \ and\ \bibinfo {author} {\bibfnamefont {Martin}\ \bibnamefont
  {Lenz}},\ }\bibfield  {title} {\enquote {\bibinfo {title} {{Fiber networks
  amplify active stress}},}\ }\href@noop {} {\bibfield  {journal} {\bibinfo
  {journal} {Proc. Natl. Acad. Sci.}\ }\textbf {\bibinfo {volume} {113}},\
  \bibinfo {pages} {2827--2832} (\bibinfo {year} {2016})}\BibitemShut {NoStop}%
\bibitem [{\citenamefont {Lenz}\ \emph {et~al.}(2012)\citenamefont {Lenz},
  \citenamefont {Thoresen}, \citenamefont {Gardel},\ and\ \citenamefont
  {Dinner}}]{lenz2012contractile}%
  \BibitemOpen
  \bibfield  {author} {\bibinfo {author} {\bibfnamefont {Martin}\ \bibnamefont
  {Lenz}}, \bibinfo {author} {\bibfnamefont {Todd}\ \bibnamefont {Thoresen}},
  \bibinfo {author} {\bibfnamefont {Margaret~L}\ \bibnamefont {Gardel}}, \ and\
  \bibinfo {author} {\bibfnamefont {Aaron~R}\ \bibnamefont {Dinner}},\
  }\bibfield  {title} {\enquote {\bibinfo {title} {{Contractile Units in
  Disordered Actomyosin Bundles Arise from F-Actin Buckling}},}\ }\href
  {\doibase 10.1103/PhysRevLett.108.238107} {\bibfield  {journal} {\bibinfo
  {journal} {Phys. Rev. Lett.}\ }\textbf {\bibinfo {volume} {108}},\ \bibinfo
  {pages} {238107} (\bibinfo {year} {2012})}\BibitemShut {NoStop}%
\bibitem [{\citenamefont {Wang}\ and\ \citenamefont
  {Wolynes}(2012)}]{wang2012active}%
  \BibitemOpen
  \bibfield  {author} {\bibinfo {author} {\bibfnamefont {Shenshen}\
  \bibnamefont {Wang}}\ and\ \bibinfo {author} {\bibfnamefont {Peter~G}\
  \bibnamefont {Wolynes}},\ }\bibfield  {title} {\enquote {\bibinfo {title}
  {{Active contractility in actomyosin networks}},}\ }\href@noop {} {\bibfield
  {journal} {\bibinfo  {journal} {Proc. Natl. Acad. Sci.}\ }\textbf {\bibinfo
  {volume} {109}},\ \bibinfo {pages} {6446--6451} (\bibinfo {year}
  {2012})}\BibitemShut {NoStop}%
\bibitem [{\citenamefont {Howard}(2002)}]{howard2002mechanics}%
  \BibitemOpen
  \bibfield  {author} {\bibinfo {author} {\bibfnamefont {J}~\bibnamefont
  {Howard}},\ }\bibfield  {title} {\enquote {\bibinfo {title} {{Mechanics of
  motor proteins}},}\ }in\ \href@noop {} {\emph {\bibinfo {booktitle} {Phys.
  bio-molecules cells. Phys. des biomol{\{}{\'{e}}{\}}cules des cellules}}}\
  (\bibinfo  {publisher} {Springer},\ \bibinfo {year} {2002})\ pp.\ \bibinfo
  {pages} {69--94}\BibitemShut {NoStop}%
\bibitem [{\citenamefont {Levine}\ and\ \citenamefont
  {MacKintosh}(2009)}]{levine2009mechanics}%
  \BibitemOpen
  \bibfield  {author} {\bibinfo {author} {\bibfnamefont {Alex~J.}\ \bibnamefont
  {Levine}}\ and\ \bibinfo {author} {\bibfnamefont {F.~C.}\ \bibnamefont
  {MacKintosh}},\ }\bibfield  {title} {\enquote {\bibinfo {title} {{The
  Mechanics and Fluctuation Spectrum of Active Gels ?}},}\ }\href {\doibase
  10.1021/jp808192w} {\bibfield  {journal} {\bibinfo  {journal} {J. Phys. Chem.
  B}\ }\textbf {\bibinfo {volume} {113}},\ \bibinfo {pages} {3820--3830}
  (\bibinfo {year} {2009})}\BibitemShut {NoStop}%
\bibitem [{\citenamefont {Storm}\ \emph {et~al.}(2005)\citenamefont {Storm},
  \citenamefont {Pastore}, \citenamefont {MacKintosh}, \citenamefont
  {Lubensky},\ and\ \citenamefont {Janmey}}]{Storm2005}%
  \BibitemOpen
  \bibfield  {author} {\bibinfo {author} {\bibfnamefont {Cornelis}\
  \bibnamefont {Storm}}, \bibinfo {author} {\bibfnamefont {Jennifer~J}\
  \bibnamefont {Pastore}}, \bibinfo {author} {\bibfnamefont {F~C}\ \bibnamefont
  {MacKintosh}}, \bibinfo {author} {\bibfnamefont {T~C}\ \bibnamefont
  {Lubensky}}, \ and\ \bibinfo {author} {\bibfnamefont {Paul~A}\ \bibnamefont
  {Janmey}},\ }\bibfield  {title} {\enquote {\bibinfo {title} {{Nonlinear
  elasticity in biological gels}},}\ }\href@noop {} {\bibfield  {journal}
  {\bibinfo  {journal} {Nature}\ }\textbf {\bibinfo {volume} {435}},\ \bibinfo
  {pages} {191--194} (\bibinfo {year} {2005})}\BibitemShut {NoStop}%
\bibitem [{\citenamefont {Gardel}(2004)}]{Gardel2004}%
  \BibitemOpen
  \bibfield  {author} {\bibinfo {author} {\bibfnamefont {M~L}\ \bibnamefont
  {Gardel}},\ }\bibfield  {title} {\enquote {\bibinfo {title} {{Elastic
  Behavior of Cross-Linked and Bundled Actin Networks}},}\ }\href {\doibase
  10.1126/science.1095087} {\bibfield  {journal} {\bibinfo  {journal}
  {Science}\ }\textbf {\bibinfo {volume} {304}},\ \bibinfo {pages} {1301--1305}
  (\bibinfo {year} {2004})}\BibitemShut {NoStop}%
\bibitem [{\citenamefont {Kasza}\ \emph {et~al.}(2009)\citenamefont {Kasza},
  \citenamefont {Koenderink}, \citenamefont {Lin}, \citenamefont {Broedersz},
  \citenamefont {Messner}, \citenamefont {Nakamura}, \citenamefont {Stossel},
  \citenamefont {{Mac Kintosh}},\ and\ \citenamefont {Weitz}}]{Kasza2009}%
  \BibitemOpen
  \bibfield  {author} {\bibinfo {author} {\bibfnamefont {K~E}\ \bibnamefont
  {Kasza}}, \bibinfo {author} {\bibfnamefont {G~H}\ \bibnamefont {Koenderink}},
  \bibinfo {author} {\bibfnamefont {Y~C}\ \bibnamefont {Lin}}, \bibinfo
  {author} {\bibfnamefont {C~P}\ \bibnamefont {Broedersz}}, \bibinfo {author}
  {\bibfnamefont {W}~\bibnamefont {Messner}}, \bibinfo {author} {\bibfnamefont
  {F}~\bibnamefont {Nakamura}}, \bibinfo {author} {\bibfnamefont {T~P}\
  \bibnamefont {Stossel}}, \bibinfo {author} {\bibfnamefont {F~C}\ \bibnamefont
  {{Mac Kintosh}}}, \ and\ \bibinfo {author} {\bibfnamefont {D~A}\ \bibnamefont
  {Weitz}},\ }\bibfield  {title} {\enquote {\bibinfo {title} {{Nonlinear
  elasticity of stiff biopolymers connected by flexible linkers}},}\ }\href
  {\doibase 10.1103/PhysRevE.79.041928} {\bibfield  {journal} {\bibinfo
  {journal} {Phys. Rev. E}\ }\textbf {\bibinfo {volume} {79}},\ \bibinfo
  {pages} {41928} (\bibinfo {year} {2009})}\BibitemShut {NoStop}%
\bibitem [{\citenamefont {Lin}\ \emph {et~al.}(2010)\citenamefont {Lin},
  \citenamefont {Yao}, \citenamefont {Broedersz}, \citenamefont {Herrmann},
  \citenamefont {MacKintosh},\ and\ \citenamefont {Weitz}}]{Lin2010}%
  \BibitemOpen
  \bibfield  {author} {\bibinfo {author} {\bibfnamefont {Yi-Chia}\ \bibnamefont
  {Lin}}, \bibinfo {author} {\bibfnamefont {Norman~Y}\ \bibnamefont {Yao}},
  \bibinfo {author} {\bibfnamefont {Chase~P}\ \bibnamefont {Broedersz}},
  \bibinfo {author} {\bibfnamefont {Harald}\ \bibnamefont {Herrmann}}, \bibinfo
  {author} {\bibfnamefont {Fred~C}\ \bibnamefont {MacKintosh}}, \ and\ \bibinfo
  {author} {\bibfnamefont {David~A}\ \bibnamefont {Weitz}},\ }\bibfield
  {title} {\enquote {\bibinfo {title} {{Origins of Elasticity in Intermediate
  Filament Networks}},}\ }\href {\doibase 10.1103/PhysRevLett.104.058101}
  {\bibfield  {journal} {\bibinfo  {journal} {Phys. Rev. Lett.}\ }\textbf
  {\bibinfo {volume} {104}},\ \bibinfo {pages} {58101} (\bibinfo {year}
  {2010})}\BibitemShut {NoStop}%
\bibitem [{\citenamefont {Shokef}\ and\ \citenamefont
  {Safran}(2012)}]{shokef2012scaling}%
  \BibitemOpen
  \bibfield  {author} {\bibinfo {author} {\bibfnamefont {Yair}\ \bibnamefont
  {Shokef}}\ and\ \bibinfo {author} {\bibfnamefont {Samuel~A}\ \bibnamefont
  {Safran}},\ }\bibfield  {title} {\enquote {\bibinfo {title} {{Scaling laws
  for the response of nonlinear elastic media with implications for cell
  mechanics}},}\ }\href@noop {} {\bibfield  {journal} {\bibinfo  {journal}
  {Phys. Rev. Lett.}\ }\textbf {\bibinfo {volume} {108}},\ \bibinfo {pages}
  {178103} (\bibinfo {year} {2012})}\BibitemShut {NoStop}%
\bibitem [{\citenamefont {Ronceray}\ and\ \citenamefont
  {Lenz}(2015)}]{ronceray2015connecting}%
  \BibitemOpen
  \bibfield  {author} {\bibinfo {author} {\bibfnamefont {Pierre}\ \bibnamefont
  {Ronceray}}\ and\ \bibinfo {author} {\bibfnamefont {Martin}\ \bibnamefont
  {Lenz}},\ }\bibfield  {title} {\enquote {\bibinfo {title} {{Connecting local
  active forces to macroscopic stress in elastic media}},}\ }\href@noop {}
  {\bibfield  {journal} {\bibinfo  {journal} {Soft Matter}\ }\textbf {\bibinfo
  {volume} {11}},\ \bibinfo {pages} {1597--1605} (\bibinfo {year}
  {2015})}\BibitemShut {NoStop}%
\bibitem [{\citenamefont {Hawkins}\ and\ \citenamefont
  {Liverpool}(2014)}]{hawkins2014stress}%
  \BibitemOpen
  \bibfield  {author} {\bibinfo {author} {\bibfnamefont {Rhoda~J}\ \bibnamefont
  {Hawkins}}\ and\ \bibinfo {author} {\bibfnamefont {Tanniemola~B}\
  \bibnamefont {Liverpool}},\ }\bibfield  {title} {\enquote {\bibinfo {title}
  {{Stress reorganization and response in active solids}},}\ }\href@noop {}
  {\bibfield  {journal} {\bibinfo  {journal} {Phys. Rev. Lett.}\ }\textbf
  {\bibinfo {volume} {113}},\ \bibinfo {pages} {28102} (\bibinfo {year}
  {2014})}\BibitemShut {NoStop}%
\bibitem [{\citenamefont {Xu}\ and\ \citenamefont
  {Safran}(2015)}]{xu2015nonlinearities}%
  \BibitemOpen
  \bibfield  {author} {\bibinfo {author} {\bibfnamefont {Xinpeng}\ \bibnamefont
  {Xu}}\ and\ \bibinfo {author} {\bibfnamefont {Samuel~A}\ \bibnamefont
  {Safran}},\ }\bibfield  {title} {\enquote {\bibinfo {title} {{Nonlinearities
  of biopolymer gels increase the range of force transmission}},}\ }\href
  {\doibase 10.1103/PhysRevE.92.032728} {\bibfield  {journal} {\bibinfo
  {journal} {Phys. Rev. E}\ }\textbf {\bibinfo {volume} {92}},\ \bibinfo
  {pages} {032728} (\bibinfo {year} {2015})}\BibitemShut {NoStop}%
\bibitem [{\citenamefont {Chen}\ and\ \citenamefont
  {Shenoy}(2011)}]{chen2011strain}%
  \BibitemOpen
  \bibfield  {author} {\bibinfo {author} {\bibfnamefont {Peng}\ \bibnamefont
  {Chen}}\ and\ \bibinfo {author} {\bibfnamefont {Vivek~B}\ \bibnamefont
  {Shenoy}},\ }\bibfield  {title} {\enquote {\bibinfo {title} {{Strain
  stiffening induced by molecular motors in active crosslinked biopolymer
  networks}},}\ }\href@noop {} {\bibfield  {journal} {\bibinfo  {journal} {Soft
  Matter}\ }\textbf {\bibinfo {volume} {7}},\ \bibinfo {pages} {355--358}
  (\bibinfo {year} {2011})}\BibitemShut {NoStop}%
\bibitem [{\citenamefont {Tee}\ \emph {et~al.}(2009)\citenamefont {Tee},
  \citenamefont {Bausch},\ and\ \citenamefont {Janmey}}]{tee2009mechanical}%
  \BibitemOpen
  \bibfield  {author} {\bibinfo {author} {\bibfnamefont {Shang-You}\
  \bibnamefont {Tee}}, \bibinfo {author} {\bibfnamefont {Andreas~R}\
  \bibnamefont {Bausch}}, \ and\ \bibinfo {author} {\bibfnamefont {Paul~A}\
  \bibnamefont {Janmey}},\ }\bibfield  {title} {\enquote {\bibinfo {title}
  {{The mechanical cell}},}\ }\href@noop {} {\bibfield  {journal} {\bibinfo
  {journal} {Curr. Biol.}\ }\textbf {\bibinfo {volume} {19}},\ \bibinfo {pages}
  {R745----R748} (\bibinfo {year} {2009})}\BibitemShut {NoStop}%
\bibitem [{\citenamefont {Lam}\ \emph {et~al.}(2011)\citenamefont {Lam},
  \citenamefont {Chaudhuri}, \citenamefont {Crow}, \citenamefont {Webster},
  \citenamefont {Li}, \citenamefont {Kita}, \citenamefont {Huang},\ and\
  \citenamefont {Fletcher}}]{lam2011mechanics}%
  \BibitemOpen
  \bibfield  {author} {\bibinfo {author} {\bibfnamefont {Wilbur~A}\
  \bibnamefont {Lam}}, \bibinfo {author} {\bibfnamefont {Ovijit}\ \bibnamefont
  {Chaudhuri}}, \bibinfo {author} {\bibfnamefont {Ailey}\ \bibnamefont {Crow}},
  \bibinfo {author} {\bibfnamefont {Kevin~D}\ \bibnamefont {Webster}}, \bibinfo
  {author} {\bibfnamefont {Tai-De}\ \bibnamefont {Li}}, \bibinfo {author}
  {\bibfnamefont {Ashley}\ \bibnamefont {Kita}}, \bibinfo {author}
  {\bibfnamefont {James}\ \bibnamefont {Huang}}, \ and\ \bibinfo {author}
  {\bibfnamefont {Daniel~A}\ \bibnamefont {Fletcher}},\ }\bibfield  {title}
  {\enquote {\bibinfo {title} {{Mechanics and contraction dynamics of single
  platelets and implications for clot stiffening}},}\ }\href@noop {} {\bibfield
   {journal} {\bibinfo  {journal} {Nat. Mater.}\ }\textbf {\bibinfo {volume}
  {10}},\ \bibinfo {pages} {61} (\bibinfo {year} {2011})}\BibitemShut {NoStop}%
\bibitem [{\citenamefont {Jansen}\ \emph {et~al.}(2013)\citenamefont {Jansen},
  \citenamefont {Bacabac}, \citenamefont {Piechocka},\ and\ \citenamefont
  {Koenderink}}]{jansen2013cells}%
  \BibitemOpen
  \bibfield  {author} {\bibinfo {author} {\bibfnamefont {Karin~A.}\
  \bibnamefont {Jansen}}, \bibinfo {author} {\bibfnamefont {Rommel~G.}\
  \bibnamefont {Bacabac}}, \bibinfo {author} {\bibfnamefont {Izabela~K.}\
  \bibnamefont {Piechocka}}, \ and\ \bibinfo {author} {\bibfnamefont
  {Gijsje~H.}\ \bibnamefont {Koenderink}},\ }\bibfield  {title} {\enquote
  {\bibinfo {title} {{Cells Actively Stiffen Fibrin Networks by Generating
  Contractile Stress}},}\ }\href {\doibase 10.1016/j.bpj.2013.10.008}
  {\bibfield  {journal} {\bibinfo  {journal} {Biophys. J.}\ }\textbf {\bibinfo
  {volume} {105}},\ \bibinfo {pages} {2240--2251} (\bibinfo {year}
  {2013})}\BibitemShut {NoStop}%
\bibitem [{\citenamefont {Toyota}\ \emph {et~al.}(2011)\citenamefont {Toyota},
  \citenamefont {Head}, \citenamefont {Schmidt},\ and\ \citenamefont
  {Mizuno}}]{toyota2011non}%
  \BibitemOpen
  \bibfield  {author} {\bibinfo {author} {\bibfnamefont {Toshihiro}\
  \bibnamefont {Toyota}}, \bibinfo {author} {\bibfnamefont {David~A}\
  \bibnamefont {Head}}, \bibinfo {author} {\bibfnamefont {Christoph~F}\
  \bibnamefont {Schmidt}}, \ and\ \bibinfo {author} {\bibfnamefont {Daisuke}\
  \bibnamefont {Mizuno}},\ }\bibfield  {title} {\enquote {\bibinfo {title}
  {{Non-Gaussian athermal fluctuations in active gels}},}\ }\href@noop {}
  {\bibfield  {journal} {\bibinfo  {journal} {Soft Matter}\ }\textbf {\bibinfo
  {volume} {7}},\ \bibinfo {pages} {3234--3239} (\bibinfo {year}
  {2011})}\BibitemShut {NoStop}%
\bibitem [{\citenamefont {Stuhrmann}\ \emph {et~al.}(2012)\citenamefont
  {Stuhrmann}, \citenamefont {e~Silva}, \citenamefont {Depken}, \citenamefont
  {MacKintosh},\ and\ \citenamefont
  {Koenderink}}]{stuhrmann2012nonequilibrium}%
  \BibitemOpen
  \bibfield  {author} {\bibinfo {author} {\bibfnamefont {Bj{\"{o}}rn}\
  \bibnamefont {Stuhrmann}}, \bibinfo {author} {\bibfnamefont {Marina~Soares}\
  \bibnamefont {e~Silva}}, \bibinfo {author} {\bibfnamefont {Martin}\
  \bibnamefont {Depken}}, \bibinfo {author} {\bibfnamefont {Frederick~C}\
  \bibnamefont {MacKintosh}}, \ and\ \bibinfo {author} {\bibfnamefont
  {Gijsje~H}\ \bibnamefont {Koenderink}},\ }\bibfield  {title} {\enquote
  {\bibinfo {title} {{Nonequilibrium fluctuations of a remodeling in vitro
  cytoskeleton}},}\ }\href@noop {} {\bibfield  {journal} {\bibinfo  {journal}
  {Phys. Rev. E}\ }\textbf {\bibinfo {volume} {86}},\ \bibinfo {pages} {20901}
  (\bibinfo {year} {2012})}\BibitemShut {NoStop}%
\bibitem [{\citenamefont {Bertrand}\ \emph {et~al.}(2012)\citenamefont
  {Bertrand}, \citenamefont {Fygenson},\ and\ \citenamefont
  {Saleh}}]{Bertrand2012}%
  \BibitemOpen
  \bibfield  {author} {\bibinfo {author} {\bibfnamefont {Olivier J~N}\
  \bibnamefont {Bertrand}}, \bibinfo {author} {\bibfnamefont {Deborah~Kuchnir}\
  \bibnamefont {Fygenson}}, \ and\ \bibinfo {author} {\bibfnamefont {Omar~A}\
  \bibnamefont {Saleh}},\ }\bibfield  {title} {\enquote {\bibinfo {title}
  {{Active, motor-driven mechanics in a DNA gel}},}\ }\href {\doibase
  10.1073/pnas.1208732109} {\bibfield  {journal} {\bibinfo  {journal} {Proc.
  Natl. Acad. Sci.}\ }\textbf {\bibinfo {volume} {109}},\ \bibinfo {pages}
  {17342--17347} (\bibinfo {year} {2012})}\BibitemShut {NoStop}%
\bibitem [{\citenamefont {Deng}\ \emph {et~al.}(2006)\citenamefont {Deng},
  \citenamefont {Trepat}, \citenamefont {Butler}, \citenamefont {Millet},
  \citenamefont {Morgan}, \citenamefont {Weitz},\ and\ \citenamefont
  {Fredberg}}]{deng2006fast}%
  \BibitemOpen
  \bibfield  {author} {\bibinfo {author} {\bibfnamefont {Linhong}\ \bibnamefont
  {Deng}}, \bibinfo {author} {\bibfnamefont {Xavier}\ \bibnamefont {Trepat}},
  \bibinfo {author} {\bibfnamefont {James~P}\ \bibnamefont {Butler}}, \bibinfo
  {author} {\bibfnamefont {Emil}\ \bibnamefont {Millet}}, \bibinfo {author}
  {\bibfnamefont {Kathleen~G}\ \bibnamefont {Morgan}}, \bibinfo {author}
  {\bibfnamefont {David~A}\ \bibnamefont {Weitz}}, \ and\ \bibinfo {author}
  {\bibfnamefont {Jeffrey~J}\ \bibnamefont {Fredberg}},\ }\bibfield  {title}
  {\enquote {\bibinfo {title} {{Fast and slow dynamics of the cytoskeleton}},}\
  }\href@noop {} {\bibfield  {journal} {\bibinfo  {journal} {Nat Mater}\
  }\textbf {\bibinfo {volume} {5}},\ \bibinfo {pages} {636--640} (\bibinfo
  {year} {2006})}\BibitemShut {NoStop}%
\bibitem [{\citenamefont {Ahmed}\ and\ \citenamefont
  {Betz}(2015)}]{ahmed2015dynamic}%
  \BibitemOpen
  \bibfield  {author} {\bibinfo {author} {\bibfnamefont {Wylie~W}\ \bibnamefont
  {Ahmed}}\ and\ \bibinfo {author} {\bibfnamefont {Timo}\ \bibnamefont
  {Betz}},\ }\bibfield  {title} {\enquote {\bibinfo {title} {{Dynamic
  cross-links tune the solid--fluid behavior of living cells}},}\ }\href@noop
  {} {\bibfield  {journal} {\bibinfo  {journal} {Proc. Natl. Acad. Sci.}\
  }\textbf {\bibinfo {volume} {112}},\ \bibinfo {pages} {6527--6528} (\bibinfo
  {year} {2015})}\BibitemShut {NoStop}%
\bibitem [{\citenamefont {Ehrlicher}\ \emph {et~al.}(2015)\citenamefont
  {Ehrlicher}, \citenamefont {Krishnan}, \citenamefont {Guo}, \citenamefont
  {Bidan}, \citenamefont {Weitz},\ and\ \citenamefont
  {Pollak}}]{ehrlicher2015alpha}%
  \BibitemOpen
  \bibfield  {author} {\bibinfo {author} {\bibfnamefont {Allen~J}\ \bibnamefont
  {Ehrlicher}}, \bibinfo {author} {\bibfnamefont {Ramaswamy}\ \bibnamefont
  {Krishnan}}, \bibinfo {author} {\bibfnamefont {Ming}\ \bibnamefont {Guo}},
  \bibinfo {author} {\bibfnamefont {C{\'{e}}cile~M}\ \bibnamefont {Bidan}},
  \bibinfo {author} {\bibfnamefont {David~A}\ \bibnamefont {Weitz}}, \ and\
  \bibinfo {author} {\bibfnamefont {Martin~R}\ \bibnamefont {Pollak}},\
  }\bibfield  {title} {\enquote {\bibinfo {title} {{Alpha-actinin binding
  kinetics modulate cellular dynamics and force generation}},}\ }\href@noop {}
  {\bibfield  {journal} {\bibinfo  {journal} {Proc. Natl. Acad. Sci.}\ }\textbf
  {\bibinfo {volume} {112}},\ \bibinfo {pages} {6619--6624} (\bibinfo {year}
  {2015})}\BibitemShut {NoStop}%
\bibitem [{\citenamefont {Yao}\ \emph {et~al.}(2013)\citenamefont {Yao},
  \citenamefont {Broedersz}, \citenamefont {Depken}, \citenamefont {Becker},
  \citenamefont {Pollak}, \citenamefont {MacKintosh},\ and\ \citenamefont
  {Weitz}}]{yao2013stress}%
  \BibitemOpen
  \bibfield  {author} {\bibinfo {author} {\bibfnamefont {Norman~Y}\
  \bibnamefont {Yao}}, \bibinfo {author} {\bibfnamefont {Chase~P}\ \bibnamefont
  {Broedersz}}, \bibinfo {author} {\bibfnamefont {Martin}\ \bibnamefont
  {Depken}}, \bibinfo {author} {\bibfnamefont {Daniel~J}\ \bibnamefont
  {Becker}}, \bibinfo {author} {\bibfnamefont {Martin~R}\ \bibnamefont
  {Pollak}}, \bibinfo {author} {\bibfnamefont {Frederick~C}\ \bibnamefont
  {MacKintosh}}, \ and\ \bibinfo {author} {\bibfnamefont {David~A}\
  \bibnamefont {Weitz}},\ }\bibfield  {title} {\enquote {\bibinfo {title}
  {{Stress-enhanced gelation: A dynamic nonlinearity of elasticity}},}\
  }\href@noop {} {\bibfield  {journal} {\bibinfo  {journal} {Phys. Rev. Lett.}\
  }\textbf {\bibinfo {volume} {110}},\ \bibinfo {pages} {18103} (\bibinfo
  {year} {2013})}\BibitemShut {NoStop}%
\bibitem [{\citenamefont {Humphrey}\ \emph {et~al.}(2002)\citenamefont
  {Humphrey}, \citenamefont {Duggan}, \citenamefont {Saha}, \citenamefont
  {Smith},\ and\ \citenamefont {Kas}}]{Humphrey2002}%
  \BibitemOpen
  \bibfield  {author} {\bibinfo {author} {\bibfnamefont {D}~\bibnamefont
  {Humphrey}}, \bibinfo {author} {\bibfnamefont {C}~\bibnamefont {Duggan}},
  \bibinfo {author} {\bibfnamefont {D}~\bibnamefont {Saha}}, \bibinfo {author}
  {\bibfnamefont {D}~\bibnamefont {Smith}}, \ and\ \bibinfo {author}
  {\bibfnamefont {J}~\bibnamefont {Kas}},\ }\bibfield  {title} {\enquote
  {\bibinfo {title} {{Active fluidization of polymer networks through molecular
  motors}},}\ }\href@noop {} {\bibfield  {journal} {\bibinfo  {journal}
  {Nature}\ }\textbf {\bibinfo {volume} {416}},\ \bibinfo {pages} {413--416}
  (\bibinfo {year} {2002})}\BibitemShut {NoStop}%
\bibitem [{\citenamefont {Fern{\'{a}}ndez}\ and\ \citenamefont
  {Ott}(2008)}]{Fernandez2008}%
  \BibitemOpen
  \bibfield  {author} {\bibinfo {author} {\bibfnamefont {Pablo}\ \bibnamefont
  {Fern{\'{a}}ndez}}\ and\ \bibinfo {author} {\bibfnamefont {Albrecht}\
  \bibnamefont {Ott}},\ }\bibfield  {title} {\enquote {\bibinfo {title}
  {{Single Cell Mechanics: Stress Stiffening and Kinematic Hardening}},}\
  }\href {\doibase 10.1103/PhysRevLett.100.238102} {\bibfield  {journal}
  {\bibinfo  {journal} {Phys. Rev. Lett.}\ }\textbf {\bibinfo {volume} {100}},\
  \bibinfo {pages} {238102} (\bibinfo {year} {2008})}\BibitemShut {NoStop}%
\bibitem [{\citenamefont {Wolff}\ \emph {et~al.}(2012)\citenamefont {Wolff},
  \citenamefont {Fern{\'{a}}ndez},\ and\ \citenamefont {Kroy}}]{Wolff2012}%
  \BibitemOpen
  \bibfield  {author} {\bibinfo {author} {\bibfnamefont {Lars}\ \bibnamefont
  {Wolff}}, \bibinfo {author} {\bibfnamefont {Pablo}\ \bibnamefont
  {Fern{\'{a}}ndez}}, \ and\ \bibinfo {author} {\bibfnamefont {Klaus}\
  \bibnamefont {Kroy}},\ }\bibfield  {title} {\enquote {\bibinfo {title}
  {{Resolving the Stiffening-Softening Paradox in Cell Mechanics}},}\ }\href
  {\doibase 10.1371/journal.pone.0040063} {\bibfield  {journal} {\bibinfo
  {journal} {PLoS One}\ }\textbf {\bibinfo {volume} {7}},\ \bibinfo {pages}
  {1--7} (\bibinfo {year} {2012})}\BibitemShut {NoStop}%
\bibitem [{\citenamefont {Krishnan}\ \emph {et~al.}(2009)\citenamefont
  {Krishnan}, \citenamefont {Park}, \citenamefont {Lin}, \citenamefont {Mead},
  \citenamefont {Jaspers}, \citenamefont {Trepat}, \citenamefont {Lenormand},
  \citenamefont {Tambe}, \citenamefont {Smolensky}, \citenamefont {Knoll},
  \citenamefont {Butler},\ and\ \citenamefont {Fredberg}}]{Krishnan2009}%
  \BibitemOpen
  \bibfield  {author} {\bibinfo {author} {\bibfnamefont {Ramaswamy}\
  \bibnamefont {Krishnan}}, \bibinfo {author} {\bibfnamefont {Chan~Young}\
  \bibnamefont {Park}}, \bibinfo {author} {\bibfnamefont {Yu-Chun}\
  \bibnamefont {Lin}}, \bibinfo {author} {\bibfnamefont {Jere}\ \bibnamefont
  {Mead}}, \bibinfo {author} {\bibfnamefont {Richard~T}\ \bibnamefont
  {Jaspers}}, \bibinfo {author} {\bibfnamefont {Xavier}\ \bibnamefont
  {Trepat}}, \bibinfo {author} {\bibfnamefont {Guillaume}\ \bibnamefont
  {Lenormand}}, \bibinfo {author} {\bibfnamefont {Dhananjay}\ \bibnamefont
  {Tambe}}, \bibinfo {author} {\bibfnamefont {Alexander~V}\ \bibnamefont
  {Smolensky}}, \bibinfo {author} {\bibfnamefont {Andrew~H}\ \bibnamefont
  {Knoll}}, \bibinfo {author} {\bibfnamefont {James~P}\ \bibnamefont {Butler}},
  \ and\ \bibinfo {author} {\bibfnamefont {Jeffrey~J}\ \bibnamefont
  {Fredberg}},\ }\bibfield  {title} {\enquote {\bibinfo {title} {{Reinforcement
  versus Fluidization in Cytoskeletal Mechanoresponsiveness}},}\ }\href
  {\doibase 10.1371/journal.pone.0005486} {\bibfield  {journal} {\bibinfo
  {journal} {PLoS One}\ }\textbf {\bibinfo {volume} {4}},\ \bibinfo {pages}
  {e5486} (\bibinfo {year} {2009})}\BibitemShut {NoStop}%
\bibitem [{\citenamefont {Trepat}\ \emph {et~al.}(2007)\citenamefont {Trepat},
  \citenamefont {Deng}, \citenamefont {An}, \citenamefont {Navajas},
  \citenamefont {Tschumperlin}, \citenamefont {Gerthoffer}, \citenamefont
  {Butler},\ and\ \citenamefont {Fredberg}}]{Trepat2007}%
  \BibitemOpen
  \bibfield  {author} {\bibinfo {author} {\bibfnamefont {Xavier}\ \bibnamefont
  {Trepat}}, \bibinfo {author} {\bibfnamefont {Linhong}\ \bibnamefont {Deng}},
  \bibinfo {author} {\bibfnamefont {Steven~S}\ \bibnamefont {An}}, \bibinfo
  {author} {\bibfnamefont {Daniel}\ \bibnamefont {Navajas}}, \bibinfo {author}
  {\bibfnamefont {Daniel~J}\ \bibnamefont {Tschumperlin}}, \bibinfo {author}
  {\bibfnamefont {William~T}\ \bibnamefont {Gerthoffer}}, \bibinfo {author}
  {\bibfnamefont {James~P}\ \bibnamefont {Butler}}, \ and\ \bibinfo {author}
  {\bibfnamefont {Jeffrey~J}\ \bibnamefont {Fredberg}},\ }\bibfield  {title}
  {\enquote {\bibinfo {title} {{Universal physical responses to stretch in the
  living cell}},}\ }\href {\doibase 10.1038/nature05824} {\bibfield  {journal}
  {\bibinfo  {journal} {Nature}\ }\textbf {\bibinfo {volume} {447}},\ \bibinfo
  {pages} {592--595} (\bibinfo {year} {2007})}\BibitemShut {NoStop}%
\bibitem [{\citenamefont {Sollich}\ \emph {et~al.}(1997)\citenamefont
  {Sollich}, \citenamefont {Lequeux}, \citenamefont {H{\'{e}}braud},\ and\
  \citenamefont {Cates}}]{Sollich1997}%
  \BibitemOpen
  \bibfield  {author} {\bibinfo {author} {\bibfnamefont {Peter}\ \bibnamefont
  {Sollich}}, \bibinfo {author} {\bibfnamefont {Fran{\c{c}}ois}\ \bibnamefont
  {Lequeux}}, \bibinfo {author} {\bibfnamefont {Pascal}\ \bibnamefont
  {H{\'{e}}braud}}, \ and\ \bibinfo {author} {\bibfnamefont {Michael~E}\
  \bibnamefont {Cates}},\ }\bibfield  {title} {\enquote {\bibinfo {title}
  {{Rheology of Soft Glassy Materials}},}\ }\href {\doibase
  10.1103/PhysRevLett.78.2020} {\bibfield  {journal} {\bibinfo  {journal}
  {Phys. Rev. Lett.}\ }\textbf {\bibinfo {volume} {78}},\ \bibinfo {pages}
  {2020--2023} (\bibinfo {year} {1997})}\BibitemShut {NoStop}%
\bibitem [{\citenamefont {Semmrich}\ \emph {et~al.}(2007)\citenamefont
  {Semmrich}, \citenamefont {Storz}, \citenamefont {Glaser}, \citenamefont
  {Merkel}, \citenamefont {Bausch},\ and\ \citenamefont {Kroy}}]{Semmrich2007}%
  \BibitemOpen
  \bibfield  {author} {\bibinfo {author} {\bibfnamefont {Christine}\
  \bibnamefont {Semmrich}}, \bibinfo {author} {\bibfnamefont {Tobias}\
  \bibnamefont {Storz}}, \bibinfo {author} {\bibfnamefont {Jens}\ \bibnamefont
  {Glaser}}, \bibinfo {author} {\bibfnamefont {Rudolf}\ \bibnamefont {Merkel}},
  \bibinfo {author} {\bibfnamefont {Andreas~R}\ \bibnamefont {Bausch}}, \ and\
  \bibinfo {author} {\bibfnamefont {Klaus}\ \bibnamefont {Kroy}},\ }\bibfield
  {title} {\enquote {\bibinfo {title} {{Glass transition and rheological
  redundancy in F-actin solutions}},}\ }\href {\doibase
  10.1073/pnas.0705513104} {\bibfield  {journal} {\bibinfo  {journal} {Proc.
  Natl. Acad. Sci.}\ }\textbf {\bibinfo {volume} {104}},\ \bibinfo {pages}
  {20199--20203} (\bibinfo {year} {2007})}\BibitemShut {NoStop}%
\bibitem [{\citenamefont {Hoffman}\ and\ \citenamefont
  {Crocker}(2009)}]{hoffman2009cell}%
  \BibitemOpen
  \bibfield  {author} {\bibinfo {author} {\bibfnamefont {Brenton~D}\
  \bibnamefont {Hoffman}}\ and\ \bibinfo {author} {\bibfnamefont {John~C}\
  \bibnamefont {Crocker}},\ }\bibfield  {title} {\enquote {\bibinfo {title}
  {{Cell mechanics: dissecting the physical responses of cells to force}},}\
  }\href@noop {} {\bibfield  {journal} {\bibinfo  {journal} {Annu. Rev. Biomed.
  Eng.}\ }\textbf {\bibinfo {volume} {11}},\ \bibinfo {pages} {259--288}
  (\bibinfo {year} {2009})}\BibitemShut {NoStop}%
\bibitem [{\citenamefont {Balland}\ \emph {et~al.}(2006)\citenamefont
  {Balland}, \citenamefont {Desprat}, \citenamefont {Icard}, \citenamefont
  {F{\'{e}}r{\'{e}}ol}, \citenamefont {Asnacios}, \citenamefont {Browaeys},
  \citenamefont {H{\'{e}}non},\ and\ \citenamefont
  {Gallet}}]{balland2006power}%
  \BibitemOpen
  \bibfield  {author} {\bibinfo {author} {\bibfnamefont {Martial}\ \bibnamefont
  {Balland}}, \bibinfo {author} {\bibfnamefont {Nicolas}\ \bibnamefont
  {Desprat}}, \bibinfo {author} {\bibfnamefont {Delphine}\ \bibnamefont
  {Icard}}, \bibinfo {author} {\bibfnamefont {Sophie}\ \bibnamefont
  {F{\'{e}}r{\'{e}}ol}}, \bibinfo {author} {\bibfnamefont {Atef}\ \bibnamefont
  {Asnacios}}, \bibinfo {author} {\bibfnamefont {Julien}\ \bibnamefont
  {Browaeys}}, \bibinfo {author} {\bibfnamefont {Sylvie}\ \bibnamefont
  {H{\'{e}}non}}, \ and\ \bibinfo {author} {\bibfnamefont {Fran{\c{c}}ois}\
  \bibnamefont {Gallet}},\ }\bibfield  {title} {\enquote {\bibinfo {title}
  {{Power laws in microrheology experiments on living cells: Comparative
  analysis and modeling}},}\ }\href@noop {} {\bibfield  {journal} {\bibinfo
  {journal} {Phys. Rev. E}\ }\textbf {\bibinfo {volume} {74}},\ \bibinfo
  {pages} {21911} (\bibinfo {year} {2006})}\BibitemShut {NoStop}%
\bibitem [{\citenamefont {Sakaue}\ and\ \citenamefont
  {Saito}(2017)}]{Sakaue2017}%
  \BibitemOpen
  \bibfield  {author} {\bibinfo {author} {\bibfnamefont {Takahiro}\
  \bibnamefont {Sakaue}}\ and\ \bibinfo {author} {\bibfnamefont {Takuya}\
  \bibnamefont {Saito}},\ }\bibfield  {title} {\enquote {\bibinfo {title}
  {{Active diffusion of model chromosomal loci driven by athermal noise}},}\
  }\href {\doibase 10.1039/C6SM00775A} {\bibfield  {journal} {\bibinfo
  {journal} {Soft Matter}\ }\textbf {\bibinfo {volume} {13}},\ \bibinfo {pages}
  {81--87} (\bibinfo {year} {2017})}\BibitemShut {NoStop}%
\bibitem [{\citenamefont {Alcaraz}\ \emph {et~al.}(2003)\citenamefont
  {Alcaraz}, \citenamefont {Buscemi}, \citenamefont {Grabulosa}, \citenamefont
  {Trepat}, \citenamefont {Fabry}, \citenamefont {Farr{\'{e}}},\ and\
  \citenamefont {Navajas}}]{Alcaraz2003}%
  \BibitemOpen
  \bibfield  {author} {\bibinfo {author} {\bibfnamefont {Jordi}\ \bibnamefont
  {Alcaraz}}, \bibinfo {author} {\bibfnamefont {Lara}\ \bibnamefont {Buscemi}},
  \bibinfo {author} {\bibfnamefont {Mireia}\ \bibnamefont {Grabulosa}},
  \bibinfo {author} {\bibfnamefont {Xavier}\ \bibnamefont {Trepat}}, \bibinfo
  {author} {\bibfnamefont {Ben}\ \bibnamefont {Fabry}}, \bibinfo {author}
  {\bibfnamefont {Ramon}\ \bibnamefont {Farr{\'{e}}}}, \ and\ \bibinfo {author}
  {\bibfnamefont {Daniel}\ \bibnamefont {Navajas}},\ }\bibfield  {title}
  {\enquote {\bibinfo {title} {{Microrheology of Human Lung Epithelial Cells
  Measured by Atomic Force Microscopy}},}\ }\href {\doibase
  10.1016/S0006-3495(03)75014-0} {\bibfield  {journal} {\bibinfo  {journal}
  {Biophys. J.}\ }\textbf {\bibinfo {volume} {84}},\ \bibinfo {pages}
  {2071--2079} (\bibinfo {year} {2003})}\BibitemShut {NoStop}%
\bibitem [{\citenamefont {Brangwynne}\ \emph
  {et~al.}(2007{\natexlab{b}})\citenamefont {Brangwynne}, \citenamefont
  {MacKintosh},\ and\ \citenamefont {Weitz}}]{brangwynne2007force}%
  \BibitemOpen
  \bibfield  {author} {\bibinfo {author} {\bibfnamefont {Clifford~P}\
  \bibnamefont {Brangwynne}}, \bibinfo {author} {\bibfnamefont {F~C}\
  \bibnamefont {MacKintosh}}, \ and\ \bibinfo {author} {\bibfnamefont
  {David~A}\ \bibnamefont {Weitz}},\ }\bibfield  {title} {\enquote {\bibinfo
  {title} {{Force fluctuations and polymerization dynamics of intracellular
  microtubules}},}\ }\href@noop {} {\bibfield  {journal} {\bibinfo  {journal}
  {Proc. Natl. Acad. Sci.}\ }\textbf {\bibinfo {volume} {104}},\ \bibinfo
  {pages} {16128--16133} (\bibinfo {year} {2007}{\natexlab{b}})}\BibitemShut
  {NoStop}%
\bibitem [{\citenamefont {Razin}\ \emph
  {et~al.}(2017{\natexlab{a}})\citenamefont {Razin}, \citenamefont {Voituriez},
  \citenamefont {Elgeti},\ and\ \citenamefont {Gov}}]{Razin2017}%
  \BibitemOpen
  \bibfield  {author} {\bibinfo {author} {\bibfnamefont {Nitzan}\ \bibnamefont
  {Razin}}, \bibinfo {author} {\bibfnamefont {Raphael}\ \bibnamefont
  {Voituriez}}, \bibinfo {author} {\bibfnamefont {Jens}\ \bibnamefont
  {Elgeti}}, \ and\ \bibinfo {author} {\bibfnamefont {Nir~S.}\ \bibnamefont
  {Gov}},\ }\bibfield  {title} {\enquote {\bibinfo {title} {{Generalized
  Archimedes' principle in active fluids}},}\ }\href {\doibase
  10.1103/PhysRevE.96.032606} {\bibfield  {journal} {\bibinfo  {journal} {Phys.
  Rev. E}\ }\textbf {\bibinfo {volume} {96}},\ \bibinfo {pages} {032606}
  (\bibinfo {year} {2017}{\natexlab{a}})},\ \Eprint
  {http://arxiv.org/abs/1703.07359} {arXiv:1703.07359} \BibitemShut {NoStop}%
\bibitem [{\citenamefont {Razin}\ \emph
  {et~al.}(2017{\natexlab{b}})\citenamefont {Razin}, \citenamefont {Voituriez},
  \citenamefont {Elgeti},\ and\ \citenamefont {Gov}}]{Razin2017a}%
  \BibitemOpen
  \bibfield  {author} {\bibinfo {author} {\bibfnamefont {Nitzan}\ \bibnamefont
  {Razin}}, \bibinfo {author} {\bibfnamefont {Raphael}\ \bibnamefont
  {Voituriez}}, \bibinfo {author} {\bibfnamefont {Jens}\ \bibnamefont
  {Elgeti}}, \ and\ \bibinfo {author} {\bibfnamefont {Nir~S.}\ \bibnamefont
  {Gov}},\ }\bibfield  {title} {\enquote {\bibinfo {title} {{Forces in
  inhomogeneous open active-particle systems}},}\ }\href
  {http://arxiv.org/abs/1708.05370} {\bibfield  {journal} {\bibinfo  {journal}
  {arXiv Prepr.}\ } (\bibinfo {year} {2017}{\natexlab{b}})},\ \Eprint
  {http://arxiv.org/abs/1708.05370} {arXiv:1708.05370} \BibitemShut {NoStop}%
\bibitem [{\citenamefont {Ahmed}\ \emph {et~al.}(2015)\citenamefont {Ahmed},
  \citenamefont {Fodor}, \citenamefont {Almonacid}, \citenamefont {Bussonnier},
  \citenamefont {Verlhac}, \citenamefont {Gov}, \citenamefont {Visco},
  \citenamefont {van Wijland},\ and\ \citenamefont {Betz}}]{ahmed2015active}%
  \BibitemOpen
  \bibfield  {author} {\bibinfo {author} {\bibfnamefont {Wylie~W}\ \bibnamefont
  {Ahmed}}, \bibinfo {author} {\bibfnamefont {Etienne}\ \bibnamefont {Fodor}},
  \bibinfo {author} {\bibfnamefont {Maria}\ \bibnamefont {Almonacid}}, \bibinfo
  {author} {\bibfnamefont {Matthias}\ \bibnamefont {Bussonnier}}, \bibinfo
  {author} {\bibfnamefont {Marie-Helene}\ \bibnamefont {Verlhac}}, \bibinfo
  {author} {\bibfnamefont {Nir~S}\ \bibnamefont {Gov}}, \bibinfo {author}
  {\bibfnamefont {Paolo}\ \bibnamefont {Visco}}, \bibinfo {author}
  {\bibfnamefont {Frederic}\ \bibnamefont {van Wijland}}, \ and\ \bibinfo
  {author} {\bibfnamefont {Timo}\ \bibnamefont {Betz}},\ }\bibfield  {title}
  {\enquote {\bibinfo {title} {{Active mechanics reveal molecular-scale force
  kinetics in living oocytes}},}\ }\href {http://arxiv.org/abs/1510.08299}
  {\bibfield  {journal} {\bibinfo  {journal} {arXiv Prepr.}\ } (\bibinfo {year}
  {2015})},\ \Eprint {http://arxiv.org/abs/1510.08299} {arXiv:1510.08299}
  \BibitemShut {NoStop}%
\bibitem [{\citenamefont {Weber}\ \emph {et~al.}(2010)\citenamefont {Weber},
  \citenamefont {Spakowitz},\ and\ \citenamefont
  {Theriot}}]{weber2010bacterial}%
  \BibitemOpen
  \bibfield  {author} {\bibinfo {author} {\bibfnamefont {Stephanie~C}\
  \bibnamefont {Weber}}, \bibinfo {author} {\bibfnamefont {Andrew~J}\
  \bibnamefont {Spakowitz}}, \ and\ \bibinfo {author} {\bibfnamefont {Julie~A}\
  \bibnamefont {Theriot}},\ }\bibfield  {title} {\enquote {\bibinfo {title}
  {{Bacterial chromosomal loci move subdiffusively through a viscoelastic
  cytoplasm}},}\ }\href@noop {} {\bibfield  {journal} {\bibinfo  {journal}
  {Phys. Rev. Lett.}\ }\textbf {\bibinfo {volume} {104}},\ \bibinfo {pages}
  {238102} (\bibinfo {year} {2010})}\BibitemShut {NoStop}%
\bibitem [{\citenamefont {Vandebroek}\ and\ \citenamefont
  {Vanderzande}(2015)}]{Vandebroek2015}%
  \BibitemOpen
  \bibfield  {author} {\bibinfo {author} {\bibfnamefont {Hans}\ \bibnamefont
  {Vandebroek}}\ and\ \bibinfo {author} {\bibfnamefont {Carlo}\ \bibnamefont
  {Vanderzande}},\ }\bibfield  {title} {\enquote {\bibinfo {title} {{Dynamics
  of a polymer in an active and viscoelastic bath}},}\ }\href {\doibase
  10.1103/PhysRevE.92.060601} {\bibfield  {journal} {\bibinfo  {journal} {Phys.
  Rev. E - Stat. Nonlinear, Soft Matter Phys.}\ }\textbf {\bibinfo {volume}
  {92}} (\bibinfo {year} {2015}),\ 10.1103/PhysRevE.92.060601},\ \Eprint
  {http://arxiv.org/abs/1507.00889} {arXiv:1507.00889} \BibitemShut {NoStop}%
\bibitem [{\citenamefont {MacKintosh}(2012)}]{mackintosh2012active}%
  \BibitemOpen
  \bibfield  {author} {\bibinfo {author} {\bibfnamefont {Fred~C}\ \bibnamefont
  {MacKintosh}},\ }\bibfield  {title} {\enquote {\bibinfo {title} {{Active
  diffusion: the erratic dance of chromosomal loci}},}\ }\href@noop {}
  {\bibfield  {journal} {\bibinfo  {journal} {Proc. Natl. Acad. Sci.}\ }\textbf
  {\bibinfo {volume} {109}},\ \bibinfo {pages} {7138--7139} (\bibinfo {year}
  {2012})}\BibitemShut {NoStop}%
\bibitem [{\citenamefont {Parry}\ \emph {et~al.}(2014)\citenamefont {Parry},
  \citenamefont {Surovtsev}, \citenamefont {Cabeen}, \citenamefont {O'Hern},
  \citenamefont {Dufresne},\ and\ \citenamefont {Jacobs-Wagner}}]{Parry2014}%
  \BibitemOpen
  \bibfield  {author} {\bibinfo {author} {\bibfnamefont {Bradley~R.}\
  \bibnamefont {Parry}}, \bibinfo {author} {\bibfnamefont {Ivan~V.}\
  \bibnamefont {Surovtsev}}, \bibinfo {author} {\bibfnamefont {Matthew~T.}\
  \bibnamefont {Cabeen}}, \bibinfo {author} {\bibfnamefont {Corey~S.}\
  \bibnamefont {O'Hern}}, \bibinfo {author} {\bibfnamefont {Eric~R.}\
  \bibnamefont {Dufresne}}, \ and\ \bibinfo {author} {\bibfnamefont
  {Christine}\ \bibnamefont {Jacobs-Wagner}},\ }\bibfield  {title} {\enquote
  {\bibinfo {title} {{The bacterial cytoplasm has glass-like properties and is
  fluidized by metabolic activity}},}\ }\href {\doibase
  10.1016/j.cell.2013.11.028} {\bibfield  {journal} {\bibinfo  {journal}
  {Cell}\ }\textbf {\bibinfo {volume} {156}},\ \bibinfo {pages} {183--194}
  (\bibinfo {year} {2014})},\ \Eprint {http://arxiv.org/abs/NIHMS150003}
  {arXiv:NIHMS150003} \BibitemShut {NoStop}%
\bibitem [{\citenamefont {Evans}(1983)}]{Evans1983}%
  \BibitemOpen
  \bibfield  {author} {\bibinfo {author} {\bibfnamefont {E.~A.}\ \bibnamefont
  {Evans}},\ }\bibfield  {title} {\enquote {\bibinfo {title} {{Bending elastic
  modulus of red blood cell membrane derived from buckling instability in
  micropipet aspiration tests}},}\ }\href {\doibase
  10.1016/S0006-3495(83)84319-7} {\bibfield  {journal} {\bibinfo  {journal}
  {Biophys. J.}\ }\textbf {\bibinfo {volume} {43}},\ \bibinfo {pages} {27--30}
  (\bibinfo {year} {1983})}\BibitemShut {NoStop}%
\bibitem [{\citenamefont {Granek}(1997)}]{granek1997}%
  \BibitemOpen
  \bibfield  {author} {\bibinfo {author} {\bibfnamefont {R.}~\bibnamefont
  {Granek}},\ }\bibfield  {title} {\enquote {\bibinfo {title} {{From
  Semi-Flexible Polymers to Membranes: Anomalous Diffusion and Reptation}},}\
  }\href {\doibase 10.1051/jp2:1997214} {\bibfield  {journal} {\bibinfo
  {journal} {J. Phys. II}\ }\textbf {\bibinfo {volume} {7}},\ \bibinfo {pages}
  {1761--1788} (\bibinfo {year} {1997})}\BibitemShut {NoStop}%
\bibitem [{\citenamefont {Granek}(2011)}]{granek2011}%
  \BibitemOpen
  \bibfield  {author} {\bibinfo {author} {\bibfnamefont {Rony}\ \bibnamefont
  {Granek}},\ }\bibfield  {title} {\enquote {\bibinfo {title} {{Membrane
  surrounded by viscoelastic continuous media: anomalous diffusion and linear
  response to force}},}\ }\href {\doibase 10.1039/c0sm01271h} {\bibfield
  {journal} {\bibinfo  {journal} {Soft Matter}\ }\textbf {\bibinfo {volume}
  {7}},\ \bibinfo {pages} {5281} (\bibinfo {year} {2011})}\BibitemShut
  {NoStop}%
\bibitem [{\citenamefont {Lin}\ \emph {et~al.}(2006)\citenamefont {Lin},
  \citenamefont {Gov},\ and\ \citenamefont {Brown}}]{Lin2006}%
  \BibitemOpen
  \bibfield  {author} {\bibinfo {author} {\bibfnamefont {Lawrence~C.L.}\
  \bibnamefont {Lin}}, \bibinfo {author} {\bibfnamefont {Nir}\ \bibnamefont
  {Gov}}, \ and\ \bibinfo {author} {\bibfnamefont {Frank~L.H.}\ \bibnamefont
  {Brown}},\ }\bibfield  {title} {\enquote {\bibinfo {title} {{Nonequilibrium
  membrane fluctuations driven by active proteins}},}\ }\href {\doibase
  10.1063/1.2166383} {\bibfield  {journal} {\bibinfo  {journal} {J. Chem.
  Phys.}\ }\textbf {\bibinfo {volume} {124}} (\bibinfo {year} {2006}),\
  10.1063/1.2166383}\BibitemShut {NoStop}%
\bibitem [{\citenamefont {Milner}\ and\ \citenamefont
  {Safran}(1987)}]{Milner1987}%
  \BibitemOpen
  \bibfield  {author} {\bibinfo {author} {\bibfnamefont {Scott~T.}\
  \bibnamefont {Milner}}\ and\ \bibinfo {author} {\bibfnamefont {S.~A.}\
  \bibnamefont {Safran}},\ }\bibfield  {title} {\enquote {\bibinfo {title}
  {{Dynamical fluctuations of droplet microemulsions and vesicles}},}\ }\href
  {\doibase 10.1103/PhysRevA.36.4371} {\bibfield  {journal} {\bibinfo
  {journal} {Phys. Rev. A}\ }\textbf {\bibinfo {volume} {36}},\ \bibinfo
  {pages} {4371--4379} (\bibinfo {year} {1987})}\BibitemShut {NoStop}%
\bibitem [{\citenamefont {Brochard}\ and\ \citenamefont
  {Lennon}(1975)}]{Brochard1975}%
  \BibitemOpen
  \bibfield  {author} {\bibinfo {author} {\bibfnamefont {F.}~\bibnamefont
  {Brochard}}\ and\ \bibinfo {author} {\bibfnamefont {J.F.}\ \bibnamefont
  {Lennon}},\ }\bibfield  {title} {\enquote {\bibinfo {title} {{Frequency
  spectrum of the flicker phenomenon in erythrocytes}},}\ }\href {\doibase
  10.1051/jphys:0197500360110103500} {\bibfield  {journal} {\bibinfo  {journal}
  {J. Phys.}\ }\textbf {\bibinfo {volume} {36}},\ \bibinfo {pages} {1035--1047}
  (\bibinfo {year} {1975})}\BibitemShut {NoStop}%
\bibitem [{\citenamefont {Strey}\ \emph {et~al.}(1995)\citenamefont {Strey},
  \citenamefont {Peterson},\ and\ \citenamefont {Sackmann}}]{Strey1995}%
  \BibitemOpen
  \bibfield  {author} {\bibinfo {author} {\bibfnamefont {H.}~\bibnamefont
  {Strey}}, \bibinfo {author} {\bibfnamefont {M.}~\bibnamefont {Peterson}}, \
  and\ \bibinfo {author} {\bibfnamefont {E.}~\bibnamefont {Sackmann}},\
  }\bibfield  {title} {\enquote {\bibinfo {title} {{Measurement of erythrocyte
  membrane elasticity by flicker eigenmode decomposition}},}\ }\href {\doibase
  10.1016/S0006-3495(95)79921-0} {\bibfield  {journal} {\bibinfo  {journal}
  {Biophys. J.}\ }\textbf {\bibinfo {volume} {69}},\ \bibinfo {pages}
  {478--488} (\bibinfo {year} {1995})}\BibitemShut {NoStop}%
\bibitem [{\citenamefont {Gov}(2004)}]{Gov2004}%
  \BibitemOpen
  \bibfield  {author} {\bibinfo {author} {\bibfnamefont {N.}~\bibnamefont
  {Gov}},\ }\bibfield  {title} {\enquote {\bibinfo {title} {{Membrane
  undulations driven by force fluctuations of active proteins}},}\ }\href
  {\doibase 10.1103/PhysRevLett.93.268104} {\bibfield  {journal} {\bibinfo
  {journal} {Phys. Rev. Lett.}\ }\textbf {\bibinfo {volume} {93}} (\bibinfo
  {year} {2004}),\ 10.1103/PhysRevLett.93.268104}\BibitemShut {NoStop}%
\bibitem [{\citenamefont {Gov}\ and\ \citenamefont {Safran}(2005)}]{Gov2005}%
  \BibitemOpen
  \bibfield  {author} {\bibinfo {author} {\bibfnamefont {N.S.}\ \bibnamefont
  {Gov}}\ and\ \bibinfo {author} {\bibfnamefont {S.A.}\ \bibnamefont
  {Safran}},\ }\bibfield  {title} {\enquote {\bibinfo {title} {{Red Blood Cell
  Membrane Fluctuations and Shape Controlled by ATP-Induced Cytoskeletal
  Defects}},}\ }\href {\doibase 10.1529/biophysj.104.045328} {\bibfield
  {journal} {\bibinfo  {journal} {Biophys. J.}\ }\textbf {\bibinfo {volume}
  {88}},\ \bibinfo {pages} {1859--1874} (\bibinfo {year} {2005})}\BibitemShut
  {NoStop}%
\bibitem [{\citenamefont {Park}\ \emph {et~al.}(2010)\citenamefont {Park},
  \citenamefont {Best}, \citenamefont {Auth}, \citenamefont {Gov},
  \citenamefont {Safran}, \citenamefont {Popescu}, \citenamefont {Suresh},\
  and\ \citenamefont {Feld}}]{Park2010}%
  \BibitemOpen
  \bibfield  {author} {\bibinfo {author} {\bibfnamefont {Y.}~\bibnamefont
  {Park}}, \bibinfo {author} {\bibfnamefont {C.~A.}\ \bibnamefont {Best}},
  \bibinfo {author} {\bibfnamefont {T.}~\bibnamefont {Auth}}, \bibinfo {author}
  {\bibfnamefont {N.~S.}\ \bibnamefont {Gov}}, \bibinfo {author} {\bibfnamefont
  {S.~A.}\ \bibnamefont {Safran}}, \bibinfo {author} {\bibfnamefont
  {G.}~\bibnamefont {Popescu}}, \bibinfo {author} {\bibfnamefont
  {S.}~\bibnamefont {Suresh}}, \ and\ \bibinfo {author} {\bibfnamefont {M.~S.}\
  \bibnamefont {Feld}},\ }\bibfield  {title} {\enquote {\bibinfo {title}
  {{Metabolic remodeling of the human red blood cell membrane}},}\ }\href
  {\doibase 10.1073/pnas.0910785107} {\bibfield  {journal} {\bibinfo  {journal}
  {Proc. Natl. Acad. Sci.}\ }\textbf {\bibinfo {volume} {107}},\ \bibinfo
  {pages} {1289--1294} (\bibinfo {year} {2010})}\BibitemShut {NoStop}%
\bibitem [{\citenamefont {Yoon}\ \emph {et~al.}(2011)\citenamefont {Yoon},
  \citenamefont {Kotar}, \citenamefont {Brown},\ and\ \citenamefont
  {Cicuta}}]{yoon2011red}%
  \BibitemOpen
  \bibfield  {author} {\bibinfo {author} {\bibfnamefont {Young~Zoon}\
  \bibnamefont {Yoon}}, \bibinfo {author} {\bibfnamefont {Jurij}\ \bibnamefont
  {Kotar}}, \bibinfo {author} {\bibfnamefont {Aidan~T}\ \bibnamefont {Brown}},
  \ and\ \bibinfo {author} {\bibfnamefont {Pietro}\ \bibnamefont {Cicuta}},\
  }\bibfield  {title} {\enquote {\bibinfo {title} {{Red blood cell dynamics:
  from spontaneous fluctuations to non-linear response}},}\ }\href@noop {}
  {\bibfield  {journal} {\bibinfo  {journal} {Soft Matter}\ }\textbf {\bibinfo
  {volume} {7}},\ \bibinfo {pages} {2042--2051} (\bibinfo {year}
  {2011})}\BibitemShut {NoStop}%
\bibitem [{\citenamefont {Rodr{\'{i}}guez-Garc{\'{i}}a}\ \emph
  {et~al.}(2016)\citenamefont {Rodr{\'{i}}guez-Garc{\'{i}}a}, \citenamefont
  {L{\'{o}}pez-Montero}, \citenamefont {Mell}, \citenamefont {Egea},
  \citenamefont {Gov},\ and\ \citenamefont {Monroy}}]{Rodriguez-Garcia2016}%
  \BibitemOpen
  \bibfield  {author} {\bibinfo {author} {\bibfnamefont {Ruddi}\ \bibnamefont
  {Rodr{\'{i}}guez-Garc{\'{i}}a}}, \bibinfo {author} {\bibfnamefont
  {Iv{\'{a}}n}\ \bibnamefont {L{\'{o}}pez-Montero}}, \bibinfo {author}
  {\bibfnamefont {Michael}\ \bibnamefont {Mell}}, \bibinfo {author}
  {\bibfnamefont {Gustavo}\ \bibnamefont {Egea}}, \bibinfo {author}
  {\bibfnamefont {Nir~S.}\ \bibnamefont {Gov}}, \ and\ \bibinfo {author}
  {\bibfnamefont {Francisco}\ \bibnamefont {Monroy}},\ }\bibfield  {title}
  {\enquote {\bibinfo {title} {{Direct Cytoskeleton Forces Cause Membrane
  Softening in Red Blood Cells}},}\ }\href {\doibase 10.1016/j.bpj.2016.08.022}
  {\bibfield  {journal} {\bibinfo  {journal} {Biophys. J.}\ }\textbf {\bibinfo
  {volume} {111}},\ \bibinfo {pages} {1101} (\bibinfo {year}
  {2016})}\BibitemShut {NoStop}%
\bibitem [{\citenamefont {Fenz}\ \emph {et~al.}(2017)\citenamefont {Fenz},
  \citenamefont {Bihr}, \citenamefont {Schmidt}, \citenamefont {Merkel},
  \citenamefont {Seifert}, \citenamefont {Sengupta},\ and\ \citenamefont
  {Smith}}]{Fenz2017}%
  \BibitemOpen
  \bibfield  {author} {\bibinfo {author} {\bibfnamefont {Susanne~F.}\
  \bibnamefont {Fenz}}, \bibinfo {author} {\bibfnamefont {Timo}\ \bibnamefont
  {Bihr}}, \bibinfo {author} {\bibfnamefont {Daniel}\ \bibnamefont {Schmidt}},
  \bibinfo {author} {\bibfnamefont {Rudolf}\ \bibnamefont {Merkel}}, \bibinfo
  {author} {\bibfnamefont {Udo}\ \bibnamefont {Seifert}}, \bibinfo {author}
  {\bibfnamefont {Kheya}\ \bibnamefont {Sengupta}}, \ and\ \bibinfo {author}
  {\bibfnamefont {Ana-Sun{\v{c}}ana}\ \bibnamefont {Smith}},\ }\bibfield
  {title} {\enquote {\bibinfo {title} {{Membrane fluctuations mediate lateral
  interaction between cadherin bonds}},}\ }\href {\doibase 10.1038/nphys4138}
  {\bibfield  {journal} {\bibinfo  {journal} {Nat. Phys.}\ } (\bibinfo {year}
  {2017}),\ 10.1038/nphys4138}\BibitemShut {NoStop}%
\bibitem [{\citenamefont {Gallet}\ \emph {et~al.}(2009)\citenamefont {Gallet},
  \citenamefont {Arcizet}, \citenamefont {Bohec},\ and\ \citenamefont
  {Richert}}]{gallet2009power}%
  \BibitemOpen
  \bibfield  {author} {\bibinfo {author} {\bibfnamefont {Fran{\c{c}}ois}\
  \bibnamefont {Gallet}}, \bibinfo {author} {\bibfnamefont {Delphine}\
  \bibnamefont {Arcizet}}, \bibinfo {author} {\bibfnamefont {Pierre}\
  \bibnamefont {Bohec}}, \ and\ \bibinfo {author} {\bibfnamefont {Alain}\
  \bibnamefont {Richert}},\ }\bibfield  {title} {\enquote {\bibinfo {title}
  {{Power spectrum of out-of-equilibrium forces in living cells: amplitude and
  frequency dependence}},}\ }\href {\doibase 10.1039/b901311c} {\bibfield
  {journal} {\bibinfo  {journal} {Soft Matter}\ }\textbf {\bibinfo {volume}
  {5}},\ \bibinfo {pages} {2947} (\bibinfo {year} {2009})}\BibitemShut
  {NoStop}%
\bibitem [{\citenamefont {Osmanovi{\'{c}}}\ and\ \citenamefont
  {Rabin}(2017)}]{Osmanovic2017}%
  \BibitemOpen
  \bibfield  {author} {\bibinfo {author} {\bibfnamefont {Dino}\ \bibnamefont
  {Osmanovi{\'{c}}}}\ and\ \bibinfo {author} {\bibfnamefont {Yitzhak}\
  \bibnamefont {Rabin}},\ }\bibfield  {title} {\enquote {\bibinfo {title}
  {{Dynamics of active Rouse chains}},}\ }\href {\doibase 10.1039/C6SM02722A}
  {\bibfield  {journal} {\bibinfo  {journal} {Soft Matter}\ }\textbf {\bibinfo
  {volume} {13}},\ \bibinfo {pages} {963--968} (\bibinfo {year} {2017})},\
  \Eprint {http://arxiv.org/abs/1608.05914} {arXiv:1608.05914} \BibitemShut
  {NoStop}%
\bibitem [{\citenamefont {Samanta}\ and\ \citenamefont
  {Chakrabarti}(2016)}]{Samanta2016}%
  \BibitemOpen
  \bibfield  {author} {\bibinfo {author} {\bibfnamefont {Nairhita}\
  \bibnamefont {Samanta}}\ and\ \bibinfo {author} {\bibfnamefont {Rajarshi}\
  \bibnamefont {Chakrabarti}},\ }\bibfield  {title} {\enquote {\bibinfo {title}
  {{Chain reconfiguration in active noise}},}\ }\href {\doibase
  10.1088/1751-8113/49/19/195601} {\bibfield  {journal} {\bibinfo  {journal}
  {J. Phys. A Math. Theor.}\ }\textbf {\bibinfo {volume} {49}},\ \bibinfo
  {pages} {195601} (\bibinfo {year} {2016})},\ \Eprint
  {http://arxiv.org/abs/1512.04056} {arXiv:1512.04056} \BibitemShut {NoStop}%
\bibitem [{\citenamefont {Romanczuk}\ \emph {et~al.}(2012)\citenamefont
  {Romanczuk}, \citenamefont {B{\"{a}}r}, \citenamefont {Ebeling},
  \citenamefont {Lindner},\ and\ \citenamefont
  {Schimansky-Geier}}]{Romanczuk2012}%
  \BibitemOpen
  \bibfield  {author} {\bibinfo {author} {\bibfnamefont {P}~\bibnamefont
  {Romanczuk}}, \bibinfo {author} {\bibfnamefont {M}~\bibnamefont {B{\"{a}}r}},
  \bibinfo {author} {\bibfnamefont {W}~\bibnamefont {Ebeling}}, \bibinfo
  {author} {\bibfnamefont {B}~\bibnamefont {Lindner}}, \ and\ \bibinfo {author}
  {\bibfnamefont {L}~\bibnamefont {Schimansky-Geier}},\ }\bibfield  {title}
  {\enquote {\bibinfo {title} {{Active Brownian particles}},}\ }\href {\doibase
  10.1140/epjst/e2012-01529-y} {\bibfield  {journal} {\bibinfo  {journal} {Eur.
  Phys. J. Spec. Top.}\ }\textbf {\bibinfo {volume} {202}},\ \bibinfo {pages}
  {1--162} (\bibinfo {year} {2012})}\BibitemShut {NoStop}%
\bibitem [{\citenamefont {Gardel}\ \emph {et~al.}(2005)\citenamefont {Gardel},
  \citenamefont {Valentine},\ and\ \citenamefont
  {Weitz}}]{gardel2005microrheology}%
  \BibitemOpen
  \bibfield  {author} {\bibinfo {author} {\bibfnamefont {Margaret~L}\
  \bibnamefont {Gardel}}, \bibinfo {author} {\bibfnamefont {Megan~T}\
  \bibnamefont {Valentine}}, \ and\ \bibinfo {author} {\bibfnamefont {David~A}\
  \bibnamefont {Weitz}},\ }\bibfield  {title} {\enquote {\bibinfo {title}
  {{Microrheology}},}\ }in\ \href@noop {} {\emph {\bibinfo {booktitle}
  {Microscale diagnostic Tech.}}}\ (\bibinfo  {publisher} {Springer},\ \bibinfo
  {year} {2005})\ pp.\ \bibinfo {pages} {1--49}\BibitemShut {NoStop}%
\bibitem [{\citenamefont {Ben-Isaac}\ \emph {et~al.}(2015)\citenamefont
  {Ben-Isaac}, \citenamefont {Fodor}, \citenamefont {Visco}, \citenamefont {van
  Wijland},\ and\ \citenamefont {Gov}}]{ben2015modeling}%
  \BibitemOpen
  \bibfield  {author} {\bibinfo {author} {\bibfnamefont {E.}~\bibnamefont
  {Ben-Isaac}}, \bibinfo {author} {\bibfnamefont {{\'{E}}}~\bibnamefont
  {Fodor}}, \bibinfo {author} {\bibfnamefont {P.}~\bibnamefont {Visco}},
  \bibinfo {author} {\bibfnamefont {F.}~\bibnamefont {van Wijland}}, \ and\
  \bibinfo {author} {\bibfnamefont {Nir~S.}\ \bibnamefont {Gov}},\ }\bibfield
  {title} {\enquote {\bibinfo {title} {{Modeling the dynamics of a tracer
  particle in an elastic active gel}},}\ }\href {\doibase
  10.1103/PhysRevE.92.012716} {\bibfield  {journal} {\bibinfo  {journal} {Phys.
  Rev. E}\ }\textbf {\bibinfo {volume} {92}},\ \bibinfo {pages} {12716}
  (\bibinfo {year} {2015})},\ \Eprint {http://arxiv.org/abs/1507.00917v1}
  {arXiv:1507.00917v1} \BibitemShut {NoStop}%
\bibitem [{\citenamefont {Doi}(2013)}]{Doi2013a}%
  \BibitemOpen
  \bibfield  {author} {\bibinfo {author} {\bibfnamefont {Masao}\ \bibnamefont
  {Doi}},\ }\href@noop {} {\emph {\bibinfo {title} {{Soft Matter Physics}}}}\
  (\bibinfo  {publisher} {Oxford University Press},\ \bibinfo {year}
  {2013})\BibitemShut {NoStop}%
\bibitem [{\citenamefont {Gardiner}(1985)}]{gardiner1985stochastic}%
  \BibitemOpen
  \bibfield  {author} {\bibinfo {author} {\bibfnamefont {Crispin}\ \bibnamefont
  {Gardiner}},\ }\bibfield  {title} {\enquote {\bibinfo {title} {{Stochastic
  methods}},}\ }\href@noop {} {\bibfield  {journal} {\bibinfo  {journal}
  {Springer Ser. Synerg. (Springer-Verlag, Berlin, 2009)}\ } (\bibinfo {year}
  {1985})}\BibitemShut {NoStop}%
\bibitem [{\citenamefont {Jarzynski}(2017)}]{Jarzynski2017}%
  \BibitemOpen
  \bibfield  {author} {\bibinfo {author} {\bibfnamefont {Christopher}\
  \bibnamefont {Jarzynski}},\ }\bibfield  {title} {\enquote {\bibinfo {title}
  {{Stochastic and Macroscopic Thermodynamics of Strongly Coupled Systems}},}\
  }\href {\doibase 10.1103/PhysRevX.7.011008} {\bibfield  {journal} {\bibinfo
  {journal} {Phys. Rev. X}\ }\textbf {\bibinfo {volume} {7}},\ \bibinfo {pages}
  {011008} (\bibinfo {year} {2017})}\BibitemShut {NoStop}%
\bibitem [{\citenamefont {Gallavotti}\ and\ \citenamefont
  {Cohen}(1995)}]{GallavPRL}%
  \BibitemOpen
  \bibfield  {author} {\bibinfo {author} {\bibfnamefont {G}~\bibnamefont
  {Gallavotti}}\ and\ \bibinfo {author} {\bibfnamefont {E~G~D}\ \bibnamefont
  {Cohen}},\ }\bibfield  {title} {\enquote {\bibinfo {title} {{Dynamical
  Ensembles in Nonequilibrium Statistical Mechanics}},}\ }\href@noop {}
  {\bibfield  {journal} {\bibinfo  {journal} {Phys. Rev. Lett.}\ }\textbf
  {\bibinfo {volume} {74}},\ \bibinfo {pages} {2694--2697} (\bibinfo {year}
  {1995})}\BibitemShut {NoStop}%
\bibitem [{\citenamefont {Jarzynski}(1996)}]{Jarzynski1997}%
  \BibitemOpen
  \bibfield  {author} {\bibinfo {author} {\bibfnamefont {C.}~\bibnamefont
  {Jarzynski}},\ }\bibfield  {title} {\enquote {\bibinfo {title} {{A
  nonequilibrium equality for free energy differences}},}\ }\href {\doibase
  10.1103/PhysRevLett.78.2690} {\bibfield  {journal} {\bibinfo  {journal}
  {Phys. Rev. Lett.}\ }\textbf {\bibinfo {volume} {78}},\ \bibinfo {pages}
  {2690--2693} (\bibinfo {year} {1996})},\ \Eprint
  {http://arxiv.org/abs/9610209} {arXiv:9610209 [cond-mat]} \BibitemShut
  {NoStop}%
\bibitem [{\citenamefont {Crooks}(1999)}]{CrooksPRE}%
  \BibitemOpen
  \bibfield  {author} {\bibinfo {author} {\bibfnamefont {Gavin~E}\ \bibnamefont
  {Crooks}},\ }\bibfield  {title} {\enquote {\bibinfo {title} {{Entropy
  production fluctuation theorem and the nonequilibrium work relation for free
  energy differences}},}\ }\href@noop {} {\bibfield  {journal} {\bibinfo
  {journal} {Phys. Rev. E}\ }\textbf {\bibinfo {volume} {60}},\ \bibinfo
  {pages} {2721--2726} (\bibinfo {year} {1999})}\BibitemShut {NoStop}%
\bibitem [{\citenamefont {Kurchan}(1998)}]{KurchanFT}%
  \BibitemOpen
  \bibfield  {author} {\bibinfo {author} {\bibfnamefont {Jorge}\ \bibnamefont
  {Kurchan}},\ }\bibfield  {title} {\enquote {\bibinfo {title} {{Fluctuation
  theorem for stochastic dynamics}},}\ }\href@noop {} {\bibfield  {journal}
  {\bibinfo  {journal} {J. Phys. A. Math. Gen.}\ }\textbf {\bibinfo {volume}
  {31}},\ \bibinfo {pages} {3719} (\bibinfo {year} {1998})}\BibitemShut
  {NoStop}%
\bibitem [{\citenamefont {Sekimoto}(1997)}]{Sekimoto1}%
  \BibitemOpen
  \bibfield  {author} {\bibinfo {author} {\bibfnamefont {K}~\bibnamefont
  {Sekimoto}},\ }\bibfield  {title} {\enquote {\bibinfo {title} {{Kinetik
  characterization of heat bath and the enrgetics of thermal ratchet
  models}},}\ }\href@noop {} {\bibfield  {journal} {\bibinfo  {journal} {J.
  Phys. Soc. Japan}\ }\textbf {\bibinfo {volume} {66}} (\bibinfo {year}
  {1997})}\BibitemShut {NoStop}%
\bibitem [{\citenamefont {Sekimoto}(1998)}]{Sekimoto2}%
  \BibitemOpen
  \bibfield  {author} {\bibinfo {author} {\bibfnamefont {K}~\bibnamefont
  {Sekimoto}},\ }\bibfield  {title} {\enquote {\bibinfo {title} {{Langevin
  equation and thermodynamics}},}\ }\href@noop {} {\bibfield  {journal}
  {\bibinfo  {journal} {Prog. Theor. Phys. Supp.}\ }\textbf {\bibinfo {volume}
  {130}} (\bibinfo {year} {1998})}\BibitemShut {NoStop}%
\bibitem [{\citenamefont {Seifert}(2005)}]{Seifert_stoch_FT}%
  \BibitemOpen
  \bibfield  {author} {\bibinfo {author} {\bibfnamefont {U}~\bibnamefont
  {Seifert}},\ }\bibfield  {title} {\enquote {\bibinfo {title} {{Entropy
  Production along a Stochastic Trajectory and an Integral Fluctuaction
  Theorem}},}\ }\href@noop {} {\bibfield  {journal} {\bibinfo  {journal} {Phys.
  Rev. Lett.}\ }\textbf {\bibinfo {volume} {95}},\ \bibinfo {pages}
  {14380--14385} (\bibinfo {year} {2005})}\BibitemShut {NoStop}%
\bibitem [{\citenamefont {Lebowitz}\ and\ \citenamefont
  {Spohn}(1999)}]{Lebowitz1999}%
  \BibitemOpen
  \bibfield  {author} {\bibinfo {author} {\bibfnamefont {Joel~L}\ \bibnamefont
  {Lebowitz}}\ and\ \bibinfo {author} {\bibfnamefont {Herbert}\ \bibnamefont
  {Spohn}},\ }\bibfield  {title} {\enquote {\bibinfo {title} {{A
  Gallavotti--Cohen-Type Symmetry in the Large Deviation Functional for
  Stochastic Dynamics}},}\ }\href@noop {} {\bibfield  {journal} {\bibinfo
  {journal} {J. Stat. Phys.}\ }\textbf {\bibinfo {volume} {95}},\ \bibinfo
  {pages} {333--365} (\bibinfo {year} {1999})}\BibitemShut {NoStop}%
\bibitem [{\citenamefont {Speck}\ \emph {et~al.}(2007)\citenamefont {Speck},
  \citenamefont {Blickle}, \citenamefont {Bechinger},\ and\ \citenamefont
  {Seifert}}]{Speck_stoch_coll}%
  \BibitemOpen
  \bibfield  {author} {\bibinfo {author} {\bibfnamefont {T}~\bibnamefont
  {Speck}}, \bibinfo {author} {\bibfnamefont {V}~\bibnamefont {Blickle}},
  \bibinfo {author} {\bibfnamefont {C}~\bibnamefont {Bechinger}}, \ and\
  \bibinfo {author} {\bibfnamefont {U}~\bibnamefont {Seifert}},\ }\bibfield
  {title} {\enquote {\bibinfo {title} {{Distribution of entropy production for
  a colloidal particle in a nonequilibrium steady state}},}\ }\href {\doibase
  10.1209/0295-5075/79/30002} {\bibfield  {journal} {\bibinfo  {journal}
  {Europhys. Lett.}\ }\textbf {\bibinfo {volume} {79}},\ \bibinfo {pages}
  {30002} (\bibinfo {year} {2007})}\BibitemShut {NoStop}%
\bibitem [{\citenamefont {Schmiedl}\ and\ \citenamefont
  {Seifert}(2007{\natexlab{a}})}]{Schmiedl2007}%
  \BibitemOpen
  \bibfield  {author} {\bibinfo {author} {\bibfnamefont {Tim}\ \bibnamefont
  {Schmiedl}}\ and\ \bibinfo {author} {\bibfnamefont {Udo}\ \bibnamefont
  {Seifert}},\ }\bibfield  {title} {\enquote {\bibinfo {title} {{Optimal
  finite-time processes in stochastic thermodynamics}},}\ }\href {\doibase
  10.1103/PhysRevLett.98.108301} {\bibfield  {journal} {\bibinfo  {journal}
  {Phys. Rev. Lett.}\ }\textbf {\bibinfo {volume} {98}} (\bibinfo {year}
  {2007}{\natexlab{a}}),\ 10.1103/PhysRevLett.98.108301},\ \Eprint
  {http://arxiv.org/abs/0701554} {arXiv:0701554 [cond-mat]} \BibitemShut
  {NoStop}%
\bibitem [{\citenamefont {Machta}(2015)}]{Machta2015}%
  \BibitemOpen
  \bibfield  {author} {\bibinfo {author} {\bibfnamefont {Benjamin~B.}\
  \bibnamefont {Machta}},\ }\bibfield  {title} {\enquote {\bibinfo {title}
  {{Dissipation Bound for Thermodynamic Control}},}\ }\href {\doibase
  10.1103/PhysRevLett.115.260603} {\bibfield  {journal} {\bibinfo  {journal}
  {Phys. Rev. Lett.}\ }\textbf {\bibinfo {volume} {115}},\ \bibinfo {pages}
  {1--5} (\bibinfo {year} {2015})},\ \Eprint {http://arxiv.org/abs/1508.04150}
  {arXiv:1508.04150} \BibitemShut {NoStop}%
\bibitem [{\citenamefont {Sivak}\ and\ \citenamefont
  {Crooks}(2012)}]{Sivak2012}%
  \BibitemOpen
  \bibfield  {author} {\bibinfo {author} {\bibfnamefont {David~A.}\
  \bibnamefont {Sivak}}\ and\ \bibinfo {author} {\bibfnamefont {Gavin~E.}\
  \bibnamefont {Crooks}},\ }\bibfield  {title} {\enquote {\bibinfo {title}
  {{Thermodynamic metrics and optimal paths}},}\ }\href {\doibase
  10.1103/PhysRevLett.108.190602} {\bibfield  {journal} {\bibinfo  {journal}
  {Phys. Rev. Lett.}\ }\textbf {\bibinfo {volume} {108}},\ \bibinfo {pages}
  {1--5} (\bibinfo {year} {2012})},\ \Eprint {http://arxiv.org/abs/1201.4166}
  {arXiv:1201.4166} \BibitemShut {NoStop}%
\bibitem [{\citenamefont {Harada}\ and\ \citenamefont
  {Sasa}(2006)}]{Harada_Sasa}%
  \BibitemOpen
  \bibfield  {author} {\bibinfo {author} {\bibfnamefont {Takahiro}\
  \bibnamefont {Harada}}\ and\ \bibinfo {author} {\bibfnamefont {Shin-ichi}\
  \bibnamefont {Sasa}},\ }\bibfield  {title} {\enquote {\bibinfo {title}
  {{Energy dissipation and violation of the fluctuation-response relation in
  nonequilibrium Langevin systems}},}\ }\href {\doibase
  10.1103/PhysRevE.73.026131} {\bibfield  {journal} {\bibinfo  {journal} {Phys.
  Rev. E}\ }\textbf {\bibinfo {volume} {73}},\ \bibinfo {pages} {026131}
  (\bibinfo {year} {2006})}\BibitemShut {NoStop}%
\bibitem [{\citenamefont {Toyabe}\ \emph {et~al.}(2010)\citenamefont {Toyabe},
  \citenamefont {Okamoto}, \citenamefont {Watanabe-Nakayama}, \citenamefont
  {Taketani}, \citenamefont {Kudo},\ and\ \citenamefont
  {Muneyuki}}]{Toyabe_ATPase}%
  \BibitemOpen
  \bibfield  {author} {\bibinfo {author} {\bibfnamefont {Shoichi}\ \bibnamefont
  {Toyabe}}, \bibinfo {author} {\bibfnamefont {Tetsuaki}\ \bibnamefont
  {Okamoto}}, \bibinfo {author} {\bibfnamefont {Takahiro}\ \bibnamefont
  {Watanabe-Nakayama}}, \bibinfo {author} {\bibfnamefont {Hiroshi}\
  \bibnamefont {Taketani}}, \bibinfo {author} {\bibfnamefont {Seishi}\
  \bibnamefont {Kudo}}, \ and\ \bibinfo {author} {\bibfnamefont {Eiro}\
  \bibnamefont {Muneyuki}},\ }\bibfield  {title} {\enquote {\bibinfo {title}
  {{Nonequilibrium Energetics of a Single F-ATPase Molecule}},}\ }\href
  {\doibase 10.1103/PhysRevLett.104.198103} {\bibfield  {journal} {\bibinfo
  {journal} {Phys. Rev. Lett.}\ }\textbf {\bibinfo {volume} {104}},\ \bibinfo
  {pages} {198103} (\bibinfo {year} {2010})}\BibitemShut {NoStop}%
\bibitem [{\citenamefont {Ariga}\ \emph {et~al.}(2017)\citenamefont {Ariga},
  \citenamefont {Tomishige},\ and\ \citenamefont {Mizuno}}]{Ariga_Kinesin}%
  \BibitemOpen
  \bibfield  {author} {\bibinfo {author} {\bibfnamefont {Takayuki}\
  \bibnamefont {Ariga}}, \bibinfo {author} {\bibfnamefont {Michio}\
  \bibnamefont {Tomishige}}, \ and\ \bibinfo {author} {\bibfnamefont {Daisuke}\
  \bibnamefont {Mizuno}},\ }\bibfield  {title} {\enquote {\bibinfo {title}
  {{Nonequilibrium Energetics of Molecular Motor Kinesin}},}\ }\href
  {http://arxiv.org/abs/1704.05302} {\bibfield  {journal} {\bibinfo  {journal}
  {Arxiv}\ } (\bibinfo {year} {2017})},\ \Eprint
  {http://arxiv.org/abs/1704.05302} {arXiv:1704.05302} \BibitemShut {NoStop}%
\bibitem [{\citenamefont {Esposito}(2012)}]{Esposito2016stochcoarsgran}%
  \BibitemOpen
  \bibfield  {author} {\bibinfo {author} {\bibfnamefont {Massimiliano}\
  \bibnamefont {Esposito}},\ }\bibfield  {title} {\enquote {\bibinfo {title}
  {{Stochastic thermodynamics under coarse graining}},}\ }\href@noop {}
  {\bibfield  {journal} {\bibinfo  {journal} {Phys. Rev. E}\ }\textbf {\bibinfo
  {volume} {85}},\ \bibinfo {pages} {41125} (\bibinfo {year}
  {2012})}\BibitemShut {NoStop}%
\bibitem [{\citenamefont {Wang}\ \emph {et~al.}(2016)\citenamefont {Wang},
  \citenamefont {Kawaguchi}, \citenamefont {Sasa},\ and\ \citenamefont
  {Tang}}]{wang2016timescalesep}%
  \BibitemOpen
  \bibfield  {author} {\bibinfo {author} {\bibfnamefont {Shou-Wen}\
  \bibnamefont {Wang}}, \bibinfo {author} {\bibfnamefont {Kyogo}\ \bibnamefont
  {Kawaguchi}}, \bibinfo {author} {\bibfnamefont {Shin-ichi}\ \bibnamefont
  {Sasa}}, \ and\ \bibinfo {author} {\bibfnamefont {Lei-Han}\ \bibnamefont
  {Tang}},\ }\bibfield  {title} {\enquote {\bibinfo {title} {{Entropy
  Production of Nanosystems with Time Scale Separation}},}\ }\href@noop {}
  {\bibfield  {journal} {\bibinfo  {journal} {Phys. Rev. Lett.}\ }\textbf
  {\bibinfo {volume} {117}},\ \bibinfo {pages} {70601} (\bibinfo {year}
  {2016})}\BibitemShut {NoStop}%
\bibitem [{\citenamefont {Rold{\'{a}}n}\ \emph {et~al.}(2015)\citenamefont
  {Rold{\'{a}}n}, \citenamefont {Neri}, \citenamefont {D{\"{o}}rpinghaus},
  \citenamefont {Meyr},\ and\ \citenamefont {J{\"{u}}licher}}]{Roldan2015}%
  \BibitemOpen
  \bibfield  {author} {\bibinfo {author} {\bibfnamefont {{\'{E}}dgar}\
  \bibnamefont {Rold{\'{a}}n}}, \bibinfo {author} {\bibfnamefont {Izaak}\
  \bibnamefont {Neri}}, \bibinfo {author} {\bibfnamefont {Meik}\ \bibnamefont
  {D{\"{o}}rpinghaus}}, \bibinfo {author} {\bibfnamefont {Heinrich}\
  \bibnamefont {Meyr}}, \ and\ \bibinfo {author} {\bibfnamefont {Frank}\
  \bibnamefont {J{\"{u}}licher}},\ }\bibfield  {title} {\enquote {\bibinfo
  {title} {{Decision Making in the Arrow of Time}},}\ }\href {\doibase
  10.1103/PhysRevLett.115.250602} {\bibfield  {journal} {\bibinfo  {journal}
  {Phys. Rev. Lett.}\ }\textbf {\bibinfo {volume} {115}},\ \bibinfo {pages}
  {250602} (\bibinfo {year} {2015})},\ \Eprint
  {http://arxiv.org/abs/1508.02018} {arXiv:1508.02018} \BibitemShut {NoStop}%
\bibitem [{\citenamefont {Neri}\ \emph {et~al.}(2017)\citenamefont {Neri},
  \citenamefont {Rold{\'{a}}n},\ and\ \citenamefont
  {J{\"{u}}licher}}]{Neri2016}%
  \BibitemOpen
  \bibfield  {author} {\bibinfo {author} {\bibfnamefont {Izaak}\ \bibnamefont
  {Neri}}, \bibinfo {author} {\bibfnamefont {{\'{E}}dgar}\ \bibnamefont
  {Rold{\'{a}}n}}, \ and\ \bibinfo {author} {\bibfnamefont {Frank}\
  \bibnamefont {J{\"{u}}licher}},\ }\bibfield  {title} {\enquote {\bibinfo
  {title} {{Statistics of Infima and Stopping Times of Entropy Production and
  Applications to Active Molecular Processes}},}\ }\href {\doibase
  10.1103/PhysRevX.7.011019} {\bibfield  {journal} {\bibinfo  {journal} {Phys.
  Rev. X}\ }\textbf {\bibinfo {volume} {7}},\ \bibinfo {pages} {011019}
  (\bibinfo {year} {2017})},\ \Eprint {http://arxiv.org/abs/1604.04159}
  {arXiv:1604.04159} \BibitemShut {NoStop}%
\bibitem [{\citenamefont {Berezhkovskii}\ \emph {et~al.}(2006)\citenamefont
  {Berezhkovskii}, \citenamefont {Hummer},\ and\ \citenamefont
  {Bezrukov}}]{Berezhkovskii2006}%
  \BibitemOpen
  \bibfield  {author} {\bibinfo {author} {\bibfnamefont {Alexander}\
  \bibnamefont {Berezhkovskii}}, \bibinfo {author} {\bibfnamefont {Gerhard}\
  \bibnamefont {Hummer}}, \ and\ \bibinfo {author} {\bibfnamefont {Sergey}\
  \bibnamefont {Bezrukov}},\ }\bibfield  {title} {\enquote {\bibinfo {title}
  {{Identity of Distributions of Direct Uphill and Downhill Translocation Times
  for Particles Traversing Membrane Channels}},}\ }\href {\doibase
  10.1103/PhysRevLett.97.020601} {\bibfield  {journal} {\bibinfo  {journal}
  {Phys. Rev. Lett.}\ }\textbf {\bibinfo {volume} {97}},\ \bibinfo {pages}
  {020601} (\bibinfo {year} {2006})}\BibitemShut {NoStop}%
\bibitem [{\citenamefont {Qian}\ and\ \citenamefont {{Sunney
  Xie}}(2006)}]{Qian2006a}%
  \BibitemOpen
  \bibfield  {author} {\bibinfo {author} {\bibfnamefont {Hong}\ \bibnamefont
  {Qian}}\ and\ \bibinfo {author} {\bibfnamefont {X.}~\bibnamefont {{Sunney
  Xie}}},\ }\bibfield  {title} {\enquote {\bibinfo {title} {{Generalized
  Haldane equation and fluctuation theorem in the steady-state cycle kinetics
  of single enzymes}},}\ }\href {\doibase 10.1103/PhysRevE.74.010902}
  {\bibfield  {journal} {\bibinfo  {journal} {Phys. Rev. E}\ }\textbf {\bibinfo
  {volume} {74}},\ \bibinfo {pages} {010902} (\bibinfo {year} {2006})},\
  \Eprint {http://arxiv.org/abs/0507659} {arXiv:0507659 [cond-mat]}
  \BibitemShut {NoStop}%
\bibitem [{\citenamefont {Kolomeisky}\ \emph {et~al.}(2005)\citenamefont
  {Kolomeisky}, \citenamefont {Stukalin},\ and\ \citenamefont
  {Popov}}]{Kolomeisky2005}%
  \BibitemOpen
  \bibfield  {author} {\bibinfo {author} {\bibfnamefont {Anatoly~B.}\
  \bibnamefont {Kolomeisky}}, \bibinfo {author} {\bibfnamefont {Evgeny~B.}\
  \bibnamefont {Stukalin}}, \ and\ \bibinfo {author} {\bibfnamefont {Alex~A.}\
  \bibnamefont {Popov}},\ }\bibfield  {title} {\enquote {\bibinfo {title}
  {{Understanding mechanochemical coupling in kinesins using first-passage-time
  processes}},}\ }\href {\doibase 10.1103/PhysRevE.71.031902} {\bibfield
  {journal} {\bibinfo  {journal} {Phys. Rev. E}\ }\textbf {\bibinfo {volume}
  {71}},\ \bibinfo {pages} {031902} (\bibinfo {year} {2005})},\ \Eprint
  {http://arxiv.org/abs/0406578} {arXiv:0406578 [cond-mat]} \BibitemShut
  {NoStop}%
\bibitem [{\citenamefont {Stern}(1977)}]{Stern1977}%
  \BibitemOpen
  \bibfield  {author} {\bibinfo {author} {\bibfnamefont {Frederick}\
  \bibnamefont {Stern}},\ }\bibfield  {title} {\enquote {\bibinfo {title} {{An
  independence in Brownian motion with constant drift}},}\ }\href@noop {}
  {\bibfield  {journal} {\bibinfo  {journal} {Ann. Probab.}\ }\textbf {\bibinfo
  {volume} {5}},\ \bibinfo {pages} {571--572} (\bibinfo {year}
  {1977})}\BibitemShut {NoStop}%
\bibitem [{\citenamefont {Loutchko}\ \emph {et~al.}(2017)\citenamefont
  {Loutchko}, \citenamefont {Eisbach},\ and\ \citenamefont
  {Mikhailov}}]{Loutchko2017}%
  \BibitemOpen
  \bibfield  {author} {\bibinfo {author} {\bibfnamefont {Dimitri}\ \bibnamefont
  {Loutchko}}, \bibinfo {author} {\bibfnamefont {Maximilian}\ \bibnamefont
  {Eisbach}}, \ and\ \bibinfo {author} {\bibfnamefont {Alexander~S.}\
  \bibnamefont {Mikhailov}},\ }\bibfield  {title} {\enquote {\bibinfo {title}
  {{Stochastic thermodynamics of a chemical nanomachine: The channeling enzyme
  tryptophan synthase}},}\ }\href {\doibase 10.1063/1.4973544} {\bibfield
  {journal} {\bibinfo  {journal} {J. Chem. Phys.}\ }\textbf {\bibinfo {volume}
  {146}} (\bibinfo {year} {2017}),\ 10.1063/1.4973544}\BibitemShut {NoStop}%
\bibitem [{\citenamefont {Berg}(2008)}]{Berg2008}%
  \BibitemOpen
  \bibfield  {author} {\bibinfo {author} {\bibfnamefont {Johannes}\
  \bibnamefont {Berg}},\ }\bibfield  {title} {\enquote {\bibinfo {title}
  {{Out-of-equilibrium dynamics of gene expression and the Jarzynski
  equality}},}\ }\href {\doibase 10.1103/PhysRevLett.100.188101} {\bibfield
  {journal} {\bibinfo  {journal} {Phys. Rev. Lett.}\ }\textbf {\bibinfo
  {volume} {100}},\ \bibinfo {pages} {1--4} (\bibinfo {year} {2008})},\ \Eprint
  {http://arxiv.org/abs/0712.0170} {arXiv:0712.0170} \BibitemShut {NoStop}%
\bibitem [{\citenamefont {Liphardt}(2002)}]{Liphardt2002}%
  \BibitemOpen
  \bibfield  {author} {\bibinfo {author} {\bibfnamefont {J.}~\bibnamefont
  {Liphardt}},\ }\bibfield  {title} {\enquote {\bibinfo {title} {{Equilibrium
  Information from Nonequilibrium Measurements in an Experimental Test of
  Jarzynski's Equality}},}\ }\href {\doibase 10.1126/science.1071152}
  {\bibfield  {journal} {\bibinfo  {journal} {Science}\ }\textbf {\bibinfo
  {volume} {296}},\ \bibinfo {pages} {1832--1835} (\bibinfo {year}
  {2002})}\BibitemShut {NoStop}%
\bibitem [{\citenamefont {Schmiedl}\ and\ \citenamefont
  {Seifert}(2007{\natexlab{b}})}]{Seifert_networks}%
  \BibitemOpen
  \bibfield  {author} {\bibinfo {author} {\bibfnamefont {Tim}\ \bibnamefont
  {Schmiedl}}\ and\ \bibinfo {author} {\bibfnamefont {Udo}\ \bibnamefont
  {Seifert}},\ }\bibfield  {title} {\enquote {\bibinfo {title} {{Stochastic
  thermodynamics of chemical reaction networks}},}\ }\href {\doibase
  10.1063/1.2428297} {\bibfield  {journal} {\bibinfo  {journal} {J. Chem.
  Phys.}\ }\textbf {\bibinfo {volume} {126}},\ \bibinfo {pages} {044101}
  (\bibinfo {year} {2007}{\natexlab{b}})}\BibitemShut {NoStop}%
\bibitem [{\citenamefont {Schmiedl}\ and\ \citenamefont
  {Seifert}(2008)}]{Seifert_motors}%
  \BibitemOpen
  \bibfield  {author} {\bibinfo {author} {\bibfnamefont {T.}~\bibnamefont
  {Schmiedl}}\ and\ \bibinfo {author} {\bibfnamefont {U}~\bibnamefont
  {Seifert}},\ }\bibfield  {title} {\enquote {\bibinfo {title} {{Efficiency of
  molecular motors at maximum power}},}\ }\href {\doibase
  10.1209/0295-5075/83/30005} {\bibfield  {journal} {\bibinfo  {journal} {EPL
  (Europhysics Lett.}\ }\textbf {\bibinfo {volume} {83}},\ \bibinfo {pages}
  {30005} (\bibinfo {year} {2008})}\BibitemShut {NoStop}%
\bibitem [{\citenamefont {Hartich}\ \emph {et~al.}(2015)\citenamefont
  {Hartich}, \citenamefont {Barato},\ and\ \citenamefont
  {Seifert}}]{Seifert_sensing}%
  \BibitemOpen
  \bibfield  {author} {\bibinfo {author} {\bibfnamefont {David}\ \bibnamefont
  {Hartich}}, \bibinfo {author} {\bibfnamefont {Andre~C}\ \bibnamefont
  {Barato}}, \ and\ \bibinfo {author} {\bibfnamefont {Udo}\ \bibnamefont
  {Seifert}},\ }\bibfield  {title} {\enquote {\bibinfo {title} {{Nonequilibrium
  sensing and its analogy to kinetic proofreading}},}\ }\href {\doibase
  10.1088/1367-2630/17/5/055026} {\bibfield  {journal} {\bibinfo  {journal}
  {New J. Phys.}\ }\textbf {\bibinfo {volume} {17}},\ \bibinfo {pages} {055026}
  (\bibinfo {year} {2015})}\BibitemShut {NoStop}%
\bibitem [{\citenamefont {England}(2013)}]{England2013}%
  \BibitemOpen
  \bibfield  {author} {\bibinfo {author} {\bibfnamefont {Jeremy~L.}\
  \bibnamefont {England}},\ }\bibfield  {title} {\enquote {\bibinfo {title}
  {{Statistical physics of self-replication}},}\ }\href {\doibase
  10.1063/1.4818538} {\bibfield  {journal} {\bibinfo  {journal} {J. Chem.
  Phys.}\ }\textbf {\bibinfo {volume} {139}} (\bibinfo {year} {2013}),\
  10.1063/1.4818538},\ \Eprint {http://arxiv.org/abs/1209.1179}
  {arXiv:1209.1179} \BibitemShut {NoStop}%
\bibitem [{\citenamefont {England}(2015)}]{England2015}%
  \BibitemOpen
  \bibfield  {author} {\bibinfo {author} {\bibfnamefont {Jeremy~L.}\
  \bibnamefont {England}},\ }\bibfield  {title} {\enquote {\bibinfo {title}
  {{Dissipative adaptation in driven self-assembly}},}\ }\href {\doibase
  10.1038/nnano.2015.250} {\bibfield  {journal} {\bibinfo  {journal} {Nat.
  Nanotechnol.}\ }\textbf {\bibinfo {volume} {10}},\ \bibinfo {pages}
  {919--923} (\bibinfo {year} {2015})},\ \Eprint
  {http://arxiv.org/abs/arXiv:1502.07086} {arXiv:arXiv:1502.07086} \BibitemShut
  {NoStop}%
\bibitem [{\citenamefont {Perunov}\ \emph {et~al.}(2016)\citenamefont
  {Perunov}, \citenamefont {Marsland},\ and\ \citenamefont
  {England}}]{Perunov2016}%
  \BibitemOpen
  \bibfield  {author} {\bibinfo {author} {\bibfnamefont {Nikolay}\ \bibnamefont
  {Perunov}}, \bibinfo {author} {\bibfnamefont {Robert~A.}\ \bibnamefont
  {Marsland}}, \ and\ \bibinfo {author} {\bibfnamefont {Jeremy~L.}\
  \bibnamefont {England}},\ }\bibfield  {title} {\enquote {\bibinfo {title}
  {{Statistical physics of adaptation}},}\ }\href {\doibase
  10.1103/PhysRevX.6.021036} {\bibfield  {journal} {\bibinfo  {journal} {Phys.
  Rev. X}\ }\textbf {\bibinfo {volume} {6}} (\bibinfo {year} {2016}),\
  10.1103/PhysRevX.6.021036},\ \Eprint {http://arxiv.org/abs/1412.1875}
  {arXiv:1412.1875} \BibitemShut {NoStop}%
\bibitem [{\citenamefont {Rouvas-Nicolis}\ and\ \citenamefont
  {Nicolis}(2009)}]{Rouvas-Nicolis2009}%
  \BibitemOpen
  \bibfield  {author} {\bibinfo {author} {\bibfnamefont {Catherine}\
  \bibnamefont {Rouvas-Nicolis}}\ and\ \bibinfo {author} {\bibfnamefont
  {Gregoire}\ \bibnamefont {Nicolis}},\ }\bibfield  {title} {\enquote {\bibinfo
  {title} {{Butterfly effect}},}\ }\href {\doibase 10.4249/scholarpedia.1720}
  {\bibfield  {journal} {\bibinfo  {journal} {Scholarpedia}\ }\textbf {\bibinfo
  {volume} {4}},\ \bibinfo {pages} {1720} (\bibinfo {year} {2009})}\BibitemShut
  {NoStop}%
\bibitem [{\citenamefont {Weiss}(2003)}]{Weiss2003}%
  \BibitemOpen
  \bibfield  {author} {\bibinfo {author} {\bibfnamefont {Jeffrey~B.}\
  \bibnamefont {Weiss}},\ }\bibfield  {title} {\enquote {\bibinfo {title}
  {{Coordinate invariance in stochastic dynamical systems}},}\ }\href {\doibase
  10.1034/j.1600-0870.2003.00014.x} {\bibfield  {journal} {\bibinfo  {journal}
  {Tellus A}\ }\textbf {\bibinfo {volume} {55}},\ \bibinfo {pages} {208--218}
  (\bibinfo {year} {2003})}\BibitemShut {NoStop}%
\bibitem [{\citenamefont {Barkai}\ and\ \citenamefont
  {Leibler}(1997)}]{Barkai1997}%
  \BibitemOpen
  \bibfield  {author} {\bibinfo {author} {\bibfnamefont {N}~\bibnamefont
  {Barkai}}\ and\ \bibinfo {author} {\bibfnamefont {S}~\bibnamefont
  {Leibler}},\ }\bibfield  {title} {\enquote {\bibinfo {title} {{Robustness in
  simple biochemical networks to transfer and process information.}}}\ }\href
  {\doibase 10.1038/43199} {\bibfield  {journal} {\bibinfo  {journal} {Nature}\
  }\textbf {\bibinfo {volume} {387}},\ \bibinfo {pages} {913--917} (\bibinfo
  {year} {1997})}\BibitemShut {NoStop}%
\bibitem [{\citenamefont {Alon}\ \emph {et~al.}(1999)\citenamefont {Alon},
  \citenamefont {Surette}, \citenamefont {Barkai},\ and\ \citenamefont
  {Leibler}}]{Leibler1999}%
  \BibitemOpen
  \bibfield  {author} {\bibinfo {author} {\bibfnamefont {U}~\bibnamefont
  {Alon}}, \bibinfo {author} {\bibfnamefont {M~G}\ \bibnamefont {Surette}},
  \bibinfo {author} {\bibfnamefont {N}~\bibnamefont {Barkai}}, \ and\ \bibinfo
  {author} {\bibfnamefont {S}~\bibnamefont {Leibler}},\ }\bibfield  {title}
  {\enquote {\bibinfo {title} {{Robustness in bacterial chemotaxis.}}}\ }\href
  {\doibase 10.1038/16483} {\bibfield  {journal} {\bibinfo  {journal} {Nature}\
  }\textbf {\bibinfo {volume} {397}},\ \bibinfo {pages} {168--171} (\bibinfo
  {year} {1999})}\BibitemShut {NoStop}%
\bibitem [{\citenamefont {Qian}(2006)}]{Qian2006}%
  \BibitemOpen
  \bibfield  {author} {\bibinfo {author} {\bibfnamefont {Hong}\ \bibnamefont
  {Qian}},\ }\bibfield  {title} {\enquote {\bibinfo {title} {{Reducing
  Intrinsic Biochemical Noise in Cells and Its Thermodynamic Limit}},}\ }\href
  {\doibase 10.1016/j.jmb.2006.07.068} {\bibfield  {journal} {\bibinfo
  {journal} {J. Mol. Biol.}\ }\textbf {\bibinfo {volume} {362}},\ \bibinfo
  {pages} {387--392} (\bibinfo {year} {2006})}\BibitemShut {NoStop}%
\bibitem [{\citenamefont {Ito}\ and\ \citenamefont {Sagawa}(2013)}]{Ito2013}%
  \BibitemOpen
  \bibfield  {author} {\bibinfo {author} {\bibfnamefont {Sosuke}\ \bibnamefont
  {Ito}}\ and\ \bibinfo {author} {\bibfnamefont {Takahiro}\ \bibnamefont
  {Sagawa}},\ }\bibfield  {title} {\enquote {\bibinfo {title} {{Information
  thermodynamics on causal networks}},}\ }\href {\doibase
  10.1103/PhysRevLett.111.180603} {\bibfield  {journal} {\bibinfo  {journal}
  {Phys. Rev. Lett.}\ }\textbf {\bibinfo {volume} {111}},\ \bibinfo {pages}
  {1--6} (\bibinfo {year} {2013})},\ \Eprint {http://arxiv.org/abs/1306.2756}
  {arXiv:1306.2756} \BibitemShut {NoStop}%
\bibitem [{\citenamefont {Sartori}\ \emph {et~al.}(2014)\citenamefont
  {Sartori}, \citenamefont {Granger}, \citenamefont {Lee},\ and\ \citenamefont
  {Horowitz}}]{Sartori2014}%
  \BibitemOpen
  \bibfield  {author} {\bibinfo {author} {\bibfnamefont {Pablo}\ \bibnamefont
  {Sartori}}, \bibinfo {author} {\bibfnamefont {L{\'{e}}o}\ \bibnamefont
  {Granger}}, \bibinfo {author} {\bibfnamefont {Chiu~Fan}\ \bibnamefont {Lee}},
  \ and\ \bibinfo {author} {\bibfnamefont {Jordan~M.}\ \bibnamefont
  {Horowitz}},\ }\bibfield  {title} {\enquote {\bibinfo {title} {{Thermodynamic
  Costs of Information Processing in Sensory Adaptation}},}\ }\href {\doibase
  10.1371/journal.pcbi.1003974} {\bibfield  {journal} {\bibinfo  {journal}
  {PLoS Comput. Biol.}\ }\textbf {\bibinfo {volume} {10}},\ \bibinfo {pages}
  {e1003974} (\bibinfo {year} {2014})},\ \Eprint
  {http://arxiv.org/abs/1404.1027} {arXiv:1404.1027} \BibitemShut {NoStop}%
\bibitem [{\citenamefont {Lang}\ \emph {et~al.}(2014)\citenamefont {Lang},
  \citenamefont {Fisher}, \citenamefont {Mora},\ and\ \citenamefont
  {Mehta}}]{Lang2014}%
  \BibitemOpen
  \bibfield  {author} {\bibinfo {author} {\bibfnamefont {Alex~H.}\ \bibnamefont
  {Lang}}, \bibinfo {author} {\bibfnamefont {Charles~K.}\ \bibnamefont
  {Fisher}}, \bibinfo {author} {\bibfnamefont {Thierry}\ \bibnamefont {Mora}},
  \ and\ \bibinfo {author} {\bibfnamefont {Pankaj}\ \bibnamefont {Mehta}},\
  }\bibfield  {title} {\enquote {\bibinfo {title} {{Thermodynamics of
  Statistical Inference by Cells}},}\ }\href {\doibase
  10.1103/PhysRevLett.113.148103} {\bibfield  {journal} {\bibinfo  {journal}
  {Phys. Rev. Lett.}\ }\textbf {\bibinfo {volume} {113}},\ \bibinfo {pages}
  {148103} (\bibinfo {year} {2014})},\ \Eprint {http://arxiv.org/abs/1405.4001}
  {arXiv:1405.4001} \BibitemShut {NoStop}%
\bibitem [{\citenamefont {Tom{\'{e}}}\ and\ \citenamefont
  {de~Oliveira}(2012)}]{Tome2012}%
  \BibitemOpen
  \bibfield  {author} {\bibinfo {author} {\bibfnamefont {T{\^{a}}nia}\
  \bibnamefont {Tom{\'{e}}}}\ and\ \bibinfo {author} {\bibfnamefont
  {M{\'{a}}rio~J.}\ \bibnamefont {de~Oliveira}},\ }\bibfield  {title} {\enquote
  {\bibinfo {title} {{Entropy Production in Nonequilibrium Systems at
  Stationary States}},}\ }\href {\doibase 10.1103/PhysRevLett.108.020601}
  {\bibfield  {journal} {\bibinfo  {journal} {Phys. Rev. Lett.}\ }\textbf
  {\bibinfo {volume} {108}},\ \bibinfo {pages} {020601} (\bibinfo {year}
  {2012})}\BibitemShut {NoStop}%
\bibitem [{\citenamefont {Stein}\ and\ \citenamefont
  {Litman}(2014)}]{Stein2014}%
  \BibitemOpen
  \bibfield  {author} {\bibinfo {author} {\bibfnamefont {Wilfred~D.}\
  \bibnamefont {Stein}}\ and\ \bibinfo {author} {\bibfnamefont {Thomas}\
  \bibnamefont {Litman}},\ }\href {\doibase 10.1016/C2012-0-07690-9} {\emph
  {\bibinfo {title} {Channels, Carriers, Pumps An Introd. to Membr. Transp.}}}\
  (\bibinfo {year} {2014})\ pp.\ \bibinfo {pages} {1--406}\BibitemShut
  {NoStop}%
\bibitem [{\citenamefont {Bezrukov}\ \emph {et~al.}(2000)\citenamefont
  {Bezrukov}, \citenamefont {Berezhkovskii}, \citenamefont {Pustovoit},\ and\
  \citenamefont {Szabo}}]{Bezrukov2000}%
  \BibitemOpen
  \bibfield  {author} {\bibinfo {author} {\bibfnamefont {Sergey~M.}\
  \bibnamefont {Bezrukov}}, \bibinfo {author} {\bibfnamefont {Alexander~M.}\
  \bibnamefont {Berezhkovskii}}, \bibinfo {author} {\bibfnamefont {Mark~A.}\
  \bibnamefont {Pustovoit}}, \ and\ \bibinfo {author} {\bibfnamefont {Attila}\
  \bibnamefont {Szabo}},\ }\bibfield  {title} {\enquote {\bibinfo {title}
  {{Particle number fluctuations in a membrane channel}},}\ }\href {\doibase
  10.1063/1.1314862} {\bibfield  {journal} {\bibinfo  {journal} {J. Chem.
  Phys.}\ }\textbf {\bibinfo {volume} {113}},\ \bibinfo {pages} {8206--8211}
  (\bibinfo {year} {2000})}\BibitemShut {NoStop}%
\bibitem [{\citenamefont {Berezhkovskii}\ and\ \citenamefont
  {Bezrukov}(2008)}]{Berezhkovskii2008}%
  \BibitemOpen
  \bibfield  {author} {\bibinfo {author} {\bibfnamefont {Alexander}\
  \bibnamefont {Berezhkovskii}}\ and\ \bibinfo {author} {\bibfnamefont
  {Sergey}\ \bibnamefont {Bezrukov}},\ }\bibfield  {title} {\enquote {\bibinfo
  {title} {{Counting Translocations of Strongly Repelling Particles through
  Single Channels: Fluctuation Theorem for Membrane Transport}},}\ }\href
  {\doibase 10.1103/PhysRevLett.100.038104} {\bibfield  {journal} {\bibinfo
  {journal} {Phys. Rev. Lett.}\ }\textbf {\bibinfo {volume} {100}},\ \bibinfo
  {pages} {038104} (\bibinfo {year} {2008})}\BibitemShut {NoStop}%
\bibitem [{\citenamefont {Gr{\"{u}}nwald}\ \emph {et~al.}(2011)\citenamefont
  {Gr{\"{u}}nwald}, \citenamefont {Singer},\ and\ \citenamefont
  {Rout}}]{Grunwald2011}%
  \BibitemOpen
  \bibfield  {author} {\bibinfo {author} {\bibfnamefont {David}\ \bibnamefont
  {Gr{\"{u}}nwald}}, \bibinfo {author} {\bibfnamefont {Robert~H}\ \bibnamefont
  {Singer}}, \ and\ \bibinfo {author} {\bibfnamefont {Michael}\ \bibnamefont
  {Rout}},\ }\bibfield  {title} {\enquote {\bibinfo {title} {{Nuclear export
  dynamics of RNA?protein complexes}},}\ }\href {\doibase
  10.1038/nature10318} {\bibfield  {journal} {\bibinfo  {journal} {Nature}\
  }\textbf {\bibinfo {volume} {475}},\ \bibinfo {pages} {333--341} (\bibinfo
  {year} {2011})},\ \Eprint {http://arxiv.org/abs/NIHMS150003}
  {arXiv:NIHMS150003} \BibitemShut {NoStop}%
\bibitem [{\citenamefont {Gingrich}\ \emph {et~al.}(2016)\citenamefont
  {Gingrich}, \citenamefont {Horowitz}, \citenamefont {Perunov},\ and\
  \citenamefont {England}}]{Gingrich2016}%
  \BibitemOpen
  \bibfield  {author} {\bibinfo {author} {\bibfnamefont {Todd~R.}\ \bibnamefont
  {Gingrich}}, \bibinfo {author} {\bibfnamefont {Jordan~M.}\ \bibnamefont
  {Horowitz}}, \bibinfo {author} {\bibfnamefont {Nikolay}\ \bibnamefont
  {Perunov}}, \ and\ \bibinfo {author} {\bibfnamefont {Jeremy~L.}\ \bibnamefont
  {England}},\ }\bibfield  {title} {\enquote {\bibinfo {title} {{Dissipation
  Bounds All Steady-State Current Fluctuations}},}\ }\href {\doibase
  10.1103/PhysRevLett.116.120601} {\bibfield  {journal} {\bibinfo  {journal}
  {Phys. Rev. Lett.}\ }\textbf {\bibinfo {volume} {116}},\ \bibinfo {pages}
  {120601} (\bibinfo {year} {2016})},\ \Eprint
  {http://arxiv.org/abs/1512.02212} {arXiv:1512.02212} \BibitemShut {NoStop}%
\bibitem [{\citenamefont {Barato}\ and\ \citenamefont
  {Seifert}(2015)}]{Barato2015}%
  \BibitemOpen
  \bibfield  {author} {\bibinfo {author} {\bibfnamefont {Andre~C.}\
  \bibnamefont {Barato}}\ and\ \bibinfo {author} {\bibfnamefont {Udo}\
  \bibnamefont {Seifert}},\ }\bibfield  {title} {\enquote {\bibinfo {title}
  {{Thermodynamic Uncertainty Relation for Biomolecular Processes}},}\ }\href
  {\doibase 10.1103/PhysRevLett.114.158101} {\bibfield  {journal} {\bibinfo
  {journal} {Phys. Rev. Lett.}\ }\textbf {\bibinfo {volume} {114}},\ \bibinfo
  {pages} {158101} (\bibinfo {year} {2015})},\ \Eprint
  {http://arxiv.org/abs/1502.0594} {arXiv:1502.0594} \BibitemShut {NoStop}%
\bibitem [{\citenamefont {Pietzonka}\ \emph {et~al.}(2016)\citenamefont
  {Pietzonka}, \citenamefont {Barato},\ and\ \citenamefont
  {Seifert}}]{Pietzonka2016}%
  \BibitemOpen
  \bibfield  {author} {\bibinfo {author} {\bibfnamefont {Patrick}\ \bibnamefont
  {Pietzonka}}, \bibinfo {author} {\bibfnamefont {Andre~C.}\ \bibnamefont
  {Barato}}, \ and\ \bibinfo {author} {\bibfnamefont {Udo}\ \bibnamefont
  {Seifert}},\ }\bibfield  {title} {\enquote {\bibinfo {title} {{Universal
  bound on the efficiency of molecular motors}},}\ }\href {\doibase
  10.1088/1742-5468/2016/12/124004} {\bibfield  {journal} {\bibinfo  {journal}
  {J. Stat. Mech. Theory Exp.}\ }\textbf {\bibinfo {volume} {2016}},\ \bibinfo
  {pages} {124004} (\bibinfo {year} {2016})},\ \Eprint
  {http://arxiv.org/abs/1609.08046} {arXiv:1609.08046} \BibitemShut {NoStop}%
\bibitem [{\citenamefont {Pietzonka}\ \emph {et~al.}(2017)\citenamefont
  {Pietzonka}, \citenamefont {Ritort},\ and\ \citenamefont
  {Seifert}}]{Pietzonka2017}%
  \BibitemOpen
  \bibfield  {author} {\bibinfo {author} {\bibfnamefont {Patrick}\ \bibnamefont
  {Pietzonka}}, \bibinfo {author} {\bibfnamefont {Felix}\ \bibnamefont
  {Ritort}}, \ and\ \bibinfo {author} {\bibfnamefont {Udo}\ \bibnamefont
  {Seifert}},\ }\bibfield  {title} {\enquote {\bibinfo {title} {{Finite-time
  generalization of the thermodynamic uncertainty relation}},}\ }\href
  {\doibase 10.1103/PhysRevE.96.012101} {\bibfield  {journal} {\bibinfo
  {journal} {Phys. Rev. E}\ }\textbf {\bibinfo {volume} {96}},\ \bibinfo
  {pages} {012101} (\bibinfo {year} {2017})},\ \Eprint
  {http://arxiv.org/abs/1702.07699} {arXiv:1702.07699} \BibitemShut {NoStop}%
\bibitem [{\citenamefont {Horowitz}\ and\ \citenamefont
  {Gingrich}(2017)}]{Horowitz2017}%
  \BibitemOpen
  \bibfield  {author} {\bibinfo {author} {\bibfnamefont {Jordan~M.}\
  \bibnamefont {Horowitz}}\ and\ \bibinfo {author} {\bibfnamefont {Todd~R.}\
  \bibnamefont {Gingrich}},\ }\bibfield  {title} {\enquote {\bibinfo {title}
  {{Proof of the finite-time thermodynamic uncertainty relation for
  steady-state currents}},}\ }\href {\doibase 10.1103/PhysRevE.96.020103}
  {\bibfield  {journal} {\bibinfo  {journal} {Phys. Rev. E}\ }\textbf {\bibinfo
  {volume} {96}},\ \bibinfo {pages} {020103} (\bibinfo {year} {2017})},\
  \Eprint {http://arxiv.org/abs/1707.03805} {arXiv:1707.03805} \BibitemShut
  {NoStop}%
\bibitem [{\citenamefont {Riedel?Kruse}\ \emph {et~al.}(2007)\citenamefont
  {Riedel?Kruse}, \citenamefont {Hilfinger}, \citenamefont {Howard},\ and\
  \citenamefont {J{\"{u}}licher}}]{RiedelKruse2007}%
  \BibitemOpen
  \bibfield  {author} {\bibinfo {author} {\bibfnamefont {Ingmar~H.}\
  \bibnamefont {Riedel?Kruse}}, \bibinfo {author} {\bibfnamefont {Andreas}\
  \bibnamefont {Hilfinger}}, \bibinfo {author} {\bibfnamefont {Jonathon}\
  \bibnamefont {Howard}}, \ and\ \bibinfo {author} {\bibfnamefont {Frank}\
  \bibnamefont {J{\"{u}}licher}},\ }\bibfield  {title} {\enquote {\bibinfo
  {title} {{How molecular motors shape the flagellar beat}},}\ }\href {\doibase
  10.2976/1.2773861} {\bibfield  {journal} {\bibinfo  {journal} {HFSP J.}\
  }\textbf {\bibinfo {volume} {1}},\ \bibinfo {pages} {192--208} (\bibinfo
  {year} {2007})}\BibitemShut {NoStop}%
\bibitem [{\citenamefont {Wan}\ and\ \citenamefont
  {Goldstein}(2014)}]{Wan2014}%
  \BibitemOpen
  \bibfield  {author} {\bibinfo {author} {\bibfnamefont {Kirsty~Y.}\
  \bibnamefont {Wan}}\ and\ \bibinfo {author} {\bibfnamefont {Raymond~E.}\
  \bibnamefont {Goldstein}},\ }\bibfield  {title} {\enquote {\bibinfo {title}
  {{Rhythmicity, recurrence, and recovery of flagellar beating}},}\ }\href
  {\doibase 10.1103/PhysRevLett.113.238103} {\bibfield  {journal} {\bibinfo
  {journal} {Phys. Rev. Lett.}\ }\textbf {\bibinfo {volume} {113}} (\bibinfo
  {year} {2014}),\ 10.1103/PhysRevLett.113.238103},\ \Eprint
  {http://arxiv.org/abs/arXiv:1406.3725v2} {arXiv:arXiv:1406.3725v2}
  \BibitemShut {NoStop}%
\bibitem [{\citenamefont {Singla}(2006)}]{Singla2006}%
  \BibitemOpen
  \bibfield  {author} {\bibinfo {author} {\bibfnamefont {V.}~\bibnamefont
  {Singla}},\ }\bibfield  {title} {\enquote {\bibinfo {title} {{The Primary
  Cilium as the Cell's Antenna: Signaling at a Sensory Organelle}},}\ }\href
  {\doibase 10.1126/science.1124534} {\bibfield  {journal} {\bibinfo  {journal}
  {Science}\ }\textbf {\bibinfo {volume} {313}},\ \bibinfo {pages} {629--633}
  (\bibinfo {year} {2006})}\BibitemShut {NoStop}%
\bibitem [{\citenamefont {Barnes}(1961)}]{Barnes1961}%
  \BibitemOpen
  \bibfield  {author} {\bibinfo {author} {\bibfnamefont {B~G}\ \bibnamefont
  {Barnes}},\ }\bibfield  {title} {\enquote {\bibinfo {title} {{Ciliated
  secretory cells in the pars distalis of the mouse hypophysis.}}}\ }\href
  {\doibase 10.1016/S0022-5320(61)80019-1} {\bibfield  {journal} {\bibinfo
  {journal} {J. Ultrastruct. Res.}\ }\textbf {\bibinfo {volume} {5}},\ \bibinfo
  {pages} {453--467} (\bibinfo {year} {1961})}\BibitemShut {NoStop}%
\bibitem [{\citenamefont {Paijmans}\ \emph {et~al.}(2017)\citenamefont
  {Paijmans}, \citenamefont {Lubensky},\ and\ \citenamefont {ten
  Wolde}}]{Paijmans2017}%
  \BibitemOpen
  \bibfield  {author} {\bibinfo {author} {\bibfnamefont {Joris}\ \bibnamefont
  {Paijmans}}, \bibinfo {author} {\bibfnamefont {David~K.}\ \bibnamefont
  {Lubensky}}, \ and\ \bibinfo {author} {\bibfnamefont {Pieter~Rein}\
  \bibnamefont {ten Wolde}},\ }\bibfield  {title} {\enquote {\bibinfo {title}
  {{A thermodynamically consistent model of the post-translational Kai
  circadian clock}},}\ }\href {\doibase 10.1371/journal.pcbi.1005415}
  {\bibfield  {journal} {\bibinfo  {journal} {PLoS Comput. Biol.}\ }\textbf
  {\bibinfo {volume} {13}} (\bibinfo {year} {2017}),\
  10.1371/journal.pcbi.1005415},\ \Eprint {http://arxiv.org/abs/1612.02715}
  {arXiv:1612.02715} \BibitemShut {NoStop}%
\bibitem [{\citenamefont {{J. C. Kimmel et
  al.}}(2017)}]{JacobC.Kimmeletal.2017}%
  \BibitemOpen
  \bibfield  {author} {\bibinfo {author} {\bibnamefont {{J. C. Kimmel et
  al.}}},\ }\bibfield  {title} {\enquote {\bibinfo {title} {{Inferring cell
  state by quantitative motility analysis reveals a dynamic state system and
  broken detailed balance}},}\ }\href@noop {} {\bibfield  {journal} {\bibinfo
  {journal} {bioRxiv:168534}\ } (\bibinfo {year} {2017})}\BibitemShut {NoStop}%
\bibitem [{\citenamefont {Lander}\ \emph {et~al.}(2012)\citenamefont {Lander},
  \citenamefont {Mehl}, \citenamefont {Blickle}, \citenamefont {Bechinger},\
  and\ \citenamefont {Seifert}}]{Lander2012}%
  \BibitemOpen
  \bibfield  {author} {\bibinfo {author} {\bibfnamefont {B.}~\bibnamefont
  {Lander}}, \bibinfo {author} {\bibfnamefont {J.}~\bibnamefont {Mehl}},
  \bibinfo {author} {\bibfnamefont {V.}~\bibnamefont {Blickle}}, \bibinfo
  {author} {\bibfnamefont {C.}~\bibnamefont {Bechinger}}, \ and\ \bibinfo
  {author} {\bibfnamefont {U.}~\bibnamefont {Seifert}},\ }\bibfield  {title}
  {\enquote {\bibinfo {title} {{Noninvasive measurement of dissipation in
  colloidal systems}},}\ }\href {\doibase 10.1103/PhysRevE.86.030401}
  {\bibfield  {journal} {\bibinfo  {journal} {Phys. Rev. E - Stat. Nonlinear,
  Soft Matter Phys.}\ }\textbf {\bibinfo {volume} {86}} (\bibinfo {year}
  {2012}),\ 10.1103/PhysRevE.86.030401},\ \Eprint
  {http://arxiv.org/abs/1209.2009} {arXiv:1209.2009} \BibitemShut {NoStop}%
\bibitem [{\citenamefont {Shannon}\ and\ \citenamefont
  {Weaver}(1949)}]{Shannon}%
  \BibitemOpen
  \bibfield  {author} {\bibinfo {author} {\bibfnamefont {Claude~Elwood}\
  \bibnamefont {Shannon}}\ and\ \bibinfo {author} {\bibfnamefont {Warren}\
  \bibnamefont {Weaver}},\ }\href@noop {} {\emph {\bibinfo {title} {{The
  mathematical theory of communication}}}}\ (\bibinfo  {publisher} {University
  of Illinois Press},\ \bibinfo {address} {Urbana},\ \bibinfo {year} {1949})\
  p.\ \bibinfo {pages} {125}\BibitemShut {NoStop}%
\bibitem [{\citenamefont {Weber}\ \emph {et~al.}(2015)\citenamefont {Weber},
  \citenamefont {Suzuki}, \citenamefont {Schaller}, \citenamefont {Aranson},
  \citenamefont {Bausch},\ and\ \citenamefont {Frey}}]{Weber2015}%
  \BibitemOpen
  \bibfield  {author} {\bibinfo {author} {\bibfnamefont {Christoph~A.}\
  \bibnamefont {Weber}}, \bibinfo {author} {\bibfnamefont {Ryo}\ \bibnamefont
  {Suzuki}}, \bibinfo {author} {\bibfnamefont {Volker}\ \bibnamefont
  {Schaller}}, \bibinfo {author} {\bibfnamefont {Igor~S.}\ \bibnamefont
  {Aranson}}, \bibinfo {author} {\bibfnamefont {Andreas~R.}\ \bibnamefont
  {Bausch}}, \ and\ \bibinfo {author} {\bibfnamefont {Erwin}\ \bibnamefont
  {Frey}},\ }\bibfield  {title} {\enquote {\bibinfo {title} {{Random bursts
  determine dynamics of active filaments}},}\ }\href {\doibase
  10.1073/pnas.1421322112} {\bibfield  {journal} {\bibinfo  {journal} {Proc.
  Natl. Acad. Sci.}\ }\textbf {\bibinfo {volume} {112}},\ \bibinfo {pages}
  {10703--10707} (\bibinfo {year} {2015})}\BibitemShut {NoStop}%
\bibitem [{\citenamefont {Everaers}\ \emph {et~al.}(2006)\citenamefont
  {Everaers}, \citenamefont {Julicher}, \citenamefont {Ajdari},\ and\
  \citenamefont {Maggs}}]{Everaers1999}%
  \BibitemOpen
  \bibfield  {author} {\bibinfo {author} {\bibfnamefont {R}~\bibnamefont
  {Everaers}}, \bibinfo {author} {\bibfnamefont {F}~\bibnamefont {Julicher}},
  \bibinfo {author} {\bibfnamefont {A}~\bibnamefont {Ajdari}}, \ and\ \bibinfo
  {author} {\bibfnamefont {A~C}\ \bibnamefont {Maggs}},\ }\bibfield  {title}
  {\enquote {\bibinfo {title} {{Dynamic Fluctuations of Semiflexible
  Polymers}},}\ }\href {\doibase 10.1103/PhysRevLett.82.3717} {\bibfield
  {journal} {\bibinfo  {journal} {Phys. Rev. E - Stat. Nonlinear Soft Matter
  Phys.}\ }\textbf {\bibinfo {volume} {73}},\ \bibinfo {pages} {021507}
  (\bibinfo {year} {2006})},\ \Eprint {http://arxiv.org/abs/9808010}
  {arXiv:9808010 [cond-mat]} \BibitemShut {NoStop}%
\bibitem [{\citenamefont {Liverpool}(2003)}]{Liverpool2003}%
  \BibitemOpen
  \bibfield  {author} {\bibinfo {author} {\bibfnamefont {Tanniemola~B.}\
  \bibnamefont {Liverpool}},\ }\bibfield  {title} {\enquote {\bibinfo {title}
  {{Anomalous fluctuations of active polar filaments}},}\ }\href {\doibase
  10.1103/PhysRevE.67.031909} {\bibfield  {journal} {\bibinfo  {journal} {Phys.
  Rev. E}\ }\textbf {\bibinfo {volume} {67}},\ \bibinfo {pages} {031909}
  (\bibinfo {year} {2003})},\ \Eprint {http://arxiv.org/abs/0208168}
  {arXiv:0208168 [cond-mat]} \BibitemShut {NoStop}%
\bibitem [{\citenamefont {Levine}\ \emph {et~al.}(2004)\citenamefont {Levine},
  \citenamefont {Liverpool},\ and\ \citenamefont {MacKintosh}}]{Levine2004}%
  \BibitemOpen
  \bibfield  {author} {\bibinfo {author} {\bibfnamefont {Alex~J.}\ \bibnamefont
  {Levine}}, \bibinfo {author} {\bibfnamefont {T.~B.}\ \bibnamefont
  {Liverpool}}, \ and\ \bibinfo {author} {\bibfnamefont {F.~C.}\ \bibnamefont
  {MacKintosh}},\ }\bibfield  {title} {\enquote {\bibinfo {title} {{Dynamics of
  rigid and flexible extended bodies in viscous films and membranes}},}\ }\href
  {\doibase 10.1103/PhysRevLett.93.038102} {\bibfield  {journal} {\bibinfo
  {journal} {Phys. Rev. Lett.}\ }\textbf {\bibinfo {volume} {93}},\ \bibinfo
  {pages} {038102--1} (\bibinfo {year} {2004})},\ \Eprint
  {http://arxiv.org/abs/0306065} {arXiv:0306065 [cond-mat]} \BibitemShut
  {NoStop}%
\bibitem [{\citenamefont {Kikuchi}\ \emph {et~al.}(2009)\citenamefont
  {Kikuchi}, \citenamefont {Ehrlicher}, \citenamefont {Koch}, \citenamefont
  {Kas}, \citenamefont {Ramaswamy},\ and\ \citenamefont
  {Rao}}]{Kikuchi24112009}%
  \BibitemOpen
  \bibfield  {author} {\bibinfo {author} {\bibfnamefont {Norio}\ \bibnamefont
  {Kikuchi}}, \bibinfo {author} {\bibfnamefont {Allen}\ \bibnamefont
  {Ehrlicher}}, \bibinfo {author} {\bibfnamefont {Daniel}\ \bibnamefont
  {Koch}}, \bibinfo {author} {\bibfnamefont {J.~A.}\ \bibnamefont {Kas}},
  \bibinfo {author} {\bibfnamefont {Sriram}\ \bibnamefont {Ramaswamy}}, \ and\
  \bibinfo {author} {\bibfnamefont {Madan}\ \bibnamefont {Rao}},\ }\bibfield
  {title} {\enquote {\bibinfo {title} {{Buckling, stiffening, and negative
  dissipation in the dynamics of a biopolymer in an active medium}},}\ }\href
  {\doibase 10.1073/pnas.0900451106} {\bibfield  {journal} {\bibinfo  {journal}
  {Proc. Natl. Acad. Sci.}\ }\textbf {\bibinfo {volume} {106}},\ \bibinfo
  {pages} {19776--19779} (\bibinfo {year} {2009})}\BibitemShut {NoStop}%
\bibitem [{\citenamefont {Loi}\ \emph {et~al.}(2011)\citenamefont {Loi},
  \citenamefont {Mossa},\ and\ \citenamefont {Cugliandolo}}]{Loi2011}%
  \BibitemOpen
  \bibfield  {author} {\bibinfo {author} {\bibfnamefont {Davide}\ \bibnamefont
  {Loi}}, \bibinfo {author} {\bibfnamefont {Stefano}\ \bibnamefont {Mossa}}, \
  and\ \bibinfo {author} {\bibfnamefont {Leticia~F.}\ \bibnamefont
  {Cugliandolo}},\ }\bibfield  {title} {\enquote {\bibinfo {title}
  {{Non-conservative forces and effective temperatures in active polymers}},}\
  }\href {\doibase 10.1039/c1sm05819c} {\bibfield  {journal} {\bibinfo
  {journal} {Soft Matter}\ }\textbf {\bibinfo {volume} {7}},\ \bibinfo {pages}
  {10193} (\bibinfo {year} {2011})},\ \Eprint {http://arxiv.org/abs/1105.0806}
  {arXiv:1105.0806} \BibitemShut {NoStop}%
\bibitem [{\citenamefont {Ghosh}\ and\ \citenamefont
  {Gov}(2014)}]{Ghosh20141065}%
  \BibitemOpen
  \bibfield  {author} {\bibinfo {author} {\bibfnamefont {A}~\bibnamefont
  {Ghosh}}\ and\ \bibinfo {author} {\bibfnamefont {N~S}\ \bibnamefont {Gov}},\
  }\bibfield  {title} {\enquote {\bibinfo {title} {{Dynamics of Active
  Semiflexible Polymers}},}\ }\href {\doibase
  http://dx.doi.org/10.1016/j.bpj.2014.07.034} {\bibfield  {journal} {\bibinfo
  {journal} {Biophys. J.}\ }\textbf {\bibinfo {volume} {107}},\ \bibinfo
  {pages} {1065--1073} (\bibinfo {year} {2014})}\BibitemShut {NoStop}%
\bibitem [{\citenamefont {Eisenstecken}\ \emph {et~al.}(2017)\citenamefont
  {Eisenstecken}, \citenamefont {Gompper},\ and\ \citenamefont
  {Winkler}}]{Eisenstecken2017}%
  \BibitemOpen
  \bibfield  {author} {\bibinfo {author} {\bibfnamefont {Thomas}\ \bibnamefont
  {Eisenstecken}}, \bibinfo {author} {\bibfnamefont {Gerhard}\ \bibnamefont
  {Gompper}}, \ and\ \bibinfo {author} {\bibfnamefont {Roland~G.}\ \bibnamefont
  {Winkler}},\ }\bibfield  {title} {\enquote {\bibinfo {title} {{Internal
  dynamics of semiflexible polymers with active noise}},}\ }\href {\doibase
  10.1063/1.4981012} {\bibfield  {journal} {\bibinfo  {journal} {J. Chem.
  Phys.}\ }\textbf {\bibinfo {volume} {146}},\ \bibinfo {pages} {154903}
  (\bibinfo {year} {2017})}\BibitemShut {NoStop}%
\bibitem [{\citenamefont {Kratky}\ and\ \citenamefont
  {Porod}(1949)}]{kratkyPorod}%
  \BibitemOpen
  \bibfield  {author} {\bibinfo {author} {\bibfnamefont {O}~\bibnamefont
  {Kratky}}\ and\ \bibinfo {author} {\bibfnamefont {G}~\bibnamefont {Porod}},\
  }\bibfield  {title} {\enquote {\bibinfo {title} {{R{\"{o}}ntgenuntersuchung
  gel{\"{o}}ster fadenmolek{\"{u}}le}},}\ }\href {\doibase
  10.1002/recl.19490681203} {\bibfield  {journal} {\bibinfo  {journal} {Recl.
  des Trav. Chim. des Pays-Bas}\ }\textbf {\bibinfo {volume} {68}},\ \bibinfo
  {pages} {1106--1122} (\bibinfo {year} {1949})}\BibitemShut {NoStop}%
\bibitem [{\citenamefont {Goldstein}\ and\ \citenamefont
  {Langer}(1995)}]{Goldstein1995}%
  \BibitemOpen
  \bibfield  {author} {\bibinfo {author} {\bibfnamefont {Raymond~E.}\
  \bibnamefont {Goldstein}}\ and\ \bibinfo {author} {\bibfnamefont
  {Stephen~A.}\ \bibnamefont {Langer}},\ }\bibfield  {title} {\enquote
  {\bibinfo {title} {{Nonlinear Dynamics of Stiff Polymers}},}\ }\href
  {\doibase 10.1103/PhysRevLett.75.1094} {\bibfield  {journal} {\bibinfo
  {journal} {Phys. Rev. Lett.}\ }\textbf {\bibinfo {volume} {75}},\ \bibinfo
  {pages} {1094--1097} (\bibinfo {year} {1995})}\BibitemShut {NoStop}%
\bibitem [{\citenamefont {Hallatschek}\ \emph {et~al.}(2007)\citenamefont
  {Hallatschek}, \citenamefont {Frey},\ and\ \citenamefont
  {Kroy}}]{Hallatschek2007}%
  \BibitemOpen
  \bibfield  {author} {\bibinfo {author} {\bibfnamefont {Oskar}\ \bibnamefont
  {Hallatschek}}, \bibinfo {author} {\bibfnamefont {Erwin}\ \bibnamefont
  {Frey}}, \ and\ \bibinfo {author} {\bibfnamefont {Klaus}\ \bibnamefont
  {Kroy}},\ }\bibfield  {title} {\enquote {\bibinfo {title} {{Tension dynamics
  in semiflexible polymers. I. Coarse-grained equations of motion}},}\ }\href
  {\doibase 10.1103/PhysRevE.75.031905} {\bibfield  {journal} {\bibinfo
  {journal} {Phys. Rev. E}\ }\textbf {\bibinfo {volume} {75}},\ \bibinfo
  {pages} {031905} (\bibinfo {year} {2007})}\BibitemShut {NoStop}%
\bibitem [{\citenamefont {Koenderink}\ \emph {et~al.}(2006)\citenamefont
  {Koenderink}, \citenamefont {Atakhorrami}, \citenamefont {MacKintosh},\ and\
  \citenamefont {Schmidt}}]{Koenderink2006}%
  \BibitemOpen
  \bibfield  {author} {\bibinfo {author} {\bibfnamefont {G.~H.}\ \bibnamefont
  {Koenderink}}, \bibinfo {author} {\bibfnamefont {M.}~\bibnamefont
  {Atakhorrami}}, \bibinfo {author} {\bibfnamefont {F.~C.}\ \bibnamefont
  {MacKintosh}}, \ and\ \bibinfo {author} {\bibfnamefont {C.~F.}\ \bibnamefont
  {Schmidt}},\ }\bibfield  {title} {\enquote {\bibinfo {title} {{High-Frequency
  Stress Relaxation in Semiflexible Polymer Solutions and Networks}},}\ }\href
  {\doibase 10.1103/PhysRevLett.96.138307} {\bibfield  {journal} {\bibinfo
  {journal} {Phys. Rev. Lett.}\ }\textbf {\bibinfo {volume} {96}},\ \bibinfo
  {pages} {138307} (\bibinfo {year} {2006})}\BibitemShut {NoStop}%
\bibitem [{\citenamefont {Mizuno}\ \emph {et~al.}(2009)\citenamefont {Mizuno},
  \citenamefont {Bacabac}, \citenamefont {Tardin}, \citenamefont {Head},\ and\
  \citenamefont {Schmidt}}]{Mizuno2009}%
  \BibitemOpen
  \bibfield  {author} {\bibinfo {author} {\bibfnamefont {Daisuke}\ \bibnamefont
  {Mizuno}}, \bibinfo {author} {\bibfnamefont {Rommel}\ \bibnamefont
  {Bacabac}}, \bibinfo {author} {\bibfnamefont {Catherine}\ \bibnamefont
  {Tardin}}, \bibinfo {author} {\bibfnamefont {David}\ \bibnamefont {Head}}, \
  and\ \bibinfo {author} {\bibfnamefont {Christoph~F}\ \bibnamefont
  {Schmidt}},\ }\bibfield  {title} {\enquote {\bibinfo {title}
  {{High-Resolution Probing of Cellular Force Transmission}},}\ }\href
  {\doibase 10.1103/PhysRevLett.102.168102} {\bibfield  {journal} {\bibinfo
  {journal} {Phys. Rev. Lett.}\ }\textbf {\bibinfo {volume} {102}},\ \bibinfo
  {pages} {168102} (\bibinfo {year} {2009})}\BibitemShut {NoStop}%
\bibitem [{\citenamefont {Yuval}\ and\ \citenamefont
  {Safran}(2013)}]{Yuval2013}%
  \BibitemOpen
  \bibfield  {author} {\bibinfo {author} {\bibfnamefont {Janni}\ \bibnamefont
  {Yuval}}\ and\ \bibinfo {author} {\bibfnamefont {Samuel~A.}\ \bibnamefont
  {Safran}},\ }\bibfield  {title} {\enquote {\bibinfo {title} {{Dynamics of
  elastic interactions in soft and biological matter}},}\ }\href {\doibase
  10.1103/PhysRevE.87.042703} {\bibfield  {journal} {\bibinfo  {journal} {Phys.
  Rev. E - Stat. Nonlinear, Soft Matter Phys.}\ }\textbf {\bibinfo {volume}
  {87}} (\bibinfo {year} {2013}),\ 10.1103/PhysRevE.87.042703}\BibitemShut
  {NoStop}%
\bibitem [{\citenamefont {Weiss}(2007)}]{WeissFluctuationProperties}%
  \BibitemOpen
  \bibfield  {author} {\bibinfo {author} {\bibfnamefont {Jeffrey~B}\
  \bibnamefont {Weiss}},\ }\bibfield  {title} {\enquote {\bibinfo {title}
  {{Fluctuation properties of steady-state Langevin systems}},}\ }\href
  {\doibase 10.1103/PhysRevE.76.061128} {\bibfield  {journal} {\bibinfo
  {journal} {Phys. Rev. E}\ }\textbf {\bibinfo {volume} {76}},\ \bibinfo
  {pages} {061128} (\bibinfo {year} {2007})}\BibitemShut {NoStop}%
\bibitem [{\citenamefont {Lim}\ \emph {et~al.}(2014)\citenamefont {Lim},
  \citenamefont {Surovtsev}, \citenamefont {Beltran}, \citenamefont {Huang},
  \citenamefont {Bewersdorf},\ and\ \citenamefont {Jacobs-Wagner}}]{Lim2014}%
  \BibitemOpen
  \bibfield  {author} {\bibinfo {author} {\bibfnamefont {Hoong~Chuin}\
  \bibnamefont {Lim}}, \bibinfo {author} {\bibfnamefont {Ivan~V.}\ \bibnamefont
  {Surovtsev}}, \bibinfo {author} {\bibfnamefont {Bruno~G.}\ \bibnamefont
  {Beltran}}, \bibinfo {author} {\bibfnamefont {Fang}\ \bibnamefont {Huang}},
  \bibinfo {author} {\bibfnamefont {Joerg}\ \bibnamefont {Bewersdorf}}, \ and\
  \bibinfo {author} {\bibfnamefont {Christine}\ \bibnamefont {Jacobs-Wagner}},\
  }\bibfield  {title} {\enquote {\bibinfo {title} {{Evidence for a DNA-relay
  mechanism in ParABS-mediated chromosome segregation}},}\ }\href {\doibase
  10.7554/eLife.02758} {\bibfield  {journal} {\bibinfo  {journal} {Elife}\
  }\textbf {\bibinfo {volume} {2014}} (\bibinfo {year} {2014}),\
  10.7554/eLife.02758}\BibitemShut {NoStop}%
\bibitem [{\citenamefont {Wang}\ \emph {et~al.}(2013)\citenamefont {Wang},
  \citenamefont {Llopis},\ and\ \citenamefont {Rudner}}]{Wang2013}%
  \BibitemOpen
  \bibfield  {author} {\bibinfo {author} {\bibfnamefont {Xindan}\ \bibnamefont
  {Wang}}, \bibinfo {author} {\bibfnamefont {Paula~Montero}\ \bibnamefont
  {Llopis}}, \ and\ \bibinfo {author} {\bibfnamefont {David~Z.}\ \bibnamefont
  {Rudner}},\ }\bibfield  {title} {\enquote {\bibinfo {title} {{Organization
  and segregation of bacterial chromosomes}},}\ }\href {\doibase
  10.1038/nrg3375} {\bibfield  {journal} {\bibinfo  {journal} {Nat. Rev.
  Genet.}\ }\textbf {\bibinfo {volume} {14}},\ \bibinfo {pages} {191--203}
  (\bibinfo {year} {2013})},\ \Eprint {http://arxiv.org/abs/NIHMS150003}
  {arXiv:NIHMS150003} \BibitemShut {NoStop}%
\bibitem [{\citenamefont {Wilhelm}\ \emph {et~al.}(2015)\citenamefont
  {Wilhelm}, \citenamefont {B{\"{u}}rmann}, \citenamefont {Minnen},
  \citenamefont {Shin}, \citenamefont {Toseland}, \citenamefont {Oh},\ and\
  \citenamefont {Gruber}}]{Wilhelm2015}%
  \BibitemOpen
  \bibfield  {author} {\bibinfo {author} {\bibfnamefont {Larissa}\ \bibnamefont
  {Wilhelm}}, \bibinfo {author} {\bibfnamefont {Frank}\ \bibnamefont
  {B{\"{u}}rmann}}, \bibinfo {author} {\bibfnamefont {Anita}\ \bibnamefont
  {Minnen}}, \bibinfo {author} {\bibfnamefont {Ho~Chul}\ \bibnamefont {Shin}},
  \bibinfo {author} {\bibfnamefont {Christopher~P.}\ \bibnamefont {Toseland}},
  \bibinfo {author} {\bibfnamefont {Byung~Ha}\ \bibnamefont {Oh}}, \ and\
  \bibinfo {author} {\bibfnamefont {Stephan}\ \bibnamefont {Gruber}},\
  }\bibfield  {title} {\enquote {\bibinfo {title} {{SMC condensin entraps
  chromosomal DNA by an ATP hydrolysis dependent loading mechanism in Bacillus
  subtilis}},}\ }\href {\doibase 10.7554/eLife.06659} {\bibfield  {journal}
  {\bibinfo  {journal} {Elife}\ }\textbf {\bibinfo {volume} {4}} (\bibinfo
  {year} {2015}),\ 10.7554/eLife.06659}\BibitemShut {NoStop}%
\bibitem [{\citenamefont {Grill}\ \emph {et~al.}(2001)\citenamefont {Grill},
  \citenamefont {G{\"{o}}nczy}, \citenamefont {Stelzer},\ and\ \citenamefont
  {Hyman}}]{Grill2001}%
  \BibitemOpen
  \bibfield  {author} {\bibinfo {author} {\bibfnamefont {S~W}\ \bibnamefont
  {Grill}}, \bibinfo {author} {\bibfnamefont {P}~\bibnamefont {G{\"{o}}nczy}},
  \bibinfo {author} {\bibfnamefont {E~H}\ \bibnamefont {Stelzer}}, \ and\
  \bibinfo {author} {\bibfnamefont {a~a}\ \bibnamefont {Hyman}},\ }\bibfield
  {title} {\enquote {\bibinfo {title} {{Polarity controls forces governing
  asymmetric spindle positioning in the Caenorhabditis elegans embryo.}}}\
  }\href {\doibase 10.1038/35054572} {\bibfield  {journal} {\bibinfo  {journal}
  {Nature}\ }\textbf {\bibinfo {volume} {409}},\ \bibinfo {pages} {630--633}
  (\bibinfo {year} {2001})}\BibitemShut {NoStop}%
\bibitem [{\citenamefont {Grill}(2003)}]{Grill2003}%
  \BibitemOpen
  \bibfield  {author} {\bibinfo {author} {\bibfnamefont {S.~W.}\ \bibnamefont
  {Grill}},\ }\bibfield  {title} {\enquote {\bibinfo {title} {{The Distribution
  of Active Force Generators Controls Mitotic Spindle Position}},}\ }\href
  {\doibase 10.1126/science.1086560} {\bibfield  {journal} {\bibinfo  {journal}
  {Science}\ }\textbf {\bibinfo {volume} {301}},\ \bibinfo {pages} {518--521}
  (\bibinfo {year} {2003})},\ \Eprint {http://arxiv.org/abs/1002.1037}
  {arXiv:1002.1037} \BibitemShut {NoStop}%
\bibitem [{\citenamefont {Pecreaux}\ \emph {et~al.}(2006)\citenamefont
  {Pecreaux}, \citenamefont {R{\"{o}}per}, \citenamefont {Kruse}, \citenamefont
  {J{\"{u}}licher}, \citenamefont {Hyman}, \citenamefont {Grill},\ and\
  \citenamefont {Howard}}]{Pecreaux2006}%
  \BibitemOpen
  \bibfield  {author} {\bibinfo {author} {\bibfnamefont {Jacques}\ \bibnamefont
  {Pecreaux}}, \bibinfo {author} {\bibfnamefont {Jens~Christian}\ \bibnamefont
  {R{\"{o}}per}}, \bibinfo {author} {\bibfnamefont {Karsten}\ \bibnamefont
  {Kruse}}, \bibinfo {author} {\bibfnamefont {Frank}\ \bibnamefont
  {J{\"{u}}licher}}, \bibinfo {author} {\bibfnamefont {Anthony~A.}\
  \bibnamefont {Hyman}}, \bibinfo {author} {\bibfnamefont {Stephan~W.}\
  \bibnamefont {Grill}}, \ and\ \bibinfo {author} {\bibfnamefont {Jonathon}\
  \bibnamefont {Howard}},\ }\bibfield  {title} {\enquote {\bibinfo {title}
  {{Spindle Oscillations during Asymmetric Cell Division Require a Threshold
  Number of Active Cortical Force Generators}},}\ }\href {\doibase
  10.1016/j.cub.2006.09.030} {\bibfield  {journal} {\bibinfo  {journal} {Curr.
  Biol.}\ }\textbf {\bibinfo {volume} {16}},\ \bibinfo {pages} {2111--2122}
  (\bibinfo {year} {2006})}\BibitemShut {NoStop}%
\bibitem [{\citenamefont {Ou}\ \emph {et~al.}(2010)\citenamefont {Ou},
  \citenamefont {Stuurman}, \citenamefont {D'Ambrosio},\ and\ \citenamefont
  {Vale}}]{Ou2010}%
  \BibitemOpen
  \bibfield  {author} {\bibinfo {author} {\bibfnamefont {G.}~\bibnamefont
  {Ou}}, \bibinfo {author} {\bibfnamefont {N.}~\bibnamefont {Stuurman}},
  \bibinfo {author} {\bibfnamefont {M.}~\bibnamefont {D'Ambrosio}}, \ and\
  \bibinfo {author} {\bibfnamefont {R.~D.}\ \bibnamefont {Vale}},\ }\bibfield
  {title} {\enquote {\bibinfo {title} {{Polarized Myosin Produces Unequal-Size
  Daughters During Asymmetric Cell Division}},}\ }\href {\doibase
  10.1126/science.1196112} {\bibfield  {journal} {\bibinfo  {journal}
  {Science}\ }\textbf {\bibinfo {volume} {330}},\ \bibinfo {pages} {677--680}
  (\bibinfo {year} {2010})},\ \Eprint {http://arxiv.org/abs/NIHMS150003}
  {arXiv:NIHMS150003} \BibitemShut {NoStop}%
\bibitem [{\citenamefont {Mayer}\ \emph {et~al.}(2010)\citenamefont {Mayer},
  \citenamefont {Depken}, \citenamefont {Bois}, \citenamefont
  {J{\"{u}}licher},\ and\ \citenamefont {Grill}}]{Mayer2010}%
  \BibitemOpen
  \bibfield  {author} {\bibinfo {author} {\bibfnamefont {Mirjam}\ \bibnamefont
  {Mayer}}, \bibinfo {author} {\bibfnamefont {Martin}\ \bibnamefont {Depken}},
  \bibinfo {author} {\bibfnamefont {Justin~S.}\ \bibnamefont {Bois}}, \bibinfo
  {author} {\bibfnamefont {Frank}\ \bibnamefont {J{\"{u}}licher}}, \ and\
  \bibinfo {author} {\bibfnamefont {Stephan~W.}\ \bibnamefont {Grill}},\
  }\bibfield  {title} {\enquote {\bibinfo {title} {{Anisotropies in cortical
  tension reveal the physical basis of polarizing cortical flows}},}\ }\href
  {\doibase 10.1038/nature09376} {\bibfield  {journal} {\bibinfo  {journal}
  {Nature}\ }\textbf {\bibinfo {volume} {467}},\ \bibinfo {pages} {617--621}
  (\bibinfo {year} {2010})},\ \Eprint {http://arxiv.org/abs/77957364208}
  {arXiv:77957364208} \BibitemShut {NoStop}%
\bibitem [{\citenamefont {Thutupalli}\ \emph {et~al.}(2015)\citenamefont
  {Thutupalli}, \citenamefont {Sun}, \citenamefont {Bunyak}, \citenamefont
  {Palaniappan},\ and\ \citenamefont {Shaevitz}}]{Thutupalli2015}%
  \BibitemOpen
  \bibfield  {author} {\bibinfo {author} {\bibfnamefont {Shashi}\ \bibnamefont
  {Thutupalli}}, \bibinfo {author} {\bibfnamefont {Mingzhai}\ \bibnamefont
  {Sun}}, \bibinfo {author} {\bibfnamefont {Filiz}\ \bibnamefont {Bunyak}},
  \bibinfo {author} {\bibfnamefont {Kannappan}\ \bibnamefont {Palaniappan}}, \
  and\ \bibinfo {author} {\bibfnamefont {Joshua~W}\ \bibnamefont {Shaevitz}},\
  }\bibfield  {title} {\enquote {\bibinfo {title} {{Directional reversals
  enable Myxococcus xanthus cells to produce collective one-dimensional streams
  during fruiting-body formation.}}}\ }\href {\doibase 10.1098/rsif.2015.0049}
  {\bibfield  {journal} {\bibinfo  {journal} {J. R. Soc. Interface}\ }\textbf
  {\bibinfo {volume} {12}},\ \bibinfo {pages} {20150049} (\bibinfo {year}
  {2015})},\ \Eprint {http://arxiv.org/abs/1410.7230} {arXiv:1410.7230}
  \BibitemShut {NoStop}%
\bibitem [{\citenamefont {Peruani}\ \emph {et~al.}(2012)\citenamefont
  {Peruani}, \citenamefont {Starru{\ss}}, \citenamefont {Jakovljevic},
  \citenamefont {S{\"{o}}gaard-Andersen}, \citenamefont {Deutsch},\ and\
  \citenamefont {B{\"{a}}r}}]{Peruani2012}%
  \BibitemOpen
  \bibfield  {author} {\bibinfo {author} {\bibfnamefont {Fernando}\
  \bibnamefont {Peruani}}, \bibinfo {author} {\bibfnamefont {J{\"{o}}rn}\
  \bibnamefont {Starru{\ss}}}, \bibinfo {author} {\bibfnamefont {Vladimir}\
  \bibnamefont {Jakovljevic}}, \bibinfo {author} {\bibfnamefont {Lotte}\
  \bibnamefont {S{\"{o}}gaard-Andersen}}, \bibinfo {author} {\bibfnamefont
  {Andreas}\ \bibnamefont {Deutsch}}, \ and\ \bibinfo {author} {\bibfnamefont
  {Markus}\ \bibnamefont {B{\"{a}}r}},\ }\bibfield  {title} {\enquote {\bibinfo
  {title} {{Collective motion and nonequilibrium cluster formation in colonies
  of gliding bacteria}},}\ }\href {\doibase 10.1103/PhysRevLett.108.098102}
  {\bibfield  {journal} {\bibinfo  {journal} {Phys. Rev. Lett.}\ }\textbf
  {\bibinfo {volume} {108}} (\bibinfo {year} {2012}),\
  10.1103/PhysRevLett.108.098102},\ \Eprint {http://arxiv.org/abs/1302.0311}
  {arXiv:1302.0311} \BibitemShut {NoStop}%
\bibitem [{\citenamefont {Frauenfelder}\ \emph {et~al.}(1999)\citenamefont
  {Frauenfelder}, \citenamefont {Wolynes},\ and\ \citenamefont
  {Austin}}]{Frauenfelder1999}%
  \BibitemOpen
  \bibfield  {author} {\bibinfo {author} {\bibfnamefont {H}~\bibnamefont
  {Frauenfelder}}, \bibinfo {author} {\bibfnamefont {P~G}\ \bibnamefont
  {Wolynes}}, \ and\ \bibinfo {author} {\bibfnamefont {R~H}\ \bibnamefont
  {Austin}},\ }\bibfield  {title} {\enquote {\bibinfo {title} {{Biological
  physics}},}\ }\href {\doibase 10.1038/467033a} {\bibfield  {journal}
  {\bibinfo  {journal} {Rev. Mod. Phys.}\ }\textbf {\bibinfo {volume} {71}},\
  \bibinfo {pages} {S419--S430} (\bibinfo {year} {1999})}\BibitemShut {NoStop}%
\bibitem [{\citenamefont {Agerschou}\ \emph {et~al.}(2017)\citenamefont
  {Agerschou}, \citenamefont {Mast},\ and\ \citenamefont
  {Braun}}]{Agerschou2017}%
  \BibitemOpen
  \bibfield  {author} {\bibinfo {author} {\bibfnamefont {Emil~Dandanell}\
  \bibnamefont {Agerschou}}, \bibinfo {author} {\bibfnamefont {Christof~B.}\
  \bibnamefont {Mast}}, \ and\ \bibinfo {author} {\bibfnamefont {Dieter}\
  \bibnamefont {Braun}},\ }\bibfield  {title} {\enquote {\bibinfo {title}
  {{Emergence of Life from Trapped Nucleotides? Non-Equilibrium Behavior of
  Oligonucleotides in Thermal Gradients}},}\ }\href {\doibase
  10.1055/s-0036-1588653} {\bibfield  {journal} {\bibinfo  {journal} {Synlett}\
  }\textbf {\bibinfo {volume} {28}},\ \bibinfo {pages} {56--63} (\bibinfo
  {year} {2017})}\BibitemShut {NoStop}%
\bibitem [{\citenamefont {Schwille}\ and\ \citenamefont
  {Diez}(2009)}]{Schwille2009}%
  \BibitemOpen
  \bibfield  {author} {\bibinfo {author} {\bibfnamefont {Petra}\ \bibnamefont
  {Schwille}}\ and\ \bibinfo {author} {\bibfnamefont {Stefan}\ \bibnamefont
  {Diez}},\ }\bibfield  {title} {\enquote {\bibinfo {title} {{Synthetic biology
  of minimal systems}},}\ }\href {\doibase 10.1080/10409230903074549}
  {\bibfield  {journal} {\bibinfo  {journal} {Crit. Rev. Biochem. Mol. Biol.}\
  }\textbf {\bibinfo {volume} {44}},\ \bibinfo {pages} {223--242} (\bibinfo
  {year} {2009})}\BibitemShut {NoStop}%
\end{thebibliography}
\end{document}